\documentclass[english]{article}
\usepackage[T1]{fontenc}
\usepackage[utf8]{inputenc}
\usepackage{geometry}
\usepackage{color}
\usepackage{babel}
\usepackage{enumitem}
\usepackage{amsmath}
\usepackage{amssymb}
\usepackage[unicode=true,pdfusetitle,
 bookmarks=true,bookmarksnumbered=false,bookmarksopen=false,
 breaklinks=false,pdfborder={0 0 0},pdfborderstyle={},backref=false,colorlinks=true]
 {hyperref}
\hypersetup{
 linkcolor=blue,citecolor=blue,urlcolor=black}

\makeatletter
\numberwithin{equation}{section}
\newcommand{\lyxaddress}[1]{
	\par {\raggedright #1
	\vspace{1.4em}
	\noindent\par}
}

\@ifundefined{date}{}{\date{}}
\makeatother

\begin{document}
\title{On exact overlaps for $\mathfrak{gl}(N)$ symmetric spin chains}
\author{Tamás Gombor}
\maketitle

\lyxaddress{\begin{center}
MTA-ELTE “Momentum” Integrable Quantum Dynamics Research Group, Department
of Theoretical Physics, Eötvös Loránd University\\
Holographic QFT Group, Wigner Research Centre for Physics, Budapest,
Hungary
\par\end{center}}
\begin{abstract}
We study the integrable two-site states of the quantum integrable
models solvable by the nested algebraic Bethe ansatz and possessing
$\mathfrak{gl}(N)$-invariant R-matrix. We investigate the overlaps
between the integrable two-site states and the wave-functions. To
find exact derivations for the factorized overlap formulas for the
nested integrable systems is a longstanding unsolved problem. In this
paper we give a derivation for a large class of the integrable states
of the $\mathfrak{gl}(N)$ symmetric spin chain. The first part of
the derivation is to calculate recurrence relations for the off-shell
overlap that uniquely fix it. Using these recursions we prove that
the normalized overlaps of the multi-particle states have factorized
forms which contain the products of the one-particle overlaps and
the ratio of the Gaudin-like determinants. We also show that the previously
proposed overlap formulas agree with our general formula.
\end{abstract}

\section{Introduction}

In recent years there has been renewed interest and relevant progress
in the calculation of the exact overlaps between the eigenstates of
the integrable models and a distinguished set of states, which are
now called integrable initial states \cite{Piroli:2017sei,Pozsgay_2019}.
These overlaps display very special features and they appear in quite
distinct parts of theoretical physics including statistical physics
and the gauge/string duality.

In statistical physics these quantities appear in the context of non-equilibrium
dynamics of the integrable models and the overlaps play a central
role in the study of the quantum quenches. In a quantum quench, a
parameter of the Hamiltonian is suddenly changed implying that the
ground-state of the pre-quenched Hamiltonian is no longer an eigenstate
of the post-quench Hamiltonian and the goal is to study the non-equilibrium
dynamics, the emergence of steady states, and their properties \cite{Essler:2016ufo}.
One of the main methods for the investigation of the steady states
is the so-called Quench Action method \cite{Caux:2013ra}, where the
knowledge of the exact overlaps is an important input. The exact overlap
formulas also played a central role in the early studies of the Generalized
Gibbs Ensemble (GGE) in interacting integrable models, see \cite{Wouters_2014,Pozsgay_2014}.

In the AdS/CFT correspondence there are at least two places where
overlaps appeared so far. The first one is the calculation of a special
class of three-point functions involving a local gauge invariant single
trace operator and two determinant operators dual to the giant gravitons.
It was shown that these three-point functions can be calculated as
an overlap between the finite volume multi-particle state and an integrable
initial state. First these quantities were investigated in the $AdS_{5}/CFT_{4}$
duality \cite{Jiang:2019xdz,Jiang:2019zig} and recently the method
was generalized to the $AdS_{4}/CFT_{3}$ duality \cite{Yang:2021hrl}.

The other, much more investigated application is the calculation of
the one-point functions in defect version of AdS/CFT. To be more precise,
in a co-dimension one defect $\mathcal{N}=4$ SYM the tree-level one-point
functions are given by the overlaps between multi-particle states
corresponding to single trace operators and special two-site states
or Matrix Product States\cite{Buhl-Mortensen:2015gfd,deLeeuw:2015hxa,deLeeuw:2016umh,DeLeeuw:2018cal,Kristjansen:2020mhn}.

In recent years several exact overlaps were determined. Based on these
results it seems that if the initial state is a so-called two-site
state and it satisfies an integrability condition \cite{Piroli:2017sei}
then the normalized overlap always can be written in the following
form\footnote{In this paper we use the notations of \cite{Hutsalyuk:2016srn,Hutsalyuk:2017tcx,Hutsalyuk:2017way,Hutsalyuk:2020dlw}
where the symbol $\mathbb{B}(\bar{u})$ without ket denotes the Bethe
state.}
\begin{equation}
{\color{blue}\frac{\left|{\color{red}\langle\Psi|}{\color{blue}\mathbb{B}(\bar{u})}\right|^{2}}{||\mathbb{B}(\bar{u})||^{2}}}={\color{red}\underbrace{\prod_{j,\nu}\mathcal{F}_{\nu}(u_{j}^{\nu})}_{\text{boundary dependent}}}\times{\color{blue}\underbrace{\frac{\det G^{+}}{\det G^{-}}}_{\text{universal}}}\label{eq:gen}
\end{equation}
where $\langle\Psi|$ is the integrable two-site state and $\mathbb{B}(\bar{u})$
is a Bethe state. We can see that the exact overlaps have factorized
form where the first term is a product over certain functions $\mathcal{F}_{\nu}(u_{j}^{\nu})$.
Physically this product can be interpreted as a product of one-particle
overlaps. The second term is a universal term which is independent
from the final state $\langle\Psi|$. It depends only on the concrete
spin chain and the Bethe roots $\bar{u}$. This universal term is
a quotient of the Gaudin-like determinants \cite{Brockmann_2014}.
There exist exact proofs of these formulas for XXX and XXZ spin chains
\cite{Brockmann_2014,Brockmann_2014odd,Foda:2015nfk}. However, these
methods work only for the defining representation and the generalization
for higher spins is not clear.

There exists an other approach for proving the factorized overlap
formulas \cite{Jiang:2020sdw}. This method uses the coordinate Bethe
Ansatz and it was used not only for the two dimensional, but also
for one of the infinite dimensional representations. Theoretically
this method can be used for any representation, but in practice one
has to handle each representation separately, since the method is
based on the coordinate Bethe Ansatz. It was a significant step forward
when the algebraic Bethe Ansatz was successfully used in the calculations
\cite{Gombor:2021uxz}. This method is based on the so called $KT$-relation,
which makes it possible to replace the creation operators of the Bethe
states with the annihilation ones, therefore it provides a recursive
method for the evaluation of the overlaps. It was the first proof
which simultaneously gave the exact overlaps for the $\mathfrak{gl}(2)$
symmetric spin chains for all spins and for all possible two-site
states.

In the numerical investigations of the nested spin chains it was observed
that the overlaps of the two-site states have the factorized form
for these models, too. However the analytical methods described above
are not directly applicable. Nevertheless, the exact overlaps can
be guessed, since in the factorized formulas only the single-particle
overlaps $\mathcal{F}_{\nu}(u_{j}^{\nu})$ are the unknown quantities.
To determine these quantities there exist two approaches which uses
coordinate Bethe Ansatz \cite{deLeeuw:2015hxa,Gombor:2020auk,Gombor:2020kgu,Jiang:2019xdz,Yang:2021hrl}
or Quantum Transfer Matrix \cite{Pozsgay_2018,Piroli:2018ksf,Piroli:2018don,deLeeuw:2019ebw}
methods. However, we emphasize that these procedures are not suitable
for proving the overlap formulas. Here we also note that the so-called
integrable matrix product states (MPS) \cite{Pozsgay_2019} are also
relevant for the applications. For these states the overlap functions
are modified: the boundary dependent part is not just a product, but
a sum of such products. For certain MPS-s there have been proposed
exact overlap formulas for the $\mathfrak{su}(2)$ \cite{Buhl-Mortensen:2015gfd},
the $\mathfrak{su}(3)$ \cite{deLeeuw:2016umh} and the $\mathfrak{so}(6)$
\cite{DeLeeuw:2018cal,deLeeuw:2019ebw} symmetric spin chains. It
was shown in \cite{deLeeuw:2019ebw} that the most MPS overlaps can
be obtained from the two-site overlaps using the representation theory
of the twisted Yangian algebras.

We emphasize that the overlap formulas (\ref{eq:gen}) are valid only
for on-shell Bethe states\footnote{The vector $\mathbb{B}(\bar{u})$ is an eigenvector of the transfer
matrix (or the Hamiltonian) if the Bethe roots $u_{j}^{\nu}$ satisfy
the Bethe Ansatz equations and we say that the Bethe vector $\mathbb{B}(\bar{u})$
is on-shell. For general set $\bar{u}$ the Bethe vector $\mathbb{B}(\bar{u})$
is called off-shell.}. For off-shell Bethe states there are available determinant formulas
as well. In \cite{Pozsgay_2014_2} it was showed that off-shell overlaps
are equivalent some partition functions and the determinant formula
of \cite{Tsuchiya_1998} gives the overlap between off-shell Bethe
states and two-site states built from diagonal K-matrices for the
XXZ spin chain. The on-shell limit of this formula is problematic
since it contains 0/0 ambiguities. By taking the proper limit, this
formula was used in \cite{Brockmann_2014} for the first proof of
the on-shell overlap (\ref{eq:gen}). In \cite{Foda:2015nfk} the
authors derived an other determinant formula for the overlap between
off-shell Bethe states and two-site states built from upper-triangular
K-matrices for the XXX spin chain. The methods of \cite{Tsuchiya_1998}
and \cite{Foda:2015nfk} were limited the spin 1/2 case. The paper
\cite{Gombor:2021uxz} showed the generalization for higher spins
but this result also only works for upper-triangular K-matrices. The
existence of a determinant formula for the general integrable two-site
states is not yet clear. In addition, it is worth to mention another
existing determinant formula for the off-shell overlaps coming from
an separation of variables (SoV) calculation \cite{Gombor:2021uxz}.
The interesting thing about this formula is that the on-shell limit
is well defined, while the homogeneous limit contains the 0/0 ambiguity.
Currently, we do not know whether off-shell determinant formulas exist
in general cases (for non upper-triangular K-matrices and spin chains
with higher ranks), and it is not the purpose of this paper to investigate
this issue.

In this paper our goal is to provide a widely applicable proof for
the two-site on-shell overlaps (\ref{eq:gen}). We concentrate on
the $\mathfrak{gl}(N)$ symmetric spin chains. The calculation of
the overlaps between the two-site states and the Bethe states is very
similar to the calculation of the norm of the Bethe states, which
in itself is a very complex task. In 1982 V. Korepin proved the Gaudin
hypothesis for a wide class of the quantum integrable models \cite{korepin-norms}.
The Gaudin hypothesis says that the square of the norm of the eigenfunction
is proportional to a Jacobian of the Bethe equations which is called
Gaudin determinant. Recently the Gaudin hypothesis was proven for
the $\mathfrak{gl}(N|M)$ symmetric spin chains \cite{Hutsalyuk:2017way}.
This approach generalizes the original Korepin argument for the $\mathfrak{gl}(N|M)$
symmetric spin chains. In this paper we combine the method of \cite{Hutsalyuk:2017way}
with the recently discovered $KT$-relation \cite{Gombor:2021uxz}
and we gain an approach which proves the factorized overlap formulas
for the $\mathfrak{gl}(N)$ symmetric spin chains for a wide class
of the two-site states. One of the main advantages of this method
is that it is based on the algebraic Bethe Ansatz, so it is independent
from the quantum space.

The paper is organized as follows. In section \ref{sec:Descr_glN}
we briefly recall the notation of the $\mathfrak{gl}(N)$ symmetric
spin chains. In section \ref{sec:Integrable-final-states} we define
the integrable two-site states and the $KT$-relations. We show that
the $KT$-relation gives a systematic method to calculate the off-shell
overlaps for the $\mathfrak{so}(N)$ and the $\mathfrak{gl}(\left\lfloor \frac{N}{2}\right\rfloor )\oplus\mathfrak{gl}(\left\lceil \frac{N}{2}\right\rceil )$
symmetric two-site states. In section \ref{sec:Main-results} we collect
our main results. In the first subsection we show various properties
which completely define the off-shell overlap formula. It is not a
closed formula, some building blocks are given only by recurrence
relations. In the second subsection we give the closed form of the
on-shell formula. In section \ref{sec:Representations-of-the} we
show the concrete representations of the integrable two-sites states
and apply our general formula for some specific models. We also show
that the previously proposed exact overlaps agree with our general
formula. All proofs of the statements of the section \ref{sec:Main-results}
can be found in the appendix.

\section{Description of $\mathfrak{gl}(N)$ symmetric spin chains\label{sec:Descr_glN}}

In this section we review the notation of quantum inverse scattering
of spin chains associated with the $\mathfrak{gl}(N)$ symmetric $R$-matrix.
In the following sections we use the sum rules and action formulas
of \cite{Hutsalyuk:2017tcx,Hutsalyuk:2020dlw} therefore we use the
conventions of these papers. The R-matrix of the models with $\mathfrak{gl}(N)$
symmetry has the form 
\begin{equation}
R(u-v)=\mathbf{I}+g(u,v)\mathbf{P},\qquad g(u,v)=\frac{c}{u-v},
\end{equation}
where $c$ is a constant and $\mathbf{I}$ and $\mathbf{P}$ are the
identity and the permutation operators in the vector space $\mathbb{C}^{N}\otimes\mathbb{C}^{N}$.
The $R$-matrix defines the $RTT$ algebra
\begin{equation}
R_{1,2}(u-v)T_{1}(u)T_{2}(v)=T_{2}(v)T_{1}(u)R_{1,2}(u-v),
\end{equation}
where 
\begin{equation}
T(u)=\sum_{i,j=1}^{N}E_{i,j}\otimes T_{i,j}(u)
\end{equation}
is the monodromy matrix and $E_{i,j}$-s are the $\mathfrak{gl}(N)$
generators. The monodromy matrix entries $T_{i,j}(u)$ generate the
famous Yangian algebra $Y(N)$ \cite{Molev:1994rs}. Using the monodromy
matrix entries the $RTT$ algebra reads as
\begin{equation}
\left[T_{i,j}(u),T_{k,l}(v)\right]=g(u,v)\left(T_{k,j}(v)T_{i,l}(u)-T_{k,j}(u)T_{i,l}(v)\right).
\end{equation}

We can define a representation of the Yangian $Y(N)$ on the quantum
space $\mathcal{H}$. A representation is highest weight if there
exists a unique pseudo-vacuum $\left|0\right\rangle \in\mathcal{H}$
such that
\begin{equation}
\begin{split}T_{i,i}\left|0\right\rangle  & =\lambda_{i}(u)\left|0\right\rangle ,\qquad\text{for }i=1,\dots,N,\\
T_{j,i}\left|0\right\rangle  & =0,\qquad\qquad\text{for }1\leq i<j\leq N.
\end{split}
\end{equation}
The $\lambda_{i}(u)$-s are the vacuum eigenvalues. One can also define
the transfer matrix as
\begin{equation}
\mathcal{T}(u)=\mathrm{tr}T(u)=\sum_{i=1}^{N}T_{i,i}(u).
\end{equation}
Since the monodromy matrix satisfies the $RTT$-relation the transfer
matrices are commuting operators
\begin{equation}
\left[\mathcal{T}(u),\mathcal{T}(v)\right]=0,
\end{equation}
therefore the eigenvectors of the transfer matrix do not dependent
on the spectral parameter $u$. We use the conventions of \cite{Hutsalyuk:2017tcx}
for the eigenvectors (Bethe vectors)
\begin{equation}
\mathcal{T}(u)\mathbb{B}(\bar{t})=\tau(u|\bar{t})\mathbb{B}(\bar{t}),
\end{equation}
where 
\begin{equation}
\bar{t}=\left\{ t_{1}^{1},\dots,t_{r_{1}}^{1};t_{1}^{2},\dots,t_{r_{2}}^{2};\dots;t_{1}^{N-1},\dots,t_{r_{N-1}}^{N-1}\right\} 
\end{equation}
is the set of Bethe roots $t_{k}^{\nu}$. The Bethe vectors are symmetric
over permutations of the parameters $t_{k}^{\nu}$ within the set
$\bar{t}^{\nu}=\left\{ t_{1}^{\nu},\dots,t_{r_{\nu}}^{\nu}\right\} $.
For generic Bethe vectors the Bethe roots $t_{k}^{\nu}$ are generic
complex numbers. If these parameters satisfy a special system of equations
(Bethe equations) then the corresponding vector becomes an eigenvector
of the transfer matrix. In this case it is called on-shell Bethe vector
and for generic Bethe roots $t_{k}^{\nu}$ it is called off-shell
Bethe vector.

In appendix \ref{sec:Off-shell-Bethe-vectors} we review the defining
equations of the off-shell Bethe vectors based on \cite{Hutsalyuk:2017tcx}.
The action formula of \cite{Hutsalyuk:2020dlw} is also shown. From
the action formula we can see that the expression $T_{i,j}(u)\mathbb{B}(\bar{t})$
can be written as a linear combination Bethe vectors. The diagonal
elements $T_{i,j}(u)$ do not change the quantum numbers $r_{1},\dots,r_{N-1}$.
The monodromy matrix entries $T_{i,j}(u)$ for $i<j$ \emph{increase}
the quantum numbers $r_{i},r_{i+1},\dots,r_{j-1}$ by one therefore
these operators are called \emph{creation operators}. The monodromy
matrix entries $T_{j,i}(u)$ for $i<j$ \emph{decrease} the quantum
numbers $r_{i},r_{i+1},\dots,r_{j-1}$ by one therefore these operators
are called \emph{annihilation operators}.

The eigenvalues of the transfer matrix can be written as \cite{Hutsalyuk:2017way}
\begin{equation}
\tau(u|\bar{t})=\sum_{i=1}^{N}\lambda_{i}(u)\prod_{k=1}^{r_{i}}\frac{u-t_{k}^{i}-c}{u-t_{k}^{i}}\prod_{k=1}^{r_{i-1}}\frac{u-t_{k}^{i-1}+c}{u-t_{k}^{i-1}}=\sum_{i=1}^{N}\lambda_{i}(u)f(\bar{t}^{i},u)f(u,\bar{t}^{i-1}),\label{eq:eig}
\end{equation}
where $r_{0}=r_{N}=0$ and we used the following notations
\begin{equation}
\begin{split}f(u,v) & =1+g(u,v)=\frac{u-v+c}{u-v},\\
f(u,\bar{t}^{i}) & =\prod_{k=1}^{r_{i}}f(u,t_{k}^{i}),\quad f(\bar{t}^{i},u)=\prod_{k=1}^{r_{i}}f(t_{k}^{i},u),\quad f(\bar{t}^{i},\bar{t}^{j})=\prod_{k=1}^{r_{i}}f(t_{k}^{i},\bar{t}^{j}).
\end{split}
\end{equation}
The Bethe equations are
\begin{equation}
\alpha_{\mu}(t_{k}^{\mu}):=\frac{\lambda_{\mu}(t_{k}^{\mu})}{\lambda_{\mu+1}(t_{k}^{\mu})}=\frac{f(t_{k}^{\mu},\bar{t}_{k}^{\mu})}{f(\bar{t}_{k}^{\mu},t_{k}^{\mu})}\frac{f(\bar{t}^{\mu+1},t_{k}^{\mu})}{f(t_{k}^{\mu},\bar{t}^{\mu-1})},\label{eq:bethe}
\end{equation}
where $\bar{t}_{k}^{\mu}=\bar{t}^{\mu}\backslash t_{k}^{\mu}$. These
equations can be rewritten as
\begin{equation}
\Phi_{k}^{(\mu)}=1,
\end{equation}
where we defined the expressions
\begin{equation}
\Phi_{k}^{(\mu)}=\alpha_{\mu}(t_{k}^{\mu})\frac{f(\bar{t}_{k}^{\mu},t_{k}^{\mu})}{f(t_{k}^{\mu},\bar{t}_{k}^{\mu})}\frac{f(t_{k}^{\mu},\bar{t}^{\mu-1})}{f(\bar{t}^{\mu+1},t_{k}^{\mu})}.
\end{equation}
One can also define the left eigenvectors of the transfer matrix
\begin{equation}
\mathbb{C}(\bar{t})\mathcal{T}(u)=\tau(u|\bar{t})\mathbb{C}(\bar{t}),
\end{equation}
and the square of the norm of the on-shell Bethe states satisfies
the Gaudin hypothesis \cite{Hutsalyuk:2017way}
\begin{equation}
\mathbb{C}(\bar{t})\mathbb{B}(\bar{t})=\frac{\prod_{\nu=1}^{N-1}\prod_{k\neq l}f(t_{l}^{\nu},t_{k}^{\nu})}{\prod_{\nu=1}^{N-2}f(\bar{t}^{\nu+1},\bar{t}^{\nu})}\det G,\label{eq:norm}
\end{equation}
where $G$ is the Gaudin matrix given by
\begin{equation}
G_{j,k}^{(\mu,\nu)}=-c\frac{\partial\log\Phi_{j}^{(\mu)}}{\partial t_{k}^{\nu}}.
\end{equation}

We can also define an other monodromy matrix which satisfy the same
$RTT$-algebra \cite{Liashyk:2018egk}. This transfer matrix can be
obtained from one of the quantum minors as
\begin{align}
\widehat{T}_{N+1-j,N+1-i}(u) & =(-1)^{i+j}t_{1,\dots,\hat{i},\dots,N}^{1,\dots,\hat{j},\dots,N}(u-c)\mathrm{qdet}(T(u))^{-1},\\
t_{b_{1},b_{2},\dots,b_{m}}^{a_{1},a_{2},\dots,a_{m}}(u) & =\sum_{p}\mathrm{sgn}(p)T_{a,b_{p(1)}}(u)T_{a,b_{p(2)}}(u-c)\dots T_{a,b_{p(m)}}(u-(m-1)c),\label{eq:Qminors}\\
\mathrm{qdet}(T(u)) & =t_{1,2,\dots,N}^{1,2,\dots,N}(u).
\end{align}
Here $\hat{i}$ an\^{d} $\hat{j}$ mean that the corresponding indices
are omitted. We call $\widehat{T}$ as twisted monodromy matrix. The
twisted monodromy matrix $\widehat{T}$ is also a highest weight representation
of $Y(N)$ with
\begin{equation}
\begin{split}\widehat{T}_{i,i}\left|0\right\rangle  & =\hat{\lambda}_{i}(u)\left|0\right\rangle ,\qquad\text{for }i=1,\dots,N,\\
\widehat{T}_{j,i}\left|0\right\rangle  & =0,\qquad\qquad\text{for }1\leq i<j\leq N,
\end{split}
\end{equation}
where
\begin{equation}
\hat{\lambda}_{i}(u)=\frac{1}{\lambda_{N-i+1}(u-(N-i)c)}\prod_{k=1}^{N-i}\frac{\lambda_{k}(u-kc)}{\lambda_{k}(u-(k-1)c)},
\end{equation}
therefore the ratios of the vacuum eigenvalues have the following
form
\begin{equation}
\hat{\alpha}_{i}(u)=\frac{\hat{\lambda}_{i}(u)}{\hat{\lambda}_{i+1}(u)}=\alpha_{N-i}(u-(N-i)c).
\end{equation}
Let $\hat{\mathbb{B}}(\bar{t})$ be the off-shell Bethe vector generated
from $\widehat{T}_{i,j}$. In \cite{Liashyk:2018egk} the connection
between the Bethe vectors $\mathbb{B}(\bar{t})$ and $\hat{\mathbb{B}}(\bar{t})$
was determined
\begin{equation}
\hat{\mathbb{B}}(\bar{t})=(-1)^{\#\bar{t}}\left(\prod_{s=1}^{N-2}f(\bar{t}^{s+1},\bar{t}^{s})\right)^{-1}\mathbb{B}(\mu(\bar{t})),\label{eq:connBhatB}
\end{equation}
where
\begin{equation}
\mu(\bar{t})=\{\bar{t}^{N-1}-c,\bar{t}^{N-2}-2c,\dots,\bar{t}^{1}-(N-1)c\}.
\end{equation}
From this identity we can obtain the eigenvalue of the twisted transfer
matrix 
\begin{align}
\widehat{\mathcal{T}}(u) & =\mathrm{tr}\widehat{T}(u)=\sum_{i=1}^{N}\widehat{T}_{i,i}(u),\\
\widehat{\mathcal{T}}(u)\mathbb{B}(\bar{t}) & =\hat{\tau}(u|\bar{t})\mathbb{B}(\bar{t}).
\end{align}
Using (\ref{eq:connBhatB}), (\ref{eq:eig}) and the fact that $\widehat{T}$
satisfies the same $RTT$-relation we can obtain that
\begin{equation}
\hat{\tau}(u|\bar{t})=\sum_{i=1}^{N}\hat{\lambda}_{i}(u)f(\bar{t}^{N-i}+(N-i)c,u)f(u,\bar{t}^{N-i+1}+(N-i+1)c).\label{eq:tweig}
\end{equation}

The twisted monodromy matrix $\widehat{T}$ is similar to the inverse
of the original monodromy matrix $T$:
\begin{equation}
V\widehat{T}^{t}(u)VT(u)=1,
\end{equation}
where $V$ is an off-diagonal $N\times N$ matrix of the auxiliary
space with the components $V_{i,j}=\delta_{i,N+1-j}$ and the superscript
t is the transposition in the auxiliary space, i.e. $\left[\widehat{T}^{t}(u)\right]_{i,j}=\widehat{T}_{j,i}(u)$.
Applying this equation to the $RTT$-relation we obtain the $R\widehat{T}T$-relation
\begin{equation}
\bar{R}_{1,2}(u-v)\widehat{T}_{1}(u)T_{2}(v)=T_{2}(v)\widehat{T}_{1}(u)\bar{R}_{1,2}(u-v),
\end{equation}
where we used the crossed $R$-matrix
\begin{equation}
\bar{R}_{1,2}(u)=V_{2}R_{1,2}^{t_{2}}(-u)V_{2}.
\end{equation}
Using the monodromy matrix entries the $R\widehat{T}T$-relation can
be written as
\begin{equation}
\left[\widehat{T}_{i,j}(u),T_{k,l}(v)\right]=g(u,v)\left(\delta_{i,N+1-k}\sum_{a=1}^{N}\widehat{T}_{a,j}(u)T_{N+1-a,l}(v)-\delta_{j,N+1-l}\sum_{a=1}^{N}T_{k,a}(v)\widehat{T}_{i,N+1-a}(u)\right).
\end{equation}

The monodromy matrix has also a useful co-product property. Let $\mathcal{H}^{(1)}$
and $\mathcal{H}^{(2)}$ be two quantum spaces and the corresponding
monodromy matrices are $T^{(1)}(u)$ and $T^{(2)}(u)$. In the tensor
product space $\mathcal{H}=\mathcal{H}^{(1)}\otimes\mathcal{H}^{(2)}$
the monodromy matrix can be written as
\begin{equation}
T_{i,j}(u)=\sum_{k=1}^{N}T_{k,j}^{(1)}(u)\otimes T_{i,k}^{(2)}(u).
\end{equation}
This monodromy matrix satisfies the $RTT$-relation and the vacuum
eigenvalues are given by
\begin{equation}
\lambda_{i}(u)=\lambda_{i}^{(1)}(u)\lambda_{i}^{(2)}(u).
\end{equation}

Let us close this section by some important algebra automorphism.
The spectral parameter shifted monodromy matrices $T(u+x)$ satisfy
the same $RTT$-algebra therefore we can introduce an algebra automorphism
\begin{equation}
\mathcal{B}_{x}:T(u)\to T(u+x).
\end{equation}
This automorphism transforms the vacuum eigenvalues and the quantum
minors as well
\begin{equation}
\mathcal{B}_{x}(\lambda_{\nu}(u))=\lambda_{\nu}(u+x),\qquad\mathcal{B}_{x}(\widehat{T}(u))=\widehat{T}(u+x).
\end{equation}

It is clear that $RTT$-relation defines different algebra for different
$N$-s. For a fixed $N$ let us use the notation $T_{i,j}^{N}(u)$
for the monodromy matrix entries to emphasize the explicit value of
$N$. There are natural embeddings for the algebra of $N-1$ to the
algebra of $N$. We can see that the $RTT$ algebra is closed for
the entries $\left\{ T_{i,j}^{N}\right\} _{i,j=1}^{N-1}$ and $\left\{ T_{i,j}^{N}\right\} _{i,j=2}^{N}$
therefore they defines the same algebra as $\left\{ T_{i,j}^{N-1}\right\} _{i,j=1}^{N-1}$.
Let us introduce the following notations for these injections
\begin{equation}
\begin{split}\mathfrak{i}^{-}:\quad & T_{i,j}^{N-1}(u)\to T_{i,j}^{N}(u),\\
\mathfrak{i}^{+}:\quad & T_{i,j}^{N-1}(u)\to T_{i+1,j+1}^{N}(u),
\end{split}
\end{equation}
where $i,j=1,\dots,N-1$. These injections act on the $\lambda$-s
as
\begin{equation}
\mathfrak{i}^{-}(\lambda_{\nu}^{N-1}(u))=\lambda_{\nu}^{N}(u),\qquad\mathfrak{i}^{+}(\lambda_{\nu}^{N-1}(u))=\lambda_{\nu+1}^{N}(u).
\end{equation}

\section{Integrable final states and $KT$-relations\label{sec:Integrable-final-states}}

In this section we introduce the so-called $KT$-relation which serves
as the defining relation for the integrable two-site states. One can
introduce the $KT$-relation in two different ways. In the first and
the second subsections we analyze the untwisted and the twisted $KT$-relations,
separately. We show that the main advantage of these relations is
that we can replace the creation operators in the Bethe vectors with
the annihilation ones. In appendix \ref{sec:Elementary} we demonstrate
this approach using the Bethe vectors with one or two Bethe roots.
We will also see that the non-vanishing overlaps for the on-shell
Bethe vectors require certain pair structures for the Bethe roots.
We close this section by the co-product property of the integrable
two-site states.

\subsection{Untwisted $KT$-relation}

Let us consider the final states $\langle\Psi|$ which satisfy the
following $KT$-relation
\begin{equation}
K_{0}(u)\langle\Psi|T_{0}(u)=\langle\Psi|T_{0}(-u)K_{0}(u),\label{eq:KT_Utw}
\end{equation}
where $K(u)$ is an invertible $N\times N$ matrix of the auxiliary
space. We call the states which satisfy the $KT$-relation as \emph{integrable
two-site states}. The motivation for the name will be clear when we
discuss their co-product property in subsection \ref{subsec:Coproduct}.
In \cite{Gombor:2021uxz} it was shown that the integrable two-site
states satisfy this relation for the $\mathfrak{gl}(2)$ symmetric
models and now we just generalize this relation to higher rank models.
In this section we do not specify $\langle\Psi|$ just require the
constraint (\ref{eq:KT_Utw}). Notice that we have not specified the
monodromy matrix neither. We only prescribe the $RTT$-relation and
the vacuum eigenvalues $\lambda_{\mu}(u)$. In section \ref{sec:Representations-of-the}
when we turn to the concrete examples we explicitly define $T(u)$
and the corresponding $\langle\Psi|$ as well.

\subsubsection{Consistency of $KT$- and $RTT$-relations and the boundary YB equation}

To obtain a consistent algebra the $KT$-relation has to be compatible
with the $RTT$-relation. Let us consider the action of two monodromy
matrices on the final state $\langle\Psi|$ as
\begin{equation}
\langle\Psi|T_{1}(u)T_{2}(v).
\end{equation}
We can reflect the two monodromy matrices on the final state in two
ways:
\begin{equation}
\begin{split}\langle\Psi|T_{1}(u)T_{2}(v) & =K_{1}^{-1}(u)\langle\Psi|T_{1}(-u)T_{2}(v)K_{1}(u)=\\
 & =K_{1}^{-1}(u)R_{1,2}^{-1}(-u-v)\langle\Psi|T_{2}(v)T_{1}(-u)R_{1,2}(-u-v)K_{1}(u)=\\
 & =K_{1}^{-1}(u)R_{1,2}^{-1}(-u-v)K_{2}^{-1}(v)\langle\Psi|T_{2}(-v)T_{1}(-u)K_{2}(v)R_{1,2}(-u-v)K_{1}(u)=\\
=K_{1}^{-1}(u) & R_{1,2}^{-1}(-u-v)K_{2}^{-1}(v)R_{12}^{-1}(u-v)\Bigl[\langle\Psi|T_{1}(-u)T_{2}(-v)\Bigr]R_{1,2}(u-v)K_{2}(v)R_{1,2}(-u-v)K_{1}(u),
\end{split}
\label{eq:T1T2v1}
\end{equation}
or
\begin{equation}
\begin{split}\langle\Psi|T_{1}(u)T_{2}(v) & =R_{1,2}^{-1}(u-v)\langle\Psi|T_{2}(v)T_{1}(u)R_{1,2}(u-v)=\\
 & =R_{1,2}^{-1}(u-v)K_{2}^{-1}(v)\langle\Psi|T_{2}(-v)T_{1}(u)K_{2}(v)R_{1,2}(u-v)=\\
 & =R_{1,2}^{-1}(u-v)K_{2}^{-1}(v)R_{1,2}^{-1}(-u-v)\langle\Psi|T_{1}(u)T_{2}(-v)R_{1,2}(-u-v)K_{2}(v)R_{1,2}(u-v)=\\
=R_{1,2}^{-1}(u-v) & K_{2}^{-1}(v)R_{1,2}^{-1}(-u-v)K_{1}^{-1}(u)\Bigl[\langle\Psi|T_{1}(-u)T_{2}(-v)\Bigr]K_{1}(u)R_{1,2}(-u-v)K_{2}(v)R_{1,2}(u-v).
\end{split}
\label{eq:T1T2v2}
\end{equation}
We can see that the equations (\ref{eq:T1T2v1}) and (\ref{eq:T1T2v2})
are consistent if the $K$-matrix satisfies the following equation
\begin{equation}
R_{1,2}(u-v)K_{2}(v)R_{1,2}(-u-v)K_{1}(u)=K_{1}(u)R_{1,2}(-u-v)K_{2}(v)R_{1,2}(u-v),
\end{equation}
which is the famous reflection or boundary Yang-Baxter equation. An
equivalent but more common form is
\begin{equation}
R_{1,2}(u-v)K_{1}(-u)R_{1,2}(u+v)K_{2}(-v)=K_{2}(-v)R_{1,2}(u+v)K_{1}(-u)R_{1,2}(u-v).
\end{equation}
The most general solution of this equation is well known \cite{Arnaudon:2004sd}.
In this paper we use the convention\footnote{The reflection equation is homogeneous in the $K$-matrix $K(u)$
therefore we can always multiply the $K$-matrix with an arbitrary
function of the spectral parameter.} 
\begin{equation}
K(u)=\frac{\mathfrak{a}}{u}\mathbf{1}+\mathcal{U},\label{eq:Kdef_UTw}
\end{equation}
where $\mathcal{U}$ is an $N\times N$ matrix with the following
constraint
\begin{equation}
\mathcal{U}^{2}=\mathbf{1}.
\end{equation}
The constant $\mathfrak{a}\in\mathbb{C}$ is a free parameter of the
$K$-matrix. One can diagonalize $\mathcal{U}$ as
\begin{equation}
G\mathcal{U}G^{-1}=\mathrm{diag}(\underbrace{1,\dots,1}_{M},\underbrace{-1,\dots,-1}_{N-M}),\label{eq:sign}
\end{equation}
where $G\in GL(N)$ and the number of $1$-s and $-1$ are $M$ and
$N-M$ respectively. We say that the matrix $\mathcal{U}$ has signature
$(M,N-M)$. We can see that the $K$-matrix commutes with a subalgebra
$\mathfrak{gl}(M)\oplus\mathfrak{gl}(N-M)$ of the original algebra
$\mathfrak{gl}(N)$ therefore we call this $K$-matrix and the corresponding
final state $\langle\Psi|$ the $\mathfrak{gl}(M)\oplus\mathfrak{gl}(N-M)$
symmetric $K$-matrix and final state.

\subsubsection{Evaluating off-shell overlaps}

The main goal of this paper is to calculate the overlap
\begin{equation}
S_{K}^{N}(\bar{t})=\left\langle \Psi\right|\mathbb{B}(\bar{t}),
\end{equation}
where we stressed that the overlap depends on the $K$-matrix and
the algebra rank $N$. The main advantage of the $KT$-relation is
that one can evaluate the overlap systematically for a special class
of $K$-matrices. In the following we show this procedure which is
based on that we can change the creation operators to the annihilation
ones. In this subsection we use the following shorthand notation for
the Bethe states
\begin{equation}
\mathbb{B}(\bar{t})\to\mathbb{B}^{\bar{r}},
\end{equation}
where we do not care about explicit values of the Bethe roots, the
only important thing is their quantum numbers $r_{1},\dots,r_{N-1}$.
In \cite{Hutsalyuk:2017tcx} a recurrence equation for the Bethe states
was derived and the their explicit form can be found in appendix \ref{sec:Off-shell-Bethe-vectors}.
Using our shorthand notation the sum formula reads as
\begin{equation}
\mathbb{B}^{r_{1},r_{2},\dots,r_{N-1}}=\sum_{k=2}^{N}T_{1,k}\sum\mathbb{B}^{r_{1}-1,\dots,r_{k-1}-1,r_{k},\dots,r_{N-1}}(\dots),\label{eq:Sum_UTW_1}
\end{equation}
where the second sum goes over different Bethe vectors with fixed
quantum numbers. The $(\dots)$ denotes the coefficients of the Bethe
vectors. We can use the component $(N,k)$ of the $KT$-relation (\ref{eq:KT_Utw})
\begin{equation}
\left\langle \Psi\right|T_{1,k}(u)=\sum_{j=1}^{N}\left\langle \Psi\right|T_{N,j}(-u)\frac{K_{j,k}(u)}{K_{N,1}(u)}-\sum_{i=2}^{N}\frac{K_{N,i}(u)}{K_{N,1}(u)}\left\langle \Psi\right|T_{i,k}(u),\label{eq:KT_UTW_1k}
\end{equation}
where we assumed that 
\begin{equation}
K_{N,1}(u)\neq0.
\end{equation}
The equation (\ref{eq:KT_UTW_1k}) can be used to change the creation
operators $T_{1,k}(u)$ to operators $T_{N,j}(-u)$ and $T_{i,k}(u)$
where $i,k=2,\dots,N$ and $j=1,\dots,N$. Combining equations (\ref{eq:Sum_UTW_1})
and (\ref{eq:KT_UTW_1k}) we obtain that
\begin{equation}
\left\langle \Psi\right|\mathbb{B}^{r_{1},\dots}=\sum_{k=2}^{N}\left(\sum_{j=1}^{N}\sum\left\langle \Psi\right|T_{N,j}\mathbb{B}^{r_{1}-1,\dots}+\sum_{i=2}^{N}\sum\left\langle \Psi\right|T_{i,k}\mathbb{B}^{r_{1}-1,\dots}\right).\label{eq:stage2}
\end{equation}
Now we can use the action formula for $T_{N,j}\mathbb{B}$ and $T_{i,k}\mathbb{B}$.
The explicit form of this formula can be found in appendix \ref{sec:Off-shell-Bethe-vectors}.
The operators $T_{N,i}$ and $T_{i,k}$ for $i>1$ do not change the
number of the first Bethe roots and the operator $T_{N,1}$ decreases
it by one. Substituting the action formula to (\ref{eq:stage2}) we
obtain that
\begin{equation}
\left\langle \Psi\right|\mathbb{B}^{r_{1},\dots}=\sum\left\langle \Psi\right|\mathbb{B}^{r_{1}-1,\dots}+\sum\left\langle \Psi\right|\mathbb{B}^{r_{1}-2,\dots}.
\end{equation}
Repeating these steps one can express the general overlap with a sum
of overlaps without the first Bethe roots
\begin{equation}
\left\langle \Psi\right|\mathbb{B}^{r_{1},\dots}=\sum\left\langle \Psi\right|\mathbb{B}^{0,\dots}.
\end{equation}

We can continue with the elimination of the $(N-1$)-th type of Bethe
roots $t_{k}^{N-1}$. Using the other recurrence formula of \cite{Hutsalyuk:2017tcx}
(its explicit form can be found in appendix \ref{sec:Off-shell-Bethe-vectors})
we obtain that
\begin{equation}
\left\langle \Psi\right|\mathbb{B}^{0,r_{2}\dots,r_{N-1}}=\sum_{k=2}^{N-1}\left\langle \Psi\right|T_{k,N}\sum\mathbb{B}^{0,r_{2}\dots,r_{k-1},r_{k}-1,\dots,r_{N-1}-1}.
\end{equation}
Using the component $(k,1)$ of the $KT$-relation (\ref{eq:KT_Utw})
\begin{equation}
\left\langle \Psi\right|T_{k,N}(-u)=\sum_{i=1}^{N}\frac{K_{k,i}(u)}{K_{N,1}(u)}\left\langle \Psi\right|T_{i,1}(u)-\sum_{j=1}^{N-1}\left\langle \Psi\right|T_{k,j}(-u)\frac{K_{j,1}(u)}{K_{N,1}(u)}.
\end{equation}
The operators $T_{j,1}$ and $T_{k,j}$ for $j<N$ do not change the
number of $(N-1)$-th Bethe roots and the operator $T_{N,1}$ decreases
it by one. It is important that these operators do not increase the
number of first Bethe roots therefore we obtain that
\begin{equation}
\left\langle \Psi\right|\mathbb{B}^{0,\dots,r_{N-1}}=\sum\mathbb{B}^{0,\dots,r_{N-1}-1}+\sum\mathbb{B}^{0,\dots,r_{N-1}-2}.
\end{equation}
Repeating these steps one can express the general overlap with a sum
of overlaps without the first and the $(N-1)$-th Bethe roots:
\begin{equation}
\left\langle \Psi\right|\mathbb{B}^{r_{1},\dots,r_{N-1}}=\sum\left\langle \Psi\right|\mathbb{B}^{0,\dots,0}.
\end{equation}
We can see that we reduced the original $\mathfrak{gl}(N)$ Bethe
state to $\mathfrak{gl}(N-2)$ Bethe states. The $\mathfrak{gl}(N-2)$
Bethe states are generated by the monodromy matrix entries $\left\{ T_{i,j}\right\} _{i,j=2}^{N-1}$.
The natural question is: Is the $KT$-relation closed for the monodromy
matrix entries $\left\{ T_{i,j}\right\} _{i,j=2}^{N-1}$? Naively,
the answer is no since the component $(a,b)$ of the $KT$-relation
(\ref{eq:KT_Utw}) reads as
\begin{multline}
\sum_{c=2}^{N-1}K_{a,c}(u)\left\langle \Psi\right|T_{c,b}(u)+K_{a,1}(u)\left\langle \Psi\right|T_{1,b}(u)+K_{a,N}(u)\left\langle \Psi\right|T_{N,b}(u)=\\
\sum_{c=2}^{N-1}\left\langle \Psi\right|T_{a,c}(-u)K_{c,b}(u)+\left\langle \Psi\right|T_{a,N}(-u)K_{N,b}(u)+\left\langle \Psi\right|T_{a,1}(-u)K_{1,b}(u).\label{eq:KTreq}
\end{multline}
where $a,b=2,\dots,N-1$. In this equation there are operators $T_{1,b}$
and $T_{a,N}$ which increase the quantum numbers $r_{1}$ and $r_{N-1}$,
respectively. This means that we can not continue the eliminations
of Bethe roots $\bar{t}^{2}$ and $\bar{t}^{N-2}$ since repeating
the previous steps we would create the first and the $(N-1)$-th Bethe
roots. We should not increase these quantum numbers therefore we have
to eliminate the operators $T_{1,b}$ and $T_{a,N}$ from equation
(\ref{eq:KTreq}).

Notice that there are two operators in (\ref{eq:KTreq}) which are
annihilates the Bethe vectors $\mathbb{B}^{0,\dots,0}$:
\begin{equation}
T_{N,b}(u)\mathbb{B}^{0,\dots,0}=T_{a,1}(u)\mathbb{B}^{0,\dots,0}=0.
\end{equation}
Let us introduce the following notation
\begin{equation}
A\cong B\quad\Longleftrightarrow\quad A\mathbb{B}^{0,\dots,0}=B\mathbb{B}^{0,\dots,0},
\end{equation}
therefore $\cong$ denotes equality in the subspace generated by the
Bethe vectors $\mathbb{B}^{0,\dots,0}$. Using this notation the previous
equation (\ref{eq:KTreq}) simplifies as
\begin{equation}
\sum_{c=2}^{N-1}K_{a,c}(u)\left\langle \Psi\right|T_{c,b}(u)+K_{a,1}(u)\left\langle \Psi\right|T_{1,b}(u)\cong\sum_{c=2}^{N-1}\left\langle \Psi\right|T_{a,c}(-u)K_{c,b}(u)+\left\langle \Psi\right|T_{a,N}(-u)K_{N,b}(u).
\end{equation}
Let us see the components $(N,b)$ and $(a,1)$ of the $KT$-relation
(\ref{eq:KT_Utw})
\begin{align}
 & \sum_{c=2}^{N-1}K_{N,c}(u)\left\langle \Psi\right|T_{c,b}(u)+K_{N,1}(u)\left\langle \Psi\right|T_{1,b}(u)\cong\left\langle \Psi\right|T_{N,N}(-u)K_{N,b}(u),\label{eq:KTreq2}\\
 & K_{a,1}(u)\left\langle \Psi\right|T_{1,1}(u)\cong\sum_{c=2}^{N-1}\left\langle \Psi\right|T_{a,c}(-u)K_{c,1}(u)+\left\langle \Psi\right|T_{a,N}(-u)K_{N,1}(u).\label{eq:KTreq3}
\end{align}
Applying the component $(N,1)$ of the $KT$-relation (\ref{eq:KT_Utw})
on the pseudo-vacuum we obtain that
\begin{equation}
K_{N,1}(u)\left\langle \Psi\right|T_{1,1}(u)\left|0\right\rangle =\left\langle \Psi\right|T_{N,N}(-u)\left|0\right\rangle K_{N,1}(u),
\end{equation}
therefore the vacuum eigenvalues have to satisfy the following constraint
\begin{equation}
\lambda_{1}(u)=\lambda_{N}(-u).
\end{equation}
Using this symmetry property we obtain that
\begin{equation}
T_{1,1}(u)\cong T_{N,N}(-u).\label{eq:KTreq4}
\end{equation}
Combining the equation (\ref{eq:KTreq}), (\ref{eq:KTreq2}),(\ref{eq:KTreq3})
and (\ref{eq:KTreq4}) we obtain the following equation on the subspace
generated by the Bethe states $\mathbb{B}^{0,\dots,0}$ 
\begin{equation}
\sum_{c=2}^{N-1}K_{a,c}^{(2)}(u)\left\langle \Psi\right|T_{c,b}(u)\cong\sum_{c=2}^{N-1}\left\langle \Psi\right|T_{a,c}(-u)K_{c,b}^{(2)}(u),\label{eq:KT2}
\end{equation}
where $a,b=2,\dots,N-1$ and
\begin{equation}
K_{a,b}^{(2)}(u)=K_{N,1}^{(1)}(u)K_{a,b}^{(1)}(u)-K_{a,1}^{(1)}(u)K_{N,b}^{(1)}(u),\qquad K^{(1)}(u)=K(u).
\end{equation}

Now we can see that the new $KT$-relation (\ref{eq:KT2}) is closed
for the $\mathfrak{gl}(N-2)$ Bethe vectors
\begin{equation}
\sum_{c=2}^{N-1}K_{a,c}^{(2)}(u)\left\langle \Psi\right|T_{c,b}^{N}(u)\mathbb{B}^{N}(\emptyset,\left\{ \bar{t}^{k}\right\} _{k=2}^{N-2},\emptyset)=\sum_{c=2}^{N-1}K_{c,b}^{(2)}(u)\left\langle \Psi\right|T_{a,c}^{N}(-u)\mathbb{B}^{N}(\emptyset,\left\{ \bar{t}^{k}\right\} _{k=2}^{N-2},\emptyset).\label{eq:UTw_embed}
\end{equation}
Let us study this embedding more precisely. Using the injections $\mathfrak{i}^{\pm}$,
the equation (\ref{eq:UTw_embed}) can be written as
\begin{multline}
\sum_{c=1}^{N-2}K_{a+1,c+1}^{(2)}(u)\left\langle \Psi\right|\left(\mathfrak{i}^{+}\circ\mathfrak{i}^{-}\right)\left(T_{c,b}^{N-2}(u)\mathbb{B}^{N-2}(\left\{ \bar{t}^{k}\right\} _{k=2}^{N-2})\right)=\\
\sum_{c=1}^{N-2}K_{c+1,b+1}^{(2)}(u)\left\langle \Psi\right|\left(\mathfrak{i}^{+}\circ\mathfrak{i}^{-}\right)\left(T_{a,c}^{N-2}(-u)\mathbb{B}^{N-2}(\left\{ \bar{t}^{k}\right\} _{k=2}^{N-2})\right),
\end{multline}
where $a,b=1,\dots,N-2$. Defining the co-vector $\left\langle \Psi^{(2)}\right|=\left\langle \Psi\right|\left(\mathfrak{i}^{+}\circ\mathfrak{i}^{-}\right)$
we obtain the relation
\begin{equation}
\sum_{c=1}^{N-2}K_{a+1,c+1}^{(2)}\left\langle \Psi^{(2)}\right|T_{c,b}^{N-2}(u)\mathbb{B}^{N-2}(\left\{ \bar{t}^{k}\right\} _{k=2}^{N-2})=\sum_{c=1}^{N-2}K_{c+1,b+1}^{(2)}(u)\left\langle \Psi^{(2)}\right|T_{a,c}^{N-2}(-u)\mathbb{B}^{N-2}(\left\{ \bar{t}^{k}\right\} _{k=2}^{N-2}),
\end{equation}
which is the $KT$-relation of the $\mathfrak{gl}(N-2)$ spin chain
with the $K$-matrix $K^{(2)}(u)$. Since 
\begin{equation}
\left\langle \Psi\right|\mathbb{B}^{N}(\emptyset,\left\{ \bar{t}^{k}\right\} _{k=2}^{N-2},\emptyset)=\left(\mathfrak{i}^{+}\circ\mathfrak{i}^{-}\right)\left(\left\langle \Psi^{(2)}\right|T_{c,b}^{N-2}(u)\mathbb{B}^{N-2}(\left\{ \bar{t}^{k}\right\} _{k=2}^{N-2})\right)
\end{equation}
the overlaps of the $\mathfrak{gl}(N)$ model where $\bar{t}^{1}=\bar{t}^{N-1}=\emptyset$
can be obtained from the overlaps of the $\mathfrak{gl}(N-2)$ model
i.e. we just obtained an embedding rule for the overlaps
\begin{equation}
S_{K}^{N}(\emptyset,\bar{t}^{2},\dots,\bar{t}^{N-2},\emptyset)=S_{K^{(2)}}^{N-2}(\bar{t}^{2},\dots,\bar{t}^{N-2})\Biggr|_{\alpha_{\nu}\to\alpha_{\nu+1}}.\label{eq:recursion_S_UTw}
\end{equation}

Repeating the previous method we can eliminate the Bethe roots $\bar{t}^{\nu}$,
therefore we obtain a recursion for the overlaps and in the end of
the day we obtain an explicit result. In the $(k+1)$-th step of the
nesting we have a $\mathfrak{gl}(N-2k)$ $K$-matrix:
\begin{equation}
\boxed{K_{a,b}^{(k+1)}(u)=K_{N+1-k,k}^{(k)}(u)K_{a,b}^{(k)}(u)-K_{a,k}^{(k)}(u)K_{N+1-k,b}^{(k)}(u),}\label{eq:reqK}
\end{equation}
where $a,b=k+1,\dots,N-k$ and we also obtain the symmetry conditions
\begin{equation}
\boxed{\lambda_{k}(u)=\lambda_{N+1-k}(-u).}\label{eq:constLamUTw}
\end{equation}
This condition implies that
\begin{equation}
\boxed{\alpha_{k}(u)=\frac{\lambda_{k}(u)}{\lambda_{k+1}(u)}=\frac{\lambda_{N+1-k}(-u)}{\lambda_{N-k}(-u)}=\frac{1}{\alpha_{N-k}(-u)}.}\label{eq:constAlpUTw}
\end{equation}
We can see that this approach gives the off-shell overlap only when
the following condition is satisfied at every level of the nesting
\begin{equation}
K_{N+1-k,k}^{(k)}(u)\neq0.
\end{equation}
It turns out this condition restricts the integrable final states
to the $\mathfrak{gl}(\frac{N}{2})\oplus\mathfrak{gl}(\frac{N}{2})$
or the $\mathfrak{gl}(\frac{N-1}{2})\oplus\mathfrak{gl}(\frac{N+1}{2})$
symmetric ones. This statement can be derived from the recurrence
equation (\ref{eq:reqK}). Let us assume that the $k$-th level $K$-matrix
has the following form
\begin{equation}
K_{a,b}^{(k)}(u)=\mathfrak{b}_{k}(\frac{\mathfrak{a}}{u}\delta_{a,b}+\mathcal{U}_{a,b}^{(k)}),
\end{equation}
where $a,b=k,\dots N+1-k$, $\mathfrak{b}_{k}\in\mathbb{C}$ and $\left(\mathcal{U}^{(k)}\right)^{2}$
i.e.

\begin{equation}
\sum_{c=k}^{N+1-k}\mathcal{U}_{a,c}^{(k)}\mathcal{U}_{c,b}^{(k)}=\delta_{a,b}.\label{eq:idVk}
\end{equation}
At first let us use the recurrence equation (\ref{eq:reqK}) 
\begin{equation}
K_{a,b}^{(k+1)}(u)=\mathfrak{b}_{k}^{2}\mathcal{U}_{N+1-k,k}^{(k)}\left(\frac{\mathfrak{a}}{u}\delta_{a,b}+\mathcal{U}_{a,b}^{(k)}-\frac{\mathcal{U}_{a,k}^{(k)}\mathcal{U}_{N+1-k,b}^{(k)}}{\mathcal{U}_{N+1-k,k}^{(k)}}\right)=\mathfrak{b}_{k+1}\left(\frac{\mathfrak{a}}{u}\delta_{a,b}+\mathcal{U}_{a,b}^{(k+1)}\right),
\end{equation}
where $a,b=k+1,\dots N-k$. Let us check that $\left(\mathcal{U}^{(k+1)}\right)^{2}=1$,
\begin{multline}
\sum_{c=k+1}^{N-k}\mathcal{U}_{a,c}^{(k+1)}\mathcal{U}_{c,b}^{(k+1)}=\\
\sum_{c=k+1}^{N-k}\left(\mathcal{U}_{a,c}^{(k)}\mathcal{U}_{c,b}^{(k)}-\frac{\mathcal{U}_{a,c}^{(k)}\mathcal{U}_{c,k}^{(k)}\mathcal{U}_{N+1-k,b}^{(k)}}{\mathcal{U}_{N+1-k,k}^{(k)}}-\frac{\mathcal{U}_{a,k}^{(k)}\mathcal{U}_{N+1-k,c}^{(k)}\mathcal{U}_{c,b}^{(k)}}{\mathcal{U}_{N+1-k,k}^{(k)}}+\frac{\mathcal{U}_{a,k}^{(k)}\mathcal{U}_{N+1-k,c}^{(k)}\mathcal{U}_{c,k}^{(k)}\mathcal{U}_{N+1-k,b}^{(k)}}{\left(\mathcal{U}_{N+1-k,k}^{(k)}\right)^{2}}\right).\label{eq:proofIdVkp}
\end{multline}
From (\ref{eq:idVk}) we obtain that
\begin{align}
\sum_{c=k+1}^{N-k}\mathcal{U}_{a,c}^{(k)}\mathcal{U}_{c,b}^{(k)} & =\delta_{a,b}-\mathcal{U}_{a,k}^{(k)}\mathcal{U}_{k,b}^{(k)}-\mathcal{U}_{a,N+1-k}^{(k)}\mathcal{U}_{N+1-k,b}^{(k)},\\
\sum_{c=k+1}^{N-k}\mathcal{U}_{a,c}^{(k)}\mathcal{U}_{c,k}^{(k)} & =-\mathcal{U}_{a,k}^{(k)}\mathcal{U}_{k,k}^{(k)}-\mathcal{U}_{a,N+1-k}^{(k)}\mathcal{U}_{N+1-k,k}^{(k)},\\
\sum_{c=k+1}^{N-k}\mathcal{U}_{N+1-k,c}^{(k)}\mathcal{U}_{c,b}^{(k)} & =-\mathcal{U}_{N+1-k,k}^{(k)}\mathcal{U}_{k,b}^{(k)}-\mathcal{U}_{N+1-k,N+1-k}^{(k)}\mathcal{U}_{N+1-k,b}^{(k)},\\
\sum_{c=k+1}^{N-k}\mathcal{U}_{N+1-k,c}^{(k)}\mathcal{U}_{c,k}^{(k)} & =-\mathcal{U}_{N+1-k,k}^{(k)}\mathcal{U}_{k,k}^{(k)}-\mathcal{U}_{N+1-k,N+1-k}^{(k)}\mathcal{U}_{N+1-k,k}^{(k)}.\label{eq:idNkk}
\end{align}
Substituting back to (\ref{eq:proofIdVkp}) we obtain that
\begin{equation}
\sum_{c=k+1}^{N-k}\mathcal{U}_{a,c}^{(k+1)}\mathcal{U}_{c,b}^{(k+1)}=\delta_{a,b}.
\end{equation}
Now let us calculate the trace of $\mathcal{U}^{(k+1)}$
\begin{equation}
\mathrm{Tr}\mathcal{U}^{(k+1)}=\sum_{a=k+1}^{N-k}\mathcal{U}_{a,a}^{(k+1)}=\sum_{a=k+1}^{N-k}\left(\mathcal{U}_{a,a}^{(k)}-\frac{\mathcal{U}_{a,k}^{(k)}\mathcal{U}_{N+1-k,a}^{(k)}}{\mathcal{U}_{N+1-k,k}^{(k)}}\right).\label{eq:proofTrV}
\end{equation}
Substituting (\ref{eq:idNkk}) to (\ref{eq:proofTrV}) we obtain that
\begin{equation}
\mathrm{Tr}\mathcal{U}^{(k+1)}=\sum_{a=k+1}^{N-k}\mathcal{U}_{a,a}^{(k)}+\mathcal{U}_{k,k}^{(k)}+\mathcal{U}_{N+1-k,N+1-k}^{(k)}=\mathrm{Tr}\mathcal{U}^{(k)}.
\end{equation}
We can see that the nesting of $K$-matrices does not change the trace
of the matrices $\mathcal{U}^{(k)}$ which means if the first level
matrix $\mathcal{U}=\mathcal{U}^{(1)}$ has signature $(M,N-M)$ then
the $k$-th level matrix $\mathcal{U}^{(k)}$ has signature $(M-k+1,N-M-k+1)$
therefore in the $k$-th step of the nesting we have $\mathfrak{gl}(M-k+1)\oplus\mathfrak{gl}(N-M-k+1)$
symmetric $K$-matrix. For $k=M+1$ the matrix $\mathcal{U}^{(M+1)}$
has signature $(0,N-2M)$ therefore the $K$-matrix $K^{(M+1)}$ is
proportional to the identity which means
\begin{equation}
K_{N-M,M+1}^{(M+1)}=0.
\end{equation}
From this argument the above procedure can be apply only for nesting
levels $k\leq M$. Since the nesting is finished at $k=\frac{N}{2}$
or $k=\frac{N-1}{2}$ for even or odd $N$, our procedure can be applied
only for $\mathfrak{gl}(\frac{N}{2})\oplus\mathfrak{gl}(\frac{N}{2})$
or $\mathfrak{gl}(\frac{N-1}{2})\oplus\mathfrak{gl}(\frac{N+1}{2})$
symmetric final states.

\subsubsection{Achiral pair structure\label{subsec:Achiral-pair-structure}}

Let us continue with the selection rules of the on-shell overlaps.
Since $K(u)$ is invertible the integrability condition 
\begin{equation}
\langle\Psi|\mathcal{T}(u)=\langle\Psi|\mathcal{T}(-u)
\end{equation}
follows from the $KT$-relation (\ref{eq:KT_Utw}). Applying this
equation to an on-shell Bethe vector we find that
\begin{equation}
\left(\tau(u|\bar{t})-\tau(-u|\bar{t})\right)\langle\Psi|\mathbb{B}(\bar{t})=0.
\end{equation}
We just obtained that the non-vanishing overlap requires that
\begin{equation}
\tau(u|\bar{t})=\tau(-u|\bar{t}).\label{eq:intcondtau}
\end{equation}
Substituting $-u$ to the eigenvalue formula (\ref{eq:eig}) we obtain
that
\begin{equation}
\tau(-u|\bar{t})=\sum_{i=1}^{N}\lambda_{i}(-u)f(\bar{t}^{i},-u)f(-u,\bar{t}^{i-1})=\sum_{i=1}^{N}\lambda_{i}(u)f(-\bar{t}^{N-i},u)f(u,-\bar{t}^{N-i+1})=\tau(u|\pi^{a}(\bar{t})),
\end{equation}
where we used that
\begin{equation}
\lambda_{i}(u)=\lambda_{N+1-i}(-u),\qquad f(u,v)=f(-v,-u),
\end{equation}
and we introduced the following map of the Bethe roots:
\begin{equation}
\pi^{a}(\bar{t}^{1},\bar{t}^{2},\dots,\bar{t}^{N-1})=\left(-\bar{t}^{N-1},-\bar{t}^{N-2},\dots,-\bar{t}^{1}\right).\label{achpair}
\end{equation}
Substituting back to (\ref{eq:intcondtau}), the condition for the
non-vanishing overlaps reads as
\begin{equation}
\tau(u|\bar{t})=\tau(u|\pi^{a}(\bar{t})).\label{eq:eig_cond}
\end{equation}
Notice that the set $\pi^{a}(\bar{t})$ also satisfies the Bethe equations
i.e. the vector $\mathbb{B}(\pi^{a}(\bar{t}))$ is also an on-shell
Bethe vector. Since different Bethe vectors have different eigenvalues\footnote{Actually this statement is precise only for the $\mathfrak{gl}(2)$
case \cite{Mukhin:2007rs}. For the $\mathfrak{gl}(N)$ case it was
proved that the Bethe algebra has non-degenerate spectrum on the subspace
of highest weight states \cite{Mukhin:2013rs} (Corollary 6.5.). The
Bethe algebra is generated by the transfer matrices $\tau_{m}(u)=\sum_{a_{1},\dots,a_{m}}t_{a_{1},\dots,a_{m}}^{a_{1},\dots,a_{m}}(u)$
coming from the quantum minors (\ref{eq:Qminors}). Using the $KT$-relation
(\ref{eq:KT_Utw}) one can derive that there exist also a generalization
of the $KT$-relation for the quantum minors 
\[
\sum_{c_{1},\dots,c_{m}}K_{a_{1},\dots,a_{m}}^{c_{1},\dots,c_{m}}(u)\langle\Psi|t_{c_{1},\dots,c_{m}}^{b_{1},\dots,b_{m}}(u)=\sum_{c_{1},\dots,c_{m}}\langle\Psi|t_{a_{1},\dots,a_{m}}^{c_{1},\dots,c_{m}}(-u)K_{c_{1},\dots,c_{m}}^{b_{1},\dots,b_{m}}(u),
\]
where $K_{a_{1},\dots,a_{m}}^{c_{1},\dots,c_{m}}(u)$ is the fused
$K$-matrix \cite{fusion-open-chains}. Using the previous argument
we can show that the equation (\ref{eq:eig_cond}) is also satisfied
for the fused transfer matrices $\tau_{m}(u|\bar{t})=\tau_{m}(u|\pi^{a}(\bar{t}))$
which generate the Bethe algebra. Since the Bethe algebra has non-degenerate
spectrum on the subspace of highest weight states we just obtained
(\ref{eq:achiral}).} we just obtained that the Bethe roots have to satisfy the selection
rule
\begin{equation}
\bar{t}=\pi^{a}(\bar{t})\label{eq:achiral}
\end{equation}
for the non-vanishing overlaps. We call the Bethe roots with condition
(\ref{eq:achiral}) as Bethe roots with \emph{achiral pair structure}.
For the achiral pair structure, every sets $\bar{t}^{\nu}$ satisfy
the condition $\bar{t}^{N-\nu}=-\bar{t}^{\nu}$ for $\nu<N/2$. For
even $N$ the $N/2$-th set of Bethe roots satisfies the condition
$\bar{t}^{\frac{N}{2}}=-\bar{t}^{\frac{N}{2}}$ therefore there exists
three disjoint subsets $\bar{t}^{+,\frac{N}{2}},\bar{t}^{-,\frac{N}{2}},\bar{t}^{0}$
for which $\bar{t}^{\frac{N}{2}}=\bar{t}^{+,\frac{N}{2}}\cup\bar{t}^{-,\frac{N}{2}}\cup\bar{t}^{0}$
and $\bar{t}^{-,\frac{N}{2}}=-\bar{t}^{+,\frac{N}{2}}$ and $\bar{t}^{0}=\emptyset$
or $\bar{t}^{0}=\{0\}$ for even or odd $r_{\frac{N}{2}}$. Using
the definitions $\bar{t}^{+,\nu}=\bar{t}^{\nu}$, $\bar{t}^{-,\nu}=\bar{t}^{N-\nu}$
for $1\leq\nu<N/2$ and $\bar{t}^{\pm}=\cup_{\nu}\bar{t}^{\pm,\nu}$
we just obtained the following decomposition of the Bethe roots

\begin{equation}
\bar{t}=\bar{t}^{+}\cup\bar{t}^{-}\cup\bar{t}^{0}.
\end{equation}

\subsection{Twisted $KT$-relation\label{subsec:Twisted--relation}}

We can also define an other $KT$-relation
\begin{equation}
K_{0}(u)\langle\Psi|T_{0}(u)=\lambda_{0}(u)\langle\Psi|\widehat{T}_{0}(-u)K_{0}(u),\label{eq:KT_Tw}
\end{equation}
where $K(u)$ is an invertible $N\times N$ matrix of the auxiliary
space and $\lambda_{0}(u)$ is a function for a proper normalization.
We call this equation as \emph{twisted} $KT$-relation. Let us repeat
the previous analysis for this twisted equation.

\subsubsection{Consistency of twisted $KT$- and $RTT$-relation and the twisted
boundary YB equation}

One can obtain the consistency condition for the twisted case in a
similar way as in the previous subsection. Repeating the same steps
as before we can derive that the consistency requires that the $K$-matrix
has to satisfy the \emph{twisted} boundary Yang-Baxter equation.
\begin{equation}
R_{1,2}(u-v)K_{1}(-u)\bar{R}_{1,2}(u+v)K_{2}(-v)=K_{2}(-v)\bar{R}_{1,2}(u+v)K_{1}(-u)R_{1,2}(u-v).
\end{equation}
The most general solutions of this equation is well known \cite{Arnaudon:2004sd}
and we use the following conventions to this general solution
\begin{equation}
K(u)=\mathcal{U}V,\label{eq:twK}
\end{equation}
where
\begin{equation}
\mathcal{U}^{t}=\pm\mathcal{U}.
\end{equation}
It is well known that the $R$-matrix has $\mathfrak{gl}(N)$ symmetry:
\begin{equation}
\begin{split}R_{1,2}(u) & =G_{1}^{-1}G_{2}^{-1}R_{1,2}(u)G_{1}G_{2},\\
\bar{R}_{1,2}(u) & =V_{1}G_{1}^{t}V_{1}G_{2}^{-1}\bar{R}_{1,2}(u)V_{1}\left(G_{1}^{t}\right)^{-1}V_{1}G_{2},
\end{split}
\end{equation}
where $G\in GL(N)$. For this $GL(N)$ transformation the $K$-matrix
transforms as
\begin{equation}
K\to GKVG^{t}V,
\end{equation}
therefore the $K$-matrix is invariant under this transformation only
when 
\begin{equation}
\mathcal{U}=G\mathcal{U}G^{t}.
\end{equation}
We can see that the $K$-matrix preserves the subalgebra $\mathfrak{so}(N)$
or $\mathfrak{sp}(N)$ of the original algebra $\mathfrak{gl}(N)$
for symmetric or anti-symmetric $\mathcal{U}$ therefore we call this
$K$-matrix and the corresponding final state $\langle\Psi|$ the
$\mathfrak{so}(N)$ or the $\mathfrak{sp}(N)$ symmetric $K$-matrix
and two-site state.

\subsubsection{Evaluating off-shell overlaps\label{subsec:EmbedTw}}

For the twisted two-site states one can calculate the overlap $\left\langle \Psi\right|\mathbb{B}(\bar{t})$
in a similar way as for the untwisted case. One has to use the recurrence
equation of the Bethe vectors (\ref{eq:rec1})
\begin{equation}
\mathbb{B}^{r_{1},\dots}=\sum_{k=2}^{N}T_{1,k}\sum\mathbb{B}^{r_{1}-1,\dots},
\end{equation}
and the component $(N,k)$ of the twisted $KT$-relation (\ref{eq:KT_Tw})
\begin{equation}
\left\langle \Psi\right|T_{1,k}(u)=\lambda_{0}(u)\sum_{j=1}^{N}\left\langle \Psi\right|\widehat{T}_{N,j}(-u)\frac{K_{j,k}}{K_{N,1}}-\sum_{i=2}^{N}\frac{K_{N,i}}{K_{N,1}}\left\langle \Psi\right|T_{i,k}(u),
\end{equation}
which can be used to change the creation operators $T_{1,k}$ to the
operators $\widehat{T}_{N,j}$ and $T_{i,k}$ where $i,k=2,\dots,N$
and $j=1,\dots,N$. The operators $\widehat{T}_{N,j}$ decrease the
number of first Bethe roots by one and the operators $\widehat{T}_{N,N}(-u)$,$T_{i,k}(u)$
do not change it. Using the actions formulas (\ref{eq:act}) and (\ref{eq:actTw})
we obtain that
\begin{equation}
\left\langle \Psi\right|\mathbb{B}^{r_{1},\dots}=\sum\left\langle \Psi\right|\mathbb{B}^{r_{1}-1,\dots}+\sum\left\langle \Psi\right|\mathbb{B}^{r_{1}-2,\dots}.
\end{equation}
When we changed the creation operators to the diagonal and the annihilation
ones we used the following assumption
\begin{equation}
K_{N,1}\neq0,
\end{equation}
which is possible only for $\mathfrak{so}(N)$ symmetric $K$-matrices
but it is obviously never true for $\mathfrak{sp}(N)$ symmetric ones.
Repeating these steps we can express the general overlap with a sum
of overlaps without the first Bethe roots
\begin{equation}
\left\langle \Psi\right|\mathbb{B}^{r_{1},\dots}=\sum\left\langle \Psi\right|\mathbb{B}^{0,\dots}.
\end{equation}

We can see that we reduced the original $\mathfrak{gl}(N)$ Bethe
state to $\mathfrak{gl}(N-1)$ ones. In appendix \ref{sec:Off-shell-Bethe-vectors}
we show that the action of operators $\left\{ T_{a,b}\right\} _{a,b=2}^{N}$
and $\left\{ \widehat{T}_{\bar{a},\bar{b}}\right\} _{\bar{a},\bar{b}=1}^{N-1}$
do not lead out of the subspace generated by the Bethe vectors $\mathbb{B}^{0,\dots}$
. Naively, the twisted $KT$-relation (\ref{eq:KT_Tw}) is not closed
for the indexes $b=2,\dots,N$ and $\bar{a}=1,\dots,N-1$ since
\begin{equation}
\sum_{c=2}^{N}K_{\bar{a},c}\left\langle \Psi\right|T_{c,b}(u)+K_{\bar{a},1}\left\langle \Psi\right|T_{1,b}(u)=\lambda_{0}(u)\sum_{\bar{c}=1}^{N-1}\left\langle \Psi\right|\widehat{T}_{\bar{a},\bar{c}}(-u)K_{\bar{c},b}+\lambda_{0}(u)\left\langle \Psi\right|\widehat{T}_{\bar{a},N}(-u)K_{N,b}.\label{eq:nestin1}
\end{equation}
We can see that the operators $T_{1,b}$ and $\widehat{T}_{\bar{a},N}$
create the first type of Bethe roots therefore we have to eliminate
them from the equation (\ref{eq:nestin1}). Let us see the components
$(N,b)$ and $(\bar{a},1)$ of the twisted $KT$-relation (\ref{eq:KT_Tw})
\begin{align}
 & \sum_{c=2}^{N}K_{N,c}\left\langle \Psi\right|T_{c,b}(u)+K_{N,1}\left\langle \Psi\right|T_{1,b}(u)\cong\lambda_{0}(u)\left\langle \Psi\right|\widehat{T}_{N,N}(-u)K_{N,b},\label{eq:nestin2}\\
 & K_{\bar{a},1}\left\langle \Psi\right|T_{1,1}(u)\cong\lambda_{0}(u)\sum_{\bar{c}=1}^{N-1}\left\langle \Psi\right|\widehat{T}_{\bar{a},\bar{c}}(-u)K_{\bar{c},1}+\lambda_{0}(u)\left\langle \Psi\right|\widehat{T}_{\bar{a},N}(-u)K_{N,1},\label{eq:nestin3}
\end{align}
where $\cong$ denotes the equality in the subspace generated by the
Bethe vectors $\mathbb{B}^{0,\dots}$. Applying the component $(N,1)$
of the twisted $KT$-relation (\ref{eq:KT_Tw}) on the pseudo-vacuum
we obtain that
\begin{equation}
K_{N,1}\left\langle \Psi\right|T_{1,1}(u)\left|0\right\rangle =\lambda_{0}(u)\left\langle \Psi\right|\widehat{T}_{N,N}(-u)\left|0\right\rangle K_{N,1},
\end{equation}
therefore the following symmetry property of the vacuum eigenvalues
have to be satisfied 
\begin{equation}
\lambda_{1}(u)=\lambda_{0}(u)\hat{\lambda}_{N}(-u).
\end{equation}
Using this symmetry property we obtain that
\begin{equation}
T_{1,1}(u)\cong\lambda_{0}(u)\widehat{T}_{N,N}(-u).\label{eq:nestin4}
\end{equation}
Combining the equations (\ref{eq:nestin1}), (\ref{eq:nestin2}),
(\ref{eq:nestin3}) and (\ref{eq:nestin4}) we obtain a new twisted
$KT$-relation on the subspace generated by the Bethe states $\mathbb{B}^{0,\dots}$
\begin{equation}
\sum_{c=2}^{N}K_{\bar{a},c}^{(2)}\left\langle \Psi\right|T_{c,b}(u)\cong\lambda_{0}(u)\sum_{\bar{c}=1}^{N-1}\left\langle \Psi\right|\widehat{T}_{\bar{a},\bar{c}}(-u)K_{\bar{c},b}^{(2)},
\end{equation}
where
\begin{equation}
K_{\bar{a},b}^{(2)}=K_{N,1}^{(1)}K_{\bar{a},b}^{(1)}-K_{\bar{a},1}^{(1)}K_{N,b}^{(1)},\qquad K^{(1)}=K,
\end{equation}
for $b=2,\dots,N$ and $\bar{a}=1,\dots,N-1$.

We can see that twisted $KT$-relation is closed for the $\mathfrak{gl}(N-1)$
Bethe vectors. 
\begin{equation}
\sum_{c=2}^{N}K_{\bar{a},c}^{(2)}\left\langle \Psi\right|T_{c,b}^{N}(u)\mathbb{B}^{N}(\emptyset,\left\{ \bar{t}^{k}\right\} _{k=2}^{N-1})=\lambda_{0}^{N}(u)\sum_{\bar{c}=1}^{N-1}K_{\bar{c},b}^{(2)}\left\langle \Psi\right|\widehat{T}_{\bar{a},\bar{c}}^{N}(-u)\mathbb{B}^{N}(\emptyset,\left\{ \bar{t}^{k}\right\} _{k=2}^{N-1}).
\end{equation}
Let us use the the injections $\mathfrak{i}^{\pm}$ \footnote{Note that the injection $\mathfrak{i}^{+}$ shifts the indices of
the untwisted monodromy matrix $T_{c,b}^{N-1}(u)\to T_{c+1,b+1}^{N}(u)$
therefore we use the range $b=1,\dots,N-1$ in the equations (\ref{eq:first})-(\ref{eq:last}).
Since we are using this shifted range, the second indexes of the $K$-matrices
also contain the shift $+1$ in the equations (\ref{eq:first})-(\ref{eq:last}).}
\begin{multline}
\sum_{d=1}^{N-1}K_{\bar{a},d+1}^{(2)}\left\langle \Psi\right|\mathfrak{i}^{+}\left(T_{d,b}^{N-1}(u)\mathbb{B}^{N-1}(\left\{ \bar{t}^{k}\right\} _{k=2}^{N-2})\right)=\\
\lambda_{0}^{N}(u)\sum_{\bar{d}=1}^{N-1}K_{\bar{d},b+1}^{(2)}\left\langle \Psi\right|\mathfrak{i}^{+}\left(\widehat{T}_{\bar{a},\bar{d}}^{N-1}(-u-c)\mathbb{B}^{N-1}(\left\{ \bar{t}^{k}\right\} _{k=2}^{N-2})\right),\label{eq:first}
\end{multline}
where $\bar{a},b=1,\dots,N-1$. Defining the co-vector $\left\langle \Psi^{(2)}\right|=\left\langle \Psi\right|\mathfrak{i}^{+}$
we obtain the relation
\begin{multline}
\sum_{d=1}^{N-1}K_{\bar{a},d+1}^{(2)}\left\langle \Psi^{(2)}\right|T_{d,b}^{N-1}(u-c/2)\mathbb{B}^{N-1}(\left\{ \bar{t}^{k}\right\} _{k=2}^{N-1})=\\
\lambda_{0}^{N-1}(u-c/2)\sum_{\bar{d}=1}^{N-1}K_{\bar{d},b+1}^{(2)}\left\langle \Psi^{(2)}\right|\widehat{T}_{\bar{a},\bar{d}}^{N-1}(-u-c/2)\mathbb{B}^{N-1}(\left\{ \bar{t}^{k}\right\} _{k=2}^{N-1}),
\end{multline}
which is almost the twisted $KT$-relation of the $\mathfrak{gl}(N-1)$
spin chain but we have extra shifts. Applying the boost operators
$\mathcal{B}_{c/2}$ we obtain the real twisted $KT$-relation 
\begin{multline}
\sum_{d=1}^{N-1}K_{\bar{a},d+1}^{(2)}\left\langle \Psi^{(2)}\right|T_{d,b}^{N-1}(u)\mathcal{B}_{c/2}\left(\mathbb{B}^{N-1}(\left\{ \bar{t}^{k}\right\} _{k=2}^{N-1})\right)=\\
\lambda_{0}^{N-1}(u)\sum_{\bar{d}=1}^{N-1}K_{\bar{d},b+1}^{(2)}\left\langle \Psi^{(2)}\right|\widehat{T}_{\bar{a},\bar{d}}^{N-1}(-u)\mathcal{B}_{c/2}\left(\mathbb{B}^{N-1}(\left\{ \bar{t}^{k}\right\} _{k=2}^{N-1})\right),\label{eq:last}
\end{multline}
therefore
\begin{equation}
S_{K^{(2)}}^{N-1}(\left\{ \bar{t}^{k}\right\} _{k=2}^{N-1})=\left\langle \Psi^{(2)}\right|\mathcal{B}_{c/2}\left(\mathbb{B}^{N-1}(\left\{ \bar{t}^{k}\right\} _{k=2}^{N-1})\right).
\end{equation}
Since 
\begin{equation}
\left\langle \Psi\right|\mathbb{B}^{N}(\emptyset,\left\{ \bar{t}^{k}\right\} _{k=2}^{N-1})=\mathfrak{i}^{+}\left(\left\langle \Psi^{(2)}\right|\mathbb{B}^{N-2}(\left\{ \bar{t}^{k}\right\} _{k=2}^{N-2})\right)
\end{equation}
the overlaps of the $\mathfrak{gl}(N)$ model where $\bar{t}^{1}=\emptyset$
can be obtained from the overlaps of the $\mathfrak{gl}(N-1)$ model
i.e. we just obtained an embedding rule for the overlaps
\begin{equation}
S_{K}^{N}(\emptyset,\bar{t}^{2},\dots,\bar{t}^{N-1})=S_{K^{(2)}}^{N-1}(\bar{t}^{2}+c/2,\dots,\bar{t}^{N-1}+c/2)\Biggr|_{\alpha_{\nu}(u)\to\alpha_{\nu+1}(u-c/2)}.\label{eq:recursion_S_Tw}
\end{equation}

Repeating the previous method we can eliminate the sets of Bethe roots
$\bar{t}^{\nu}$, therefore we obtain a recursion for the overlaps
and in the end of the day we obtain an explicit result. In the $(k+1)$-th
step of the nesting we have a $\mathfrak{gl}(N-k)$ $K$-matrix:
\begin{equation}
\boxed{K_{\bar{a},b}^{(k+1)}=K_{N+1-k,k}^{(k)}K_{\bar{a},b}^{(k)}-K_{\bar{a},k}^{(k)}K_{N+1-k,b}^{(k)},}\label{eq:reqK-1}
\end{equation}
where $\bar{a}=1,\dots,N-k$, $b=k+1,\dots,N$ and we also obtain
the symmetry conditions
\begin{equation}
\boxed{\lambda_{k}(u)=\lambda_{0}(u)\hat{\lambda}_{N+1-k}(-u).}\label{eq:constLamTw}
\end{equation}
These conditions imply that
\begin{equation}
\boxed{\alpha_{k}(u)=\frac{\lambda_{k}(u)}{\lambda_{k+1}(u)}=\frac{\hat{\lambda}_{N+1-k}(-u)}{\hat{\lambda}_{N-k}(-u)}=\frac{1}{\hat{\alpha}_{N-k}(-u)}=\frac{1}{\alpha_{k}(-u-kc)}.}\label{eq:constAlpTw}
\end{equation}
We can see that this procedure gives the off-shell overlap only when
the following condition is satisfied at every level of the nesting
\begin{equation}
K_{N+1-k,k}^{(k)}\neq0.
\end{equation}

We can give an alternative recursion for the overlaps which is based
on the other recurrence equation of the Bethe states (\ref{eq:rec2}).
This recursion eliminates the Bethe root $\bar{t}^{N-1}$ at the first
step. We can also use a $KT$-relation in the subspace generated by
the Bethe vectors $\mathbb{B}^{\dots,0}$:
\begin{equation}
\sum_{c=1}^{N-1}\bar{K}_{\bar{a},c}^{(2)}\left\langle \Psi\right|T_{c,b}(u)\cong\lambda_{0}(u)\sum_{\bar{c}=2}^{N}\left\langle \Psi\right|\widehat{T}_{\bar{a},\bar{c}}(-u)\bar{K}_{\bar{c},b}^{(2)},
\end{equation}
where
\begin{equation}
\bar{K}_{\bar{a},b}^{(2)}=K_{N,1}^{(1)}K_{\bar{a},b}^{(1)}-K_{\bar{a},1}^{(1)}K_{N,b}^{(1)},\qquad\bar{K}^{(1)}=K,
\end{equation}
for $b=1,\dots,N-1$ and $\bar{a}=2,\dots,N$. The existence of this
$KT$-equation means that we can get an other embedding rule for the
overlaps
\begin{equation}
S_{K}^{N}(\bar{t}^{1},\dots,\bar{t}^{N-2},\emptyset)=S_{\bar{K}^{(2)}}^{N-1}(\bar{t}^{1},\dots,\bar{t}^{N-2}).\label{eq:recursion_S_Tw-1}
\end{equation}

\subsubsection{Chiral pair structure\label{subsec:Chiral-pair-structure}}

Let us continue with the selection rules of the on-shell overlaps.
Since the $K$-matrix is invertible the integrability condition 
\begin{equation}
\langle\Psi|\mathcal{T}(u)=\lambda_{0}(u)\langle\Psi|\widehat{\mathcal{T}}(-u)
\end{equation}
follows from the twisted $KT$-relation (\ref{eq:KT_Tw}). Applying
this equation to an on-shell Bethe vector we find that
\begin{equation}
\left(\tau(u|\bar{t})-\lambda_{0}(u)\hat{\tau}(-u|\bar{t})\right)\langle\Psi|\mathbb{B}(\bar{t})=0.
\end{equation}
The non-vanishing overlap requires that
\begin{equation}
\tau(u|\bar{t})=\lambda_{0}(u)\hat{\tau}(-u|\bar{t}).\label{eq:intcondtau-1}
\end{equation}
The eigenvalue $\hat{\tau}(-u|\bar{t})$ can be written as (\ref{eq:tweig})
\begin{multline}
\lambda_{0}(u)\hat{\tau}(-u|\bar{t})=\sum_{i=1}^{N}\lambda_{0}(u)\hat{\lambda}_{i}(-u)f(\bar{t}^{N-i}+(N-i)c,-u)f(-u,\bar{t}^{N-i+1}+(N-i+1)c)=\\
=\sum_{i=1}^{N}\lambda_{i}(u)f(-\bar{t}^{i}-ic,u)f(u,-\bar{t}^{i-1}-(i-1)c)=\tau(u|\pi^{c}(\bar{t})),
\end{multline}
where we used that
\begin{equation}
\lambda_{i}(u)=\lambda_{0}(u)\hat{\lambda}_{N+1-i}(-u),\qquad f(u,v)=f(-v,-u).
\end{equation}
and we introduced the following map of Bethe roots:
\begin{equation}
\pi^{c}(\bar{t}^{1},\bar{t}^{2},\dots,\bar{t}^{N-1})=\left(-\bar{t}^{1}-c,-\bar{t}^{2}-2c,\dots,-\bar{t}^{N-1}-(N-1)c\right).\label{eq:chpair}
\end{equation}
Substituting back to (\ref{eq:intcondtau-1}) the condition for non-vanishing
overlaps read as
\begin{equation}
\tau(u|\bar{t})=\tau(u|\pi^{c}(\bar{t})).
\end{equation}
Notice that the set $\pi^{c}(\bar{t})$ also satisfies the Bethe equations
i.e. the vector $\mathbb{B}(\pi^{c}(\bar{t}))$ is also an on-shell
Bethe vector. Since different Bethe vectors have different eigenvalues,
we just obtain that the Bethe roots have to satisfy the selection
rule
\begin{equation}
\bar{t}=\pi^{c}(\bar{t})\label{eq:achiral-1}
\end{equation}
for the non-vanishing overlaps. We call the Bethe roots with condition
(\ref{eq:achiral-1}) as Bethe roots with \emph{chiral pair structure}.
For chiral pair structure every sets $\bar{t}^{\nu}$ satisfy the
condition
\begin{equation}
\bar{t}^{\nu}=-\bar{t}^{\nu}-\nu c.
\end{equation}
For the chiral pair structure there exist three disjoint subsets $\bar{t}^{+\nu},\bar{t}^{-,\nu},\bar{t}^{0,\nu}$
for which $\bar{t}^{\nu}=\bar{t}^{+,\nu}\cup\bar{t}^{-,\nu}\cup\bar{t}^{0,\nu}$
and $\bar{t}^{-,\nu}=-\bar{t}^{+,\nu}-\nu c$ and $\bar{t}^{0,\nu}=\emptyset$
or $\bar{t}^{0,\nu}=\{-\frac{\nu c}{2}\}$ for even or odd $r_{\nu}$.
Using the definitions $\bar{t}^{\pm}=\cup_{\nu}\bar{t}^{\pm,\nu}$
and $\bar{t}^{0}=\cup_{\nu}\bar{t}^{0,\nu}$ we can obtain the following
decomposition of the Bethe roots
\begin{equation}
\bar{t}=\bar{t}^{+}\cup\bar{t}^{-}\cup\bar{t}^{0}.
\end{equation}

\subsection{Co-product structure of the integrable states\label{subsec:Coproduct}}

Let $\mathcal{H}^{(1)},\mathcal{H}^{(2)}$ be quantum spaces and the
corresponding monodromy matrices are $T^{(1)}(u),T^{(1)}(u)$. Let
$\left\langle \Psi^{(1)}\right|\in\mathcal{H}^{(1)}$ and $\left\langle \Psi^{(2)}\right|\in\mathcal{H}^{(2)}$
be integrable final states with the same untwisted $K$-matrix i.e.
they satisfy the same $KT$-relation
\begin{equation}
K_{0}(u)\left\langle \Psi^{(i)}\right|T_{0}^{(i)}(u)=\left\langle \Psi^{(i)}\right|T_{0}^{(i)}(-u)K_{0}(u),
\end{equation}
for $i=1,2$. Let us define the tensor product quantum space $\mathcal{H}=\mathcal{H}^{(1)}\otimes\mathcal{H}^{(2)}$
for which the monodromy matrix is 
\begin{equation}
T_{i,j}(u)=\sum_{k=1}^{N}T_{k,j}^{(1)}(u)\otimes T_{i,k}^{(2)}(u).
\end{equation}
It is easy to show that the co-vector 
\begin{equation}
\left\langle \Psi\right|=\left\langle \Psi^{(1)}\right|\otimes\left\langle \Psi^{(2)}\right|\label{eq:Psi_Factor}
\end{equation}
is an integrable two-site state with the same $K$-matrix i.e. it
satisfies the $KT$-relation
\begin{equation}
K_{0}(u)\left\langle \Psi\right|T_{0}(u)=\left\langle \Psi\right|T_{0}(-u)K_{0}(u).
\end{equation}
We can also show that the same is true for the twisted cases.

In the previous subsections we saw for the $\mathfrak{so}(N)$ and
$\mathfrak{gl}(\left\lfloor \frac{N}{2}\right\rfloor )\otimes\mathfrak{gl}(\left\lceil \frac{N}{2}\right\rceil )$
symmetric cases that fixing the $K$-matrix and the vacuum eigenvalues
$\lambda_{i}(u)$ the overlaps $\left\langle \Psi\right|\mathbb{B}(\bar{t})$
are completely fixed therefore the co-vector $\left\langle \Psi\right|\in\mathcal{H}^{*}$
is unique for a fixed $K$-matrix and $\lambda_{i}$-s and normalization
$\langle\Psi|0\rangle=1$.

\section{Main results\label{sec:Main-results}}

In the previous section we schematically showed an approach which
can be used to calculate any off-shell overlaps for $\mathfrak{gl}(\left\lfloor \frac{N}{2}\right\rfloor )\oplus\mathfrak{gl}(\left\lceil \frac{N}{2}\right\rceil )$
or $\mathfrak{so}(N)$ symmetric two-site states. In this section,
we put this method into practice. Our goal is to find a concrete recurrence
equation for the off-shell overlap which is suitable for the derivation
of the on-shell formulas. The structure of the calculation is very
similar to the calculation of the scalar products of the Bethe states
\cite{Hutsalyuk:2017way}. In this section we show the main statements
and the derivations can be found in the appendix.

\subsection{Properties of the off-shell overlaps}

We choose the normalization of the two-site state as
\begin{equation}
\langle\Psi|0\rangle=1,
\end{equation}
and our goal is to calculate the off-shell overlaps
\begin{equation}
S_{K}^{N}(\bar{t})=\left\langle \Psi\right|\mathbb{B}(\bar{t}).
\end{equation}

\subsubsection*{Step 1}

It turns out that the vacuum eigenvalue dependence of the off-shell
overlaps can be written as
\begin{equation}
S_{K}^{N}(\bar{t})=\sum_{\mathrm{part}}W_{K}^{N}(\bar{t}_{\textsc{i}}|\bar{t}_{\textsc{ii}})\prod_{\nu=1}^{N-1}\alpha_{\nu}(\bar{t}_{\textsc{i}}^{\nu}),\label{eq:sumFormula}
\end{equation}
where the sets of the Bethe roots $\bar{t}^{k}$ are divided into
two disjoint subsets for which $\bar{t}^{k}=\bar{t}_{\textsc{i}}^{k}\cup\bar{t}_{\textsc{ii}}^{k}$.
The sum is taken over all possible partitions of this type. Since
our calculation based on algebraic Bethe ansatz the rational coefficients
$W_{K}^{N}(\bar{t}_{\textsc{i}}|\bar{t}_{\textsc{ii}})$ do not depend
on vacuum eigenvalues. The weights $W_{K}$ are completely determined
only from the components of the $R$- and $K$-matrices. The derivation
of the sum formula (\ref{eq:sumFormula}) can be found in appendix
\ref{sec:DSumFormula}.

\subsubsection*{Step 2}

Just like in the calculation of the scalar products \cite{Hutsalyuk:2017tcx}
we can define highest coefficients (HC) of the overlap formula as
\begin{equation}
Z_{K}^{N}(\bar{t}):=W_{K}^{N}(\bar{t}|\emptyset),\qquad\bar{Z}_{K}^{N}(\bar{t}):=W_{K}^{N}(\emptyset|\bar{t}).
\end{equation}
It turns out that a weight $W_{K}^{N}$ of a general partition is
completely defined by the HC-s as

\begin{equation}
W_{K}^{N}(\bar{t}_{\textsc{i}}|\bar{t}_{\textsc{ii}})=\frac{\prod_{\nu=1}^{N-1}f(\bar{t}_{\textsc{ii}}^{\nu},\bar{t}_{\textsc{i}}^{\nu})}{\prod_{\nu=1}^{N-2}f(\bar{t}_{\textsc{ii}}^{\nu+1},\bar{t}_{\textsc{i}}^{\nu})}Z_{K}^{N}(\bar{t}_{\textsc{i}})\bar{Z}_{K}^{N}(\bar{t}_{\textsc{ii}}).\label{eq:WZZ}
\end{equation}
We can also derive a connection between HC-s as
\begin{equation}
Z_{K}^{N}(\bar{t})=\bar{Z}_{\Pi(K)}^{N}(\pi^{a}(\bar{t}))\label{eq:connHCUTw}
\end{equation}
for $\mathfrak{gl}(\left\lfloor \frac{N}{2}\right\rfloor )\oplus\mathfrak{gl}(\left\lceil \frac{N}{2}\right\rceil )$
symmetric case and

\begin{equation}
Z_{K}^{N}(\bar{t})=(-1)^{\#\bar{t}}\left(\prod_{s=1}^{N-2}f(\bar{t}^{s+1},\bar{t}^{s})\right)^{-1}\bar{Z}_{K}^{N}(\pi^{c}(\bar{t})))\label{eq:connHCTw}
\end{equation}
for $\mathfrak{so}(N)$ symmetric case where we used the notations
(\ref{eq:achiral}), (\ref{eq:chpair}) and
\begin{equation}
\Pi(K(u))=VK^{t}(u)V,
\end{equation}
The derivations of the equations (\ref{eq:WZZ}), (\ref{eq:connHCUTw})
and (\ref{eq:connHCTw}) can be found in appendix \ref{sec:ConnectionsWZ}.

Notice that combining the equation (\ref{eq:WZZ}) with the embedding
formulas (\ref{eq:recursion_S_UTw}) and (\ref{eq:recursion_S_Tw})
we obtain that
\begin{equation}
\bar{Z}_{K}^{N}(\emptyset,\bar{t}^{2},\dots,\bar{t}^{N-2},\emptyset)=\bar{Z}_{K^{(2)}}^{N-2}(\bar{t}^{2},\dots,\bar{t}^{N-2})\label{eq:embedZ_UTw}
\end{equation}
for the untwisted case and
\begin{equation}
\bar{Z}_{K}^{N}(\emptyset,\bar{t}^{2},\dots,\bar{t}^{N-1})=\bar{Z}_{K^{(2)}}^{N-1}(\bar{t}^{2}+c/2,\dots,\bar{t}^{N-1}+c/2)\label{eq:embedZ_Tw}
\end{equation}
for the twisted case.

\subsubsection*{Step 3}

The last step is to find a recurrence relation for the highest coefficient
$\bar{Z}(\bar{t})$ which completely defines the off-shell overlap
according to the results of the previous steps.

For the $\mathfrak{gl}(\left\lfloor \frac{N}{2}\right\rfloor )\oplus\mathfrak{gl}(\left\lceil \frac{N}{2}\right\rceil )$
symmetric case the recurrence equation of the HC can be written as

\begin{multline}
\bar{Z}_{K}^{N}(\{z,\bar{t}^{1}\},\left\{ \bar{t}^{s}\right\} _{s=2}^{N-1})=\\
F_{K}^{(1)}(z)\sum_{\mathrm{part}}\bar{Z}_{K}^{N}(\{\bar{\omega}_{\textsc{ii}}^{s}\}_{s=1}^{N-2},\bar{t}_{\textsc{ii}}^{N-1})\prod_{s=1}^{N-2}\frac{f(\bar{\omega}_{\textsc{i}}^{s},\bar{\omega}_{\textsc{ii}}^{s})}{h(\bar{\omega}_{\textsc{i}}^{s},\bar{\omega}_{\textsc{i}}^{s-1})f(\bar{\omega}_{\textsc{i}}^{s},\bar{\omega}_{\textsc{ii}}^{s-1})}\frac{f(\bar{t}_{\textsc{i}}^{N-1},\bar{t}_{\textsc{ii}}^{N-1})f(\bar{t}^{N-1},-z)}{h(\bar{t}_{\textsc{i}}^{N-1},\bar{\omega}_{\textsc{i}}^{N-2})f(\bar{t}_{\textsc{i}}^{N-1},\bar{\omega}_{\textsc{ii}}^{N-2})f(\bar{t}^{2},z)}-\\
-\sum_{i=2}^{N}\frac{K_{N,i}(z)}{K_{N,1}(z)}\sum_{\mathrm{part}}\bar{Z}_{K}^{N}(\bar{w}_{\textsc{ii}}^{1},\{\bar{t}_{\textsc{ii}}^{s}\}_{s=2}^{i-1},\{\bar{t}^{s}\}_{s=i}^{N-1})\frac{f(\bar{w}_{\textsc{i}}^{1},\bar{w}_{\textsc{ii}}^{1})}{h(\bar{w}_{\textsc{i}}^{1},z)}\frac{f(\bar{t}_{\textsc{i}}^{2},\bar{t}_{\textsc{ii}}^{2})}{h(\bar{t}_{\textsc{i}}^{2},\bar{w}_{\textsc{i}}^{1})f(\bar{t}_{\textsc{i}}^{2},\bar{w}_{\textsc{ii}}^{1})}\prod_{s=3}^{i-1}\frac{f(\bar{t}_{\textsc{i}}^{s},\bar{t}_{\textsc{ii}}^{s})}{h(\bar{t}_{\textsc{i}}^{s},\bar{t}_{\textsc{i}}^{s-1})f(\bar{t}_{\textsc{i}}^{s},\bar{t}_{\textsc{ii}}^{s-1})},\label{eq:recursionZ_UTw1}
\end{multline}
where we defined the sets $\bar{w}^{1}=\left\{ z,\bar{t}^{1}\right\} $
and $\bar{\omega}^{\nu}=\left\{ -z,\bar{t}^{\nu}\right\} $ for $\nu=1,\dots,N-2$.
We set by definition $\bar{\omega}^{0}=\{-z\}$. The sum in the second
line goes over all the partitions of $\bar{t}^{N-1}=\bar{t}_{\textsc{i}}^{N-1}\cup\bar{t}_{\textsc{ii}}^{N-1}$
and $\bar{\omega}^{s}=\bar{\omega}_{\textsc{i}}^{s}\cup\bar{\omega}_{\textsc{ii}}^{s}$
where $\#\bar{t}_{\textsc{i}}^{N-1}=\#\bar{\omega}_{\textsc{i}}^{s}=1$
for $s=0,\dots,N-2$. The sum in the third line goes over all the
partitions of $\bar{w}^{1}=\bar{w}_{\textsc{i}}^{1}\cup\bar{w}_{\textsc{ii}}^{1}$
and $\bar{t}^{s}=\bar{t}_{\textsc{i}}^{s}\cup\bar{t}_{\textsc{ii}}^{s}$
where $\#\bar{w}_{\textsc{i}}^{1}=\#\bar{t}_{\textsc{i}}^{s}=1$ for
$s=2,\dots,i-1$. We also defined the one-particle overlap function
$F_{K}^{(1)}(z)$ as
\begin{equation}
F_{K}^{(1)}(z):=\frac{K_{N-1,2}(z)}{K_{N,1}(z)}-\frac{K_{N,2}(z)}{K_{N,1}(z)}\frac{K_{N-1,1}(z)}{K_{N,1}(z)}.\label{eq:defFK}
\end{equation}
We can see that the recurrence equation (\ref{eq:recursionZ_UTw1})
can be use to eliminate Bethe roots $\bar{t}^{1}$ from the HC and
it is important that this recursion does not increase the number of
$\bar{t}^{N-1}$.

We also have a recursion for the last Bethe roots:
\begin{multline}
\bar{Z}_{K}^{N}(\left\{ \bar{t}^{k}\right\} _{k=1}^{N-2},\{z,\bar{t}^{N-1}\})=\sum_{i=1}^{N}\sum_{j=1}^{N-1}\frac{K_{j,i}(-z)}{K_{N,1}(-z)}f(\bar{t}^{1},-z)\sum_{\mathrm{part}(\bar{t})}\bar{Z}_{K}^{N}(\left\{ \bar{t}_{\textsc{ii}}^{k}\right\} _{k=1}^{N-2},\bar{t}^{N-1})\times\\
\prod_{s=1}^{i-1}\frac{f(\bar{t}_{\textsc{i}}^{s},\bar{t}_{\textsc{ii}}^{s})}{h(\bar{t}_{\textsc{i}}^{s},\bar{t}_{\textsc{i}}^{s-1})f(\bar{t}_{\textsc{i}}^{s},\bar{t}_{\textsc{ii}}^{s-1})}\frac{\prod_{\nu=j}^{N-2}g(\bar{t}_{\textsc{iii}}^{\nu+1},\bar{t}_{\textsc{iii}}^{\nu})f(\bar{t}_{\textsc{iii}}^{\nu},\bar{t}_{\textsc{ii}}^{\nu})f(\bar{t}_{\textsc{iii}}^{\nu},\bar{t}_{\textsc{i}}^{\nu})}{\prod_{\nu=j}^{N-1}f(\bar{t}_{\textsc{iii}}^{\nu},\bar{t}^{\nu-1})},\label{eq:recursionZ_UTw2}
\end{multline}
where the sum in the first line goes over the partitions $\bar{t}^{\nu}=\bar{t}_{\textsc{i}}^{\nu}\cup\bar{t}_{\textsc{ii}}^{\nu}\cup\bar{t}_{\textsc{iii}}^{\nu}$
for $\nu=1,\dots,N-2$ where 
\begin{align*}
\#\bar{t}_{\textsc{i}}^{\nu} & =1,\quad\text{for \ensuremath{\nu<i}},\qquad\#\bar{t}_{\textsc{i}}^{\nu}=0,\quad\text{for \ensuremath{i\leq\nu,}}\\
\#\bar{t}_{\textsc{iii}}^{\nu} & =0,\quad\text{for \ensuremath{\nu<j}},\qquad\#\bar{t}_{\textsc{iii}}^{\nu}=1,\quad\text{for \ensuremath{j\leq\nu,}}
\end{align*}
and we set by definition $\bar{t}_{\textsc{iii}}^{N-1}=\{z\},\bar{t}_{\textsc{ii}}^{N-1}=\bar{t}^{N-1}$
and $\bar{t}_{\textsc{i}}^{0}=\{-z\}$. We can see that the recurrence
equation (\ref{eq:recursionZ_UTw2}) can be use to eliminate Bethe
roots $\bar{t}^{N-1}$ from the HC and it is important that this recursion
does not increase the number of $\bar{t}^{1}$. We can see that the
recurrence equations (\ref{eq:recursionZ_UTw1}) and (\ref{eq:recursionZ_UTw2})
with the embedding formula (\ref{eq:embedZ_UTw}) and the initial
condition $\bar{Z}(\emptyset)=1$ completely define the highest coefficients
for the untwisted case.

For the $\mathfrak{so}(N)$ symmetric case the recurrence relation
of the HC can be written as

\begin{multline}
\bar{Z}_{K}^{N}(\{z,\bar{t}^{1}\},\left\{ \bar{t}^{s}\right\} _{s=2}^{N-1})=F_{K}^{(1)}\frac{f(\bar{t}^{1},-z-c)}{f(\bar{t}^{2},-z-c)f(\bar{t}^{2},z)}\sum_{\mathrm{part}}\bar{Z}_{K}^{N}(\bar{t}_{\textsc{ii}}^{1},\{\bar{t}^{s}\}_{s=2}^{N-1})\frac{f(\bar{t}_{\textsc{i}}^{1},\bar{t}_{\textsc{ii}}^{1})}{h(\bar{t}_{\textsc{i}}^{1},-z-c)}-\\
-\sum_{i=2}^{N}\frac{K_{N,i}}{K_{N,1}}\sum_{\mathrm{part}}\bar{Z}_{K}^{N}(\bar{w}_{\textsc{ii}}^{1},\{\bar{t}_{\textsc{ii}}^{s}\}_{s=2}^{i-1},\{\bar{t}^{s}\}_{s=i}^{N-1})\frac{f(\bar{w}_{\textsc{i}}^{1},\bar{w}_{\textsc{ii}}^{1})}{h(\bar{w}_{\textsc{i}}^{1},z)}\frac{f(\bar{t}_{\textsc{i}}^{2},\bar{t}_{\textsc{ii}}^{2})}{h(\bar{t}_{\textsc{i}}^{2},\bar{w}_{\textsc{i}}^{1})f(\bar{t}_{\textsc{i}}^{2},\bar{w}_{\textsc{ii}}^{1})}\prod_{s=3}^{i-1}\frac{f(\bar{t}_{\textsc{i}}^{s},\bar{t}_{\textsc{ii}}^{s})}{h(\bar{t}_{\textsc{i}}^{s},\bar{t}_{\textsc{i}}^{s-1})f(\bar{t}_{\textsc{i}}^{s},\bar{t}_{\textsc{ii}}^{s-1})},\label{eq:recursionZ_Tw}
\end{multline}
where we defined the set $\bar{w}^{1}=\left\{ z,\bar{t}^{1}\right\} $
and the one-particle overlap as
\begin{equation}
F_{K}^{(1)}:=\left(\frac{K_{N,2}}{K_{N,1}}\right)^{2}-\frac{K_{N-1,2}}{K_{N,1}}.
\end{equation}
The sum of the first line is taken over all possible partitions $\bar{t}^{1}=\bar{t}_{\textsc{i}}^{1}\cup\bar{t}_{\textsc{ii}}^{1}$
where $\#\bar{t}_{\textsc{i}}^{1}=1$ and the sum of the second line
is taken over all possible partitions $\bar{w}^{1}=\{z,\bar{t}^{1}\}=\bar{w}_{\textsc{i}}^{1}\cup\bar{w}_{\textsc{ii}}^{1}$,
$\bar{t}^{s}=\bar{t}_{\textsc{i}}^{s}\cup\bar{t}_{\textsc{ii}}^{s}$
where $\#\bar{w}_{\textsc{i}}^{1}=\#\bar{t}_{\textsc{i}}^{s}=1$ for
$s=2,\dots,i-1$. We can see that the recurrence equations (\ref{eq:recursionZ_Tw})
with the embedding formula (\ref{eq:embedZ_Tw}) and the initial condition
$\bar{Z}(\emptyset)=1$ completely define the highest coefficients
for the twisted case.

The derivation of the recurrence equations (\ref{eq:recursionZ_UTw1}),
(\ref{eq:recursionZ_UTw2}) and (\ref{eq:recursionZ_Tw}) can be found
in appendix \ref{sec:RecursionsHC}. Notice that the equations of
the steps 1,2,3 completely determine the off-shell overlaps.

\subsubsection*{Step 4}

We saw that the non-vanishing on-shell overlaps require certain pair
structures for the Bethe roots. One can easily see that the HC-s have
poles in the pair structure limit. It turns out that the residues
of these poles are crucial for the on-shell limit.

For the $\mathfrak{gl}(\left\lfloor \frac{N}{2}\right\rfloor )\oplus\mathfrak{gl}(\left\lceil \frac{N}{2}\right\rceil )$
symmetric case the HC-s have the following poles at the limit $t_{k}^{\nu}+t_{l}^{N-\nu}\to0$
\begin{equation}
\bar{Z}_{K}^{N}(\bar{t})=\frac{c}{t_{k}^{\nu}+t_{l}^{N-\nu}}\tilde{F}_{K}^{(\nu)}(t_{k}^{\nu})\frac{f(\bar{\tau}^{\nu},t_{k}^{\nu})}{f(\bar{\tau}^{\nu+1},t_{k}^{\nu})}\frac{f(\bar{\tau}^{N-\nu},-t_{k}^{\nu})}{f(\bar{\tau}^{N-\nu+1},-t_{k}^{\nu})}\bar{Z}_{K}^{N}(\bar{\tau})+reg\label{eq:pole_Utw}
\end{equation}
for $\nu=1,\dots,\left\lfloor N/2\right\rfloor $ where $\bar{\tau}=\bar{t}\backslash\{t_{k}^{\nu},t_{l}^{N-\nu}\}$.
We also defined a re-normalized one-particle overlap functions $\tilde{F}_{K}^{(\nu)}(z):=F_{K}^{(\nu)}(z)$
for $\nu\neq\frac{N-1}{2}$ and
\begin{align}
\tilde{F}_{K}^{(\frac{N-1}{2})}(z) & :=\frac{F_{K}^{(\frac{N-1}{2})}(z)}{f(-z,z)},
\end{align}
where

\begin{equation}
\boxed{F_{K}^{(\nu)}(z):=\frac{K_{N-\nu,\nu+1}^{(\nu)}(z)}{K_{N+1-\nu,\nu}^{(\nu)}(z)}-\frac{K_{N+1-\nu,\nu+1}^{(\nu)}(z)}{K_{N+1-\nu,\nu}^{(\nu)}(z)}\frac{K_{N-\nu,\nu}^{(\nu)}(z)}{K_{N+1-\nu,\nu}^{(\nu)}(z)}.}\label{eq:onepartUTw}
\end{equation}

For the $\mathfrak{so}(N)$ symmetric case the HC-s have the following
poles at the limit $t_{k}^{\nu}+t_{l}^{\nu}+\nu c\to0$
\begin{equation}
\bar{Z}_{K}^{N}(\bar{t})=\frac{c}{t_{k}^{\nu}+t_{l}^{\nu}+\nu c}F_{K}^{(\nu)}\frac{f(\bar{\tau}^{\nu},t_{k}^{\nu})}{f(\bar{t}^{\nu+1},t_{k}^{\nu})}\frac{f(\bar{\tau}^{\nu},-t_{k}^{\nu}-\nu c)}{f(\bar{t}^{\nu+1},-t_{k}^{\nu}-\nu c)}\bar{Z}_{K}^{N}(\bar{\tau})+reg.,\label{eq:pole_Tw}
\end{equation}
for $\nu=1,\dots,N-1$ where $\bar{\tau}=\bar{t}\backslash\{t_{k}^{\nu},t_{l}^{\nu}\}$
and
\begin{equation}
\boxed{F_{K}^{(k)}:=\left(\frac{K_{N+1-k,k+1}^{(k)}}{K_{N+1-k,k}^{(k)}}\right)^{2}-\frac{K_{N-k,k+1}^{(k)}}{K_{N+1-k,k}^{(k)}}.}\label{eq:onepartTw}
\end{equation}
The derivation of the residues (\ref{eq:pole_Utw}) and (\ref{eq:pole_Tw})
can be found in appendix \ref{sec:Poles}.

\subsection{On-shell limit}

Let us continue with the on-shell limit of the overlaps. The calculation
of this section is based on the derivation of the norm of the Bethe
states \cite{Hutsalyuk:2017way}. We emphasize that the strategy is
the same as the derivation of \cite{Hutsalyuk:2017way}, we only have
to modify some technical details.

In section \ref{sec:Integrable-final-states} we saw that the non-vanishing
overlaps require pair structures $\bar{t}=\bar{t}^{+}\cup\bar{t}^{-}\cup\bar{t}^{0}$
in the on-shell limit. In this subsection we restrict ourselves to
the cases where $\bar{t}^{0}=\emptyset$ therefore we dived the set
of the Bethe roots as $\bar{t}=\bar{t}^{+}\cup\bar{t}^{-}$. These
new sets are defined for the untwisted case as
\begin{equation}
\begin{split}t_{k}^{+,\nu} & =t_{k}^{\nu},\quad t_{k}^{-,\nu}=t_{k}^{N-\nu},\quad\nu<N/2,\\
t_{k}^{+,N/2} & =t_{k}^{N/2},\quad t_{k}^{-,N/2}=t_{k+\frac{r_{N/2}}{2}}^{N/2},\quad k=1,\dots,\frac{r_{N/2}}{2},
\end{split}
\end{equation}
and 
\begin{equation}
t_{k}^{+,\nu}=t_{k}^{\nu},\quad t_{k}^{-,\nu}=t_{k+\frac{r_{\nu}}{2}}^{\nu},\quad k=1,\dots,\frac{r_{\nu}}{2}
\end{equation}
for the twisted case. The pair structure limit is $\bar{t}^{-}\to-\bar{t}^{+}$
for untwisted case and $\bar{t}^{-,\nu}\to-\bar{t}^{+,\nu}-\nu c$
for the twisted case. Let us introduce a common notation for the pair
structure limit $\bar{t}^{-}\to\hat{\pi}\left(\bar{t}^{+}\right)$
where $\hat{\pi}\left(\bar{t}^{+}\right)=-\bar{t}^{+}$ or $\hat{\pi}\left(\bar{t}^{+,\nu}\right)=-\bar{t}^{+,\nu}-\nu c$
for the achiral or the chiral pair structures\footnote{With our earlier notations (\ref{eq:achiral}) and (\ref{eq:chpair}),
$\hat{\pi}=\pi^{a}$ or $\hat{\pi}=\pi^{c}$ for achiral or the chiral
pair structures, respectively.}.

\subsubsection{Gaudin-like determinants\label{subsec:Gaudin-like-determinants}}

We show that the on-shell overlaps are proportional to the Gaudin-like
determinant $\det G^{+}$ where the Gaudin-like matrix is defined
as
\begin{equation}
G_{j,k}^{+,(\mu,\nu)}=-c\left(\frac{\partial}{\partial t_{k}^{+,\nu}}+\frac{\partial}{\partial t_{k}^{-,\nu}}\right)\log\Phi_{j}^{+,(\mu)}\Biggr|_{\bar{t}^{-}=\hat{\pi}\left(\bar{t}^{+}\right)},\label{eq:Gdef}
\end{equation}
where 
\begin{equation}
\Phi_{k}^{+,(\mu)}=\alpha_{\mu}(t_{k}^{+,\mu})\frac{f(\bar{t}_{k}^{\mu},t_{k}^{+,\mu})}{f(t_{k}^{+,\mu},\bar{t}_{k}^{\mu})}\frac{f(t_{k}^{+,\mu},\bar{t}^{\mu-1})}{f(\bar{t}^{\mu+1},t_{k}^{+,\mu})}.
\end{equation}
It is crucial that we first take the derivative in (\ref{eq:Gdef})
and only after take the pair structure limit. Using the symmetry properties
of the $\alpha-s$ (\ref{eq:constAlpUTw}) and (\ref{eq:constAlpTw}),
we can convince ourselves that the Bethe equation for $t_{k}^{-,\mu}$
is completely equivalent with the Bethe equation for $t_{k}^{+,\mu}$
which is $\Phi_{k}^{+,(\mu)}=1$. The diagonal elements of this matrix
contains derivatives of $\log\alpha$-s for which we use the following
notation
\begin{equation}
X_{j}^{+,\mu}=-c\frac{d}{du}\log\alpha_{\mu}^{+}(u)\Biggr|_{u=t_{j}^{+,\mu}}.\label{eq:Xvar}
\end{equation}
We can see that for a specific model the variables $X_{j}^{+,\mu}$
are functions of the Bethe roots. Let us define a more general case
where the variables $X_{j}^{+,\mu}$ and $t_{j}^{+,\mu}$ are independent
which gives us a more general version of the Gaudin determinant where
we do not impose (\ref{eq:Xvar})
\begin{equation}
\mathbf{F}^{(\mathbf{r}^{+})}(\bar{X}^{+},\bar{t}^{+})=\det G^{+}.
\end{equation}
This function depends on two sets of variables $\bar{X}^{+}=\cup_{\nu}\bar{X}^{+,\nu}$
and $\bar{t}^{+}=\cup_{\nu}\bar{t}^{+,\nu}$. The superscript denotes
the number of Bethe roots as $\mathbf{r}^{+}=\#\bar{X}^{+}=\#\bar{t}^{+}$.

The function $\mathbf{F}^{(\mathbf{r}^{+})}$ obeys the following
Korepin criteria.

\paragraph{Korepin criteria.}
\begin{enumerate}[label=(\roman*)]
\item \label{enum:prop1}The function $\mathbf{F}^{(\mathbf{r}^{+})}(\bar{X}^{+},\bar{t}^{+})$
is symmetric over the replacement of the pairs $(X_{j}^{+,\mu},t_{j}^{+,\mu})\leftrightarrow(X_{k}^{+,\mu},t_{k}^{+,\mu})$.
\item \label{enum:prop2}It is linear function of each $X_{j}^{+,\mu}$.
\item \label{enum:prop3}$\mathbf{F}^{(1)}(X_{1}^{+,\nu},t_{1}^{+,\nu})=X_{1}^{+,\nu}$.
\item \label{enum:prop4}The coefficient of $X_{j}^{+,\mu}$ is given by
the function $\mathbf{F}^{(\mathbf{r}^{+}-1)}$ with modified parameters
$X_{k}^{+,\nu}$
\begin{equation}
\frac{\partial\mathbf{F}^{(\mathbf{r}^{+})}(\bar{X}^{+},\bar{t}^{+})}{\partial X_{j}^{+,\mu}}=\mathbf{F}^{(\mathbf{r}^{+}-1)}(\bar{X}^{+,mod}\backslash X_{j}^{+,\mu,mod},\bar{t}^{+}\backslash t_{j}^{+,\mu}),\label{eq:derivFX}
\end{equation}
where the original variables $X_{k}^{+,\nu}$ should be replaced by
$X_{k}^{+,\nu,mod}$ which are defined in the next subsection.
\item \label{enum:prop5}$\mathbf{F}^{(\mathbf{r}^{+})}(\bar{X}^{+},\bar{t}^{+})=0$,
if all $X_{j}^{+,\mu}=0$.
\end{enumerate}
The properties \ref{enum:prop1},\ref{enum:prop2},\ref{enum:prop4}
are obvious. The property \ref{enum:prop3} is also obvious for the
untwisted case where $\nu<\left\lfloor N/2\right\rfloor $. Let us
see the $\nu=N/2$ case where we have two Bethe roots $t_{1}^{N/2}$
and $t_{2}^{N/2}$ before the pair structure limit. The Bethe equation
reads as 
\begin{equation}
\Phi_{1}^{+,(N/2)}=\alpha_{N/2}(t_{1}^{+,N/2})\frac{f(t_{1}^{-,N/2},t_{1}^{+,N/2})}{f(t_{1}^{+,N/2},t_{1}^{-,N/2})}.
\end{equation}
Substituting back to the definition of the function $\mathbf{F}^{(1)}$
we obtain that
\begin{equation}
\mathbf{F}^{(1)}(X_{1}^{+,N/2},t_{1}^{+,N/2})=-c\left(\frac{\partial}{\partial t_{1}^{+,N/2}}+\frac{\partial}{\partial t_{1}^{-,N/2}}\right)\log\Phi_{1}^{+,(N/2)}.
\end{equation}
Taking the derivatives the function $\mathbf{F}^{(1)}$ is simplified
as
\begin{equation}
\mathbf{F}^{(1)}(X_{1}^{+,N/2},t_{1}^{+,N/2})=X_{1}^{+,N/2},
\end{equation}
where we used the following identity
\begin{equation}
\left(\frac{\partial}{\partial u}+\frac{\partial}{\partial v}\right)\log f(u,v)\Biggr|_{u=-v}=0.\label{eq:idDlogf}
\end{equation}
For $\nu=\frac{N-1}{2}$ the Bethe equation reads as 
\begin{equation}
\Phi_{1}^{+,(\frac{N-1}{2})}=\alpha_{\frac{N-1}{2}}(t_{1}^{+,\frac{N-1}{2}})\frac{1}{f(t_{1}^{-,\frac{N-1}{2}},t_{1}^{+,\frac{N-1}{2}})}.
\end{equation}
Substituting to the definition of the function $\mathbf{F}^{(1)}$
we obtain that
\begin{equation}
\mathbf{F}^{(1)}(X_{1}^{+,\frac{N-1}{2}},t_{1}^{+,\frac{N-1}{2}})=X_{1}^{+,\frac{N-1}{2}},
\end{equation}
where we used (\ref{eq:idDlogf}).

For the twisted case where we have two Bethe roots $t_{1}^{\nu}$
and $t_{2}^{\nu}$ the Bethe equation reads as 
\begin{equation}
\Phi_{1}^{+,(\nu)}=\alpha_{\nu}(t_{1}^{+,\nu})\frac{f(t_{1}^{-,\nu},t_{1}^{+,\nu})}{f(t_{1}^{+,\nu},t_{1}^{-,\nu})}.
\end{equation}
Substituting back to the definition of the function $\mathbf{F}^{(1)}$
we obtain that
\begin{equation}
\mathbf{F}^{(1)}(X_{1}^{+,\nu},t_{1}^{+,\nu})=-c\left(\frac{\partial}{\partial t_{1}^{+,\nu}}+\frac{\partial}{\partial t_{1}^{-,\nu}}\right)\log\Phi_{1}^{+,(\nu)}.
\end{equation}
Taking the derivatives, the function $\mathbf{F}^{(1)}$ is simplified
as
\begin{equation}
\mathbf{F}^{(1)}(X_{1}^{+,\nu},t_{1}^{+,\nu})=X_{1}^{+,\nu},
\end{equation}
where we used the following identity
\begin{equation}
\left(\frac{\partial}{\partial u}+\frac{\partial}{\partial v}\right)\log f(u,v)\Biggr|_{u=-v-\nu c}=0.
\end{equation}
We just finished the proof of the property \ref{enum:prop3}.

For the proof of the property \ref{enum:prop5} let us calculate the
product of Bethe equations for the achiral pair structure
\begin{equation}
\prod_{\mu,j}\Phi_{j}^{+,(\mu)}=\frac{f(\bar{t}^{-,\frac{N}{2}},\bar{t}^{+,\frac{N}{2}})}{f(\bar{t}^{+,\frac{N}{2}},\bar{t}^{-,\frac{N}{2}})}\frac{1}{f(\bar{t}^{-,\frac{N}{2}},\bar{t}^{+,\frac{N}{2}-1})f(\bar{t}^{-,\frac{N}{2}-1},\bar{t}^{+,\frac{N}{2}})}\prod_{\mu}\alpha_{\mu}(\bar{t}^{+,\mu})
\end{equation}
for even $N$ and
\begin{equation}
\prod_{\mu,j}\Phi_{j}^{+,(\mu)}=\frac{1}{f(\bar{t}^{-,\frac{N-1}{2}},\bar{t}^{+,\frac{N-1}{2}})}\prod_{\mu}\alpha_{\mu}(\bar{t}^{+,\mu})
\end{equation}
for odd $N$. Using the identities (\ref{eq:idDlogf}) and
\begin{equation}
\left(\frac{\partial}{\partial u}+\frac{\partial}{\partial v}\right)\log\left(f(u,w)f(-w,v)\right)\Biggr|_{u=-v}=0,
\end{equation}
we can see that sum of the rows of the Gaudin matrix is
\begin{equation}
\sum_{\mu}\sum_{j}G_{j,k}^{+,(\mu,\nu)}=X_{k}^{+,\nu},
\end{equation}
therefore if all $X_{j}^{+,\mu}=0$ then the rows of the Gaudin matrix
are not linearly independent, and thus, $\det G^{+}=0$. In an analogous
way one can show that the property \ref{enum:prop5} is also satisfied
for the chiral pair structure.

Above we just showed that the functions $\mathbf{F}^{(\mathbf{r}^{+})}(\bar{X}^{+},\bar{t}^{+})$
satisfy the Korepin criteria. The reverse statement can be also proven
easily and this proof is the same as the proof of Proposition 4.1.
in \cite{Hutsalyuk:2017way}.

\subsubsection{Definitions of the modified $X$-s\label{subsec:modX}}

In the equation (\ref{eq:derivFX}) the modification is trivial for
$|\nu-\mu|>1$ i.e. $X_{k}^{+,\nu,mod}=X_{k}^{+,\nu}$ for $|\nu-\mu|>1$.
For the other cases we define the modified $X_{j}^{+,\nu}$-s, separately.

For the twisted case the modified $X_{j}^{+,\nu}$-s in the equation
(\ref{eq:derivFX}) are defined as
\begin{equation}
\begin{split}X_{k}^{+,\mu,mod} & =X_{k}^{+,\mu}-c\frac{d}{du}\log\frac{f(t_{j}^{+,\mu},u)}{f(u,t_{j}^{+,\mu})}\frac{f(-t_{j}^{+,\mu}-\mu c,u)}{f(u,-t_{j}^{+,\mu}-\mu c)}\Biggr|_{u=t_{k}^{+,\mu}},\\
X_{k}^{+,\mu+1,mod} & =X_{k}^{+,\mu+1}-c\frac{d}{du}\log f(u,t_{j}^{+,\mu})f(u,-t_{j}^{+,\mu}-\mu c)\Biggr|_{u=t_{k}^{+,\mu+1}},\\
X_{k}^{+,\mu-1,mod} & =X_{k}^{+,\mu-1}+c\frac{d}{du}\log f(t_{j}^{+,\mu},u)f(-t_{j}^{+,\mu}-\mu c,u)\Biggr|_{u=t_{k}^{+,\mu-1}}.
\end{split}
\end{equation}
 For the untwisted case where $N$ is even the modified $X_{j}^{+,\mu}$-s
in the equation (\ref{eq:derivFX}) are defined as
\begin{equation}
\begin{split}X_{k}^{+,\mu,mod} & =X_{k}^{+,\mu}-c\frac{d}{du}\log\frac{f(t_{j}^{+,\mu},u)}{f(u,t_{j}^{+,\mu})}\Biggr|_{u=t_{k}^{+,\mu}},\\
X_{k}^{+,\mu+1,mod} & =X_{k}^{+,\mu+1}-c\frac{d}{du}\log f(u,t_{j}^{+,\mu})\Biggr|_{u=t_{k}^{+,\mu+1}},\\
X_{k}^{+,\mu-1,mod} & =X_{k}^{+,\mu-1}+c\frac{d}{du}\log f(t_{j}^{+,\mu},u)\Biggr|_{u=t_{k}^{+,\mu-1}},
\end{split}
\end{equation}
for $\mu=1,\dots,\frac{N}{2}-2$ and
\begin{equation}
\begin{split}X_{k}^{+,\frac{N}{2}-1,mod} & =X_{k}^{+,\frac{N}{2}-1}-c\frac{d}{du}\log\frac{f(t_{j}^{+,\frac{N}{2}-1},u)}{f(u,t_{j}^{+,\frac{N}{2}-1})}\Biggr|_{u=t_{k}^{+,\frac{N}{2}-1}},\\
X_{k}^{+,\frac{N}{2},mod} & =X_{k}^{+,\frac{N}{2}}-c\frac{d}{du}\log\frac{f(u,t_{j}^{+,\frac{N}{2}-1})}{f(-t_{j}^{+,\frac{N}{2}-1},u)}\Biggr|_{u=t_{k}^{+,\frac{N}{2}}},\\
X_{k}^{+,\frac{N}{2}-2,mod} & =X_{k}^{+\frac{N}{2}-2}+c\frac{d}{du}\log f(t_{j}^{+,\frac{N}{2}-1},u)\Biggr|_{u=t_{k}^{+,\frac{N}{2}-2}},
\end{split}
\end{equation}
for $\mu=\frac{N}{2}-1$ and
\begin{equation}
\begin{split}X_{k}^{+,\frac{N}{2},mod} & =X_{k}^{+,\frac{N}{2}}-c\frac{d}{du}\log\frac{f(t_{j}^{+,\frac{N}{2}},u)}{f(u,t_{j}^{+,\frac{N}{2}})}\frac{f(-t_{j}^{+,\frac{N}{2}},u)}{f(u,-t_{j}^{+,\frac{N}{2}})}\Biggr|_{u=t_{k}^{+,\frac{N}{2}}},\\
X_{k}^{+,\frac{N}{2}-1,mod} & =X_{k}^{+\frac{N}{2}-1}+c\frac{d}{du}\log f(t_{j}^{+,\frac{N}{2}},u)\log f(-t_{j}^{+,\frac{N}{2}},u)\Biggr|_{u=t_{k}^{+,\frac{N}{2}-1}},
\end{split}
\end{equation}
for $\mu=\frac{N}{2}$.

For the odd $N$ case the modified $X_{j}^{+,\mu}$-s in the equation
(\ref{eq:derivFX}) are defined as
\begin{equation}
\begin{split}X_{k}^{+,\mu,mod} & =X_{k}^{+,\mu}-c\frac{d}{du}\log\frac{f(t_{j}^{+,\mu},u)}{f(u,t_{j}^{+,\mu})}\Biggr|_{u=t_{k}^{+,\mu}},\\
X_{k}^{+,\mu+1,mod} & =X_{k}^{+,\mu+1}-c\frac{d}{du}\log f(u,t_{j}^{+,\mu})\Biggr|_{u=t_{k}^{+,\mu+1}},\\
X_{k}^{+,\mu-1,mod} & =X_{k}^{+,\mu-1}+c\frac{d}{du}\log f(t_{j}^{+,\mu},u)\Biggr|_{u=t_{k}^{+,\mu-1}},
\end{split}
\end{equation}
for $\mu=1,\dots,\frac{N-3}{2}$ and
\begin{equation}
\begin{split}X_{k}^{+,\frac{N-1}{2},mod} & =X_{k}^{+,\frac{N-1}{2}}-c\frac{d}{du}\log\frac{f(t_{j}^{+,\frac{N-1}{2}},u)}{f(u,t_{j}^{+,\frac{N-1}{2}})f(-t_{j}^{+,\frac{N-1}{2}},u)}\Biggr|_{u=t_{k}^{+,\frac{N-1}{2}}},\\
X_{k}^{+,\frac{N-3}{2},mod} & =X_{k}^{+,\frac{N-3}{2}}+c\frac{d}{du}\log f(t_{j}^{+,\frac{N-1}{2}},u)\Biggr|_{u=t_{k}^{+,\frac{N-3}{2}}},
\end{split}
\end{equation}
for $\mu=\frac{N-1}{2}$.

\subsubsection{On-shell formulas}

It turns out that the results of the previous subsection are enough
to derive the closed form of the on-shell overlaps. In appendix \ref{sec:On-shell-limit}
we show that a re-normalized version of the overlap in the on-shell
limit satisfies the Korepin criteria therefore it is equal to the
Gaudin-like determinant.

The on-shell formula can be written as
\begin{equation}
S_{K}^{N}(\bar{t})=\prod_{\nu=1}^{\frac{N-1}{2}}\left[F_{K}^{(\nu)}\right]^{r_{\nu}/2}\times\prod_{\nu=1}^{N-1}\prod_{k\neq l}f(t_{l}^{+,\nu},t_{k}^{+,\nu})\prod_{k<l}f(t_{l}^{+,\nu},-t_{k}^{+,\nu}-\nu c)f(-t_{k}^{+,\nu}-\nu c,t_{l}^{+,\nu})\det G^{+}\label{eq:onshell1}
\end{equation}
for the twisted and

\begin{equation}
S_{K}^{N}(\bar{t})=\prod_{\nu=1}^{\frac{N-1}{2}}\tilde{F}_{K}^{(\nu)}(\bar{t}^{+,\nu})\times\frac{\prod_{\nu=1}^{\frac{N-1}{2}}\prod_{k\neq l}f(t_{l}^{+,\nu},t_{k}^{+,\nu})}{\prod_{\nu=1}^{\frac{N-3}{2}}f(\bar{t}^{+,\nu+1},\bar{t}^{+,\nu})\prod_{k<l}f(-t_{l}^{+,\frac{N-1}{2}},t_{k}^{+,\frac{N-1}{2}})}\det G^{+}\label{eq:onshell2}
\end{equation}
for the untwisted two-site states where $N$ is odd and
\begin{equation}
S_{K}^{N}(\bar{t})=\prod_{\nu=1}^{\frac{N}{2}}F_{K}^{(\nu)}(\bar{t}^{+,\nu})\times\frac{\prod_{\nu=1}^{\frac{N}{2}}\prod_{k\neq l}f(t_{l}^{+,\nu},t_{k}^{+,\nu})\prod_{k<l}f(-t_{l}^{+,\frac{N}{2}},t_{k}^{+,\frac{N}{2}})f(t_{l}^{+,\frac{N}{2}},-t_{k}^{+,\frac{N}{2}})}{\prod_{\nu=1}^{\frac{N}{2}-1}f(\bar{t}^{+,\nu+1},\bar{t}^{+,\nu})f(-\bar{t}^{+,\frac{N}{2}},\bar{t}^{+,\frac{N}{2}-1})}\det G^{+}\label{eq:onshell3}
\end{equation}
where $N$ is even.

Our final goal is to calculate normalized overlap formula. Let us
see the explicit form of the scalar product $\mathbb{C}(\bar{t})\mathbb{B}(\bar{t})$
(\ref{sec:On-shell-limit}) in the pair structure limit. The Gaudin
matrix has the following block form w.r.t. the decomposition $\bar{t}=\bar{t}^{+}\cup\bar{t}^{-}$

\begin{equation}
G=\left(\begin{array}{cc}
A^{++} & A^{+-}\\
A^{-+} & A^{++}
\end{array}\right),
\end{equation}
where we introduced the following matrices
\begin{equation}
\begin{split}A_{j,k}^{++,(\mu,\nu)} & =-c\frac{\partial}{\partial t_{k}^{+,\nu}}\log\Phi_{j}^{+,(\mu)}\Biggr|_{\bar{t}^{-}=\hat{\pi}\left(\bar{t}^{+}\right)},\qquad A_{j,k}^{+-,(\mu,\nu)}=-c\frac{\partial}{\partial t_{k}^{-,\nu}}\log\Phi_{j}^{+,(\mu)}\Biggr|_{\bar{t}^{-}=\hat{\pi}\left(\bar{t}^{+}\right)},\\
A_{j,k}^{-+,(\mu,\nu)} & =-c\frac{\partial}{\partial t_{k}^{+,\nu}}\log\Phi_{j}^{-,(\mu)}\Biggr|_{\bar{t}^{-}=\hat{\pi}\left(\bar{t}^{+}\right)},\qquad A_{j,k}^{--,(\mu,\nu)}=-c\frac{\partial}{\partial t_{k}^{-,\nu}}\log\Phi_{j}^{-,(\mu)}\Biggr|_{\bar{t}^{-}=\hat{\pi}\left(\bar{t}^{+}\right)}.
\end{split}
\end{equation}
Applying the identities
\begin{equation}
\begin{split}\frac{\partial}{\partial t_{k}^{-,\nu}}\log\Phi_{j}^{-,(\mu)}\Biggr|_{\bar{t}^{-}=\hat{\pi}\left(\bar{t}^{+}\right)}=\frac{\partial}{\partial t_{k}^{+,\nu}}\log\Phi_{j}^{+,(\mu)}\Biggr|_{\bar{t}^{-}=\hat{\pi}\left(\bar{t}^{+}\right)},\\
\frac{\partial}{\partial t_{k}^{+,\nu}}\log\Phi_{j}^{-,(\mu)}\Biggr|_{\bar{t}^{-}=\hat{\pi}\left(\bar{t}^{+}\right)}=\frac{\partial}{\partial t_{k}^{-,\nu}}\log\Phi_{j}^{+,(\mu)}\Biggr|_{\bar{t}^{-}=\hat{\pi}\left(\bar{t}^{+}\right)},
\end{split}
\end{equation}
the Gaudin determinant simplifies as
\begin{equation}
\det G=\left|\begin{array}{cc}
A^{++} & A^{+-}\\
A^{-+} & A^{--}
\end{array}\right|=\left|\begin{array}{cc}
A^{++} & A^{+-}\\
A^{+-} & A^{++}
\end{array}\right|=\left|\begin{array}{cc}
A^{++}+A^{+-} & A^{+-}\\
0 & A^{++}-A^{+-}
\end{array}\right|=\det G^{+}\det G^{-},\label{eq:factorDetG}
\end{equation}
where
\begin{equation}
G_{j,k}^{\pm,(\mu,\nu)}=-c\left(\frac{\partial}{\partial t_{k}^{+,\nu}}\pm\frac{\partial}{\partial t_{k}^{-,\nu}}\right)\log\Phi_{j}^{+,(\mu)}\Biggr|_{\bar{t}^{-}=\hat{\pi}\left(\bar{t}^{+}\right)}.
\end{equation}
We can see that the determinant $\det G^{+}$ in (\ref{eq:factorDetG})
agrees with the previously defined Gaudin-like determinant (\ref{eq:Gdef})
which appears in the on-shell overlap formulas (\ref{eq:onshell1}-\ref{eq:onshell3}).
Applying this factorization property of the Gaudin determinant we
obtain that the scalar product for the achiral pair structures reads
as
\begin{equation}
\mathbb{C}(\bar{t})\mathbb{B}(\bar{t})=\frac{\left[\prod_{\nu=1}^{\frac{N-1}{2}}\prod_{k\neq l}f(t_{l}^{+,\nu},t_{k}^{+,\nu})\right]^{2}}{\left[\prod_{\nu=1}^{\frac{N-3}{2}}f(\bar{t}^{+,\nu+1},\bar{t}^{+,\nu})\right]^{2}f(-\bar{t}^{+,\frac{N-1}{2}},\bar{t}^{+,\frac{N-1}{2}})}\det G^{+}\det G^{-},\label{eq:scalar1}
\end{equation}
where $N$ is odd and
\begin{equation}
\mathbb{C}(\bar{t})\mathbb{B}(\bar{t})=\frac{\left[\prod_{\nu=1}^{\frac{N}{2}}\prod_{k\neq l}f(t_{l}^{+,\nu},t_{k}^{+,\nu})\right]^{2}\left[\prod_{k,l}f(-t_{l}^{+,\frac{N}{2}},t_{k}^{+,\frac{N}{2}})f(t_{l}^{+,\frac{N}{2}},-t_{k}^{+,\frac{N}{2}})\right]}{\left[\prod_{\nu=1}^{\frac{N}{2}-1}f(\bar{t}^{+,\nu+1},\bar{t}^{+,\nu})f(-\bar{t}^{+,\frac{N}{2}},\bar{t}^{+,\frac{N}{2}-1})\right]^{2}}\det G^{+}\det G^{-},\label{eq:scalar2}
\end{equation}
where $N$ is even. For the chiral pair structure we obtain that
\begin{equation}
\mathbb{C}(\bar{t})\mathbb{B}(\bar{t})=\prod_{\nu=1}^{N-1}\prod_{k\neq l}f(t_{l}^{+,\nu},t_{k}^{+,\nu})^{2}\prod_{k,l}f(t_{l}^{+,\nu},-t_{k}^{+,\nu}-\nu c)f(-t_{l}^{+,\nu}-\nu c,t_{k}^{+,\nu})\det G^{+}\det G^{-}.\label{eq:scalar3}
\end{equation}
We can see that the denominator of the original scalar product (\ref{eq:norm})
is fully eliminated for the chiral pair structure since
\begin{multline}
f(\bar{t}^{\nu+1},\bar{t}^{\nu})=f(\bar{t}^{+,\nu+1},\bar{t}^{+,\nu})f(\bar{t}^{+,\nu+1},\bar{t}^{-,\nu})f(\bar{t}^{-,\nu+1},\bar{t}^{+,\nu})f(\bar{t}^{-,\nu+1},\bar{t}^{-,\nu})=\\
f(\bar{t}^{+,\nu+1},\bar{t}^{+,\nu})f(\bar{t}^{+,\nu+1},-\bar{t}^{+,\nu}-\nu c)f(-\bar{t}^{+,\nu+1}-(\nu+1)c,\bar{t}^{+,\nu})f(-\bar{t}^{+,\nu+1}-(\nu+1)c,-\bar{t}^{+,\nu}-\nu c)=\\
f(\bar{t}^{+,\nu+1},\bar{t}^{+,\nu})f(\bar{t}^{+,\nu+1},-\bar{t}^{+,\nu}-\nu c)f(-\bar{t}^{+,\nu}-(\nu+1)c,\bar{t}^{+,\nu+1})f(\bar{t}^{+,\nu},\bar{t}^{+,\nu+1}+c)=1
\end{multline}
where we used the identities $f(u,v)=f(-v,-u)$, $f(u+x,v)=f(u,v-x)$
and $f(u,v+c)f(v,u)=1$.

Combining the on-shell overlaps (\ref{eq:onshell1}-\ref{eq:onshell3})
with the scalar products (\ref{eq:scalar1}-\ref{eq:scalar3}) our
final result i.e. the normalized on-shell overlaps can be written
as
\begin{equation}
\boxed{\frac{\mathbb{C}(\bar{t})|\Psi\rangle\langle\Psi|\mathbb{B}(\bar{t})}{\mathbb{C}(\bar{t})\mathbb{B}(\bar{t})}=\prod_{\nu=1}^{n^{+}}\mathcal{F}_{K}^{(\nu)}(\bar{t}^{+,\nu})\frac{\det G^{+}}{\det G^{-}},}\label{eq:normOv}
\end{equation}
where $n^{+}=N-1,\left\lfloor \frac{N}{2}\right\rfloor $ for the
twisted and untwisted two-site states. The one-particle overlaps reads
as
\begin{equation}
\boxed{\mathcal{F}_{K}^{(\nu)}(u)=\frac{\left(F_{K}^{(\nu)}\right)^{2}}{f(u,-u-\nu c)f(-u-\nu c,u)}}\label{eq:FK1}
\end{equation}
for the $\mathfrak{so}(N)$ symmetric and
\begin{equation}
\boxed{\begin{split}\mathcal{F}_{K}^{(\nu)}(u) & =\left(F_{K}^{(\nu)}(u)\right)^{2},\qquad\nu<\left\lfloor \frac{N}{2}\right\rfloor \\
\mathcal{F}_{K}^{(\frac{N-1}{2})}(u) & =\frac{\left(F_{K}^{(\frac{N-1}{2})}(u)\right)^{2}}{f(-u,u)},\\
F_{K}^{(\frac{N}{2})}(u) & =\frac{\left(F_{K}^{(\frac{N}{2})}(u)\right)^{2}}{f(-u,u)f(u,-u)}
\end{split}
}\label{eq:FK2}
\end{equation}
for the $\mathfrak{gl}(\left\lfloor \frac{N}{2}\right\rfloor )\oplus\mathfrak{gl}(\left\lceil \frac{N}{2}\right\rceil )$
symmetric final states.

\subsubsection{Odd quantum numbers}

We close this section with the overlaps where the subset $\bar{t}^{0}$
is not empty. Let us start with the untwisted case when $N$ is even\footnote{For odd $N$ there is no set $\bar{t}^{0}$ for the untwisted case.}.
In this case the subsets are defined as
\begin{equation}
\begin{split}t_{k}^{+,\nu} & =t_{k}^{\nu},\quad t_{k}^{-,\nu}=t_{k}^{N-\nu},\quad\nu<N/2,\\
t_{k}^{+,N/2} & =t_{k}^{N/2},\quad t_{k}^{-,N/2}=t_{k+\frac{r_{N/2}-1}{2}}^{N/2},\quad k=1,\dots,\frac{r_{N/2}-1}{2}\\
\bar{t}^{0} & =\{t_{r_{N/2}}^{N/2}\}.
\end{split}
,
\end{equation}
The pair structure limit is $\bar{t}^{-}\to-\bar{t}^{+}$and $\bar{t}^{0}\to\{0\}$.

We can show that the Gaudin determinant is also factorized for this
pair structure. The Gaudin matrix has the following block structure
\begin{equation}
G=\left(\begin{array}{ccc}
A^{++} & A^{+0} & A^{+-}\\
A^{0+} & A^{00} & A^{0-}\\
A^{-+} & A^{-0} & A^{--}
\end{array}\right),
\end{equation}
where we defined the following matrices
\begin{equation}
\begin{split}A_{j}^{+0,(\mu)} & =-c\frac{\partial}{\partial t_{r_{N/2}}^{N/2}}\log\Phi_{j}^{+,(\mu)},\qquad A_{j}^{-0,(\mu)}=-c\frac{\partial}{\partial t_{r_{N/2}}^{N/2}}\log\Phi_{j}^{-,(\mu)},\\
A_{k}^{0+,(\nu)} & =-c\frac{\partial}{\partial t_{k}^{+,\nu}}\log\Phi_{r_{N/2}}^{(N/2)},\qquad A_{k}^{0-,(\nu)}=-c\frac{\partial}{\partial t_{k}^{-,\nu}}\log\Phi_{r_{N/2}}^{(N/2)},\\
A^{00} & =-c\frac{\partial}{\partial t_{r_{N/2}}^{N/2}}\log\Phi_{r_{N/2}}^{(N/2)}.
\end{split}
\label{eq:A0}
\end{equation}
Applying the identities
\begin{equation}
\frac{\partial}{\partial t_{r_{N/2}}^{N/2}}\log\Phi_{j}^{-,(\mu)}=\frac{\partial}{\partial t_{r_{N/2}}^{N/2}}\log\Phi_{j}^{+,(\mu)},\qquad\frac{\partial}{\partial t_{k}^{-,\nu}}\log\Phi_{r_{N/2}}^{(N/2)}=\frac{\partial}{\partial t_{k}^{+,\nu}}\log\Phi_{r_{N/2}}^{(N/2)},
\end{equation}
the Gaudin determinant simplifies as
\begin{equation}
\det G=\left|\begin{array}{ccc}
A^{++} & A^{+0} & A^{+-}\\
A^{0+} & A^{00} & A^{0+}\\
A^{+-} & A^{+0} & A^{++}
\end{array}\right|=\left|\begin{array}{ccc}
A^{++}+A^{+-} & A^{+0} & A^{+-}\\
2A^{0+} & A^{00} & A^{0+}\\
0 & 0 & A^{++}-A^{+-}
\end{array}\right|=\det G^{+}\det G^{-},
\end{equation}
where
\begin{equation}
\det G^{+}=\left|\begin{array}{cc}
A^{++}+A^{+-} & A^{+0}\\
2A^{0+} & A^{00}
\end{array}\right|,\quad\det G^{-}=\left|A^{++}-A^{+-}\right|.\label{eq:degGodd}
\end{equation}
In the equations (\ref{eq:A0}-\ref{eq:degGodd}) we assumed that
the achiral pair structure is satisfied. The scalar product for this
specific pair structure reads as
\begin{multline}
\mathbb{C}(\bar{t})\mathbb{B}(\bar{t})=\\
\frac{\left[\prod_{\nu=1}^{\frac{N}{2}}\prod_{k\neq l}f(t_{l}^{+,\nu},t_{k}^{+,\nu})\right]^{2}\left[\prod_{k,l}f(-t_{l}^{+,\frac{N}{2}},t_{k}^{+,\frac{N}{2}})f(t_{l}^{+,\frac{N}{2}},-t_{k}^{+,\frac{N}{2}})\right]\left[f(\bar{t}^{+,\frac{N}{2}},0)f(0,\bar{t}^{+,\frac{N}{2}})\right]^{2}}{\left[\prod_{\nu=1}^{\frac{N}{2}-1}f(\bar{t}^{+,\nu+1},\bar{t}^{+,\nu})f(-\bar{t}^{+,\frac{N}{2}},\bar{t}^{+,\frac{N}{2}-1})\right]^{2}\left[f(0,\bar{t}^{+,\frac{N}{2}-1})\right]^{2}}\times\\
\det G^{+}\det G^{-}.
\end{multline}

We can generalize the previous calculations of the on-shell overlaps
for this pair structure and we just show the final results. It turns
out that the on-shell overlap is proportional to the Gaudin-like determinant
$\det G^{+}$ (\ref{eq:degGodd}):
\begin{multline}
S_{K}^{N}(\bar{t})=F_{K}^{0}\prod_{\nu=1}^{\frac{N}{2}}F_{K}^{(\nu)}(\bar{t}^{+,\nu})\times\\
\frac{\prod_{\nu=1}^{\frac{N}{2}}\prod_{k\neq l}f(t_{l}^{+,\nu},t_{k}^{+,\nu})\prod_{k<l}f(-t_{l}^{+,\frac{N}{2}},t_{k}^{+,\frac{N}{2}})f(t_{l}^{+,\frac{N}{2}},-t_{k}^{+,\frac{N}{2}})f(\bar{t}^{+,\frac{N}{2}},0)f(0,\bar{t}^{+,\frac{N}{2}})}{\prod_{\nu=1}^{\frac{N}{2}-1}f(\bar{t}^{+,\nu+1},\bar{t}^{+,\nu})f(-\bar{t}^{+,\frac{N}{2}},\bar{t}^{+,\frac{N}{2}-1})f(0,\bar{t}^{+,\frac{N}{2}-1})}\det G^{+},
\end{multline}
where
\begin{equation}
F_{K}^{0}=\frac{\mathfrak{a}}{c\mathcal{U}_{\frac{N}{2}+1,\frac{N}{2}}^{(\frac{N}{2})}}.
\end{equation}
This extra multiplicative factor comes form the extra Bethe root $t_{r_{N/2}}^{N/2}$
and it is equal to the residue of the normalized $K$-matrix in the
$N/2$-th nesting step
\begin{equation}
F_{K}^{0}=\underset{u=0}{\mathrm{Res}}\frac{K_{\frac{N}{2}+1,\frac{N}{2}+1}^{(\frac{N}{2})}(u)}{K_{\frac{N}{2}+1,\frac{N}{2}}^{(\frac{N}{2})}(u)}.
\end{equation}
We can see that the overlap is non-vanishing only when this residue
is non-vanishing.

Substituting back, we obtain that the normalized on-shell overlap
can be written as
\begin{equation}
\boxed{\frac{\mathbb{C}(\bar{t})|\Psi\rangle\langle\Psi|\mathbb{B}(\bar{t})}{\mathbb{C}(\bar{t})\mathbb{B}(\bar{t})}=\left(F_{K}^{0}\right)^{2}\prod_{\nu=1}^{n^{+}}\mathcal{F}_{K}^{(\nu)}(\bar{t}^{+,\nu})\frac{\det G^{+}}{\det G^{-}}.}
\end{equation}

We close this section by some comments on the $\mathfrak{so}(N)$
symmetric case when $\bar{t}^{0}\neq\emptyset$. In appendix \ref{sec:Elementary}
we show that the on-shell overlaps always vanish for $\mathbf{r}=1$.
This is a consequence of the fact that the $\mathfrak{so}(N)$ symmetric
$K$-matrices are spectral parameter independent (see equation (\ref{eq:twK}))
and no poles appear in the HC-s when we take the $t_{1}^{\nu}\to-\nu c/2$
limit. This observation suggests that the on-shell overlaps are vanishing
for all $\bar{t}^{0}\neq\emptyset$ configurations. For the particular
examples which were investigated previously \cite{Piroli:2018ksf,Piroli:2018don,deLeeuw:2019ebw,Kristjansen:2020mhn}
shows that this conjecture is true but the proof for the general cases
is postponed to a later work.

\section{Representations of the monodromy matrices and integrable states\label{sec:Representations-of-the}}

In this section we apply our general formula (\ref{eq:normOv}) for
concrete spin chains and final states. At first we build the monodromy
matrix from the Lax-operators.

Let us consider the following the Lax operator
\begin{equation}
\mathcal{L}^{\Lambda}(u)=u\mathbf{1}+c\sum_{i,j=1}^{N}E_{i,j}\otimes E_{j,i}^{\Lambda}\in\mathbb{C}^{N}\otimes\mathrm{End}\mathcal{V}_{\Lambda},
\end{equation}
where $\mathcal{V}_{\Lambda}$ is a finite dimensional representation
of the Lie-algebra $\mathfrak{gl}(N)$ belongs to the $N$-tuple $\Lambda=(\Lambda_{1},\dots,\Lambda_{N})$
where $\Lambda_{i}-\Lambda_{j}\in\mathbb{N}$ for $i<j$ and $E_{i,j}^{\Lambda}\in\mathrm{End}\mathcal{V}_{\Lambda}$
are the corresponding $\mathfrak{gl}(N)$ generators for which
\begin{equation}
\begin{split}E_{i,i}^{\Lambda}|0\rangle^{\Lambda} & =\Lambda_{i}|0\rangle^{\Lambda},\quad i=1,\dots,N,\\
E_{j,i}^{\Lambda}|0\rangle^{\Lambda} & =0,\quad1\leq i<j\leq N,
\end{split}
\end{equation}
where $|0\rangle^{\Lambda}\in\mathcal{V}_{\Lambda}$ is the highest
weight state of representation $\Lambda=(\Lambda_{1},\dots,\Lambda_{N})$.
The Lax operators satisfy the $RTT$-relation. Due to the co-product
property and the boost automorphism we can build monodromy matrices
from these Lax operators which satisfy the $RTT$-relation,
\begin{equation}
T_{a}(u)=\mathcal{L}_{a,L}^{\Lambda^{(L)}}(u-\xi_{L})\dots\mathcal{L}_{a,1}^{\Lambda^{(1)}}(u-\xi_{1}),
\end{equation}
where we introduced $L$ inhomogeneities $\xi_{i}\in\mathbb{C}$ and
representations $\Lambda^{(i)}$ for $i=1,\dots,L$. Now the monodromy
matrix entries act on the quantum space $\mathcal{H}=\mathcal{V}_{\Lambda^{(1)}}\otimes\dots\otimes\mathcal{V}_{\Lambda^{(L)}}$.
The pseudo-vacuum for this monodromy matrix is
\begin{equation}
|0\rangle=|0\rangle^{\Lambda^{(1)}}\otimes\dots\otimes|0\rangle^{\Lambda^{(L)}}\in\mathcal{H},
\end{equation}
and the vacuum eigenvalues are
\begin{equation}
\lambda_{i}(u)=\prod_{j=1}^{L}(u-\xi_{j}+c\Lambda_{i}^{(j)}).
\end{equation}

\subsection{Explicit forms of untwisted two-site states}

In section \ref{sec:Integrable-final-states} we saw that to obtain
integrable two-site states for the untwisted $KT$-relation we have
to restrict our parameter space for which the vacuum eigenvalues satisfy
the symmetry property (\ref{eq:constLamUTw})
\begin{equation}
\lambda_{i}(u)=\lambda_{N+1-i}(-u).
\end{equation}
This equation is satisfied if 
\begin{equation}
\xi_{2j-1}=-\xi_{2j},\qquad\Lambda_{j}^{(2j-1)}=-\Lambda_{N+1-j}^{(2j)}.
\end{equation}
Let us define the monodromy matrix as
\[
T_{a}(u)=\mathcal{L}_{a,2L}^{\bar{\Lambda}^{(L)}}(u+\theta_{L})\mathcal{L}_{a,2L-1}^{\Lambda^{(L)}}(u-\theta_{L})\dots\mathcal{L}_{a,2}^{\bar{\Lambda}^{(1)}}(u+\theta_{1})\mathcal{L}_{a,1}^{\Lambda^{(1)}}(u-\theta_{1}),
\]
where we used the notation $\bar{\Lambda}=(-\Lambda_{N},\dots,-\Lambda_{1})$.
Clearly $\dim\mathcal{V}_{\bar{\Lambda}}=\dim\mathcal{V}_{\Lambda}$.
The generators of the representation $\bar{\Lambda}$ is chosen as
\begin{equation}
E_{i,j}^{\bar{\Lambda}}=-E_{N+1-j,N+1-i}^{\Lambda}.
\end{equation}
Since the finite dimensional representations are unique there exist
a matrix $V^{\Lambda}\in\mathrm{End}\mathcal{V}_{\Lambda}$ which
defines the following similarity transformation
\begin{equation}
E_{i,j}^{\Lambda}=V^{\Lambda}\left(E_{N+1-j,N+1-i}^{\Lambda}\right)^{t}V^{\Lambda}.
\end{equation}
The following connection between the Lax-operators will be useful
\begin{equation}
\mathcal{L}_{a,j}^{\bar{\Lambda}}(u)=-V_{a}\mathcal{L}_{a,j}^{\Lambda}(-u)^{t_{a}}V_{a}=-V_{j}^{\Lambda}\mathcal{L}_{a,j}^{\Lambda}(-u)^{t_{j}}V_{j}^{\Lambda}=u\mathbf{1}-c\sum_{i,j=1}^{N}E_{i,j}\otimes E_{N+1-i,N+1-j}^{\Lambda}.\label{eq:cross}
\end{equation}
 The vacuum eigenvalues are
\begin{equation}
\lambda_{i}(u)=\prod_{j=1}^{L}(u-\theta_{j}+c\Lambda_{i}^{(j)})(u+\theta_{j}-c\Lambda_{N+1-i}^{(j)}),
\end{equation}
which clearly satisfy the untwisted constraint $\lambda_{i}(u)=\lambda_{N+1-i}(-u)$.

Let us define a reflection matrix $K^{\Lambda}(u)\in\mathrm{End}\mathcal{V}_{\Lambda}$
which satisfies the reflection equation 
\begin{equation}
\mathcal{L}_{2,1}^{\Lambda}(u-v)K_{1}^{\Lambda}(-u)\mathcal{L}_{2,1}^{\Lambda}(u+v)K_{2}(-v)=K_{2}(-v)\mathcal{L}_{2,1}^{\Lambda}(u+v)K_{1}^{\Lambda}(-u)\mathcal{L}_{2,1}^{\Lambda}(u-v).
\end{equation}
We can also use the matrix elements $K(u)=\sum_{i,k}K_{i,k}(u)E_{i,k}$,
$K^{\Lambda}(u)=\sum_{i,k}K_{i,k}^{\Lambda}(u)E_{i,k}^{\Lambda}$
and $\mathcal{L}^{\Lambda}(u)=\sum_{i,j,k,l}\mathcal{L}^{\Lambda}(u)_{i,j}^{k,l}E_{i,k}\otimes E_{j,l}^{\Lambda}$
for which the last equation reads as
\begin{multline*}
\sum_{a_{1},a_{2},b_{1},b_{2}}\mathcal{L}^{\Lambda}(u-v)_{i_{1},j_{1}}^{a_{1},b_{1}}K_{b_{1},b_{2}}^{\Lambda}(-u)\mathcal{L}^{\Lambda}(u+v)_{a_{1},b_{2}}^{a_{2},j_{2}}K_{a_{2},i_{2}}(-v)=\\
\sum_{a_{1},a_{2},b_{1},b_{2}}K_{i_{1},a_{1}}(-v)\mathcal{L}^{\Lambda}(u+v)_{a_{1},j_{1}}^{a_{2},b_{1}}K_{b_{1},b_{2}}^{\Lambda}(-u)\mathcal{L}^{\Lambda}(u-v)_{a_{2},b_{2}}^{i_{2},j_{2}}.
\end{multline*}
An equivalent form of this equation is
\begin{equation}
\vec{K}_{4,3}(u)V_{4}\vec{K}_{2,1}^{\Lambda}(\theta)V_{2}^{\Lambda}\mathcal{L}_{3,2}^{\bar{\Lambda}}(u+\theta)\mathcal{L}_{3,1}^{\Lambda}(u-\theta)=\vec{K}_{4,3}(u)V_{4}\vec{K}_{2,1}^{\Lambda}(\theta)V_{2}^{\Lambda}\mathcal{L}_{4,1}^{\bar{\Lambda}}(u+\theta)\mathcal{L}_{4,2}^{\Lambda}(u-\theta),\label{eq:KKLL}
\end{equation}
where we used the connection between the Lax operators (\ref{eq:cross})
and introduced the following notations
\begin{equation}
\vec{K}=\sum_{i,j=1}^{N}K_{i,j}\vec{e}_{i}^{*}\otimes\vec{e}_{j}^{*},\qquad\vec{K}^{\Lambda}=\text{\ensuremath{\sum}}_{i,j=1}^{\dim\mathcal{V}_{\Lambda}}K_{i,j}^{\Lambda}\vec{e}_{i}^{\Lambda*}\otimes\vec{e}_{j}^{\Lambda*},
\end{equation}
where $\vec{e}_{i}^{*}$ and $\vec{e}_{i}^{\Lambda*}$ are the canonical
basis vectors in the dual space of $\mathbb{C}^{N}$ and $\mathcal{V}_{\Lambda}$,
respectively.

Let us define the two-site final state as
\begin{equation}
\langle\Psi|=\vec{K}_{2,1}^{\Lambda^{(1)}}(\theta_{1})V_{2}^{\Lambda}\dots\vec{K}_{2L,2L-1}^{\Lambda^{(L)}}(\theta_{L})V_{2L}^{\Lambda}.
\end{equation}
Using the relation (\ref{eq:KKLL}) we can show that the following
equation is satisfied
\begin{equation}
\vec{K}_{b,a}(u)\langle\Psi|T_{a}(u)=\vec{K}_{b,a}(u)V_{b}\langle\Psi|T_{b}^{\Pi}(u)V_{b},\label{eq:KvecT}
\end{equation}
where we introduced the space reflected monodromy matrix
\begin{equation}
T_{b}^{\Pi}(u)=\mathcal{L}_{b,1}^{\bar{\Lambda}^{(1)}}(u+\theta_{1})\mathcal{L}_{b,2}^{\Lambda^{(1)}}(u-\theta_{1})\dots\mathcal{L}_{b,2L-1}^{\bar{\Lambda}^{(L)}}(u+\theta_{L})\mathcal{L}_{a,2L}^{\Lambda^{(L)}}(u-\theta_{L}).
\end{equation}
It is easy to show that
\begin{equation}
T_{b}^{\Pi}(u)=V_{b}T_{b}^{t_{b}}(-u)V_{b}.
\end{equation}
Substituting back to (\ref{eq:KvecT}) we obtain the untwisted $KT$-relation
\begin{equation}
K_{0}(u)\langle\Psi|T_{0}(u)=\langle\Psi|T_{0}(-u)K_{0}(u).
\end{equation}

\subsection{Explicit forms of twisted two-site states}

For the twisted case we restrict ourselves to the representations
with rectangular Young diagram i.e. $(s,r):=\Lambda$ where $\Lambda_{i}=s$
for $i\leq r$ and $\Lambda_{i}=0$ for $i>r$. Let us define the
monodromy matrix as
\begin{equation}
T_{a}(u)=\mathcal{L}_{a,2L}^{(s_{L},r_{L})}(u+\theta_{L}-cs_{L}+cr_{L})\mathcal{L}_{a,2L-1}^{(s_{L},r_{L})}(u-\theta_{L})\dots\mathcal{L}_{a,2}^{(s_{1},r_{1})}(u+\theta_{1}-cs_{1}+cr_{1})\mathcal{L}_{a,1}^{(s_{1},r_{1})}(u-\theta_{1}).
\end{equation}
The vacuum eigenvalues are
\begin{equation}
\begin{split}\lambda_{i}(u) & =\prod_{j=1}^{L}(u-\theta_{j}+cs_{j})(u+\theta_{j}+cr_{j}),\quad i\leq r,\\
\lambda_{i}(u) & =\prod_{j=1}^{L}(u-\theta_{j})(u+\theta_{j}-cs_{j}+cr_{j}),\quad i>r,
\end{split}
\end{equation}
and
\begin{equation}
\begin{split}\hat{\lambda}_{i}(u) & =\prod_{j=1}^{L}\frac{1}{(u-\theta_{j}+cs_{j})(u+\theta_{j}+cr_{j})},\quad N-r<i,\\
\hat{\lambda}_{i}(u) & =\prod_{j=1}^{L}\frac{1}{(u-\theta_{j}+cs_{j})(u+\theta_{j}+cr_{j})}\frac{(u-\theta_{j}+cs_{j}-cr_{j})(u+\theta_{j})}{(u-\theta_{j}-cr_{j})(u+\theta_{j}-cs_{j})},\quad i\leq N-r,
\end{split}
\end{equation}
We can see that these functions satisfy the condition (\ref{eq:constLamTw})
\begin{equation}
\lambda_{i}(u)=\lambda_{0}(u)\hat{\lambda}_{N+1-i}(-u),
\end{equation}
where
\begin{equation}
\lambda_{0}(u)=\prod_{j=1}^{L}(u^{2}-(\theta_{j}-cs_{j})^{2})(u^{2}-(\theta_{j}+cr_{j})^{2}).
\end{equation}
The twisted monodromy can be written as
\begin{multline}
\widehat{T}_{a}(u)=\frac{1}{\lambda_{0}(u)}\times\\
\bar{\mathcal{L}}_{a,2L}^{(s_{L},N-r_{L})}(u+\theta_{L}-cs_{L})\mathcal{\bar{L}}_{a,2L-1}^{(s_{L},N-r_{L})}(u-\theta_{L}-cr_{L})\dots\bar{\mathcal{L}}_{a,2}^{(s_{1},N-r_{1})}(u+\theta_{1}-cs_{L})\bar{\mathcal{L}}_{a,1}^{(s_{1},N-r_{1})}(u-\theta_{1}-cr_{1}),
\end{multline}
where
\begin{multline}
\bar{\mathcal{L}}_{a,j}^{(s,N-r)}(u)=-V_{a}\mathcal{L}_{a,j}^{(s,r)}(-u-cs)^{t_{a}}V_{a}=-V_{j}^{(s,r)}\mathcal{L}_{a,j}^{(s,r)}(-u-cs)^{t_{j}}V_{j}^{(s,r)}=\\
=(u+cs)\mathbf{1}-c\sum_{i,j=1}^{N}E_{i,j}\otimes E_{N+1-i,N+1-j}^{(s,r)}.
\end{multline}
Let us define a reflection matrix $K^{(s,r)}(u)\in\mathrm{End}\mathcal{V}_{(s,r)}$
which satisfies the reflection equation 
\begin{equation}
\mathcal{L}_{21}^{(s,r)}(u-v)K_{1}^{(s,r)}(-u)\mathcal{\bar{L}}_{21}^{(s,N-r)}(u+v-cr)K_{2}(-v)=K_{2}(-v)\mathcal{\bar{L}}_{21}^{(s,N-r)}(u+v-cr)K_{1}^{(s,r)}(-u)\mathcal{L}_{21}^{(s,r)}(u-v).
\end{equation}
There is an alternative form of this equation
\begin{multline}
\vec{K}_{4,3}(u)V_{4}\vec{K}_{2,1}^{(s,r)}(\theta)V_{2}^{(s,r)}\mathcal{L}_{3,2}^{(s,r)}(u+\theta-cs+cr)\mathcal{L}_{3,1}^{(s,r)}(u-\theta)=\\
\vec{K}_{4,3}(u)V_{4}\vec{K}_{2,1}^{(s,r)}(\theta)V_{2}^{(s,r)}\mathcal{L}_{4,1}^{(s,r)}(u+\theta-cs+cr)\mathcal{L}_{4,2}^{(s,r)}(u-\theta).\label{eq:KKLL-1}
\end{multline}
Let us define the two-site final state as
\begin{equation}
\langle\Psi|=\vec{K}_{2,1}^{\Lambda^{(1)}}(\theta_{1})V_{2}^{\Lambda}\dots\vec{K}_{2L,2L-1}^{\Lambda^{(L)}}(\theta_{L})V_{2L}^{\Lambda}.
\end{equation}
Using the relation (\ref{eq:KKLL-1}) we can show that the following
equation is satisfied
\begin{equation}
\vec{K}_{b,a}(u)\langle\Psi|T_{a}(u)=\lambda_{0}(u)\vec{K}_{b,a}(u)V_{b}\langle\Psi|T_{b}^{\Pi}(u)V_{b},\label{eq:KvecT-1}
\end{equation}
where we defined the space reflected monodromy matrix
\begin{equation}
T_{b}^{\Pi}(u)=\bar{\mathcal{L}}_{b,1}^{(s_{1},r_{1})}(u+\theta_{1}-cs_{L}+cr_{1})\bar{\mathcal{L}}_{b,2}^{(s_{1},r_{1})}(u-\theta_{1})\dots\mathcal{\bar{L}}_{b,2L-1}^{(s_{L},r_{L})}(u+\theta_{L}-cs_{L}+cr_{L})\bar{\mathcal{L}}_{b,2L}^{(s_{L},r_{L})}(u-\theta_{L}).
\end{equation}
It is easy to show that
\begin{equation}
T_{b}^{\Pi}(u)=\lambda_{0}(u)V_{b}\widehat{T}_{b}^{t_{b}}(-u)V_{b}.
\end{equation}
Substituting back to (\ref{eq:KvecT-1}) we obtain the twisted $KT$-relation
\begin{equation}
K_{0}(u)\langle\Psi|T_{0}(u)=\lambda_{0}(u)\langle\Psi|\widehat{T}_{0}(-u)K_{0}(u).
\end{equation}

\subsection{Examples}

In this subsection we apply our general formula for $\mathfrak{gl}(3)$
and $\mathfrak{gl}(4)$ symmetric models.

\subsubsection*{$\mathfrak{gl}(3)$ models}

Let us apply our on-shell overlap formula for the $\mathfrak{so}(3)$
symmetric final state. In \cite{Piroli:2018ksf,Piroli:2018don} the
authors proposed a formula for the defining representation i.e. $s_{i}=r_{i}=1$.
Their convention for the final state was
\begin{equation}
\langle\Psi|=\vec{\psi}\otimes\dots\otimes\vec{\psi},
\end{equation}
where
\begin{equation}
\psi=\left(\begin{array}{ccc}
\kappa_{1,1} & \kappa_{1,2} & \kappa_{1,3}\\
\kappa_{1,2} & \kappa_{2,2} & \kappa_{2,3}\\
\kappa_{1,3} & \kappa_{2,3} & \kappa_{3,3}
\end{array}\right).
\end{equation}
To get the corresponding $K$-matrix we have to solve the equation
(\ref{eq:KKLL-1}) with $\psi=\vec{K}_{2,1}^{(1,0,0)}(\theta)V_{2}^{(1,0,0)}$
and the solution is
\begin{equation}
K=\left(\begin{array}{ccc}
\kappa_{1,3} & \kappa_{2,3} & \kappa_{3,3}\\
\kappa_{1,2} & \kappa_{2,2} & \kappa_{2,3}\\
\kappa_{1,1} & \kappa_{1,2} & \kappa_{1,3}
\end{array}\right).
\end{equation}
Substituting back to (\ref{eq:onepartTw}) we obtain the one-particle
overlaps
\begin{equation}
F_{K}^{(1)}=\frac{\kappa_{1,2}^{2}-\kappa_{1,1}\kappa_{2,2}}{\kappa_{1,1}^{2}},\qquad F_{K}^{(2)}=\frac{-\kappa_{1,1}\det\psi}{(\kappa_{1,1}\kappa_{2,2}-\kappa_{1,2}^{2})^{2}},
\end{equation}
therefore the on-shell overlap can be written as
\begin{equation}
\frac{\mathbb{C}(\bar{t})|\Psi\rangle\langle\Psi|\mathbb{B}(\bar{t})}{\mathbb{C}(\bar{t})\mathbb{B}(\bar{t})}=\langle\Psi|0\rangle^{2}\mathcal{F}_{K}^{(\nu)}(\bar{t}^{+,1})\mathcal{F}_{K}^{(\nu)}(\bar{t}^{+,1})\frac{\det G^{+}}{\det G^{-}}.
\end{equation}
Using the convention $c=i$ and $u_{k}=t_{k}^{1}+i/2$, $v_{k}=t_{k}^{2}+i$
and $N=r_{1}$, $M=r_{2}$ the on-shell overlap reads as
\begin{equation}
\frac{\mathbb{C}(\bar{u},\bar{v})|\Psi\rangle\langle\Psi|\mathbb{B}(\bar{u},\bar{v})}{\mathbb{C}(\bar{u},\bar{v})\mathbb{B}(\bar{u},\bar{v})}=\Gamma(\left\{ \kappa_{i,j}\right\} )\prod_{k=1}^{N/2}h(u_{k}^{+})\prod_{k=1}^{M/2}h(v_{k}^{+})\frac{\det G^{+}}{\det G^{-}},
\end{equation}
where
\begin{equation}
\Gamma(\left\{ \kappa_{i,j}\right\} )=\kappa_{1,1}^{2L}\left(F_{K}^{(1)}\right)^{N}\left(F_{K}^{(2)}\right)^{M}=\kappa_{1,1}^{2L-2N+M}(\kappa_{1,1}\kappa_{2,2}-\kappa_{1,2}^{2})^{N-2M}(\det\psi)^{M},
\end{equation}
and
\begin{equation}
h(u)=\frac{u^{2}}{u^{2}+1/4}.\label{eq:hdef}
\end{equation}
We can see that our result is in complete agreement of the conjecture
of \cite{Piroli:2018ksf,Piroli:2018don}.

\subsubsection*{$\mathfrak{gl}(4)$ models in representation $s_{k}=1$, $r_{k}=2$}

Now we calculate the overlaps for the representations $s_{k}=1$,
$r_{k}=2$ which is the defining representation of the $\mathfrak{so}(6)$
algebra which is isomorphic to $\mathfrak{gl}(4)$. We use the following
convention for the generators of the representation $(s,r)=(1,2)=(1,1,0,0)${\tiny{}
\begin{equation}
\begin{split}E_{1,1}^{(1,2)}=\left(\begin{array}{cccccc}
1 & 0 & 0 & 0 & 0 & 0\\
0 & 1 & 0 & 0 & 0 & 0\\
0 & 0 & 1 & 0 & 0 & 0\\
0 & 0 & 0 & 0 & 0 & 0\\
0 & 0 & 0 & 0 & 0 & 0\\
0 & 0 & 0 & 0 & 0 & 0
\end{array}\right),\qquad E_{2,2}^{(1,2)}=\left(\begin{array}{cccccc}
1 & 0 & 0 & 0 & 0 & 0\\
0 & 0 & 0 & 0 & 0 & 0\\
0 & 0 & 0 & 0 & 0 & 0\\
0 & 0 & 0 & 1 & 0 & 0\\
0 & 0 & 0 & 0 & 1 & 0\\
0 & 0 & 0 & 0 & 0 & 0
\end{array}\right),\\
E_{3,3}^{(1,2)}=\left(\begin{array}{cccccc}
0 & 0 & 0 & 0 & 0 & 0\\
0 & 1 & 0 & 0 & 0 & 0\\
0 & 0 & 0 & 0 & 0 & 0\\
0 & 0 & 0 & 1 & 0 & 0\\
0 & 0 & 0 & 0 & 0 & 0\\
0 & 0 & 0 & 0 & 0 & 1
\end{array}\right),\qquad E_{4,4}^{(1,2)}=\left(\begin{array}{cccccc}
0 & 0 & 0 & 0 & 0 & 0\\
0 & 0 & 0 & 0 & 0 & 0\\
0 & 0 & 1 & 0 & 0 & 0\\
0 & 0 & 0 & 0 & 0 & 0\\
0 & 0 & 0 & 0 & 1 & 0\\
0 & 0 & 0 & 0 & 0 & 1
\end{array}\right),\\
E_{1,2}^{(1,2)}=\left(\begin{array}{cccccc}
0 & 0 & 0 & 0 & 0 & 0\\
0 & 0 & 0 & 1 & 0 & 0\\
0 & 0 & 0 & 0 & -1 & 0\\
0 & 0 & 0 & 0 & 0 & 0\\
0 & 0 & 0 & 0 & 0 & 0\\
0 & 0 & 0 & 0 & 0 & 0
\end{array}\right),\quad E_{2,3}^{(1,2)}=\left(\begin{array}{cccccc}
0 & 1 & 0 & 0 & 0 & 0\\
0 & 0 & 0 & 0 & 0 & 0\\
0 & 0 & 0 & 0 & 0 & 0\\
0 & 0 & 0 & 0 & 0 & 0\\
0 & 0 & 0 & 0 & 0 & -1\\
0 & 0 & 0 & 0 & 0 & 0
\end{array}\right),\quad E_{3,4}^{(1,2)}= & \left(\begin{array}{cccccc}
0 & 0 & 0 & 0 & 0 & 0\\
0 & 0 & 1 & 0 & 0 & 0\\
0 & 0 & 0 & 0 & 0 & 0\\
0 & 0 & 0 & 0 & -1 & 0\\
0 & 0 & 0 & 0 & 0 & 0\\
0 & 0 & 0 & 0 & 0 & 0
\end{array}\right),
\end{split}
\end{equation}
}where the rest of the generators follow from the commutation rules
and the property $E_{j,i}^{(1,2)}=\left(E_{i,j}^{(1,2)}\right)^{t}$.
We also use the notation $(Z,Y,X,\bar{X},\bar{Y},\bar{Z})$ for the
basis vectors of the six dimensional representation therefore the
weights are
\begin{equation}
\begin{split}Z & \to(1,1,0,0),\qquad\bar{Z}\to(0,0,1,1),\\
Y & \to(1,0,1,0),\qquad\bar{Y}\to(0,1,0,1),\\
X & \to(1,0,0,1),\qquad\bar{X}\to(0,1,1,0).
\end{split}
\end{equation}
In this convention the $V^{(1,2)}$ matrix reads as
\begin{equation}
V^{(1,2)}=\left(\begin{array}{cccccc}
0 & 0 & 0 & 0 & 0 & 1\\
0 & 0 & 0 & 0 & -1 & 0\\
0 & 0 & 1 & 0 & 0 & 0\\
0 & 0 & 0 & 1 & 0 & 0\\
0 & -1 & 0 & 0 & 0 & 0\\
1 & 0 & 0 & 0 & 0 & 0
\end{array}\right).
\end{equation}
Using these matrices we can define the Lax operators $\mathcal{L}^{(1,1,0,0)}(u)$
and $\mathcal{L}^{\overline{(1,1,0,0)}}(u)$. Since $\overline{(1,1,0,0)}=(0,0,-1,-1)\cong(1,1,0,0)$
these two Lax matrices are similar.
\begin{equation}
\mathcal{L}_{a,j}^{\overline{(1,1,0,0)}}(u)=W_{j}\mathcal{L}_{a,j}^{(1,1,0,0)}(u-c)W_{j},
\end{equation}
where $W_{i,j}=V_{i,7-j}^{(1,2)}$. It also means that the untwisted
\begin{equation}
T_{a}^{utw}(u|\bar{\theta}^{u})=\mathcal{L}_{a,2L}^{\bar{\Lambda}}(u+\theta_{L}^{u})\mathcal{L}_{a,2L-1}^{\Lambda}(u-\theta_{L}^{u})\dots\mathcal{L}_{a,2}^{\bar{\Lambda}}(u+\theta_{1}^{u})\mathcal{L}_{a,1}^{\Lambda}(u-\theta_{1}^{u}),\label{eq:transferUTw}
\end{equation}
and twisted monodromy matrices
\begin{equation}
T_{a}^{tw}(u|\bar{\theta}^{t})=\mathcal{L}_{a,2L}^{(1,2)}(u+\theta_{L}^{t}+c)\mathcal{L}_{a,2L-1}^{(1,2)}(u-\theta_{L}^{t})\dots\mathcal{L}_{a,2}^{(1,2)}(u+\theta_{1}^{t}+c)\mathcal{L}_{a,1}^{(1,2)}(u-\theta_{1}^{t})
\end{equation}
are similar for $\Lambda=(1,1,0,0)=(1,2)$ i.e.
\begin{equation}
T_{a}^{utw}(u|\bar{\theta}^{u})=\mathcal{W}T_{a}^{tw}(u-c|\bar{\theta}^{u}-c)\mathcal{W},\label{eq:conn}
\end{equation}
where $\mathcal{W}=W_{2}W_{4}\dots W_{2L}$. In the following we use
the notations $\mathcal{L}_{a,j}(u)=\mathcal{L}_{a,j}^{(1,2)}(u)=\mathcal{L}_{a,j}^{(1,1,0,0)}(u)$
and $\bar{\mathcal{L}}_{a,j}(u)=\mathcal{L}_{a,j}^{\overline{(1,1,0,0)}}(u)$.

Let us turn to the integrable states. The possible residual symmetries
are $\mathfrak{so}(4)$,$\mathfrak{sp}(4)$ for the twisted and $\mathfrak{gl}(1)\oplus\mathfrak{gl}(3)$
and $\mathfrak{gl}(2)\oplus\mathfrak{gl}(2)$ for the untwisted classes.
Our overlap formula works for the $\mathfrak{so}(4)$ and the $\mathfrak{gl}(2)\oplus\mathfrak{gl}(2)$
cases. The papers \cite{deLeeuw:2019ebw}, \cite{Jiang:2019xdz} propose
overlap formulas for two-site states for this spin chain. Now we compere
our general overlap formulas to these conjectures.

Let us start with the paper \cite{deLeeuw:2019ebw} where the two-site
state is
\begin{equation}
\langle\Psi|=\vec{\psi}^{\otimes L},
\end{equation}
where
\begin{equation}
\vec{\psi}=Z\otimes Z+X\otimes X+Y\otimes Y+\bar{Z}\otimes\bar{Z}+\bar{X}\otimes\bar{X}+\bar{Y}\otimes\bar{Y}.
\end{equation}
In \cite{Kristjansen:2020mhn} it was shown that the overlaps of this
two-side state describe the tree level one-point functions of a defect
version of $\mathcal{N}=4$ SYM in the scalar sector. This two-site
state has symmetry $\mathfrak{so}(4)\cong\mathfrak{so}(3)\oplus\mathfrak{so}(3)$
therefore it belongs to the twisted class. For the twisted monodromy
matrix the $\alpha$-s are
\begin{equation}
\begin{split}\alpha_{1}(u) & =1,\\
\alpha_{2}(u) & =\prod_{j=1}^{L}\frac{(u-\theta_{j}^{t}+c)(u+\theta_{j}^{t}+2c)}{(u-\theta_{j}^{t})(u+\theta_{j}^{t}+c)}\\
\alpha_{3}(u) & =1.
\end{split}
,
\end{equation}
 Using the notations $w_{k}=t_{k}^{1}+c/2$, $u_{k}=t_{k}^{2}+c$,
$v_{k}=t_{k}^{3}+3c/2$ and fixing the inhomogeneities $\theta_{j}^{t}=-c/2$
the Bethe equations read
\begin{align}
1 & =\Phi_{k}^{(w)}=\prod_{l\neq k}^{K_{w}}\frac{w_{k}-w_{l}-c}{w_{k}-w_{l}+c}\prod_{l=1}^{M}\frac{w_{k}-u_{l}+c/2}{w_{k}-u_{l}-c/2},\label{eq:BEw}\\
1 & =\Phi_{k}^{(u)}=\left(\frac{u_{k}+c/2}{u_{k}-c/2}\right)^{2L}\prod_{l\neq k}^{M}\frac{u_{k}-u_{l}-c}{u_{k}-u_{l}+c}\prod_{l=1}^{K_{v}}\frac{u_{k}-v_{l}+c/2}{u_{k}-v_{l}-c/2}\prod_{l=1}^{K_{w}}\frac{u_{k}-w_{l}+c/2}{u_{k}-w_{l}-c/2},\label{eq:BEu}\\
1 & =\Phi_{k}^{(v)}=\prod_{l\neq k}^{K_{v}}\frac{v_{k}-v_{l}-c}{v_{k}-v_{l}+c}\prod_{l=1}^{M}\frac{v_{k}-u_{l}+c/2}{v_{k}-u_{l}-c/2},\label{eq:BEv}
\end{align}
which agrees with the convention of \cite{deLeeuw:2019ebw} if $c=i$.
To get the corresponding $K$-matrix we have to solve the equation
(\ref{eq:KKLL-1})
\begin{equation}
\vec{K}_{4,3}(u)V_{4}\psi_{12}\mathcal{L}_{3,2}(u+c/2)\mathcal{L}_{3,1}(u+c/2)=\vec{K}_{4,3}(u)V_{4}\psi_{12}\mathcal{L}_{4,1}(u+c/2)\mathcal{L}_{4,2}(u+c/2),
\end{equation}
and the solution is
\begin{equation}
K=\left(\begin{array}{cccc}
0 & 0 & 0 & 1\\
0 & 0 & 1 & 0\\
0 & 1 & 0 & 0\\
1 & 0 & 0 & 0
\end{array}\right).
\end{equation}
For this $K$-matrix the one-particle overlap functions are
\begin{equation}
F_{K}^{(1)}=F_{K}^{(2)}=F_{K}^{(3)}=-1,
\end{equation}
which means that the on-shell overlap reads as
\begin{equation}
\frac{\mathbb{C}(\bar{u},\bar{v},\bar{w})|\Psi\rangle\langle\Psi|\mathbb{B}(\bar{u},\bar{v},\bar{w})}{\mathbb{C}(\bar{u},\bar{v},\bar{w})\mathbb{B}(\bar{u},\bar{v},\bar{w})}=\prod_{k=1}^{M/2}h(u_{k}^{+})\prod_{k=1}^{K_{v}/2}h(v_{k}^{+})\prod_{k=1}^{K_{w}/2}h(v_{k}^{+})\frac{\det G^{+}}{\det G^{-}},
\end{equation}
where $h(u)$ is defined in (\ref{eq:hdef}). We can see that this
result is in complete agreement with the conjecture of \cite{deLeeuw:2019ebw}.

Let us continue with the proposed overlap formula of \cite{Jiang:2019xdz}.
The final state is
\begin{equation}
\langle\Psi|=\vec{\psi}\otimes\dots\otimes\vec{\psi},
\end{equation}
where
\begin{equation}
\vec{\psi}=(Z+(Y-\bar{Y})+\bar{Z})\otimes(x^{2}Z+x(Y-\bar{Y})+\bar{Z}).
\end{equation}
This state has $\mathfrak{gl}(2)\oplus\mathfrak{gl}(2)$ symmetry
therefore it belongs to the untwisted class. The Bethe state $\mathbb{B}^{JKV}$
in \cite{Jiang:2019xdz} belongs to a translation invariant transfer
matrix but our untwisted transfer matrix (\ref{eq:transferUTw}) is
obviously not translation invariant (see equation (\ref{eq:conn}))
therefore the convention for the Bethe states in \cite{Jiang:2019xdz}
are different by a rotation
\begin{equation}
\mathbb{B}^{JKV}=\mathcal{W}\mathbb{B},
\end{equation}
therefore 
\begin{equation}
\langle\Psi|\mathbb{B}^{JKV}=\langle\Psi|\mathcal{W}\mathbb{B}=\langle\Psi'|\mathbb{B},
\end{equation}
where $\langle\Psi'|$ is built from the two-site operator $\psi'_{12}=\psi_{12}W_{2}$.

For the untwisted monodromy matrix the $\alpha$-s are 
\begin{equation}
\begin{split}\alpha_{1}(u) & =1,\\
\alpha_{2}(u) & =\prod_{j=1}^{L}\frac{(u-\theta_{j}^{u}+c)(u+\theta_{j}^{u})}{(u-\theta_{j}^{u})(u+\theta_{j}^{u}-c)}\\
\alpha_{3}(u) & =1.
\end{split}
,
\end{equation}
Using the notations $w_{k}=t_{k}^{1}-c/2$, $u_{k}=t_{k}^{2}$, $v_{k}=t_{k}^{3}+c/2$
and fixing the inhomogeneities $\theta_{j}^{u}=c/2$, the Bethe equations
read as (\ref{eq:BEw}-\ref{eq:BEv}) which agrees with the convention
of \cite{Jiang:2019xdz} if $c=i$.

To get the corresponding $K$-matrix we have to solve the equation
(\ref{eq:KKLL})
\begin{equation}
\vec{K}_{4,3}(u)V_{4}\psi'_{12}\bar{\mathcal{L}}_{3,2}(u+c/2)\mathcal{L}_{3,1}(u-c/2)=\vec{K}_{4,3}(u)V_{4}\psi'_{12}\mathcal{\bar{\mathcal{L}}}_{4,1}(u+c/2)\mathcal{L}_{4,2}(u-c/2),
\end{equation}
and the solution is 
\begin{equation}
K(u)=\frac{c}{2}\mathbf{1}+u\left(\begin{array}{cccc}
\frac{x+1}{x-1} & 0 & 0 & -\frac{2x}{x+1}\\
0 & \frac{x+1}{x-1} & \frac{2x}{x-1} & 0\\
0 & \frac{-2}{x+1} & \frac{x+1}{1-x} & 0\\
\frac{2}{x+1} & 0 & 0 & \frac{x+1}{1-x}
\end{array}\right).
\end{equation}
Substituting back to (\ref{eq:onepartUTw}) we obtain the one-particle
overlap functions
\begin{equation}
F_{K}^{(1)}(u)=-1,\qquad F_{K}^{(2)}(u)=(x-1)^{2}\frac{1}{4}\frac{u^{2}-(c/2)^{2}}{u^{2}},
\end{equation}
therefore
\begin{equation}
\mathcal{F}_{K}^{(1)}(u)=1,\qquad\mathcal{F}_{K}^{(2)}(u)=(x-1)^{4}\frac{1}{16}\frac{u^{2}-(c/2)^{2}}{u^{2}}.
\end{equation}
Fixing $c=i$ the on-shell overlap reads as
\begin{equation}
\frac{\mathbb{C}(\bar{u},\bar{v},\bar{w})|\Psi'\rangle\langle\Psi'|\mathbb{B}(\bar{u},\bar{v},\bar{w})}{\mathbb{C}(\bar{u},\bar{v},\bar{w})\mathbb{B}(\bar{u},\bar{v},\bar{w})}=(x-1)^{2M}\left(\frac{1}{2}\right)^{2M}\prod_{k=1}^{M/2}\frac{u_{k}^{2}+1/4}{u_{k}^{2}}\frac{\det G^{+}}{\det G^{-}},
\end{equation}
which agrees with the proposed formula of \cite{Jiang:2019xdz}.

\subsubsection*{Alternating $\mathfrak{gl}(4)$ model}

The ABJM theory in the scalar sector can be described as an alternating
$\mathfrak{gl}(4)$ spin chain where the quantum space is $\left(\Lambda\otimes\bar{\Lambda}\right)^{\otimes L}$
where $\Lambda=(1,1)=(1,0,0,0)$ which is compatible with our untwisted
monodromy matrix. In \cite{Yang:2021hrl} the authors proposed an
overlap for the two-site state
\begin{equation}
\langle\Psi|=\vec{\psi}\otimes\dots\otimes\vec{\psi},
\end{equation}
where
\begin{equation}
\psi=\left(\begin{array}{cccc}
1 & 0 & 0 & 1\\
0 & 0 & 0 & 0\\
0 & 0 & 0 & 0\\
1 & 0 & 0 & 1
\end{array}\right).
\end{equation}
 For the untwisted monodromy matrix the $\alpha$-s are 
\begin{equation}
\begin{split}\alpha_{1}(u) & =\prod_{i=1}^{L}\frac{(u-\theta_{j}+c)}{(u-\theta_{j})},\\
\alpha_{2}(u) & =1,\\
\alpha_{3}(u) & =\prod_{i=1}^{L}\frac{(u+\theta_{j})}{(u+\theta_{j}-c)}.
\end{split}
\end{equation}
Using the notations $u_{k}=t_{k}^{1}-c/2$, $w_{k}=t_{k}^{2}$, $v_{k}=t_{k}^{3}+c/2$
and fixing the inhomogeneities $\theta_{j}=c$ the Bethe equations
read as
\begin{align}
1 & =\Phi_{k}^{(u)}=\left(\frac{u_{k}+c/2}{u_{k}-c/2}\right)^{L}\prod_{l\neq k}^{K_{u}}\frac{u_{k}-u_{l}-c}{u_{k}-u_{l}+c}\prod_{l=1}^{K_{w}}\frac{u_{k}-w_{l}+c/2}{u_{k}-w_{l}-c/2},\label{eq:BEw-1}\\
1 & =\Phi_{k}^{(w)}=\prod_{l\neq k}^{K_{w}}\frac{w_{k}-w_{l}-c}{w_{k}-w_{l}+c}\prod_{l=1}^{K_{u}}\frac{w_{k}-u_{l}+c/2}{w_{k}-u_{l}-c/2}\prod_{l=1}^{K_{v}}\frac{w_{k}-v_{l}+c/2}{w_{k}-v_{l}-c/2},\label{eq:BEu-1}\\
1 & =\Phi_{k}^{(v)}=\left(\frac{v_{k}+c/2}{v_{k}-c/2}\right)^{L}\prod_{l\neq k}^{K_{v}}\frac{v_{k}-v_{l}-c}{v_{k}-v_{l}+c}\prod_{l=1}^{K_{w}}\frac{v_{k}-w_{l}+c/2}{v_{k}-w_{l}-c/2},\label{eq:BEv-1}
\end{align}
which agrees with the convention in \cite{Yang:2021hrl} if $c=i$.
To get the corresponding $K$-matrix we have to solve the equation
(\ref{eq:KKLL-1})
\begin{equation}
\vec{K}_{4,3}(u)V_{4}\psi_{12}\bar{\mathcal{L}}_{3,2}(u+c)\mathcal{L}_{3,1}(u-c)=\vec{K}_{4,3}(u)V_{4}\psi_{12}\bar{\mathcal{L}}_{4,1}(u+c)\mathcal{L}_{4,2}(u-c),
\end{equation}
where
\begin{equation}
\mathcal{L}(u)=u\mathbf{1}+c\sum_{i,j=1}^{4}E_{i,j}\otimes E_{j,i},\quad\mathcal{\bar{L}}(u)=u\mathbf{1}-c\sum_{i,j=1}^{4}E_{i,j}\otimes E_{5-i,5-j},
\end{equation}
and the solution is
\begin{equation}
K(u)=c\mathbf{1}+u\left(\begin{array}{cccc}
0 & 0 & 0 & 1\\
0 & -1 & 0 & 0\\
0 & 0 & -1 & 0\\
1 & 0 & 0 & 0
\end{array}\right).
\end{equation}
Unfortunately, this $K$-matrix has $\mathfrak{gl}(1)\oplus\mathfrak{gl}(3)$
symmetry therefore our overlap formula can not be applied for this
final state.

Our formula works only for the $\mathfrak{gl}(2)\oplus\mathfrak{gl}(2)$
symmetric $K$-matrix, for example
\begin{equation}
K(u)=c\mathfrak{a}\mathbf{1}+u\left(\begin{array}{cccc}
0 & 0 & 0 & 1\\
0 & 0 & 1 & 0\\
0 & 1 & 0 & 0\\
1 & 0 & 0 & 0
\end{array}\right).
\end{equation}
For this $K$-matrix the two-site operator is $\psi_{1,2}\sim K_{2,1}(c)V_{2}$
therefore this $K$-matrix belongs the two-site state
\begin{equation}
\langle\Psi'|=\psi'\otimes\dots\otimes\psi',
\end{equation}
where
\begin{equation}
\psi'=\left(\begin{array}{cccc}
1 & 0 & 0 & \mathfrak{a}\\
0 & 1 & \mathfrak{a} & 0\\
0 & \mathfrak{a} & 1 & 0\\
\mathfrak{a} & 0 & 0 & 1
\end{array}\right).
\end{equation}
For this two-site state the on-shell overlap reads as
\begin{equation}
\frac{\mathbb{C}(\bar{u},\bar{w},\bar{v})|\Psi'\rangle\langle\Psi'|\mathbb{B}(\bar{u},\bar{w},\bar{v})}{\mathbb{C}(\bar{u},\bar{w},\bar{v})\mathbb{B}(\bar{u},\bar{w},\bar{v})}=\prod_{k=1}^{K_{w}/2}\frac{(w_{k}^{2}-c^{2}\mathfrak{a}^{2})^{2}}{w_{k}^{2}(w_{k}^{2}+1/4)}\frac{\det G^{+}}{\det G^{-}}.
\end{equation}
It would be interesting to find a physical observable of the ABJM
theory which can be described by this overlap. Perhaps these type
of overlaps would appear for the one-point functions of a defect version
of ABJM.

\section{Conclusions}

In this paper we applied the algebraic Bethe Ansatz method for the
calculation of the overlaps between the integrable two-site states
and the Bethe states of the $\mathfrak{gl}(N)$ symmetric spin chains.
The main advantage of this approach is that it is independent from
the quantum space. Our main results are the equations (\ref{eq:normOv}-\ref{eq:FK2})
which give the exact overlaps for all $\mathfrak{so}(N)$ or $\mathfrak{gl}(\left\lfloor \frac{N}{2}\right\rfloor )\otimes\mathfrak{gl}(\left\lceil \frac{N}{2}\right\rceil )$
symmetric two-site states. This is the first time when the factorized
overlaps are exactly proven for nested systems. We also applied our
general formula to concrete two-site states where the overlap formulas
were previously proposed and we found that the results are perfectly
matched, which proves the earlier conjectures.

There are lots of open questions.
\begin{itemize}
\item Our result is not complete even for the $\mathfrak{gl}(N)$ symmetric
spin chains since it does not work for the $\mathfrak{sp}(N)$ and
the $\mathfrak{gl}(M)\otimes\mathfrak{gl}(N-M)$ symmetric two-site
states where $M<\left\lfloor \frac{N}{2}\right\rfloor $. It would
be nice to find an extension of the method of this paper which contains
these remaining cases.
\item Our method is based on the $KT$-relation and the action formulas
of the monodromy matrix entries on the off-shell Bethe states. Recently
these action formulas were generalized to the quantum integrable models
associated with $U_{q}(\mathfrak{gl}(N))$-invariant $R$-matrices
\cite{Liashyk:2021tth}. Using these new results it would be interesting
to generalize our method to the calculation of the exact overlaps
of the two-site states for the trigonometric integrable models.
\item For the applications in the AdS/CFT duality, it would be interesting
to generalize our results to $\mathfrak{gl}(N|M)$ symmetric spin
chains. For the context of the $AdS_{5}/CFT_{4}$ duality the relevant
case is the $\mathfrak{gl}(4|4)$. The tree level one-point function
for a certain defect is given by the overlap of an $\mathfrak{osp}(4|4)$
symmetric two-site state and the exact overlap formula was proposed
in \cite{Kristjansen:2020vbe} based on \cite{Gombor:2020kgu}. It
would be nice to give an exact proof for this formula. In the context
of the three-point functions $\langle\mathcal{D}\mathcal{D}\mathcal{O}\rangle$
where $\mathcal{D},\mathcal{O}$ are determinant and single trace
operators \cite{Jiang:2019zig} the relevant symmetry class is the
$\mathfrak{gl}(2|2)\oplus\mathfrak{gl}(2|2)$ and it would be nice
to prove the overlap formula which gives the tree level three-point
functions for the full spectrum of $AdS_{5}/CFT_{4}$.
\item It would be also interesting to combine our $KT$--relation with
the recently developed SoV framework of the $\mathfrak{gl}(N)$ spin
chains \cite{Ryan:2018fyo,Maillet:2018bim}. In the $\mathfrak{gl}(2)$
case it was already done in \cite{Gombor:2021uxz} and a Vandermonde-like
determinant formula was derived for the overlap using the SoV approach.
Recently an analogous but more general formula was proposed for the
higher rank cases in \cite{Cavaglia:2021mft}, which is compatible
with the result of \cite{Gombor:2021uxz} for the $\mathfrak{gl}(2)$
spin chain. Maybe our $KT$-relation could help for the proof at the
higher rank cases.
\item Finally it is worth to mention that it was conjectured that the fermionic
and bosonic dualities of the $Q$-system could fix the on-shell overlaps
\cite{Kristjansen:2020vbe,Kristjansen:2021xno}. Maybe our general
formulas and other formulas (calculated from the above mentioned generalizations)
could help to decide whether this conjecture is true or not.
\end{itemize}

\section*{Acknowledgments}

I thank Balázs Pozsgay, Arthur Hutsalyuk and Zoltán Bajnok for the
useful discussions and the NKFIH grant K134946 for support.

\appendix

\section{Off-shell Bethe vectors\label{sec:Off-shell-Bethe-vectors}}

In this section we review the recurrence and action formulas of the
off-shell Bethe vectors which are used in our derivations of the overlaps.
The formulas of this section can be found in \cite{Hutsalyuk:2017tcx,Hutsalyuk:2017way,Liashyk:2018egk,Hutsalyuk:2020dlw}.

The off-shell Bethe vectors can be calculated from the following sum
formula \cite{Hutsalyuk:2017tcx}

\begin{equation}
\mathbb{B}(\{z,\bar{t}^{1}\},\left\{ \bar{t}^{k}\right\} _{k=2}^{N-1})=\sum_{j=2}^{N}\frac{T_{1,j}(z)}{\lambda_{2}(z)}\sum_{\mathrm{part}(\bar{t})}\mathbb{B}(\bar{t}^{1},\left\{ \bar{t}_{\textsc{ii}}^{k}\right\} _{k=2}^{j-1},\left\{ \bar{t}^{k}\right\} _{k=j}^{N-1})\frac{\prod_{\nu=2}^{j-1}\alpha_{\nu}(\bar{t}_{\textsc{i}}^{\nu})g(\bar{t}_{\textsc{i}}^{\nu},\bar{t}_{\textsc{i}}^{\nu-1})f(\bar{t}_{\textsc{ii}}^{\nu},\bar{t}_{\textsc{i}}^{\nu})}{\prod_{\nu=1}^{j-1}f(\bar{t}^{\nu+1},\bar{t}_{\textsc{i}}^{\nu})},\label{eq:rec1}
\end{equation}
where the sum goes over all the possible partitions $\bar{t}^{\nu}=\bar{t}_{\textsc{i}}^{\nu}\cup\bar{t}_{\textsc{ii}}^{\nu}$
for $\nu=2,\dots,j-1$ where $\bar{t}_{\textsc{i}}^{\nu},\bar{t}_{\textsc{ii}}^{\nu}$
are disjoint subsets and $\#\bar{t}_{\textsc{i}}^{\nu}=1$. We set
by definition $\bar{t}_{\textsc{i}}^{1}=\{z\}$ and $\bar{t}^{N}=\emptyset$.

There is another sum formula
\begin{equation}
\mathbb{B}(\left\{ \bar{t}^{k}\right\} _{k=1}^{N-2},\{z,\bar{t}^{N-1}\})=\sum_{j=1}^{N-1}\frac{T_{j,N}(z)}{\lambda_{N}(z)}\sum_{\mathrm{part}(\bar{t})}\mathbb{B}(\left\{ \bar{t}^{k}\right\} _{k=1}^{j-1},\left\{ \bar{t}_{\textsc{ii}}^{k}\right\} _{k=j}^{N-2},\bar{t}^{N-1})\frac{\prod_{\nu=j}^{N-2}g(\bar{t}_{\textsc{i}}^{\nu+1},\bar{t}_{\textsc{i}}^{\nu})f(\bar{t}_{\textsc{i}}^{\nu},\bar{t}_{\textsc{ii}}^{\nu})}{\prod_{\nu=j}^{N-1}f(\bar{t}_{\textsc{i}}^{\nu},\bar{t}^{\nu-1})},\label{eq:rec2}
\end{equation}
where the sum goes over all the possible partitions $\bar{t}^{\nu}=\bar{t}_{\textsc{i}}^{\nu}\cup\bar{t}_{\textsc{ii}}^{\nu}$
for $\nu=j,\dots,N-2$ where $\bar{t}_{\textsc{i}}^{\nu},\bar{t}_{\textsc{ii}}^{\nu}$
are disjoint subsets and $\#\bar{t}_{\textsc{i}}^{\nu}=1$. We set
by definition $\bar{t}_{\textsc{i}}^{N-1}=\{z\}$ and $\bar{t}^{0}=\emptyset$.

These recursions can be used to eliminate Bethe roots $\bar{t}^{1}$
and $\bar{t}^{N-1}$ to obtain off-shell Bethe vectors $\mathbb{B}(\emptyset,\left\{ \bar{t}^{k}\right\} _{k=2}^{N-1})$
and $\mathbb{B}(\left\{ \bar{t}^{k}\right\} _{k=1}^{N-2},\emptyset)$
which can be obtained using the algebra injections $\mathfrak{i}^{\pm}$
\begin{equation}
\begin{split}\mathbb{B}^{N}(\emptyset,\left\{ \bar{t}^{k}\right\} _{k=2}^{N-1}) & =\mathfrak{i}^{+}\left(\mathbb{B}^{N-1}(\left\{ \bar{t}^{k}\right\} _{k=2}^{N-1})\right),\\
\mathbb{B}^{N}(\left\{ \bar{t}^{k}\right\} _{k=1}^{N-2},\emptyset) & =\mathfrak{i}^{-}\left(\mathbb{B}^{N-1}(\left\{ \bar{t}^{k}\right\} _{k=1}^{N-2})\right).
\end{split}
\end{equation}
We can also see that the boosted off-shell vector can be written as
\begin{equation}
\mathcal{B}_{x}\mathbb{B}(\bar{t})=\mathbb{B}(\bar{t}+x).
\end{equation}

We also use the following action formula for the off-shell Bethe vectors
\cite{Hutsalyuk:2020dlw}

\begin{multline}
T_{i,j}(z)\mathbb{B}(\bar{t})=\lambda_{N}(z)\sum_{\mathrm{part}(\bar{w})}\mathbb{B}(\bar{w}_{\textsc{ii}})\frac{\prod_{s=j}^{i-1}f(\bar{w}_{\textsc{i}}^{s},\bar{w}_{\textsc{iii}}^{s})}{\prod_{s=j}^{i-2}f(\bar{w}_{\textsc{i}}^{s+1},\bar{w}_{\textsc{iii}}^{s})}\times\\
\prod_{s=1}^{i-1}\frac{f(\bar{w}_{\textsc{i}}^{s},\bar{w}_{\textsc{ii}}^{s})}{h(\bar{w}_{\textsc{i}}^{s},\bar{w}_{\textsc{i}}^{s-1})f(\bar{w}_{\textsc{i}}^{s},\bar{w}_{\textsc{ii}}^{s-1})}\prod_{s=j}^{N-1}\frac{\alpha_{s}(\bar{w}_{\textsc{iii}}^{s})f(\bar{w}_{\textsc{ii}}^{s},\bar{w}_{\textsc{iii}}^{s})}{h(\bar{w}_{\textsc{iii}}^{s+1},\bar{w}_{\textsc{iii}}^{s})f(\bar{w}_{\textsc{ii}}^{s+1},\bar{w}_{\textsc{iii}}^{s})},\label{eq:act}
\end{multline}
where $\bar{w}^{\nu}=\{z,\bar{t}^{\nu}\}$. The sum goes over all
the partitions of $\bar{w}^{\nu}=\bar{w}_{\textsc{i}}^{\nu}\cup\bar{w}_{\textsc{ii}}^{\nu}\cup\bar{w}_{\textsc{iii}}^{\nu}$
where $\bar{w}_{\textsc{i}}^{\nu},\bar{w}_{\textsc{ii}}^{\nu},\bar{w}_{\textsc{iii}}^{\nu}$
are disjoint sets for a fixed $\nu$ and $\#\bar{w}_{\textsc{i}}^{\nu}=\Theta(i-1-\nu)$,
$\#\bar{w}_{\textsc{iii}}^{\nu}=\Theta(\nu-j)$. We also set $\bar{w}_{\textsc{i}}^{0}=\bar{w}_{\textsc{iii}}^{N}=\{z\}$
and $\bar{w}_{\textsc{ii}}^{0}=\bar{w}_{\textsc{iii}}^{0}=\bar{w}_{\textsc{i}}^{N}=\bar{w}_{\textsc{ii}}^{N}=\emptyset$.
We also used the unit step function $\Theta(k)$ which is defined
as $\Theta(k)=1$ for $k\geq0$ and $\Theta(k)=0$ for $k<0$.

We can see that the diagonal elements $T_{i,i}(u)$ do not change
the quantum numbers $r_{j}$. The creation operators i.e. $T_{i,j}(u)$
where $i<j$ increase the quantum numbers $r_{i},r_{i+1},\dots,r_{j-1}$
by one. The annihilation operators i.e. $T_{j,i}(u)$ where $i<j$
decrease the quantum numbers $r_{i},r_{i+1},\dots,r_{j-1}$ by one.

The operators $T_{i,j}$ where $i,j<N$ act on the off-shell state
with $r_{N-1}=0$ as
\begin{multline}
T_{i,j}(z)\mathbb{B}(\left\{ \bar{t}^{k}\right\} _{k=1}^{N-2},\emptyset)=\lambda_{N-1}(z)\sum_{\mathrm{part}(\bar{w})}\mathbb{B}(\left\{ \bar{w}_{\textsc{ii}}^{k}\right\} _{k=1}^{N-2},\emptyset)\frac{\prod_{s=j}^{i-1}f(\bar{w}_{\textsc{i}}^{s},\bar{w}_{\textsc{iii}}^{s})}{\prod_{s=j}^{i-2}f(\bar{w}_{\textsc{i}}^{s+1},\bar{w}_{\textsc{iii}}^{s})}\times\\
\prod_{s=1}^{i-1}\frac{f(\bar{w}_{\textsc{i}}^{s},\bar{w}_{\textsc{ii}}^{s})}{h(\bar{w}_{\textsc{i}}^{s},\bar{w}_{\textsc{i}}^{s-1})f(\bar{w}_{\textsc{i}}^{s},\bar{w}_{\textsc{ii}}^{s-1})}\prod_{s=j}^{N-2}\frac{\alpha_{s}(\bar{w}_{\textsc{iii}}^{s})f(\bar{w}_{\textsc{ii}}^{s},\bar{w}_{\textsc{iii}}^{s})}{h(\bar{w}_{\textsc{iii}}^{s+1},\bar{w}_{\textsc{iii}}^{s})f(\bar{w}_{\textsc{ii}}^{s+1},\bar{w}_{\textsc{iii}}^{s})},
\end{multline}
which means
\begin{equation}
T_{i,j}^{N}(z)\mathbb{B}^{N}(\left\{ \bar{t}^{k}\right\} _{k=1}^{N-2},\emptyset)=\mathfrak{i}^{-}(T_{i,j}^{N-1}(z)\mathbb{B}^{N-1}(\left\{ \bar{t}^{k}\right\} _{k=1}^{N-2}))
\end{equation}
for $i,j=1,\dots,N-1$. In an analogous way we can show that
\begin{equation}
T_{i+1,j+1}^{N}(z)\mathbb{B}^{N}(\emptyset,\left\{ \bar{t}^{k}\right\} _{k=2}^{N-1})=\mathfrak{i}^{+}(T_{i,j}^{N-1}(z)\mathbb{B}^{N-1}(\left\{ \bar{t}^{k}\right\} _{k=2}^{N-1})).
\end{equation}

In some parts of the derivations it is more convenient to use the
following re-normalized Bethe states
\begin{equation}
\tilde{\mathbb{B}}(\bar{t})=\prod_{\nu=1}^{N-1}\lambda_{\nu+1}(\bar{t}^{\nu})\mathbb{B}(\bar{t}),\label{eq:renOffShell}
\end{equation}
for which the recurrence and action formulas for can be written as
\begin{equation}
\tilde{\mathbb{B}}(\{z,\bar{t}^{1}\},\left\{ \bar{t}^{k}\right\} _{k=2}^{N-1})=\sum_{j=2}^{N}T_{1,j}(z)\sum_{\mathrm{part}(\bar{t})}\tilde{\mathbb{B}}(\bar{t}^{1},\left\{ \bar{t}_{\textsc{ii}}^{k}\right\} _{k=2}^{j-1},\left\{ \bar{t}^{k}\right\} _{k=j}^{N-1})\frac{\prod_{\nu=2}^{j-1}\lambda_{\nu}(\bar{t}_{\textsc{i}}^{\nu})g(\bar{t}_{\textsc{i}}^{\nu},\bar{t}_{\textsc{i}}^{\nu-1})f(\bar{t}_{\textsc{ii}}^{\nu},\bar{t}_{\textsc{i}}^{\nu})}{\prod_{\nu=1}^{j-1}f(\bar{t}^{\nu+1},\bar{t}_{\textsc{i}}^{\nu})},\label{eq:rec1-1}
\end{equation}
\begin{multline}
\tilde{\mathbb{B}}(\left\{ \bar{t}^{k}\right\} _{k=1}^{N-2},\{z,\bar{t}^{N-1}\})=\sum_{j=1}^{N-1}T_{j,N}(z)\sum_{\mathrm{part}(\bar{t})}\tilde{\mathbb{B}}(\left\{ \bar{t}^{k}\right\} _{k=1}^{j-1},\left\{ \bar{t}_{\textsc{ii}}^{k}\right\} _{k=j}^{N-2},\bar{t}^{N-1})\times\\
\frac{\prod_{\nu=j}^{N-2}\lambda_{\nu+1}(\bar{t}_{\textsc{i}}^{\nu})g(\bar{t}_{\textsc{i}}^{\nu+1},\bar{t}_{\textsc{i}}^{\nu})f(\bar{t}_{\textsc{i}}^{\nu},\bar{t}_{\textsc{ii}}^{\nu})}{\prod_{\nu=j}^{N-1}f(\bar{t}_{\textsc{i}}^{\nu},\bar{t}^{\nu-1})},\label{eq:rec2-1}
\end{multline}
and
\begin{multline}
T_{i,j}(z)\tilde{\mathbb{B}}(\bar{t})=\sum_{\mathrm{part}(\bar{w})}\frac{\prod_{s=1}^{i-1}\lambda_{s+1}(\bar{w}_{\textsc{i}}^{s})\prod_{s=j}^{N-1}\lambda_{s}(\bar{w}_{\textsc{iii}}^{s})}{\prod_{s=1}^{N-2}\lambda_{s+1}(z)}\tilde{\mathbb{B}}(\bar{w}_{\textsc{ii}})\\
\frac{\prod_{s=j}^{i-1}f(\bar{w}_{\textsc{i}}^{s},\bar{w}_{\textsc{iii}}^{s})}{\prod_{s=j}^{i-2}f(\bar{w}_{\textsc{i}}^{s+1},\bar{w}_{\textsc{iii}}^{s})}\prod_{s=1}^{i-1}\frac{f(\bar{w}_{\textsc{i}}^{s},\bar{w}_{\textsc{ii}}^{s})}{h(\bar{w}_{\textsc{i}}^{s},\bar{w}_{\textsc{i}}^{s-1})f(\bar{w}_{\textsc{i}}^{s},\bar{w}_{\textsc{ii}}^{s-1})}\prod_{s=j}^{N-1}\frac{f(\bar{w}_{\textsc{ii}}^{s},\bar{w}_{\textsc{iii}}^{s})}{h(\bar{w}_{\textsc{iii}}^{s+1},\bar{w}_{\textsc{iii}}^{s})f(\bar{w}_{\textsc{ii}}^{s+1},\bar{w}_{\textsc{iii}}^{s})}.
\end{multline}

We will also use the co-product formula of the off-shell Bethe vectors.
Let $\mathcal{H}^{(1)},\mathcal{H}^{(2)}$ be two quantum spaces for
which $\mathcal{H}=\mathcal{H}^{(1)}\otimes\mathcal{H}^{(2)}$ and
the corresponding off-shell states are $\tilde{\mathbb{B}}^{(1)}(\bar{t}),\tilde{\mathbb{B}}^{(2)}(\bar{t})$.
The co-product formula reads as\cite{Hutsalyuk:2016srn}
\begin{equation}
\tilde{\mathbb{B}}(\bar{t})=\sum_{\mathrm{part}(\bar{t})}\frac{\prod_{\nu=1}^{N-1}\lambda_{\nu}^{(2)}(\bar{t}_{\mathrm{i}}^{\nu})\lambda_{\nu+1}^{(1)}(\bar{t}_{\mathrm{ii}}^{\nu})f(\bar{t}_{\mathrm{ii}}^{\nu},\bar{t}_{\mathrm{i}}^{\nu})}{\prod_{\nu=1}^{N-2}f(\bar{t}_{\mathrm{ii}}^{\nu+1},\bar{t}_{\mathrm{i}}^{\nu})}\mathbb{\tilde{\mathbb{B}}}^{(1)}(\bar{t}_{\mathrm{i}})\otimes\tilde{\mathbb{B}}^{(2)}(\bar{t}_{\mathrm{ii}}),\label{eq:coproduct}
\end{equation}
where the sum goes over all the possible partitions $\bar{t}^{\nu}=\bar{t}_{\mathrm{i}}^{\nu}\cup\bar{t}_{\mathrm{ii}}^{\nu}$
where $\bar{t}_{\mathrm{i}}^{\nu},\bar{t}_{\mathrm{ii}}^{\nu}$ are
disjoint subsets.

We also need the action of the twisted monodromy matrix on the off-shell
Bethe vectors. Combining the identity (\ref{eq:connBhatB}) with the
action formula (\ref{eq:act}) we can obtain the following action
formula
\begin{multline}
\widehat{T}_{i,j}(z)\mathbb{B}(\bar{t})=(-1)^{\#\bar{t}}\prod_{s=2}^{N-1}f(\bar{t}^{s-1}-c,\bar{t}^{s})\widehat{T}_{i,j}(z)\hat{\mathbb{B}}(\mu^{-1}(\bar{t}))=\\
(-1)^{\#\bar{t}}\prod_{s=2}^{N-1}f(\bar{t}^{s-1}-c,\bar{t}^{s})\hat{\lambda}_{N}(z)\sum_{\mathrm{part}(\bar{w})}\hat{\mathbb{B}}(\mu^{-1}(\bar{w}_{\textsc{ii}}))\frac{\prod_{s=N-i+1}^{N-j}f(\bar{w}_{\textsc{i}}^{s},\bar{w}_{\textsc{iii}}^{s})}{\prod_{s=N-i+2}^{N-j}f(\bar{w}_{\textsc{i}}^{s-1}-c,\bar{w}_{\textsc{iii}}^{s})}\times\\
\prod_{s=N-i+1}^{N-1}\frac{f(\bar{w}_{\textsc{i}}^{s},\bar{w}_{\textsc{ii}}^{s})}{h(\bar{w}_{\textsc{i}}^{s},\bar{w}_{\textsc{i}}^{s+1}+c)f(\bar{w}_{\textsc{i}}^{s},\bar{w}_{\textsc{ii}}^{s+1}+c)}\prod_{s=1}^{N-j}\frac{\hat{\alpha}_{N-s}(\bar{w}_{\textsc{iii}}^{s}+sc)f(\bar{w}_{\textsc{ii}}^{s},\bar{w}_{\textsc{iii}}^{s})}{h(\bar{w}_{\textsc{iii}}^{s-1}-c,\bar{w}_{\textsc{iii}}^{s})f(\bar{w}_{\textsc{ii}}^{s-1}-c,\bar{w}_{\textsc{iii}}^{s})},
\end{multline}
where $\bar{w}^{s}=\left\{ z-sc,\bar{t}^{s}\right\} $. Applying the
identity (\ref{eq:connBhatB}) again we obtain that
\begin{multline}
\widehat{T}_{i,j}(z)\mathbb{B}(\bar{t})=(-1)^{i-j}\hat{\lambda}_{N}(z)\sum_{\mathrm{part}(\bar{w})}\mathbb{B}(\bar{w}_{\textsc{ii}})\frac{\prod_{s=2}^{N-1}f(\bar{t}^{s-1}-c,\bar{t}^{s})}{\prod_{s=2}^{N-1}f(\bar{w}_{\textsc{ii}}^{s-1}-c,\bar{w}_{\textsc{ii}}^{s})}\frac{\prod_{s=N-i+1}^{N-j}f(\bar{w}_{\textsc{i}}^{s},\bar{w}_{\textsc{iii}}^{s})}{\prod_{s=N-i+2}^{N-j}f(\bar{w}_{\textsc{i}}^{s-1}-c,\bar{w}_{\textsc{iii}}^{s})}\times\\
\prod_{s=N-i+1}^{N-1}\frac{f(\bar{w}_{\textsc{i}}^{s},\bar{w}_{\textsc{ii}}^{s})}{h(\bar{w}_{\textsc{i}}^{s},\bar{w}_{\textsc{i}}^{s+1}+c)f(\bar{w}_{\textsc{i}}^{s},\bar{w}_{\textsc{ii}}^{s+1}+c)}\prod_{s=1}^{N-j}\frac{\alpha_{s}(\bar{w}_{\textsc{iii}}^{s})f(\bar{w}_{\textsc{ii}}^{s},\bar{w}_{\textsc{iii}}^{s})}{h(\bar{w}_{\textsc{iii}}^{s-1}-c,\bar{w}_{\textsc{iii}}^{s})f(\bar{w}_{\textsc{ii}}^{s-1}-c,\bar{w}_{\textsc{iii}}^{s})},\label{eq:actTw}
\end{multline}
where $\bar{w}^{\nu}=\{z-sc,\bar{t}^{\nu}\}$. The sum goes over all
the partition of $\bar{w}^{\nu}=\bar{w}_{\textsc{i}}^{\nu}\cup\bar{w}_{\textsc{ii}}^{\nu}\cup\bar{w}_{\textsc{iii}}^{\nu}$
where $\bar{w}_{\textsc{i}}^{\nu},\bar{w}_{\textsc{ii}}^{\nu},\bar{w}_{\textsc{iii}}^{\nu}$
are disjoint sets for a fixed $\nu$ and $\#\bar{w}_{\textsc{i}}^{\nu}=\Theta(\nu-N-1+i)$,
$\#\bar{w}_{\textsc{iii}}^{\nu}=\Theta(N-j-\nu)$. We also set $\bar{w}_{\textsc{iii}}^{0}=\{z\}$,
$\bar{w}_{\textsc{i}}^{N}=\{z-Nc\}$ and $\bar{w}_{\textsc{i}}^{0}=\bar{w}_{\textsc{ii}}^{0}=\bar{w}_{\textsc{ii}}^{N}=\bar{w}_{\textsc{iii}}^{N}=\emptyset$.

We can see that the diagonal elements $\widehat{T}_{i,i}(u)$ do not
change the quantum numbers $r_{j}$. The creation operators i.e. $\widehat{T}_{i,j}(u)$
where $i<j$ increase the quantum numbers $r_{N-j+1},r_{N-j+2},\dots,r_{N-i}$
by one. The annihilation operators i.e. $\widehat{T}_{j,i}(u)$ where
$i<j$ decrease the quantum numbers $r_{N-j+1},r_{N-j+2},\dots,r_{N-i}$
by one.

It is easy to show that
\begin{equation}
\begin{split}\widehat{T}_{i+1,j+1}^{N}(z)\mathbb{B}^{N}(\left\{ \bar{t}^{k}\right\} _{k=1}^{N-2},\emptyset) & =\mathfrak{i}^{-}(\widehat{T}_{i,j}^{N-1}(z)\mathbb{B}^{N-1}(\left\{ \bar{t}^{k}\right\} _{k=1}^{N-2})),\\
\widehat{T}_{i,j}^{N}(z)\mathbb{B}^{N}(\emptyset,\left\{ \bar{t}^{k}\right\} _{k=2}^{N-1}) & =\mathfrak{i}^{+}(\widehat{T}_{i,j}^{N-1}(z-c)\mathbb{B}^{N-1}(\left\{ \bar{t}^{k}\right\} _{k=1}^{N-2})),
\end{split}
\end{equation}
where $i,j=1,\dots,N-1$.

\section{Elementary overlaps\label{sec:Elementary}}

In this section we calculate some explicit overlaps. We concentrate
on the cases where the numbers of Bethe roots are one or two.

\subsection{Untwisted case}

Let us start with the untwisted case. In this subsection the calculations
are based on the components $(N,2)$ and $(N-1,1)$ of the $KT$-relation
(\ref{eq:KT_Utw})
\begin{multline}
\langle\Psi|T_{1,2}(u)=\frac{K_{N,2}(u)}{K_{N,1}(u)}\left(\langle\Psi|T_{N,N}(-u)-\langle\Psi|T_{2,2}(u)\right)+\\
\sum_{j=1}^{N-1}\frac{K_{j,2}(u)}{K_{N,1}(u)}\langle\Psi|T_{N,j}(-u)-\sum_{i=3}^{N}\frac{K_{N,i}(u)}{K_{N,1}(u)}\langle\Psi|T_{i,2}(u),\label{eq:KT12}
\end{multline}
and
\begin{multline}
\langle\Psi|T_{N-1,N}(u)=\frac{K_{N-1,1}(-u)}{K_{N,1}(-u)}\left(\langle\Psi|T_{1,1}(-u)-\langle\Psi|T_{N-1,N-1}(u)\right)+\\
\sum_{i=2}^{N}\frac{K_{N-1,i}(-u)}{K_{N,1}(-u)}\langle\Psi|T_{i,1}(-u)-\sum_{j=1}^{N-2}\frac{K_{j,1}(-u)}{K_{N,1}(-u)}\langle\Psi|T_{N-1,j}(u).\label{eq:KTN1}
\end{multline}
We can see that these equations can be used to change the creation
operators $T_{1,2}(u),T_{N-1,N}(u)$ to diagonal or annihilation operators.

\subsubsection*{$N=2$}

For $N=2$ the $K$-matrix can be parameterized as (\ref{eq:Kdef_UTw})
\begin{equation}
K(u)=\frac{\mathfrak{a}}{u}+\mathcal{U}=\left(\begin{array}{cc}
\frac{\mathfrak{a}}{u}+\mathcal{U}_{1,1} & \mathcal{U}_{1,2}\\
\mathcal{U}_{2,1} & \frac{\mathfrak{a}}{u}+\mathcal{U}_{2,2}
\end{array}\right).
\end{equation}
At first let us calculate the following overlap
\begin{equation}
\langle\Psi|T_{1,2}(u)\left|0\right\rangle =\frac{K_{2,2}(u)}{K_{2,1}(u)}\langle\Psi|\left(T_{2,2}(-u)-T_{2,2}(u)\right)\left|0\right\rangle =\frac{K_{2,2}(u)}{K_{2,1}(u)}(\lambda_{1}(u)-\lambda_{2}(u)),\label{eq:onemagnonUTw}
\end{equation}
where we used the $KT$-relation (\ref{eq:KT12}) and the symmetry
property (\ref{eq:constLamUTw}). Using the equation (\ref{eq:onemagnonUTw})
we obtain the following overlap
\begin{equation}
\langle\Psi|\mathbb{B}(\{t_{1}^{1}\})=\frac{K_{2,2}(t_{1}^{1})}{K_{2,1}(t_{1}^{1})}(\alpha_{1}(t_{1}^{1})-1)=\frac{\mathfrak{a}+\mathcal{U}_{2,2}t_{1}^{1}}{\mathcal{U}_{2,1}t_{1}^{1}}(\alpha_{1}(t_{1}^{1})-1).
\end{equation}
We can see that this overlap vanishes taking the on-shell limit $\alpha_{1}(t_{1}^{1})=1$.
We can get non-vanishing on-shell overlap if we take $t_{1}^{1}=0$
for which the overlap simplifies as
\begin{equation}
\langle\Psi|\mathbb{B}(\{0\})=\frac{\mathfrak{a}}{\mathcal{U}_{2,1}}\alpha'_{1}(0).
\end{equation}
Let us continue with the $r_{1}=2$ case for which the Bethe vector
is
\begin{equation}
\mathbb{B}(\{t_{1}^{1},t_{2}^{1}\})=\frac{T_{1,2}(t_{1}^{1})}{\lambda_{2}(t_{1}^{1})}\frac{T_{1,2}(t_{2}^{1})}{\lambda_{2}(t_{2}^{1})}\left|0\right\rangle .
\end{equation}
Let us calculate the overlap
\begin{equation}
\langle\Psi|\mathbb{B}(\{t_{1}^{1},t_{2}^{1}\})=\frac{K_{2,2}(t_{1}^{1})}{K_{2,1}(t_{1}^{1})}\langle\Psi|\left(\frac{T_{2,2}(-t_{1}^{1})}{\lambda_{2}(t_{1}^{1})}-\frac{T_{2,2}(t_{1}^{1})}{\lambda_{2}(t_{1}^{1})}\right)\frac{T_{1,2}(t_{2}^{1})}{\lambda_{2}(t_{2}^{1})}\left|0\right\rangle +\frac{K_{1,2}(t_{1}^{1})}{K_{2,1}(t_{1}^{1})}\langle\Psi|\frac{T_{2,1}(-t_{1}^{1})}{\lambda_{2}(t_{1}^{1})}\frac{T_{1,2}(t_{2}^{1})}{\lambda_{2}(t_{2}^{1})}\left|0\right\rangle ,
\end{equation}
where we used the $KT$-equation (\ref{eq:KT12}). Using the $RTT$-relations
\begin{equation}
\begin{split}T_{2,2}(u)T_{1,2}(v) & =f(u,v)T_{1,2}(v)T_{2,2}(u)-g(u,v)T_{1,2}(u)T_{2,2}(v),\\
T_{2,1}(u)T_{1,2}(v) & =T_{1,2}(v)T_{2,1}(u)+g(u,v)\left(T_{1,1}(v)T_{2,2}(u)-T_{1,1}(u)T_{2,2}(v)\right),
\end{split}
\end{equation}
we obtain that
\begin{multline}
\langle\Psi|\mathbb{B}(\{t_{1}^{1},t_{2}^{1}\})=\\
\frac{K_{2,2}(t_{1}^{1})}{K_{2,1}(t_{1}^{1})}\left(\frac{K_{2,2}(t_{2}^{1})}{K_{2,1}(t_{2}^{1})}f(-t_{1}^{1},t_{2}^{1})\alpha_{1}(t_{1}^{1})(\alpha_{1}(t_{2}^{1})-1)-\frac{K_{2,2}(-t_{1}^{1})}{K_{2,1}(-t_{1}^{1})}g(-t_{1}^{1},t_{2}^{1})(1-\alpha_{1}(t_{1}^{1}))\right)-\\
-\frac{K_{2,2}(t_{1}^{1})}{K_{2,1}(t_{1}^{1})}\left(\frac{K_{2,2}(t_{2}^{1})}{K_{2,1}(t_{2}^{1})}f(t_{1}^{1},t_{2}^{1})(\alpha_{1}(t_{2}^{1})-1)-\frac{K_{2,2}(t_{1}^{1})}{K_{2,1}(t_{1}^{1})}g(t_{1}^{1},t_{2}^{1})(\alpha_{1}(t_{1}^{1})-1)\right)+\\
\frac{K_{1,2}(t_{1}^{1})}{K_{2,1}(t_{1}^{1})}g(-t_{1}^{1},t_{2}^{1})(\alpha_{1}(t_{1}^{1})\alpha_{1}(t_{2}^{1})-1),
\end{multline}
where we used the one particle overlap formula (\ref{eq:onemagnonUTw})
and the symmetry property (\ref{eq:constLamUTw}). Let us collect
the $\alpha$ terms.
\begin{multline}
\langle\Psi|\mathbb{B}(\{t_{1}^{1},t_{2}^{1}\})=\left(\frac{K_{2,2}(t_{1}^{1})}{K_{2,1}(t_{1}^{1})}\frac{K_{2,2}(t_{2}^{1})}{K_{2,1}(t_{2}^{1})}f(-t_{1}^{1},t_{2}^{1})+\frac{K_{1,2}(t_{1}^{1})}{K_{2,1}(t_{1}^{1})}g(-t_{1}^{1},t_{2}^{1})\right)\alpha_{1}(t_{2}^{1})\alpha_{1}(t_{1}^{1})-\\
-\frac{K_{2,2}(t_{1}^{1})}{K_{2,1}(t_{1}^{1})}\frac{K_{2,2}(t_{2}^{1})}{K_{2,1}(t_{2}^{1})}f(t_{1}^{1},t_{2}^{1})\alpha_{1}(t_{2}^{1})-\\
\left(\frac{K_{2,2}(t_{1}^{1})}{K_{2,1}(t_{1}^{1})}\frac{K_{2,2}(t_{2}^{1})}{K_{2,1}(t_{2}^{1})}f(-t_{1}^{1},t_{2}^{1})-\frac{K_{2,2}(t_{1}^{1})}{K_{2,1}(t_{1}^{1})}\left(\frac{K_{2,2}(-t_{1}^{1})}{K_{2,1}(-t_{1}^{1})}g(-t_{1}^{1},t_{2}^{1})+\frac{K_{2,2}(t_{1}^{1})}{K_{2,1}(t_{1}^{1})}g(t_{1}^{1},t_{2}^{1})\right)\right)\alpha_{1}(t_{1}^{1})+\\
+\left(\frac{K_{2,2}(t_{1}^{1})}{K_{2,1}(t_{1}^{1})}\left(\frac{K_{2,2}(t_{2}^{1})}{K_{2,1}(t_{2}^{1})}f(t_{1}^{1},t_{2}^{1})-\left(\frac{K_{2,2}(-t_{1}^{1})}{K_{2,1}(-t_{1}^{1})}g(-t_{1}^{1},t_{2}^{1})+\frac{K_{2,2}(t_{1}^{1})}{K_{2,1}(t_{1}^{1})}g(t_{1}^{1},t_{2}^{1})\right)\right)-\frac{K_{1,2}(t_{1}^{1})}{K_{2,1}(t_{1}^{1})}g(-t_{1}^{1},t_{2}^{1})\right).
\end{multline}
Using the explicit parametrization of the $K$-matrix we can obtain
the following identity
\begin{equation}
\frac{K_{2,2}(-t_{1}^{1})}{K_{2,1}(-t_{1}^{1})}g(-t_{1}^{1},t_{2}^{1})+\frac{K_{2,2}(t_{1}^{1})}{K_{2,1}(t_{1}^{1})}g(t_{1}^{1},t_{2}^{1})=\frac{K_{2,2}(t_{2}^{1})}{K_{2,1}(t_{2}^{1})}\left(g(-t_{1}^{1},t_{2}^{1})+g(t_{1}^{1},t_{2}^{1})\right).
\end{equation}
Substituting back, the overlap simplifies as
\begin{align}
\langle\Psi|\mathbb{B}(\{t_{1}^{1},t_{2}^{1}\}) & =\left(\frac{K_{2,2}(t_{1}^{1})}{K_{2,1}(t_{1}^{1})}\frac{K_{2,2}(t_{2}^{1})}{K_{2,1}(t_{2}^{1})}f(-t_{1}^{1},t_{2}^{1})+\frac{K_{1,2}(t_{1}^{1})}{K_{2,1}(t_{1}^{1})}g(-t_{1}^{1},t_{2}^{1})\right)\alpha_{1}(t_{2}^{1})\alpha_{1}(t_{1}^{1})-\nonumber \\
 & -\frac{K_{2,2}(t_{1}^{1})}{K_{2,1}(t_{1}^{1})}\frac{K_{2,2}(t_{2}^{1})}{K_{2,1}(t_{2}^{1})}f(t_{1}^{1},t_{2}^{1})\alpha_{1}(t_{2}^{1})-\frac{K_{2,2}(t_{1}^{1})}{K_{2,1}(t_{1}^{1})}\frac{K_{2,2}(t_{2}^{1})}{K_{2,1}(t_{2}^{1})}f(-t_{1}^{1},-t_{2}^{1})\alpha_{1}(t_{1}^{1})+\nonumber \\
 & +\left(\frac{K_{2,2}(t_{1}^{1})}{K_{2,1}(t_{1}^{1})}\frac{K_{2,2}(t_{2}^{1})}{K_{2,1}(t_{2}^{1})}f(t_{1}^{1},-t_{2}^{1})+\frac{K_{1,2}(t_{1}^{1})}{K_{2,1}(t_{1}^{1})}g(t_{1}^{1},-t_{2}^{1})\right).
\end{align}
We can check that if we first take the on-shell limit
\begin{equation}
\alpha_{1}(t_{1}^{1})=\frac{f(t_{1}^{1},t_{2}^{1})}{f(t_{2}^{1},t_{1}^{1})},\qquad\alpha_{1}(t_{2}^{1})=\frac{f(t_{2}^{1},t_{1}^{1})}{f(t_{1}^{1},t_{2}^{1})},
\end{equation}
then the overlap is vanishing. For non-vanishing overlap we have to
take the $t_{2}^{1}\to-t_{1}^{1}$ limit first:
\begin{multline}
\langle\Psi|\mathbb{B}(\{t_{1}^{1},-t_{1}^{1}\})=\left(\frac{K_{1,2}(t_{1}^{1})}{K_{2,1}(t_{1}^{1})}-\frac{K_{2,2}(t_{1}^{1})}{K_{2,1}(t_{1}^{1})}\frac{K_{1,1}(t_{1}^{1})}{K_{2,1}(t_{1}^{1})}\right)X_{1}^{+,1}+\\
+\frac{K_{2,2}(t_{1}^{1})}{K_{2,1}(t_{1}^{1})}\frac{K_{1,1}(t_{1}^{1})}{K_{2,1}(t_{1}^{1})}\left(\frac{f(t_{1}^{1},-t_{1}^{1})}{\alpha_{1}(t_{1}^{1})}+f(-t_{1}^{1},t_{1}^{1})\alpha_{1}(t_{1}^{1})-2\right),
\end{multline}
where we used the symmetry property (\ref{eq:constAlpUTw}) and the
identity
\begin{equation}
\frac{K_{2,2}(-u)}{K_{2,1}(-u)}=-\frac{K_{1,1}(u)}{K_{2,1}(u)},
\end{equation}
and we introduced the notation
\begin{equation}
X_{1}^{+,1}=-c\frac{d}{du}\log\alpha_{1}(u)\Biggr|_{u=t_{1}^{1}}.
\end{equation}
Now we can take the on-shell limit therefore we just obtained the
non-vanishing on-shell formula
\begin{equation}
\langle\Psi|\mathbb{B}(\{t_{1}^{1},-t_{1}^{1}\})=F_{K}^{(1)}(t_{1}^{1})X_{1}^{+,1},
\end{equation}
where we used the notation (\ref{eq:defFK}) for $N=2$.

\subsubsection*{$N=3$}

We can do the same calculation for $N=3$. Let us start with the overlap
\begin{equation}
\langle\Psi|T_{2,3}(u)\left|0\right\rangle =\frac{K_{2,1}(-u)}{K_{3,1}(-u)}\langle\Psi|T_{1,1}(-u)-T_{2,2}(u)\left|0\right\rangle =\frac{K_{2,1}(-u)}{K_{3,1}(-u)}(\lambda_{3}(u)-\lambda_{2}(u)),\label{eq:onemagnonN3}
\end{equation}
where we used the $KT$-relation (\ref{eq:KTN1}) and the symmetry
property (\ref{eq:constLamUTw}). Let us continue with the state $r_{1}=r_{2}=1$
which has the following form
\begin{equation}
\mathbb{B}(\{t_{1}^{1}\},\{t_{1}^{2}\})=\frac{1}{f(t_{1}^{2},t_{1}^{1})}\left(\frac{T_{1,2}(t_{1}^{1})}{\lambda_{2}(t_{1}^{1})}\frac{T_{2,3}(t_{1}^{2})}{\lambda_{3}(t_{1}^{2})}+g(t_{1}^{2},t_{1}^{1})\alpha_{2}(t_{1}^{2})\frac{T_{1,3}(t_{1}^{1})}{\lambda_{2}(t_{1}^{1})}\right)\left|0\right\rangle .
\end{equation}
Let us calculate the overlap
\begin{multline}
\langle\Psi|\frac{T_{1,2}(t_{1}^{1})}{\lambda_{2}(t_{1}^{1})}\frac{T_{2,3}(t_{1}^{2})}{\lambda_{3}(t_{1}^{2})}\left|0\right\rangle =\frac{K_{3,2}(t_{1}^{1})}{K_{3,1}(t_{1}^{1})}\langle\Psi|\left(\frac{T_{3,3}(-t_{1}^{1})}{\lambda_{2}(t_{1}^{1})}-\frac{T_{2,2}(t_{1}^{1})}{\lambda_{2}(t_{1}^{1})}\right)\frac{T_{2,3}(t_{1}^{2})}{\lambda_{3}(t_{1}^{2})}\left|0\right\rangle +\\
\frac{K_{2,2}(t_{1}^{1})}{K_{3,1}(t_{1}^{1})}\langle\Psi|\frac{T_{3,2}(-t_{1}^{1})}{\lambda_{2}(t_{1}^{1})}\frac{T_{2,3}(t_{1}^{2})}{\lambda_{3}(t_{1}^{2})}\left|0\right\rangle -\frac{K_{3,3}(t_{1}^{1})}{K_{3,1}(t_{1}^{1})}\langle\Psi|\frac{T_{3,2}(t_{1}^{1})}{\lambda_{2}(t_{1}^{1})}\frac{T_{2,3}(t_{1}^{2})}{\lambda_{3}(t_{1}^{2})}\left|0\right\rangle ,
\end{multline}
where we used the $KT$-relation (\ref{eq:KT12}). Using the $RTT$-relations
\begin{align}
T_{3,3}(u)T_{2,3}(v) & =f(u,v)T_{2,3}(v)T_{3,3}(u)-g(u,v)T_{2,3}(u)T_{3,3}(v),\nonumber \\
T_{2,2}(u)T_{2,3}(v) & =f(v,u)T_{2,3}(v)T_{2,2}(u)+g(u,v)T_{2,3}(u)T_{2,2}(v),\\
T_{3,2}(u)T_{2,3}(v) & =T_{2,3}(v)T_{3,2}(u)+g(u,v)\left(T_{2,2}(v)T_{3,3}(u)-T_{2,2}(u)T_{3,3}(v)\right),\nonumber 
\end{align}
we obtain that
\begin{multline}
\langle\Psi|\frac{T_{1,2}(t_{1}^{1})}{\lambda_{2}(t_{1}^{1})}\frac{T_{2,3}(t_{1}^{2})}{\lambda_{3}(t_{1}^{2})}\left|0\right\rangle =\\
\frac{K_{3,2}(t_{1}^{1})}{K_{3,1}(t_{1}^{1})}\left(\frac{K_{2,1}(-t_{1}^{2})}{K_{3,1}(-t_{1}^{2})}f(-t_{1}^{1},t_{1}^{2})\alpha_{1}(t_{1}^{1})(1-\alpha_{2}(t_{1}^{2}))-\frac{K_{2,1}(t_{1}^{1})}{K_{3,1}(t_{1}^{1})}g(-t_{1}^{1},t_{1}^{2})(\alpha_{1}(t_{1}^{1})-1)\right)-\\
-\frac{K_{3,2}(t_{1}^{1})}{K_{3,1}(t_{1}^{1})}\left(\frac{K_{2,1}(-t_{1}^{2})}{K_{3,1}(-t_{1}^{2})}f(t_{1}^{2},t_{1}^{1})(1-\alpha_{2}(t_{1}^{2}))+\frac{K_{2,1}(-t_{1}^{1})}{K_{3,1}(-t_{1}^{1})}g(t_{1}^{1},t_{1}^{2})\alpha_{2}(t_{1}^{2})(\frac{1}{\alpha_{2}(t_{1}^{1})}-1)\right)+\\
+\frac{K_{2,2}(t_{1}^{1})}{K_{3,1}(t_{1}^{1})}g(-t_{1}^{1},t_{1}^{2})(\alpha_{1}(t_{1}^{1})\alpha_{2}(t_{1}^{2})-1)-\frac{K_{3,3}(t_{1}^{1})}{K_{3,1}(t_{1}^{1})}g(t_{1}^{1},t_{1}^{2})(\alpha_{2}(t_{1}^{2})\frac{1}{\alpha_{2}(t_{1}^{1})}-1),
\end{multline}
where we used the one-magnon overlap (\ref{eq:onemagnonN3}) and the
symmetry property (\ref{eq:constLamUTw}). Let us continue with the
overlap
\begin{multline}
\langle\Psi|\frac{T_{1,3}(t_{1}^{1})}{\lambda_{2}(t_{1}^{1})}\left|0\right\rangle =\frac{K_{3,3}(t_{1}^{1})}{K_{3,1}(t_{1}^{1})}\langle\Psi|\left(\frac{T_{3,3}(-t_{1}^{1})}{\lambda_{2}(t_{1}^{1})}-\frac{T_{3,3}(t_{1}^{1})}{\lambda_{2}(t_{1}^{1})}\right)\left|0\right\rangle -\frac{K_{3,2}(t_{1}^{1})}{K_{3,1}(t_{1}^{1})}\langle\Psi|\frac{T_{2,3}(t_{1}^{1})}{\lambda_{2}(t_{1}^{1})}\left|0\right\rangle =\\
\frac{K_{3,3}(t_{1}^{1})}{K_{3,1}(t_{1}^{1})}\left(\alpha_{1}(t_{1}^{1})-\frac{1}{\alpha_{2}(t_{1}^{1})}\right)-\frac{K_{3,2}(t_{1}^{1})}{K_{3,1}(t_{1}^{1})}\frac{K_{2,1}(-t_{1}^{1})}{K_{3,1}(-t_{1}^{1})}\left(\frac{1}{\alpha_{2}(t_{1}^{1})}-1\right),
\end{multline}
where we used the component $(3,1)$ of the $KT$-relation and the
symmetry property (\ref{eq:constLamUTw}). Substituting back, we obtain
that
\begin{multline}
f(t_{1}^{2},t_{1}^{1})\langle\Psi|\mathbb{B}(\{t_{1}^{1}\},\{t_{1}^{2}\})=\\
\left(\frac{K_{2,2}(t_{1}^{1})}{K_{3,1}(t_{1}^{1})}g(-t_{1}^{1},t_{1}^{2})-\frac{K_{3,2}(t_{1}^{1})}{K_{3,1}(t_{1}^{1})}\frac{K_{2,1}(-t_{1}^{2})}{K_{3,1}(-t_{1}^{2})}f(-t_{1}^{1},t_{1}^{2})+\frac{K_{3,3}(t_{1}^{1})}{K_{3,1}(t_{1}^{1})}g(t_{1}^{2},t_{1}^{1})\right)\alpha_{1}(t_{1}^{1})\alpha_{2}(t_{1}^{2})+\\
+\frac{K_{3,2}(t_{1}^{1})}{K_{3,1}(t_{1}^{1})}\frac{K_{2,1}(t_{1}^{2})}{K_{3,1}(t_{1}^{2})}\alpha_{1}(t_{1}^{1})+\left(\frac{K_{3,2}(t_{1}^{1})}{K_{3,1}(t_{1}^{1})}\frac{K_{2,1}(-t_{1}^{2})}{K_{3,1}(-t_{1}^{2})}f(t_{1}^{2},t_{1}^{1})\right)\alpha_{2}(t_{1}^{2})-\\
\left(\frac{K_{2,2}(t_{1}^{1})}{K_{3,1}(t_{1}^{1})}g(-t_{1}^{1},t_{1}^{2})+\frac{K_{3,3}(t_{1}^{1})}{K_{3,1}(t_{1}^{1})}g(t_{1}^{2},t_{1}^{1})+\frac{K_{3,2}(t_{1}^{1})}{K_{3,1}(t_{1}^{1})}\frac{K_{2,1}(-t_{1}^{2})}{K_{3,1}(-t_{1}^{2})}\left(f(t_{1}^{2},t_{1}^{1})-g(-t_{1}^{1},t_{1}^{2})\right)\right).
\end{multline}
We emphasize that the ratio of two non-diagonal matrix element of
the $K$-matrix is independent on argument (see the explicit form
(\ref{eq:Kdef_UTw})). We can check that the overlap vanishing in
the on-shell limit
\begin{equation}
\alpha_{1}(t_{1}^{1})=f(t_{1}^{2},t_{1}^{1}),\qquad\alpha_{2}(t_{1}^{2})=\frac{1}{f(t_{1}^{2},t_{1}^{1})}.
\end{equation}
For a non-vanishing overlap, at first we have to take the $t_{1}^{2}\to-t_{1}^{1}$
limit
\begin{multline}
f(-t_{1}^{1},t_{1}^{1})\langle\Psi|\mathbb{B}(\{t_{1}^{1}\},\{-t_{1}^{1}\})=\left(\frac{K_{2,2}(t_{1}^{1})}{K_{3,1}(t_{1}^{1})}-\frac{K_{3,2}(t_{1}^{1})}{K_{3,1}(t_{1}^{1})}\frac{K_{2,1}(t_{1}^{1})}{K_{3,1}(t_{1}^{1})}\right)X_{1}^{+,1}+\\
+\frac{K_{3,2}(t_{1}^{1})}{K_{3,1}(t_{1}^{1})}\frac{K_{2,1}(t_{1}^{1})}{K_{3,1}(t_{1}^{1})}\left(\alpha_{1}(t_{1}^{1})+f(-t_{1}^{1},t_{1}^{1})\frac{1}{\alpha_{1}(t_{1}^{1})}-1-f(-t_{1}^{1},t_{1}^{1})\right),
\end{multline}
where we used the symmetry property (\ref{eq:constAlpUTw}). Taking
the on-shell limit we obtain that
\begin{equation}
\langle\Psi|\mathbb{B}(\{t_{1}^{1}\},\{-t_{1}^{1}\})=\frac{F_{K}^{(1)}(t_{1}^{1})}{f(-t_{1}^{1},t_{1}^{1})}X_{1}^{+,1},
\end{equation}
where we used the notation (\ref{eq:defFK}) for $N=3$.

\subsubsection*{General $N$}

We can also calculate these overlaps for $N>3$. Let us start with
the overlap
\begin{equation}
\langle\Psi|T_{N-1,N}(u)\left|0\right\rangle =\frac{K_{N-1,1}(-u)}{K_{N,1}(-u)}\langle\Psi|T_{1,1}(-u)-T_{N-1,N-1}(u)\left|0\right\rangle =\frac{K_{N-1,1}(-u)}{K_{N,1}(-u)}(\lambda_{N}(u)-\lambda_{N-1}(u)),\label{eq:onemagnonNN}
\end{equation}
where we used the equation (\ref{eq:KTN1}) and the symmetry property
(\ref{eq:constLamUTw}). Let us continue with the state $r_{1}=r_{N-1}=1$
which reads as
\begin{equation}
\mathbb{B}(\{t_{1}^{1}\},\emptyset,\left\{ t_{1}^{N-1}\right\} )=\left(\frac{T_{1,2}(t_{1}^{1})}{\lambda_{2}(t_{1}^{1})}\frac{T_{N-1,N}(t_{1}^{N-1})}{\lambda_{N}(t_{1}^{N-1})}\right)\left|0\right\rangle .
\end{equation}
Now the overlap is
\begin{multline}
\langle\Psi|\mathbb{B}(\{t_{1}^{1}\},\emptyset,\left\{ t_{1}^{N-1}\right\} )=\frac{K_{N,2}(t_{1}^{1})}{K_{N,1}(t_{1}^{1})}\langle\Psi|\left(\frac{T_{N,N}(-t_{1}^{1})}{\lambda_{2}(t_{1}^{1})}-\frac{T_{2,2}(t_{1}^{1})}{\lambda_{2}(t_{1}^{1})}\right)\frac{T_{N-1,N}(t_{1}^{N-1})}{\lambda_{N}(t_{1}^{N-1})}\left|0\right\rangle +\\
+\frac{K_{N-1,2}(t_{1}^{1})}{K_{N,1}(t_{1}^{1})}\langle\Psi|\frac{T_{N,N-1}(-t_{1}^{1})}{\lambda_{2}(t_{1}^{1})}\frac{T_{N-1,N}(t_{1}^{N-1})}{\lambda_{N}(t_{1}^{N-1})}\left|0\right\rangle ,
\end{multline}
where we used the KT-relation (\ref{eq:KT12}). Using the $RTT$-relations
\begin{align}
T_{N,N}(u)T_{N-1,N}(v) & =f(u,v)T_{N-1,N}(v)T_{N,N}(u)-g(u,v)T_{N-1,N}(u)T_{N,N}(v),\nonumber \\
\left[T_{N-1,N}(u),T_{2,2}(v)\right] & =g(u,v)\left(T_{2,N}(v)T_{N-1,2}(u)-T_{2,N}(u)T_{N-1,2}(v)\right),\\
\left[T_{N,N-1}(u),T_{N-1,N}(v)\right] & =g(u,v)\left(T_{N-1,N-1}(v)T_{N,N}(u)-T_{N-1,N-1}(u)T_{N,N}(v)\right),\nonumber 
\end{align}
we obtain that
\begin{multline}
\langle\Psi|\mathbb{B}(\{t_{1}^{1}\},\emptyset,\left\{ t_{1}^{N-1}\right\} )=\\
\left(\frac{K_{N-1,2}(t_{1}^{1})}{K_{N,1}(t_{1}^{1})}g(-t_{1}^{1},t_{1}^{N-1})-\frac{K_{N,2}(t_{1}^{1})}{K_{N,1}(t_{1}^{1})}\frac{K_{N-1,1}(-t_{1}^{N-1})}{K_{N,1}(-t_{1}^{N-1})}f(-t_{1}^{1},t_{1}^{N-1})\right)\left(\alpha_{1}(t_{1}^{1})\alpha_{N-1}(t_{1}^{N-1})-1\right)+\\
+\frac{K_{N,2}(t_{1}^{1})}{K_{N,1}(t_{1}^{1})}\frac{K_{N-1,1}(-t_{1}^{N-1})}{K_{N,1}(-t_{1}^{N-1})}\left(\alpha_{1}(t_{1}^{1})-\alpha_{N-1}(t_{1}^{N-1})\right),
\end{multline}
where we used the one-magnon overlap (\ref{eq:onemagnonNN}), the
symmetry property (\ref{eq:constLamUTw}) and the identity
\begin{equation}
\frac{K_{N-1,1}(u)}{K_{N,1}(u)}=\frac{K_{N-1,1}(v)}{K_{N,1}(v)},
\end{equation}
for $N>3$. We can see that the overlap is vanishing for the on-shell
limit
\begin{equation}
\alpha_{1}(t_{1}^{1})=1,\qquad\alpha_{1}(t_{1}^{N-1})=1.
\end{equation}
For a non-vanishing overlap, at first we have to take the $t_{1}^{N-1}\to-t_{1}^{1}$
limit.
\begin{multline}
\langle\Psi|\mathbb{B}(\{t_{1}^{1}\},\emptyset,\left\{ t_{1}^{N-1}\right\} )=\left(\frac{K_{N-1,2}(t_{1}^{1})}{K_{N,1}(t_{1}^{1})}-\frac{K_{N,2}(t_{1}^{1})}{K_{N,1}(t_{1}^{1})}\frac{K_{N-1,1}(t_{1}^{1})}{K_{N,1}(t_{1}^{1})}\right)X_{1}^{+,1}(t_{1}^{1})+\\
+\frac{K_{N,2}(t_{1}^{1})}{K_{N,1}(t_{1}^{1})}\frac{K_{N-1,1}(t_{1}^{1})}{K_{N,1}(t_{1}^{1})}\left(\alpha_{1}(t_{1}^{1})-1/\alpha_{1}(t_{1}^{1})\right),
\end{multline}
where we used the symmetry property (\ref{eq:constAlpUTw}). Now we
can take the on-shell limit 
\begin{equation}
\langle\Psi|\mathbb{B}(\{t_{1}^{1}\},\emptyset^{\times(N-3)},\left\{ -t_{1}^{1}\right\} )=F_{K}^{(1)}(t_{1}^{1})X_{1}^{+,1}.
\end{equation}
Using the embedding rule (\ref{eq:recursion_S_UTw}) we can also obtain
the following on-shell formulas
\begin{equation}
\langle\Psi|\mathbb{B}(\emptyset^{\times(k-1)},\{t_{1}^{k}\},\emptyset^{\times(N-1-2k)},\left\{ -t_{1}^{k}\right\} ,\emptyset^{\times(k-1)})=\tilde{F}_{K}^{(k)}(t_{1}^{k})X_{1}^{+,k},
\end{equation}
for $k<N/2$ and
\begin{equation}
\langle\Psi|\mathbb{B}(\emptyset^{\times(\frac{N}{2}-1)},\left\{ t_{1}^{\frac{N}{2}},-t_{1}^{\frac{N}{2}}\right\} ,\emptyset^{\times(\frac{N}{2}-1)})=F_{K}^{(\frac{N}{2})}(t_{1}^{\frac{N}{2}})X_{1}^{+,\frac{N}{2}},
\end{equation}
where we used the notation (\ref{eq:onepartUTw}).

\subsection{Twisted case}

As we saw in subsection \ref{subsec:Twisted--relation} our method
works only for the $\mathfrak{so}(N)$ symmetric $K$-matrix which
has explicit form (\ref{eq:twK}) therefore $K=VK^{T}V.$ For $N=2$
it means that $K_{1,1}=K_{2,2}$.

For $N=2$ the twisted $KT$-relation (\ref{eq:KT_Tw}) can be written
as
\begin{equation}
\langle\Psi|T_{1,2}(u)=\frac{K_{2,2}}{K_{2,1}}\left(\lambda_{0}(u)\langle\Psi|\widehat{T}_{2,2}(-u)-\langle\Psi|T_{2,2}(u)\right)+\frac{K_{1,2}}{K_{2,1}}\lambda_{0}(u)\langle\Psi|\widehat{T}_{2,1}(-u).\label{eq:KTN2}
\end{equation}
At first let us calculate the following overlaps for the twisted case
\begin{equation}
\langle\Psi|T_{1,2}(u)\left|0\right\rangle =\frac{K_{2,2}}{K_{2,1}}\langle\Psi|\left(\lambda_{0}(u)\widehat{T}_{2,2}(-u)-T_{2,2}(u)\right)\left|0\right\rangle =\frac{K_{2,2}}{K_{2,1}}(\lambda_{1}(u)-\lambda_{2}(u)),
\end{equation}
and
\begin{equation}
\lambda_{0}(-u)\langle\Psi|\widehat{T}_{1,2}(u)\left|0\right\rangle =\frac{K_{1,1}}{K_{2,1}}\langle\Psi|\left(T_{1,1}(-u)-\lambda_{0}(-u)\widehat{T}_{1,1}(u)\right)\left|0\right\rangle =\frac{K_{1,1}}{K_{2,1}}(\lambda_{1}(-u)-\lambda_{2}(-u)),
\end{equation}
where we used the $KT$-relation (\ref{eq:KTN2}) and the symmetry
property (\ref{eq:constLamTw}). The Bethe vector with $r_{1}=2$
is
\begin{equation}
\mathbb{B}(\{t_{1}^{1},t_{2}^{1}\})=\frac{T_{1,2}(t_{1}^{1})}{\lambda_{2}(t_{1}^{1})}\frac{T_{1,2}(t_{2}^{1})}{\lambda_{2}(t_{2}^{1})}\left|0\right\rangle .
\end{equation}
Let us calculate the overlap
\begin{multline}
\langle\Psi|\mathbb{B}(\{t_{1}^{1},t_{2}^{1}\})=\frac{K_{2,2}}{K_{2,1}}\langle\Psi|\left(\frac{\lambda_{0}(t_{1}^{1})\widehat{T}_{2,2}(-t_{1}^{1})}{\lambda_{2}(t_{1}^{1})}-\frac{T_{2,2}(t_{1}^{1})}{\lambda_{2}(t_{1}^{1})}\right)\frac{T_{1,2}(t_{2}^{1})}{\lambda_{2}(t_{2}^{1})}\left|0\right\rangle +\\
+\frac{K_{1,2}}{K_{2,1}}\langle\Psi|\frac{\lambda_{0}(t_{1}^{1})\widehat{T}_{2,1}(-t_{1}^{1})}{\lambda_{2}(t_{1}^{1})}\frac{T_{1,2}(t_{2}^{1})}{\lambda_{2}(t_{2}^{1})}\left|0\right\rangle ,
\end{multline}
where we used the $KT$-relation (\ref{eq:KTN2}). Using the $R\widehat{T}T$-relation
\begin{equation}
\begin{split}\widehat{T}_{2,2}(u)T_{1,2}(v) & =f(u-c,v)T_{1,2}(v)\widehat{T}_{2,2}(u)+g(u-c,v)\widehat{T}_{1,2}(u)T_{2,2}(v),\\
\widehat{T}_{21}(u)T_{12}(v) & =T_{1,2}(v)\widehat{T}_{2,1}(u)-g(u-c,v)\left(T_{1,1}(v)\widehat{T}_{2,2}(u)-\widehat{T}_{1,1}(u)T_{2,2}(v)\right),
\end{split}
\end{equation}
 we obtain that
\begin{multline}
\langle\Psi|\mathbb{B}(\{t_{1}^{1},t_{2}^{1}\})=\left(\left(\frac{K_{2,2}}{K_{2,1}}\right)^{2}f(-t_{1}^{1}-c,t_{2}^{1})-\frac{K_{1,2}}{K_{2,1}}g(-t_{1}^{1}-c,t_{2}^{1})\right)\alpha_{1}(t_{2}^{1})\alpha_{1}(t_{1}^{1})-\\
-\left(\frac{K_{2,2}}{K_{2,1}}\right)^{2}f(t_{1}^{1},t_{2}^{1})\alpha_{1}(t_{2}^{1})-\left(\frac{K_{2,2}}{K_{2,1}}\right)^{2}f(t_{2}^{1},t_{1}^{1})\alpha_{1}(t_{1}^{1})+\\
+\left(\left(\frac{K_{2,2}}{K_{2,1}}\right)^{2}f(t_{2}^{1},-t_{1}^{1}-c)-\frac{K_{1,2}}{K_{2,1}}g(t_{2}^{1},-t_{1}^{1}-c)\right),
\end{multline}
where we used the symmetry property (\ref{eq:constLamTw}) and the
equality $K_{1,1}=K_{2,2}$. We can see that the overlap is vanishing
for the on-shell limit
\begin{equation}
\alpha_{1}(t_{1}^{1})=\frac{f(t_{1}^{1},t_{2}^{1})}{f(t_{2}^{1},t_{1}^{1})},\qquad\alpha_{1}(t_{2}^{1})=\frac{f(t_{2}^{1},t_{1}^{1})}{f(t_{1}^{1},t_{2}^{1})}.
\end{equation}
For a non-vanishing overlap, at first we have to take the $t_{2}^{1}\to-t_{1}^{1}-c$
limit.
\begin{equation}
\langle\Psi|\mathbb{B}(\{t_{1}^{1},t_{2}^{1}\})=\left(\left(\frac{K_{2,2}}{K_{2,1}}\right)^{2}-\frac{K_{1,2}}{K_{2,1}}\right)X_{1}^{+,1}-\left(\frac{K_{2,2}}{K_{2,1}}\right)^{2}\left(\frac{f(t_{1}^{1},-t_{1}^{1}-c)}{\alpha_{1}(t_{1}^{1})}+f(-t_{1}^{1}-c,t_{1}^{1})\alpha_{1}(t_{1}^{1})-2\right),
\end{equation}
where we used the symmetry property (\ref{eq:constAlpTw}). Now we
can take the on-shell limit and we obtain that
\begin{equation}
\langle\Psi|\mathbb{B}(\{t_{1}^{1},t_{2}^{1}\})=F_{K}^{(1)}X_{1}^{+,1},
\end{equation}
where we used the notation (\ref{eq:onepartTw}). Using the embedding
rules (\ref{eq:recursion_S_Tw}) and (\ref{eq:recursion_S_Tw-1})
we can also obtain the following on-shell formulas
\begin{equation}
\langle\Psi|\mathbb{B}(\emptyset^{\times(k-1)},\{t_{1}^{k},-t_{1}^{k}-kc\},\emptyset^{\times(N-1-k)})=F_{K}^{(k)}X_{1}^{+,k}.
\end{equation}

\section{Derivation of the sum formula\label{sec:DSumFormula}}

In this section we derive the sum formula (\ref{eq:sumFormula}).
This formula is universal for both the untwisted and the twisted two-site
states but we derive it separately.

\subsection{Twisted case}

We start with the twisted case. We want to prove the sum formula 
\begin{equation}
S_{K}^{N}(\bar{t})=\sum_{\mathrm{part}(\bar{t})}W_{K}^{N}(\bar{t}_{\textsc{i}}|\bar{t}_{\textsc{ii}})\prod_{\nu=1}^{N-1}\alpha_{\nu}(\bar{t}_{\textsc{i}}^{\nu}),
\end{equation}
where the weights $W_{K}^{N}(\bar{t}_{\textsc{i}}|\bar{t}_{\textsc{ii}})$
do not depend on $\alpha$. In the derivation we only care about $\alpha$
dependence of the overlap therefore we use the notation $(\dots)$
for the $\alpha$ independent coefficients. Using this notation the
sum formula takes the form
\begin{equation}
S_{K}^{N}(\bar{t})=\sum_{\mathrm{part}(\bar{t})}\prod_{\nu=1}^{N-1}\alpha_{\nu}(\bar{t}_{\textsc{i}}^{\nu})(\dots).\label{eq:sumrule_Tw}
\end{equation}

The derivation based on an observation of the subsection \ref{subsec:EmbedTw}
where we saw that there exists an embedding rule for the overlap function:
\begin{equation}
S_{K}^{N}(\emptyset,\bar{t}^{2},\dots,\bar{t}^{N-1})=S_{K^{(2)}}^{N-1}(\bar{t}^{2}+c/2,\dots,\bar{t}^{N-1}+c/2)\Biggr|_{\alpha_{\nu}(u)\to\alpha_{\nu+1}(u-c/2)},
\end{equation}
therefore we can prove the sum rule (\ref{eq:sumrule_Tw}) by induction
over $N$.

Let us start with $N=2$. Using the explicit form of the Bethe vector
$\mathbb{B}(\bar{t}^{1})$ we obtain that
\begin{equation}
\langle\Psi|\mathbb{B}(\{z,\bar{t}^{1}\})=\langle\Psi|\frac{T_{1,2}(z)}{\lambda_{2}(z)}\mathbb{B}(\bar{t}^{1}).\label{eq:recN2-1}
\end{equation}
We can apply the twisted $KT$-relation (\ref{eq:KT_Tw})
\begin{equation}
K_{2,1}\left\langle \Psi\right|T_{1,2}(u)=\lambda_{0}(u)\left\langle \Psi\right|\widehat{T}_{2,2}(-u)K_{2,2}-K_{2,2}\left\langle \Psi\right|T_{2,2}(u)+\lambda_{0}(u)\left\langle \Psi\right|\widehat{T}_{2,1}(-u)K_{1,2}.\label{eq:KTtemp}
\end{equation}
Using the symmetry property (\ref{eq:constLamTw}), the action formulas
(\ref{eq:act}) and (\ref{eq:actTw}) of the operators of the r.h.s.
read as
\begin{equation}
\begin{split}\frac{\lambda_{0}(z)\widehat{T}_{2,2}(-z)}{\lambda_{2}(z)}\mathbb{B}(\bar{t}^{1}) & =\alpha_{1}(z)\sum_{\bar{\omega}^{1}=\bar{\omega}_{\textsc{i}}^{1}\cup\bar{\omega}_{\textsc{ii}}^{1}}\mathbb{B}(\bar{\omega}_{\textsc{ii}}^{1})\frac{f(\bar{\omega}_{\textsc{i}}^{1},\bar{\omega}_{\textsc{ii}}^{1})}{h(\bar{\omega}_{\textsc{i}}^{1},-z-c)},\\
\frac{\lambda_{0}(z)\widehat{T}_{2,1}(-z)}{\lambda_{2}(z)}\mathbb{B}(\bar{t}^{1}) & =\alpha_{1}(z)\sum_{\bar{\omega}^{1}=\bar{\omega}_{\textsc{i}}^{1}\cup\bar{\omega}_{\textsc{ii}}^{1}\cup\bar{\omega}_{\textsc{iii}}^{1}}\mathbb{B}(\bar{\omega}_{\textsc{ii}}^{1})\alpha_{1}(\bar{\omega}_{\textsc{iii}}^{1})\frac{f(\bar{\omega}_{\textsc{i}}^{1},\bar{\omega}_{\textsc{ii}}^{1})}{h(\bar{\omega}_{\textsc{i}}^{1},-z-c)}\frac{f(\bar{\omega}_{\textsc{ii}}^{1},\bar{\omega}_{\textsc{iii}}^{1})}{h(-z-c,\bar{\omega}_{\textsc{iii}}^{1})}\\
\frac{T_{2,2}(z)}{\lambda_{2}(z)}\mathbb{B}(\bar{t}^{1}) & =\sum_{\bar{w}^{1}=\bar{w}_{\textsc{i}}^{1}\cup\bar{w}_{\textsc{ii}}^{1}}\mathbb{B}(\bar{w}_{\textsc{ii}}^{1})\frac{f(\bar{w}_{\textsc{i}}^{1},\bar{w}_{\textsc{ii}}^{1})}{h(\bar{w}_{\textsc{i}}^{1},z)},
\end{split}
,\label{eq:actN2}
\end{equation}
where $\bar{\omega}^{1}=\{-z-c,\bar{t}^{1}\}$ and $\bar{w}^{1}=\{z,\bar{t}^{1}\}$.
The sums goes over the partitions where $\#\bar{\omega}_{\textsc{i}}^{1}=\#\bar{\omega}_{\textsc{iii}}^{1}=\#\bar{w}_{\textsc{i}}^{1}=1$.
Substituting back (\ref{eq:KTtemp}) and (\ref{eq:actN2}) to (\ref{eq:recN2-1}),
we obtain a recurrence equation for the overlap
\begin{align}
S_{K}^{N}(\{z,\bar{t}^{1}\}) & =\frac{K_{2,2}}{K_{2,1}}\left(\alpha_{1}(z)\sum_{\bar{\omega}^{1}=\bar{\omega}_{\textsc{i}}^{1}\cup\bar{\omega}_{\textsc{ii}}^{1}}S_{K}^{N}(\bar{\omega}_{\textsc{ii}}^{1})\frac{f(\bar{\omega}_{\textsc{i}}^{1},\bar{\omega}_{\textsc{ii}}^{1})}{h(\bar{\omega}_{\textsc{i}}^{1},-z-c)}-\sum_{\bar{w}^{1}=\bar{w}_{\textsc{i}}^{1}\cup\bar{w}_{\textsc{ii}}^{1}}S_{K}^{N}(\bar{w}_{\textsc{ii}}^{1})\frac{f(\bar{w}_{\textsc{i}}^{1},\bar{w}_{\textsc{ii}}^{1})}{h(\bar{w}_{\textsc{i}}^{1},z)}\right)+\nonumber \\
 & +\frac{K_{1,2}}{K_{2,1}}\alpha_{1}(z)\sum_{\bar{\omega}^{1}=\bar{\omega}_{\textsc{i}}^{1}\cup\bar{\omega}_{\textsc{ii}}^{1}\cup\bar{\omega}_{\textsc{iii}}^{1}}S_{K}^{N}(\bar{\omega}_{\textsc{ii}}^{1})\alpha_{1}(\bar{\omega}_{\textsc{iii}}^{1})\frac{f(\bar{\omega}_{\textsc{i}}^{1},\bar{\omega}_{\textsc{ii}}^{1})}{h(\bar{\omega}_{\textsc{i}}^{1},-z-c)}\frac{f(\bar{\omega}_{\textsc{ii}}^{1},\bar{\omega}_{\textsc{iii}}^{1})}{h(-z-c,\bar{\omega}_{\textsc{iii}}^{1})}.\label{eq:recN2}
\end{align}
For $\bar{t}^{1}=\emptyset$ we obtain that
\begin{equation}
S_{K}^{N}(\{z\})=\frac{K_{2,2}}{K_{2,1}}(\alpha_{1}(z)-1),
\end{equation}
therefore the sum formula is satisfied for $r_{1}=1$. Let us assume
that the sum formula is satisfied for $r_{1}$ or less Bethe roots.
Applying this induction assumption to the recurrence equation (\ref{eq:recN2})
we obtain that
\begin{equation}
S_{K}^{N}(\{z,\bar{t}^{1}\})=\alpha_{1}(z)\sum_{\mathrm{part}(\bar{\omega}^{1})}\alpha_{1}(\bar{\omega}_{\textsc{i}}^{1})(\dots)+\sum_{\mathrm{part}(\bar{w}^{1})}\alpha_{1}(\bar{w}_{\textsc{i}}^{1})(\dots),\label{eq:temp-1}
\end{equation}
where the first sum goes over all partitions $\bar{\omega}^{1}=\bar{\omega}_{\textsc{i}}^{1}\cup\bar{\omega}_{\textsc{ii}}^{1}$
and the second sum goes over all partitions $\bar{w}^{1}=\bar{w}_{\textsc{i}}^{1}\cup\bar{w}_{\textsc{ii}}^{1}$.
Since $\bar{\omega}_{\textsc{i}}^{1}\subset\{-z-c,\bar{t}^{1}\}$
the first sum can be written as
\begin{multline}
\alpha_{1}(z)\sum_{\mathrm{part}(\bar{\omega}^{1})}\alpha_{1}(\bar{\omega}_{\textsc{i}}^{1})(\dots)=\alpha_{1}(z)\left(\sum_{\mathrm{part}(\bar{t}^{1})}\alpha_{1}(\bar{t}_{\textsc{i}}^{1})(\dots)+\sum_{\mathrm{part}(\bar{t}^{1})}\alpha_{1}(-z-c)\alpha_{1}(\bar{t}_{\textsc{i}}^{1})(\dots)\right)=\\
=\alpha_{1}(z)\sum_{\mathrm{part}(\bar{t}^{1})}\alpha_{1}(\bar{t}_{\textsc{i}}^{1})(\dots)+\sum_{\mathrm{part}(\bar{t}^{1})}\alpha_{1}(\bar{t}_{\textsc{i}}^{1})(\dots),
\end{multline}
where we used the symmetry property (\ref{eq:constAlpTw}). Substituting
back to (\ref{eq:temp-1}) we obtain that 
\begin{equation}
S_{K}^{N}(\{z,\bar{t}^{1}\})=\alpha_{1}(z)\sum_{\mathrm{part}}\alpha_{1}(\bar{t}_{\textsc{i}}^{1})(\dots)+\sum_{\mathrm{part}}\alpha_{1}(\bar{t}_{\textsc{i}}^{1})(\dots).
\end{equation}
We can see that we just showed that the sum formula for $r_{1}+1$
Bethe roots follows from the sum formula for $r_{1}$ or less Bethe
roots therefore we just proved the sum formula for $N=2$.

Let us assume that the sum rule is true up to $N-1$ i.e. the overlap
for $r_{1}=0$ reads as
\begin{equation}
S_{K}^{N}(\emptyset,\left\{ \bar{t}^{k}\right\} _{k=2}^{N-1})=\sum_{\mathrm{part}}\prod_{\nu=2}^{N-1}\alpha_{\nu}(\bar{t}_{\textsc{i}}^{\nu})\times(\dots),\label{eq:indAssumTw}
\end{equation}
where the sum goes over the partition of $\bar{t}^{k}$ for $k=2,\dots,N-1$.

At first let us use the recurrence formula (\ref{eq:rec1})
\begin{equation}
\mathbb{B}(\{z,\bar{t}^{1}\},\left\{ \bar{t}^{k}\right\} _{k=2}^{N-1})=\sum_{j=2}^{N}\frac{T_{1,j}(z)}{\lambda_{2}(z)}\sum_{\mathrm{part}}\mathbb{B}(\bar{t}^{1},\left\{ \bar{t}_{\textsc{ii}}^{k}\right\} _{k=2}^{j-1},\left\{ \bar{t}^{k}\right\} _{k=j}^{N-1})\prod_{\nu=2}^{j-1}\alpha_{\nu}(\bar{t}_{\textsc{i}}^{\nu})\times(\dots).
\end{equation}
After that we can use the component $(N,j)$ of the $KT$-relation
(\ref{eq:KT_Tw})
\begin{equation}
K_{N,1}\left\langle \Psi\right|T_{1,j}(u)=\lambda_{0}(u)\sum_{k=1}^{N}\left\langle \Psi\right|\widehat{T}_{N,k}(-u)K_{k,j}-\sum_{i=2}^{N}K_{N,i}\left\langle \Psi\right|T_{i,j}(u),
\end{equation}
to change the creation operators $T_{1,j}(z)$ in the recurrence formula
to diagonal or annihilation operators w.r.t the first level Bethe
roots. Using the symmetry property (\ref{eq:constLamTw}), the action
formulas (\ref{eq:act}) and (\ref{eq:actTw}) of the operators of
the r.h.s. read as
\begin{equation}
\begin{split}\frac{\lambda_{0}(z)}{\lambda_{2}(z)}\widehat{T}_{N,k}(-z)\mathbb{B}(\bar{t}) & =\alpha_{1}(z)\sum_{\mathrm{part}(\bar{\omega})}\mathbb{B}(\bar{\omega}_{\textsc{ii}})\prod_{s=1}^{N-k}\alpha_{s}(\bar{\omega}_{\textsc{iii}}^{s})\times(\dots),\\
\frac{1}{\lambda_{2}(z)}T_{i,j}(z)\mathbb{B}(\bar{t}) & =\frac{1}{\prod_{s=2}^{N-1}\alpha_{s}(z)}\sum_{\mathrm{part}(\bar{w})}\mathbb{B}(\bar{w}_{\textsc{ii}})\prod_{s=j}^{N-1}\alpha_{s}(\bar{w}_{\textsc{iii}}^{s})\times(\dots),
\end{split}
\end{equation}
where $\bar{\omega}^{\nu}=\{-z-\nu c,\bar{t}^{\nu}\}$ and $\bar{w}^{\nu}=\{z,\bar{t}^{\nu}\}$.
We can see that in these formulas the $\alpha$-dependent terms are
$\alpha_{s}(t_{k}^{s})$, $\alpha_{s}(z)$ therefore we obtain the
following recursion for the overlap
\begin{equation}
\langle\Psi|\mathbb{B}(\{z,\bar{t}^{1}\},\left\{ \bar{t}^{k}\right\} _{k=2}^{N-1})=\sum_{\mathrm{part}}\langle\Psi|\mathbb{B}(\bar{w}_{\textsc{ii}})\prod_{s=2}^{N-1}\alpha_{s}(\bar{t}_{\textsc{i}}^{s})\times(\alpha_{\nu}(z),\dots),\label{eq:rectemp}
\end{equation}
where $\bar{w}_{\textsc{ii}}^{\nu}\subset\{z,-z-\nu c,\bar{t}_{\textsc{ii}}^{\nu}\}$
and $\bar{t}^{\nu}=\bar{t}_{\textsc{i}}^{\nu}\cup\bar{t}_{\textsc{ii}}^{\nu}$
for which $\bar{t}_{\textsc{i}}^{\nu},\bar{t}_{\textsc{ii}}^{\nu}$
are disjoint subsets and $\bar{w}_{\textsc{ii}}^{1}\leq r_{1}$. In
the equation above we used the notation $(\alpha_{\nu}(z),\dots)$
for the terms which depend only on $\alpha_{\nu}(z)$ but do not depend
on any $\alpha_{s}(t_{k}^{\nu})$. Applying the recurrence relation
(\ref{eq:rectemp}) on the overlap $\langle\Psi|\mathbb{B}(\bar{t})$
we can eliminate the Bethe roots $\bar{t}^{1}$ as
\begin{equation}
\langle\Psi|\mathbb{B}(\bar{t}^{1},\left\{ \bar{t}^{k}\right\} _{k=2}^{N-1})=\sum_{\mathrm{part}}\langle\Psi|\mathbb{B}(\emptyset,\left\{ \bar{w}_{\textsc{ii}}^{k}\right\} _{k=2}^{N-1})\prod_{s=2}^{N-1}\alpha_{s}(\bar{t}_{\textsc{i}}^{s})\times(\alpha_{\nu}(t_{k}^{1}),\dots),
\end{equation}
where $\bar{w}_{\textsc{ii}}^{\nu}\subset\{\bar{t}^{1},-\bar{t}^{1}-\nu c,\bar{t}_{\textsc{ii}}^{\nu}\}$
and we used the notation $(\alpha_{\nu}(t_{k}^{1}),\dots)$ for the
terms which depend only on $\alpha_{\nu}(t_{k}^{1})$ but not on $\alpha_{\nu}(t_{k}^{s})$
for $s>1$. Applying (\ref{eq:indAssumTw}) in the r.h.s we obtain
that
\begin{equation}
\langle\Psi|\mathbb{B}(\bar{t}^{1},\left\{ \bar{t}^{k}\right\} _{k=2}^{N-1})=\sum_{\mathrm{part}}\prod_{s=2}^{N-1}\alpha_{s}(\bar{t}_{\textsc{i}}^{s})\times(\alpha_{\nu}(t_{k}^{1}),\dots),\label{eq:sumv1}
\end{equation}
where we used the symmetry property (\ref{eq:constAlpTw}). We just
showed that the overlap $\langle\Psi|\mathbb{B}(\bar{t}^{1},\left\{ \bar{t}^{k}\right\} _{k=2}^{N-1})$
has the correct $\alpha$-dependence w.r.t the arguments $\bar{t}^{k}$,
where $k=2,\dots,N-1$.

It remains to prove that the $\alpha$-dependence of the overlap w.r.t
the arguments $\bar{t}^{1}$ is given by the sum rule (\ref{eq:sumrule_Tw}).
It can be done using the other embedding of the overlaps (\ref{eq:recursion_S_Tw-1})
i.e.
\begin{equation}
S_{K}^{N}(\bar{t}^{1},\dots,\bar{t}^{N-2},\emptyset)=S_{\bar{K}^{(2)}}^{N-1}(\bar{t}^{1},\dots,\bar{t}^{N-2}).
\end{equation}
Applying the induction assumption we obtain that
\begin{equation}
S_{K}^{N}(\left\{ \bar{t}^{k}\right\} _{k=1}^{N-2},\emptyset)=\sum_{\mathrm{part}}\prod_{\nu=1}^{N-2}\alpha_{\nu}(\bar{t}_{\textsc{i}}^{\nu})\times(\dots).\label{eq:indAssumTw2}
\end{equation}
Using the second recursion for the Bethe vector (\ref{eq:rec2}),
the twisted $KT$-relation (\ref{eq:KT_Tw}), the action rules (\ref{eq:act}),(\ref{eq:actTw})
and repeating the considerations above we can obtain that
\begin{equation}
\langle\Psi|\mathbb{B}(\left\{ \bar{t}^{k}\right\} _{k=1}^{N-2},\bar{t}^{N-1})=\sum_{\mathrm{part}}\langle\Psi|\mathbb{B}(\left\{ \bar{w}_{\textsc{ii}}^{k}\right\} _{k=2}^{N-1},\emptyset)\prod_{s=1}^{N-2}\alpha_{s}(\bar{t}_{\textsc{i}}^{s})\times(\alpha_{\nu}(t_{k}^{N-1}),\dots),\label{eq:otherrec}
\end{equation}
where $\bar{w}_{\textsc{ii}}^{\nu}\subset\{\bar{t}^{N-1},-\bar{t}^{N-1}-\nu c,\bar{t}_{\textsc{ii}}^{\nu}\}$
and $\bar{t}^{\nu}=\bar{t}_{\textsc{i}}^{\nu}\cup\bar{t}_{\textsc{ii}}^{\nu}$
for which $\bar{t}_{\textsc{i}}^{\nu},\bar{t}_{\textsc{ii}}^{\nu}$
are disjoint subsets for $\nu=1,\dots,N-2$. Substituting (\ref{eq:indAssumTw2})
to (\ref{eq:otherrec}) we find that

\begin{equation}
S_{K}^{N}(\left\{ \bar{t}^{k}\right\} _{k=1}^{N-2},\bar{t}^{N-1})=\sum_{\mathrm{part}}\prod_{\nu=1}^{N-2}\alpha_{\nu}(\bar{t}_{\textsc{i}}^{\nu})\times(\alpha_{\nu}(t_{k}^{N-1}),\dots),\label{eq:sumv2}
\end{equation}
where we used the symmetry property (\ref{eq:constAlpTw}). Combining
the equation (\ref{eq:sumv1}) with (\ref{eq:sumv2}) we obtain the
sum formula (\ref{eq:sumFormula}).

\subsection{Untwisted case}

In this subsection we prove the sum formula
\begin{equation}
S_{K}^{N}(\bar{t})=\sum_{\mathrm{part}(\bar{t})}\prod_{\nu=1}^{N-1}\alpha_{\nu}(\bar{t}_{\textsc{i}}^{\nu})(\dots),\label{eq:sumrule_UTw}
\end{equation}
for the untwisted two-site states. The untwisted case is more complicated
than the twisted one since in the previous derivation we could use
two type of embedding rules but for the untwisted overlaps we only
have one type of embedding (\ref{eq:recursion_S_UTw})
\begin{equation}
S_{K}^{N}(\emptyset,\bar{t}^{2},\dots,\bar{t}^{N-2},\emptyset)=S_{\bar{K}^{(2)}}^{N-2}(\bar{t}^{2},\dots,\bar{t}^{N-2})\Biggr|_{\alpha_{\nu}\to\alpha_{\nu+1}}.\label{eq:embed_UTw}
\end{equation}
At first we prove that if the equations
\begin{align}
S_{K}^{N}(\bar{t}^{1},\dots,\bar{t}^{k-1},\emptyset,\dots,\emptyset) & =\sum_{\mathrm{part}(\bar{t})}\prod_{\nu=1}^{k-1}\alpha_{\nu}(\bar{t}_{\textsc{i}}^{\nu})\times(\dots),\label{eq:induction_assum}\\
S_{K}^{N}(\emptyset,\bar{t}^{2},\dots,\bar{t}^{k},\emptyset,\dots,\emptyset) & =\sum_{\mathrm{part}(\bar{t})}\prod_{\nu=2}^{k}\alpha_{\nu}(\bar{t}_{\textsc{i}}^{\nu})\times(\dots)\label{eq:conditiont2}
\end{align}
are satisfied for a fixed $k$ then the following equation is also
satisfied
\begin{equation}
S_{K}^{N}(\bar{t}^{1},\bar{t}^{2},\dots,\bar{t}^{k},\emptyset,\dots,\emptyset)=\sum_{\mathrm{part}(\bar{t})}\prod_{\nu=1}^{k}\alpha_{\nu}(\bar{t}_{\textsc{i}}^{\nu})\times(\dots).\label{eq:iter}
\end{equation}

Let us use the recurrence formula (\ref{eq:rec1})
\begin{equation}
\mathbb{B}(\{z,\bar{t}^{1}\},\left\{ \bar{t}^{s}\right\} _{s=2}^{k},\emptyset^{N-k-1})=\sum_{j=2}^{k+1}\frac{T_{1,j}(z)}{\lambda_{2}(z)}\sum\mathbb{B}(\bar{t}^{1},\left\{ \bar{t}_{\textsc{ii}}^{s}\right\} _{s=2}^{j-1},\left\{ \bar{t}^{s}\right\} _{s=j}^{k},\emptyset^{N-k-2})\prod_{\nu=2}^{j-1}\alpha_{\nu}(t_{\textsc{i}}^{\nu})\times(\dots),
\end{equation}
and the $KT$-relation (\ref{eq:KT_Utw})
\begin{equation}
K_{N,1}(u)\left\langle \Psi\right|T_{1,j}(u)=\sum_{i=1}^{N}\left\langle \Psi\right|T_{N,i}(-u)K_{i,j}(u)-\sum_{i=2}^{N}K_{N,i}(u)\left\langle \Psi\right|T_{i,j}(u),
\end{equation}
and the action formulas (\ref{eq:act})
\begin{equation}
\begin{split}\frac{T_{N,i}(-z)}{\lambda_{2}(z)}\mathbb{B}(\left\{ \bar{t}^{s}\right\} _{s=2}^{k},\emptyset^{N-k-1}) & =\alpha_{1}(z)\sum_{\mathrm{part}}\mathbb{B}(\left\{ \bar{\omega}_{\textsc{ii}}^{s}\right\} _{s=2}^{k},\emptyset^{N-k-1})\prod_{s=i}^{k}\alpha_{s}(\bar{\omega}_{\textsc{iii}}^{s})\times(\dots),\\
\frac{T_{i,j}(z)}{\lambda_{2}(z)}\mathbb{B}(\left\{ \bar{t}^{s}\right\} _{s=2}^{k},\emptyset^{N-k-1}) & =\frac{1}{\prod_{s=2}^{N-1}\alpha_{s}(z)}\sum_{\mathrm{part}}\mathbb{B}(\left\{ \bar{w}_{\textsc{ii}}^{s}\right\} _{s=2}^{k},\emptyset^{N-k-1})\prod_{s=j}^{k}\alpha_{s}(\bar{w}_{\textsc{iii}}^{s})\times(\dots),
\end{split}
\end{equation}
where we used the symmetry property (\ref{eq:constLamUTw}). Note
that the r.h.s of the first line vanishes for $i<N$ if $k<N-1$.
We can see that in these formulas the $\alpha$-dependent terms are
$\alpha_{s}(t_{k}^{s})$ and $\alpha_{s}(z)$ therefore combining
them we obtain the following recursion for the overlap
\begin{equation}
\langle\Psi|\mathbb{B}(\{z,\bar{t}^{1}\},\left\{ \bar{t}^{s}\right\} _{s=2}^{k},\emptyset^{N-k-1})=\sum_{\mathrm{part}}\langle\Psi|\mathbb{B}(\left\{ \bar{w}_{\textsc{ii}}^{s}\right\} _{s=2}^{k},\emptyset^{N-k-1})\prod_{s=2}^{k}\alpha_{s}(\bar{t}_{\textsc{i}}^{s})\times(\alpha_{\nu}(z),\dots),
\end{equation}
where $\bar{w}_{\textsc{ii}}^{\nu}\subset\{z,-z,\bar{t}_{\textsc{ii}}^{\nu}\}$
and $\bar{t}=\bar{t}_{\textsc{i}}\cup\bar{t}_{\textsc{ii}}$ where
$\bar{w}_{\textsc{ii}}^{1}\leq r_{1}$. Applying this recursion for
the overlap we can eliminate the Bethe roots $\bar{t}^{1}$ as
\begin{equation}
\langle\Psi|\mathbb{B}(\bar{t}^{1},\left\{ \bar{t}^{s}\right\} _{s=2}^{k},\emptyset^{N-k-1})=\sum_{\mathrm{part}}\langle\Psi|\mathbb{B}(\emptyset,\left\{ \bar{w}_{\textsc{ii}}^{s}\right\} _{s=2}^{k},\emptyset^{N-k-1})\prod_{s=2}^{k}\alpha_{s}(\bar{t}_{\textsc{i}}^{s})\times(\alpha_{\nu}(t_{k}^{1}),\dots),
\end{equation}
where $\bar{w}_{\textsc{ii}}^{\nu}\subset\{\bar{t}^{1},-\bar{t}^{1},\bar{t}_{\textsc{ii}}^{\nu}\}$.
Applying (\ref{eq:conditiont2}) in the r.h.s, we obtain that
\begin{equation}
S_{K}^{N}(\bar{t}^{1},\left\{ \bar{t}^{s}\right\} _{s=2}^{k},\emptyset^{N-k-1})=\sum_{\mathrm{part}}\prod_{s=2}^{k}\alpha_{s}(\bar{t}_{\textsc{i}}^{s})\times(\alpha_{\nu}(t_{k}^{1}),\dots).\label{eq:finalform1}
\end{equation}

Let us continue with the other recursion of the Bethe vectors (\ref{eq:rec2})

\begin{equation}
\mathbb{B}(\left\{ \bar{t}^{s}\right\} _{s=1}^{k-1},\{z,\bar{t}^{k}\},\emptyset^{N-k-1})=\sum_{i=1}^{k}\frac{T_{i,k+1}(z)}{\lambda_{k+1}(z)}\sum_{\mathrm{part}}\mathbb{B}(\left\{ \bar{t}^{s}\right\} _{s=1}^{i-1},\left\{ \bar{t}_{\textsc{ii}}^{s}\right\} _{s=i}^{k-1},\bar{t}^{k},\emptyset^{N-k-1})\times(\dots).\label{eq:proofSumRec}
\end{equation}
Let us try to use the $KT$-relation (\ref{eq:KT_Utw}) to change
the creation operators $T_{i,k+1}(z)$ to the annihilation operator
\begin{equation}
\left\langle \Psi\right|T_{i,k+1}(-u)K_{k+1,1}(u)=\sum_{j=1}^{N}K_{i,j}(u)\left\langle \Psi\right|T_{j,1}(u)-\sum_{j=1}^{k}\left\langle \Psi\right|T_{i,j}(-u)K_{j,1}(u)-\sum_{j=k+2}^{N}\left\langle \Psi\right|T_{i,j}(-u)K_{j,1}(u).
\end{equation}
We can see that the last term would create $\bar{t}^{s}$ Bethe roots
for $s>k$ therefore we can not use the equation above. It turns out
that we can eliminate these terms. Let us start with the component
$(a,b)$ of the $KT$-relation (\ref{eq:KT_Utw})
\begin{equation}
\sum_{c=1}^{N-1}K_{a,c}(u)\left\langle \Psi\right|T_{c,b}(u)+K_{a,N}(u)\left\langle \Psi\right|T_{N,b}(u)=\sum_{c=1}^{N-1}\left\langle \Psi\right|T_{a,c}(-u)K_{c,b}(u)+\left\langle \Psi\right|\underline{T_{a,N}(-u)}K_{N,b}(u).
\end{equation}
We can see that the last term is a creation operator but we can eliminate
this term by using the component $(a,1)$ of the $KT$-relation (\ref{eq:KT_Utw})
\begin{align}
\sum_{c=1}^{N-1}K_{a,c}(u)\left\langle \Psi\right|T_{c,1}(u)+K_{a,N}(u)\left\langle \Psi\right|T_{N,1}(u) & =\sum_{c=1}^{N-1}\left\langle \Psi\right|T_{a,c}(-u)K_{c,1}(u)+\left\langle \Psi\right|\underline{T_{a,N}(-u)}K_{N,1}(u).
\end{align}
Making a proper substitution we can obtain the following equation

\begin{equation}
\sum_{c=1}^{N-1}K_{a,c}(u)\left\langle \Psi\right|T_{c,b}(u)+C_{a,b}^{(1)}(u)=\sum_{c=1}^{N-1}\left\langle \Psi\right|T_{a,c}(-u)\tilde{K}_{c,b}^{(1)}(u),
\end{equation}
where
\begin{align}
C_{a,b}^{(1)}(u) & =K_{a,N}(u)\left\langle \Psi\right|T_{N,b}(u)-\left(\sum_{c=1}^{N-1}K_{a,c}(u)\left\langle \Psi\right|T_{c,1}(u)+K_{a,N}(u)\left\langle \Psi\right|T_{N,1}(u)\right)\frac{K_{N,b}(u)}{K_{N,1}(u)},\\
\tilde{K}_{a,b}^{(1)}(u) & =K_{a,b}(u)-\frac{K_{a,1}(u)K_{N,b}(u)}{K_{N,1}(u)}.
\end{align}
We can see that the expression $C_{a,b}^{(1)}(u)$ contains only diagonal
and annihilation operators. Repeating this procedure we can obtain
the following equations
\begin{equation}
\sum_{c=1}^{N-l}K_{a,c}(u)\left\langle \Psi\right|T_{c,b}(u)+C_{a,b}^{(l)}(u)=\sum_{c=1}^{N-l}\left\langle \Psi\right|T_{a,c}(-u)\tilde{K}_{c,b}^{(l)}(u),\label{eq:goodRecursion}
\end{equation}
where
\begin{align}
C_{a,b}^{(k+1)}(u) & =C_{a,b}^{(k)}(u)+K_{a,N-k}(u)\left\langle \Psi\right|T_{N-k,b}(u)-\nonumber \\
 & -\left(\sum_{c=1}^{N-k-1}K_{a,c}(u)\left\langle \Psi\right|T_{c,1}(u)+K_{a,N-k}(u)\left\langle \Psi\right|T_{N-k,1}(u)+C_{a,1}^{(k)}(u)\right)\frac{\tilde{K}_{N-k,b}^{(k)}(u)}{\tilde{K}_{N-k,1}^{(k)}(u)},\\
\tilde{K}_{a,b}^{(k+1)}(u) & =\tilde{K}_{a,b}^{(k)}(u)-\frac{\tilde{K}_{a,1}^{(k)}(u)\tilde{K}_{N-k,b}^{(k)}(u)}{\tilde{K}_{N-k,1}^{(k)}(u)}.
\end{align}
We can see that the the expression $C_{a,b}^{(k)}(u)$ contains only
diagonal and annihilation operators for $b\leq N-k$. The matrices
$\tilde{K}_{a,b}^{(k)}(u)$ are just a re-normalized versions of $K_{a,b}^{(k)}(u)$
(\ref{eq:reqK}) for which we already see that their matrix elements
can be chosen to be not zero. Let us use the component $(i,1)$ of
the (\ref{eq:goodRecursion}) for $l=N-k-1$
\begin{equation}
\left\langle \Psi\right|T_{i,k+1}(-u)\tilde{K}_{k+1,1}^{(N-k-1)}(u)=\sum_{c=1}^{k+1}K_{i,c}(u)\left\langle \Psi\right|T_{c,1}(u)+C_{i,1}^{(N-k-1)}(u)-\sum_{c=1}^{k}\left\langle \Psi\right|T_{i,c}(-u)\tilde{K}_{c,1}^{(N-k-1)}(u).\label{eq:goodKT}
\end{equation}
We can see that now we can change the operators $T_{i,k+1}(u)$ to
diagonal and annihilation operators and operators $T_{i,c}(u)$ but
they do not increase the quantum numbers $r_{s}$ for $s\geq k$.
Substituting (\ref{eq:goodKT}) and the action formula (\ref{eq:act})
to (\ref{eq:proofSumRec}) we obtain that
\begin{equation}
S_{K}^{N}(\left\{ \bar{t}^{s}\right\} _{s=1}^{k-1},\{z,\bar{t}^{k}\},\emptyset^{N-k-1})=\sum_{\mathrm{part}}S_{K}^{N}(\left\{ \bar{w}_{\textsc{ii}}^{s}\right\} _{s=1}^{k},\emptyset^{N-k-1})\prod_{s=1}^{k-1}\alpha_{s}(\bar{t}_{\textsc{i}}^{s})\times(\alpha_{\nu}(z),\dots),
\end{equation}
where $\bar{w}_{\textsc{ii}}^{\nu}\subset\{z,-z,\bar{t}_{\textsc{ii}}^{\nu}\}$
and $\bar{t}=\bar{t}_{\textsc{i}}\cup\bar{t}_{\textsc{ii}}$ where
$\bar{w}_{\textsc{ii}}^{k}\leq r_{k}$. Using this recurrence equation
we can eliminate all the Bethe roots $\bar{t}^{k}$ and we obtain
that
\[
S_{K}^{N}(\left\{ \bar{t}^{s}\right\} _{s=1}^{k-1},\bar{t}^{k},\emptyset^{N-k-1})=\sum_{\mathrm{part}}S_{K}^{N}(\left\{ \bar{w}_{\textsc{ii}}^{s}\right\} _{s=1}^{k-1},\emptyset^{N-k})\prod_{s=1}^{k-1}\alpha_{s}(\bar{t}_{\textsc{i}}^{s})\times(\alpha_{\nu}(t_{l}^{k}),\dots),
\]
where $\bar{w}_{\textsc{ii}}^{\nu}\subset\{\bar{t}^{k},-\bar{t}^{k},\bar{t}_{\textsc{ii}}^{\nu}\}$.
Using the condition (\ref{eq:induction_assum}) we obtain that
\begin{equation}
S_{K}^{N}(\left\{ \bar{t}^{s}\right\} _{s=1}^{k-1},\bar{t}^{k},\emptyset^{N-k-1})=\sum_{\mathrm{part}}\prod_{s=1}^{k-1}\alpha_{s}(\bar{t}_{\textsc{i}}^{s})\times(\alpha_{\nu}(t_{l}^{k}),\dots).\label{eq:finalform2}
\end{equation}
Combining (\ref{eq:finalform1}) and (\ref{eq:finalform2}) we just
proved that the equation (\ref{eq:iter}) follows from (\ref{eq:induction_assum})
and (\ref{eq:conditiont2}).

In an completely analogous way, one can also show that the equation
\begin{equation}
S_{K}^{N}(\emptyset^{k},\left\{ \bar{t}^{s}\right\} _{s=k+1}^{N-1})=\sum_{\mathrm{part}}\prod_{s=k+1}^{N-1}\alpha_{s}(\bar{t}_{\textsc{i}}^{s})\times(\dots)
\end{equation}
follows from the equations
\begin{align}
S_{K}^{N}(\emptyset^{k+1},\left\{ \bar{t}^{s}\right\} _{s=k+2}^{N-1}) & =\sum_{\mathrm{part}}\prod_{s=k+2}^{N-1}\alpha_{s}(\bar{t}_{\textsc{i}}^{s})\times(\dots),\\
S_{K}^{N}(\emptyset^{k},\left\{ \bar{t}^{s}\right\} _{s=k+1}^{N-2},\emptyset) & =\sum_{\mathrm{part}}\prod_{s=k+1}^{N-2}\alpha_{s}(\bar{t}_{\textsc{i}}^{s})\times(\dots).
\end{align}

Let us continue with the overlap $\langle\Psi|\mathbb{B}(\bar{t}^{1},\emptyset^{N-2})$.
Using the explicit form of the Bethe vector we obtain that
\begin{equation}
\langle\Psi|\mathbb{B}(\{z,\bar{t}^{1}\},\emptyset^{N-2})=\langle\Psi|\frac{T_{1,2}(z)}{\lambda_{2}(z)}\mathbb{B}(\bar{t}^{1},\emptyset^{N-2}).
\end{equation}
We can apply the $KT$-relation (\ref{eq:KT_Utw})
\begin{multline}
K_{N,1}(u)\left\langle \Psi\right|T_{1,2}(u)=\left\langle \Psi\right|T_{N,N}(-u)K_{N,2}(u)-K_{N,2}(u)\left\langle \Psi\right|T_{2,2}(u)+\\
+\sum_{i=1}^{N-1}\left\langle \Psi\right|T_{N,i}(-u)K_{i,2}(u)-\sum_{i=3}^{N}K_{N,i}(u)\left\langle \Psi\right|T_{i,2}(u).
\end{multline}
The operators in the second line are annihilate the operator $\mathbb{B}(\bar{t}^{1},\emptyset^{N-2})$.
The actions of the remaining operators read as
\begin{equation}
\begin{split}\frac{T_{N,N}(-z)}{\lambda_{2}(z)}\mathbb{B}(\bar{t}^{1},\emptyset^{N-2}) & =\alpha_{1}(z)\mathbb{B}(\bar{t}^{1},\emptyset^{N-2}),\\
\frac{T_{2,2}(z)}{\lambda_{2}(z)}\mathbb{B}(\bar{t}^{1},\emptyset^{N-2}) & =\sum_{\mathrm{part}(\bar{w}^{1})}\mathbb{B}(\bar{w}_{\textsc{ii}}^{1},\emptyset^{N-2})\frac{f(\bar{w}_{\textsc{i}}^{1},\bar{w}_{\textsc{ii}}^{1})}{h(\bar{w}_{\textsc{i}}^{1},z)},
\end{split}
\end{equation}
where $\bar{w}^{1}=\{z,\bar{t}^{1}\}.$ Substituting back, we obtain
a recurrence equation for the overlap
\begin{equation}
S_{K}^{N}(\{z,\bar{t}^{1}\},\emptyset^{N-2})=\frac{K_{N,2}(u)}{K_{N,1}(u)}\left(\alpha_{1}(z)S_{K}^{N}(\bar{t}^{1},\emptyset^{N-2})-\sum S_{K}^{N}(\bar{w}_{\textsc{ii}}^{1},\emptyset^{N-2})\frac{f(\bar{w}_{\textsc{i}}^{1},\bar{w}_{\textsc{ii}}^{1})}{h(\bar{w}_{\textsc{i}}^{1},z)}\right).
\end{equation}
This recursion with the initial condition $S_{K}^{N}(\emptyset,\emptyset^{N-2})=1$
is completely defines the overlap $S_{K}^{N}(\bar{t}^{1},\emptyset^{N-2})$
and its $\alpha$-dependence. We can see that the overlap has the
form
\begin{equation}
S_{K}^{N}(\bar{t}^{1},\emptyset^{N-2})=\sum_{\mathrm{part}}\alpha_{1}(\bar{t}_{\textsc{i}}^{1})\times(\dots).\label{eq:sum1}
\end{equation}
In an analogous way, we can show that
\begin{equation}
S_{K}^{N}(\emptyset^{N-2},\bar{t}^{N-1})=\sum_{\mathrm{part}}\alpha_{N-1}(\bar{t}_{\textsc{i}}^{N-1})\times(\dots).\label{eq:sumN1}
\end{equation}

Let us take the overlap
\begin{equation}
S_{K}^{N}(\emptyset,\bar{t}^{2},\emptyset^{N-3})=S_{K^{(2)}}^{N-2}(\bar{t}^{2},\emptyset^{N-4})\Biggr|_{\alpha_{\nu}\to\alpha_{\nu+1}}=\sum_{\mathrm{part}}\alpha_{2}(\bar{t}_{\textsc{i}}^{2})\times(\dots),\label{eq:sum2}
\end{equation}
where we used the embedding rule (\ref{eq:embed_UTw}) and the equation
(\ref{eq:sum1}). We already showed that the expression 
\begin{equation}
S_{K}^{N}(\bar{t}^{1},\bar{t}^{2},\emptyset^{N-3})=\sum_{\mathrm{part}}\alpha_{1}(\bar{t}_{\textsc{i}}^{1})\alpha_{2}(\bar{t}_{\textsc{i}}^{2})\times(\dots)
\end{equation}
follows from (\ref{eq:sum1}) and (\ref{eq:sum2}) . Continuing this
recursion we can show that
\begin{equation}
S_{K}^{N}(\left\{ \bar{t}^{s}\right\} _{s=1}^{N-2},\emptyset)=\sum_{\mathrm{part}}\prod_{s=1}^{N-2}\alpha_{s}(\bar{t}_{\textsc{i}}^{s})\times(\dots).\label{eq:sum11}
\end{equation}
Starting with (\ref{eq:sumN1}) we can also show that
\begin{equation}
S_{K}^{N}(\emptyset,\left\{ \bar{t}^{s}\right\} _{s=2}^{N-1})=\sum_{\mathrm{part}}\prod_{s=2}^{N-1}\alpha_{s}(\bar{t}_{\textsc{i}}^{s})\times(\dots),\label{eq:sum12}
\end{equation}
but we already proved that the equation 
\begin{equation}
S_{K}^{N}(\left\{ \bar{t}^{s}\right\} _{s=1}^{N-1})=\sum\prod_{s=1}^{N-1}\alpha_{s}(\bar{t}_{\textsc{i}}^{s})\times(\dots)
\end{equation}
follows from (\ref{eq:sum11}),(\ref{eq:sum12}) and that is what
we wanted to show.

\section{Connections between the weights $W_{K}^{N}$ and the highest coefficients
$Z_{K}^{N},\bar{Z}_{K}^{N}$\label{sec:ConnectionsWZ}}

The derivation of (\ref{eq:WZZ}) can be done for the untwisted and
twisted cases, simultaneously. Let us define two quantum spaces as
$\mathcal{H}^{(1)},\mathcal{H}^{(2)}$ with the corresponding monodromy
matrices $T^{(1)},T^{(2)}$. Let us assume that $\langle\Psi^{(1)}|\in\mathcal{H}^{(1)}$
and $\langle\Psi^{(2)}|\in\mathcal{H}^{(2)}$ are integrable two-site
states with the same $K$-matrix. Let us define the tensor product
quantum space $\mathcal{H}=\mathcal{H}^{(1)}\otimes\mathcal{H}^{(2)}$.
In subsection \ref{subsec:Coproduct} we saw that the integrable two-site
state of quantum space $\mathcal{H}$ is factorized as $\left\langle \Psi\right|=\langle\Psi^{(1)}|\otimes\langle\Psi^{(2)}|$
(see (\ref{eq:Psi_Factor})) Let us choose the normalization as $\langle\Psi|0\rangle^{(i)}=1$
for $i=1,2$.

In this section we use the co-product formula for the Bethe states
(\ref{eq:coproduct}) 
\begin{equation}
\tilde{\mathbb{B}}(\bar{t})=\sum_{\mathrm{part}}\frac{\prod_{\nu=1}^{N-1}\lambda_{\nu}^{(2)}(\bar{t}_{\mathrm{i}}^{\nu})\lambda_{\nu+1}^{(1)}(\bar{t}_{\mathrm{ii}}^{\nu})f(\bar{t}_{\mathrm{ii}}^{\nu},\bar{t}_{\mathrm{i}}^{\nu})}{\prod_{\nu=1}^{N-2}f(\bar{t}_{\mathrm{ii}}^{\nu+1},\bar{t}_{\mathrm{i}}^{\nu})}\mathbb{\tilde{\mathbb{B}}}^{(1)}(\bar{t}_{\mathrm{i}})\otimes\tilde{\mathbb{B}}^{(2)}(\bar{t}_{\mathrm{ii}}),
\end{equation}
where the sum is taken over all the partitions. Let us fix a particular
partition $\bar{t}=\bar{t}_{\textsc{i}}\cup\bar{t}_{\textsc{ii}}$
and choose the highest weights as
\begin{equation}
\begin{split}\lambda_{\nu+1}^{(1)}(u) & =0,\qquad\text{for all }u\in\bar{t}_{\textsc{i}}^{\nu},\\
\lambda_{\nu}^{(2)}(u) & =0,\qquad\text{for all }u\in\bar{t}_{\textsc{ii}}^{\nu},
\end{split}
\end{equation}
therefore in the sum rule of the overlap and the co-product formula
there is only one non-vanishing term:
\begin{align}
\langle\Psi|\tilde{\mathbb{B}}(\bar{t})= & W_{K}^{N}(\bar{t}_{\textsc{i}}|\bar{t}_{\textsc{ii}})\prod_{\nu=1}^{N-1}\lambda_{\nu}(\bar{t}_{\textsc{i}}^{\nu})\lambda_{\nu+1}(\bar{t}_{\textsc{ii}}^{\nu}),\label{eq:sumterm}\\
\tilde{\mathbb{B}}(\bar{t})= & \frac{\prod_{\nu=1}^{N-1}\lambda_{\nu}^{(2)}(\bar{t}_{\textsc{i}}^{\nu})\lambda_{\nu+1}^{(1)}(\bar{t}_{\textsc{ii}}^{\nu})f(\bar{t}_{\textsc{ii}}^{\nu},\bar{t}_{\textsc{i}}^{\nu})}{\prod_{\nu=1}^{N-2}f(\bar{t}_{\textsc{ii}}^{\nu+1},\bar{t}_{\textsc{i}}^{\nu})}\tilde{\mathbb{B}}^{(1)}(\bar{t}_{\textsc{i}})\otimes\tilde{\mathbb{B}}^{(2)}(\bar{t}_{\textsc{ii}}).\label{eq:coprodterm}
\end{align}
Substituting (\ref{eq:coprodterm}) to the equation (\ref{eq:sumterm})
we obtain that
\begin{equation}
\langle\Psi|\tilde{\mathbb{B}}(\bar{t})=\frac{\prod_{\nu=1}^{N-1}\lambda_{\nu}^{(2)}(\bar{t}_{\textsc{i}}^{\nu})\lambda_{\nu+1}^{(1)}(\bar{t}_{\textsc{ii}}^{\nu})f(\bar{t}_{\textsc{ii}}^{\nu},\bar{t}_{\textsc{i}}^{\nu})}{\prod_{\nu=1}^{N-2}f(\bar{t}_{\textsc{ii}}^{\nu+1},\bar{t}_{\textsc{i}}^{\nu})}\left(\langle\Psi^{(1)}|\tilde{\mathbb{B}}^{(1)}(\bar{t}_{\textsc{i}})\right)\left(\langle\Psi^{(2)}|\tilde{\mathbb{B}}^{(2)}(\bar{t}_{\textsc{ii}})\right).
\end{equation}
Applying the sum formula (\ref{eq:sumterm}) on the l.h.s. and the
other two sum formulas
\begin{align}
\langle\Psi^{(1)}|\tilde{\mathbb{B}}^{(1)}(\bar{t}_{\textsc{i}}) & =W_{K}^{N}(\bar{t}_{\textsc{i}}|\emptyset)\prod_{\nu=1}^{N-1}\lambda_{\nu}^{(1)}(\bar{t}_{\textsc{i}}^{\nu})=Z_{K}^{N}(\bar{t}_{\textsc{i}})\prod_{\nu=1}^{N-1}\lambda_{\nu}^{(1)}(\bar{t}_{\textsc{i}}^{\nu}),\\
\langle\Psi^{(2)}|\tilde{\mathbb{B}}^{(2)}(\bar{t}_{\textsc{ii}}) & =W_{K}^{N}(\emptyset|\bar{t}_{\textsc{ii}})\prod_{\nu=1}^{N-1}\lambda_{\nu+1}^{(2)}(\bar{t}_{\textsc{ii}}^{\nu})=\bar{Z}_{K}^{N}(\bar{t}_{\textsc{ii}})\prod_{\nu=1}^{N-1}\lambda_{\nu+1}^{(2)}(\bar{t}_{\textsc{ii}}^{\nu}),
\end{align}
on r.h.s. we obtain that
\begin{equation}
W_{K}^{N}(\bar{t}_{\textsc{i}}|\bar{t}_{\textsc{ii}})=\frac{\prod_{\nu=1}^{N-1}f(\bar{t}_{\textsc{ii}}^{\nu},\bar{t}_{\textsc{i}}^{\nu})}{\prod_{\nu=1}^{N-2}f(\bar{t}_{\textsc{ii}}^{\nu+1},\bar{t}_{\textsc{i}}^{\nu})}Z_{K}^{N}(\bar{t}_{\textsc{i}})\bar{Z}_{K}^{N}(\bar{t}_{\textsc{ii}}).\label{eq:WZZbar}
\end{equation}

Now we continue with the derivations of the connections between the
highest coefficients $Z_{K}^{N}$ and $\bar{Z}_{K}^{N}$. We do these
derivations separately for the twisted and untwisted cases.

\subsection{Untwisted case}

Let us use the following algebra automorphism 
\begin{equation}
\Pi:T(u)\to VT(-u)^{t}V,
\end{equation}
where $V_{i,j}=\delta_{i,N+1-j}$ . Applying this automorphism to
the $KT$-relation (\ref{eq:KT_Utw})
\begin{equation}
K(u)\left\langle \Psi_{\Pi}\right|\Pi(T(u))=\left\langle \Psi_{\Pi}\right|\Pi(T(-u))K(u),
\end{equation}
 we can obtain that
\begin{equation}
K^{\Pi}(u)\left\langle \Psi_{\Pi}\right|T(u)=\left\langle \Psi_{\Pi}\right|T(-u)K^{\Pi}(u),\label{eq:KTmod}
\end{equation}
where we defined a transformed $K$-matrix
\begin{equation}
K^{\Pi}(u)=VK^{t}(u)V.
\end{equation}
Since
\begin{equation}
\Pi(T_{i,j}(u))\left|0\right\rangle =T_{N+1-j,N+1-i}(-u)\left|0\right\rangle =0,
\end{equation}
the pseudo-vacuum is invariant
\begin{equation}
\Pi(\left|0\right\rangle )=\left|0\right\rangle .
\end{equation}
In our convention the vacuum eigenvalues are invariant under the transformation
$\Pi$:
\begin{equation}
\Pi(T_{i,i}(u))\left|0\right\rangle =\lambda_{N+1-i}(-u)\left|0\right\rangle =\lambda_{i}(u)\left|0\right\rangle .
\end{equation}
We can apply the transformation $\Pi$ on the overlap
\begin{equation}
\langle\Psi|\tilde{\mathbb{B}}(\bar{t})=\langle\Psi_{\Pi}|\Pi(\tilde{\mathbb{B}}(\bar{t})).
\end{equation}
In the following we prove that
\begin{equation}
\Pi(\tilde{\mathbb{B}}(\bar{t}))=\tilde{\mathbb{B}}(\pi^{a}(\bar{t})),\label{eq:refB}
\end{equation}
where we used the definitions (\ref{eq:achiral}).Let us apply $\Pi$
to recurrence relation (\ref{eq:rec1-1})
\begin{multline}
\Pi(\tilde{\mathbb{B}}(\{z,\bar{t}^{1}\},\left\{ \bar{t}^{k}\right\} _{k=2}^{N-1}))=\sum_{j=2}^{N}T_{N+1-j,N}(-z)\sum_{\mathrm{part}}\Pi(\tilde{\mathbb{B}}(\bar{t}^{1},\left\{ \bar{t}_{\textsc{ii}}^{k}\right\} _{k=2}^{j-1},\left\{ \bar{t}^{k}\right\} _{k=j}^{N-1}))\times\\
\frac{\prod_{\nu=2}^{j-1}\lambda_{\nu}(\bar{t}_{\textsc{i}}^{\nu})g(\bar{t}_{\textsc{i}}^{\nu},\bar{t}_{\textsc{i}}^{\nu-1})f(\bar{t}_{\textsc{ii}}^{\nu},\bar{t}_{\textsc{i}}^{\nu})}{\prod_{\nu=1}^{j-1}f(\bar{t}^{\nu+1},\bar{t}_{\textsc{i}}^{\nu})}.
\end{multline}
Let us assume (\ref{eq:refB}) is satisfied for $r_{1}$ or less first
Bethe roots. Using this induction assumption, we can apply (\ref{eq:refB})
in the r.h.s therefore
\begin{multline}
\Pi(\tilde{\mathbb{B}}(\{z,\bar{t}^{1}\},\left\{ \bar{t}^{k}\right\} _{k=2}^{N-1}))=\sum_{j=1}^{N-1}T_{j,N}(-z)\sum_{\mathrm{part}}\tilde{\mathbb{B}}(\left\{ -\bar{t}^{N-k}\right\} _{k=1}^{j-1},\left\{ -\bar{t}_{\textsc{ii}}^{N-k}\right\} _{k=j}^{N-2},-\bar{t}^{1})\times\\
\frac{\prod_{\nu=2}^{N-j}\lambda_{\nu}(\bar{t}_{\textsc{i}}^{\nu})g(\bar{t}_{\textsc{i}}^{\nu},\bar{t}_{\textsc{i}}^{\nu-1})f(\bar{t}_{\textsc{ii}}^{\nu},\bar{t}_{\textsc{i}}^{\nu})}{\prod_{\nu=1}^{N-j}f(\bar{t}^{\nu+1},\bar{t}_{\textsc{i}}^{\nu})}.
\end{multline}
The term in the second line can be written as
\begin{multline}
\frac{\prod_{\nu=2}^{N-j}\lambda_{\nu}(\bar{t}_{\textsc{i}}^{\nu})g(\bar{t}_{\textsc{i}}^{\nu},\bar{t}_{\textsc{i}}^{\nu-1})f(\bar{t}_{\textsc{ii}}^{\nu},\bar{t}_{\textsc{i}}^{\nu})}{\prod_{\nu=1}^{N-j}f(\bar{t}^{\nu+1},\bar{t}_{\textsc{i}}^{\nu})}=\frac{\prod_{\nu=2}^{N-j}\lambda_{N+1-\nu}(-\bar{t}_{\textsc{i}}^{\nu})g(-\bar{t}_{\textsc{i}}^{\nu-1},-\bar{t}_{\textsc{i}}^{\nu})f(-\bar{t}_{\textsc{i}}^{\nu},-\bar{t}_{\textsc{ii}}^{\nu})}{\prod_{\nu=1}^{N-j}f(-\bar{t}_{\textsc{i}}^{\nu},-\bar{t}^{\nu+1})}=\\
\frac{\prod_{\nu=j}^{N-2}\lambda_{\nu+1}(-\bar{t}_{\textsc{i}}^{N-\nu})g(-\bar{t}_{\textsc{i}}^{N-\nu-1},-\bar{t}_{\textsc{i}}^{N-\nu})f(-\bar{t}_{\textsc{i}}^{N-\nu},-\bar{t}_{\textsc{ii}}^{N-\nu})}{\prod_{\nu=j}^{N-1}f(-\bar{t}_{\textsc{i}}^{N-\nu},-\bar{t}^{N-\nu+1})}.
\end{multline}
Plugging back and using the recurrence relation (\ref{eq:rec2-1})
in the r.h.s., we obtain that
\begin{equation}
\Pi(\tilde{\mathbb{B}}(\{z,\bar{t}^{1}\},\left\{ \bar{t}^{k}\right\} _{k=2}^{N-1}))=\tilde{\mathbb{B}}(\left\{ -\bar{t}^{k}\right\} _{k=N-1}^{2},\{-z,-\bar{t}^{1}\}),
\end{equation}
therefore we just proved (\ref{eq:refB}).

Using (\ref{eq:refB}) in the overlap we obtain that
\begin{equation}
\langle\Psi|\tilde{\mathbb{B}}(\bar{t})=\langle\Psi_{\Pi}|\tilde{\mathbb{B}}(\pi^{a}(\bar{t})).
\end{equation}
Since the $\langle\Psi_{\Pi}|$ satisfies a $KT$-relation (\ref{eq:KTmod})
we can use the sum formula (\ref{eq:sumFormula}) but we have to change
the $K$-matrix elements as $K_{i,j}(u)\to K_{i,j}^{\Pi}(u)=K_{N+1-j,N+1-j}(u)$
therefore
\begin{multline}
\langle\Psi_{\Pi}|\tilde{\mathbb{B}}(\pi^{a}(\bar{t}))=\sum_{\mathrm{part}}W_{\Pi(K)}^{N}(\pi^{a}(\bar{t}_{\textsc{i}})|\pi^{a}(\bar{t}_{\textsc{ii}}))\prod_{\nu=1}^{N-1}\lambda_{\nu}(-\bar{t}_{\textsc{i}}^{N-\nu})\lambda_{\nu+1}(-\bar{t}_{\textsc{ii}}^{N-\nu})=\\
\sum_{\mathrm{part}}W_{\Pi(K)}^{N}(\pi^{a}(\bar{t}_{\textsc{i}})|\pi^{a}(\bar{t}_{\textsc{ii}}))\prod_{\nu=1}^{N-1}\lambda_{\nu+1}(\bar{t}_{\textsc{i}}^{\nu})\lambda_{\nu}(\bar{t}_{\textsc{ii}}^{\nu}),
\end{multline}
which means
\begin{equation}
W_{K}^{N}(\bar{t}_{\textsc{i}}|\bar{t}_{\textsc{ii}})=W_{\Pi(K)}^{N}(\pi^{a}(\bar{t}_{\textsc{ii}})|\pi^{a}(\bar{t}_{\textsc{i}})),
\end{equation}
therefore the HC-s satisfy the following relation 
\begin{equation}
Z_{K}^{N}(\bar{t})=\bar{Z}_{\Pi(K)}^{N}(\pi^{a}(\bar{t})).\label{eq:UTwZZbar}
\end{equation}

\subsection{Twisted case}

Let us use the following algebra automorphism 
\begin{equation}
\Pi:T(u)\to\lambda_{0}(u)V\widehat{T}(-u)^{t}V,
\end{equation}
where $V_{i,j}=\delta_{i,N+1-j}$. From the definition of the twisted
monodromy matrix we can see that 
\begin{equation}
\Pi(\widehat{T}(u))=\frac{1}{\lambda_{0}(-u)}VT(-u)^{t}V.
\end{equation}
Applying this automorphism on the twisted $KT$-relation (\ref{eq:KT_Tw})
we obtain that
\begin{equation}
K^{\Pi}\left\langle \Psi_{\Pi}\right|T(u)=\lambda_{0}(u)\left\langle \Psi_{\Pi}\right|\widehat{T}(-u)K^{\Pi},
\end{equation}
where
\begin{equation}
K^{\Pi}=VK^{t}V.
\end{equation}
Since we concentrate on the $\mathfrak{so}(N)$ symmetric $K$-matrices
which have the explicit forms
\begin{equation}
K=\mathcal{U}V,\qquad\mathcal{U}=\mathcal{U}^{t},
\end{equation}
the $K$-matrix is invariant under the transformation $\Pi$ 
\begin{equation}
K^{\Pi}=K.
\end{equation}
The transformed monodromy matrix acts on the reference state as
\begin{equation}
\Pi(T_{i,j}(u))\left|0\right\rangle =\lambda_{0}(u)\widehat{T}_{N+1-j,N+1-j}(-u)\left|0\right\rangle =0,
\end{equation}
therefore the pseudo-vacuum is invariant under the transformation
$\Pi$:
\begin{equation}
\Pi(\left|0\right\rangle )=\left|0\right\rangle .
\end{equation}
The vacuum eigenvalues are invariant under $\Pi$
\begin{equation}
\Pi(T_{i,i}(u))\left|0\right\rangle =\lambda_{0}(u)\hat{\lambda}_{N+1-i}(-u)\left|0\right\rangle =\lambda_{i}(u)\left|0\right\rangle ,
\end{equation}
where we used the constraint (\ref{eq:constLamTw}). We can apply
this transformation on the overlap
\begin{equation}
\langle\Psi|\tilde{\mathbb{B}}(\bar{t})=\langle\Psi_{\Pi}|\Pi(\tilde{\mathbb{B}}(\bar{t})).
\end{equation}
In the following we prove that
\begin{equation}
\Pi(\tilde{\mathbb{B}}(\bar{t}))=\prod_{\nu=1}^{N-1}\lambda_{0}(\bar{t}^{\nu})\tilde{\hat{\mathbb{B}}}(\pi^{a}(\bar{t})),\label{eq:refB-1}
\end{equation}
where we used the definitions (\ref{eq:achiral}) and
\begin{equation}
\tilde{\hat{\mathbb{B}}}(\bar{t})=\prod_{\nu=1}^{N-1}\hat{\lambda}_{\nu+1}(\bar{t}^{\nu})\hat{\mathbb{B}}(\bar{t}).\label{eq:renBhat}
\end{equation}
Let us apply $\Pi$ to the recurrence relation (\ref{eq:rec1-1})
\begin{multline}
\Pi(\tilde{\mathbb{B}}(\{z,\bar{t}^{1}\},\left\{ \bar{t}^{k}\right\} _{k=2}^{N-1}))=\sum_{j=2}^{N}\lambda_{0}(z)\hat{T}_{N+1-j,N}(-z)\sum_{\mathrm{part}}\Pi(\tilde{\mathbb{B}}(\bar{t}^{1},\left\{ \bar{t}_{\textsc{ii}}^{k}\right\} _{k=2}^{j-1},\left\{ \bar{t}^{k}\right\} _{k=j}^{N-1}))\times\\
\frac{\prod_{\nu=2}^{j-1}\lambda_{\nu}(\bar{t}_{\textsc{i}}^{\nu})g(\bar{t}_{\textsc{i}}^{\nu},\bar{t}_{\textsc{i}}^{\nu-1})f(\bar{t}_{\textsc{ii}}^{\nu},\bar{t}_{\textsc{i}}^{\nu})}{\prod_{\nu=1}^{j-1}f(\bar{t}^{\nu+1},\bar{t}_{\textsc{i}}^{\nu})}.
\end{multline}
Let us assume (\ref{eq:refB-1}) is satisfied for $r_{1}$ or less
first Bethe roots. Using this induction assumption we can apply (\ref{eq:refB-1})
in the r.h.s therefore
\begin{multline}
\Pi(\tilde{\mathbb{B}}(\{z,\bar{t}^{1}\},\left\{ \bar{t}^{k}\right\} _{k=2}^{N-1}))=\sum_{j=1}^{N-1}\lambda_{0}(z)\widehat{T}_{j,N}(-z)\sum_{\mathrm{part}}\frac{\prod_{\nu=1}^{N-1}\lambda_{0}(\bar{t}^{\nu})}{\prod_{\nu=2}^{N-j}\lambda_{0}(\bar{t}_{\textsc{i}}^{\nu})}\tilde{\hat{\mathbb{B}}}(\left\{ -\bar{t}^{N-k}\right\} _{k=1}^{j-1},\left\{ -\bar{t}_{\textsc{ii}}^{N-k}\right\} _{k=j}^{N-2},-\bar{t}^{1})\times\\
\frac{\prod_{\nu=2}^{N-j}\lambda_{\nu}(\bar{t}_{\textsc{i}}^{\nu})g(\bar{t}_{\textsc{i}}^{\nu},\bar{t}_{\textsc{i}}^{\nu-1})f(\bar{t}_{\textsc{ii}}^{\nu},\bar{t}_{\textsc{i}}^{\nu})}{\prod_{\nu=1}^{N-j}f(\bar{t}^{\nu+1},\bar{t}_{\textsc{i}}^{\nu})}.
\end{multline}
The term in the second line can be written as
\begin{multline}
\frac{\prod_{\nu=2}^{N-j}\lambda_{\nu}(\bar{t}_{\textsc{i}}^{\nu})g(\bar{t}_{\textsc{i}}^{\nu},\bar{t}_{\textsc{i}}^{\nu-1})f(\bar{t}_{\textsc{ii}}^{\nu},\bar{t}_{\textsc{i}}^{\nu})}{\prod_{\nu=1}^{N-j}f(\bar{t}^{\nu+1},\bar{t}_{\textsc{i}}^{\nu})}=\frac{\prod_{\nu=2}^{N-j}\lambda_{0}(\bar{t}_{\textsc{i}}^{\nu})\hat{\lambda}_{N+1-\nu}(-\bar{t}_{\textsc{i}}^{\nu})g(-\bar{t}_{\textsc{i}}^{\nu-1},-\bar{t}_{\textsc{i}}^{\nu})f(-\bar{t}_{\textsc{i}}^{\nu},-\bar{t}_{\textsc{ii}}^{\nu})}{\prod_{\nu=1}^{N-j}f(-\bar{t}_{\textsc{i}}^{\nu},-\bar{t}^{\nu+1})}=\\
\prod_{\nu=2}^{N-j}\lambda_{0}(\bar{t}_{\textsc{i}}^{\nu})\frac{\prod_{\nu=j}^{N-2}\hat{\lambda}_{\nu+1}(-\bar{t}_{\textsc{i}}^{N-\nu})g(-\bar{t}_{\textsc{i}}^{N-\nu-1},-\bar{t}_{\textsc{i}}^{N-\nu})f(-\bar{t}_{\textsc{i}}^{N-\nu},-\bar{t}_{\textsc{ii}}^{N-\nu})}{\prod_{\nu=j}^{N-1}f(-\bar{t}_{\textsc{i}}^{N-\nu},-\bar{t}^{N-\nu+1})}.
\end{multline}
Plugging back and using the recurrence relation (\ref{eq:rec2-1})
we obtain that
\begin{equation}
\Pi(\tilde{\mathbb{B}}(\{z,\bar{t}^{1}\},\left\{ \bar{t}^{k}\right\} _{k=2}^{N-1}))=\lambda_{0}(z)\prod_{\nu=1}^{N-1}\lambda_{0}(\bar{t}^{\nu})\tilde{\hat{\mathbb{B}}}(\left\{ -\bar{t}^{k}\right\} _{k=N-1}^{2},\{-z,-\bar{t}^{1}\}),
\end{equation}
therefore we just proved (\ref{eq:refB-1}).

Using (\ref{eq:refB-1}), the overlap can be written as
\begin{equation}
\langle\Psi|\tilde{\mathbb{B}}(\bar{t})=\prod_{\nu=1}^{N-1}\lambda_{0}(\bar{t}^{\nu})\langle\Psi_{\Pi}|\tilde{\hat{\mathbb{B}}}(\pi^{a}(\bar{t})).\label{eq:temp}
\end{equation}
Since the $\langle\Psi_{\Pi}|$ satisfies the twisted $KT$-relation
with the $K$-matrix $K$ we can use the sum formula (\ref{eq:sumFormula}).
At first let us use (\ref{eq:connBhatB}) and (\ref{eq:renBhat})
in the r.h.s.:
\begin{equation}
\prod_{\nu=1}^{N-1}\lambda_{0}(\bar{t}^{\nu})\langle\Psi_{\Pi}|\tilde{\hat{\mathbb{B}}}(\pi^{a}(\bar{t}))=A\prod_{\nu=1}^{N-1}\lambda_{0}(\bar{t}^{\nu})\hat{\lambda}_{\nu+1}(-\bar{t}^{N-\nu})\langle\Psi_{\Pi}|\mathbb{B}(\mu(\pi^{a}(\bar{t})))=A\prod_{\nu=1}^{N-1}\lambda_{\nu}(\bar{t}^{\nu})\langle\Psi_{\Pi}|\mathbb{B}(\mu(\pi^{a}(\bar{t}))),
\end{equation}
where we used the symmetry property (\ref{eq:constLamTw}) and
\begin{equation}
A=(-1)^{\#\bar{t}}\left(\prod_{s=1}^{N-2}f(\bar{t}^{s+1},\bar{t}^{s})\right)^{-1}.
\end{equation}
We can also notice that $\mu(\pi^{a}(\bar{t}))=\mu^{c}(\bar{t})$
(see the definition (\ref{eq:chpair})). Now we can substitute the
sum formula (\ref{eq:sumFormula}):
\begin{equation}
\prod_{\nu=1}^{N-1}\lambda_{0}(\bar{t}^{\nu})\langle\Psi_{\Pi}|\tilde{\hat{\mathbb{B}}}(\pi^{a}(\bar{t}))=A\sum_{\mathrm{part}}W_{K}^{N}(\pi^{c}(\bar{t}_{\textsc{i}})|\pi^{c}(\bar{t}_{\textsc{ii}}))\prod_{\nu=1}^{N-1}\lambda_{\nu}(\bar{t}^{\nu})\alpha_{\nu}(-\bar{t}_{\textsc{i}}^{\nu}-\nu c).
\end{equation}
After some rearrangements we get
\begin{equation}
\prod_{\nu=1}^{N-1}\lambda_{0}(\bar{t}^{\nu})\langle\Psi_{\Pi}|\tilde{\hat{\mathbb{B}}}(\pi^{a}(\bar{t}))=A\sum_{\mathrm{part}}W_{K}^{N}(\pi^{c}(\bar{t}_{\textsc{i}})|\pi^{c}(\bar{t}_{\textsc{ii}}))\prod_{\nu=1}^{N-1}\lambda_{\nu}(\bar{t}_{\textsc{ii}}^{\nu})\lambda_{\nu+1}(\bar{t}_{\textsc{i}}^{\nu}).
\end{equation}
Substituting back to (\ref{eq:temp}) we obtain an identity for the
weights $W_{K}^{N}$: 
\begin{equation}
W_{K}^{N}(\bar{t}_{\textsc{i}}|\bar{t}_{\textsc{ii}})=(-1)^{\#\bar{t}}\left(\prod_{s=1}^{N-2}f(\bar{t}^{s+1},\bar{t}^{s})\right)^{-1}W_{K}^{N}(\pi^{c}(\bar{t}_{\textsc{ii}})|\pi^{c}(\bar{t}_{\textsc{i}})),
\end{equation}
therefore the HC-s satisfy the following relation 
\begin{equation}
Z_{K}^{N}(\bar{t})=(-1)^{\#\bar{t}}\left(\prod_{s=1}^{N-2}f(\bar{t}^{s+1},\bar{t}^{s})\right)^{-1}\bar{Z}_{K}^{N}(\pi^{c}(\bar{t})).\label{eq:TwZZbar}
\end{equation}

\section{Recursions for the HC-s\label{sec:RecursionsHC}}

From the sum formula (\ref{eq:sumFormula}) we can see that the $\alpha$
independent term in the overlap is just the HC $\bar{Z}^{N}(\bar{t})$.
In this section we want to derive the recurrence relations (\ref{eq:recursionZ_UTw1}),(\ref{eq:recursionZ_UTw2})
and (\ref{eq:recursionZ_Tw}) for these quantities therefore we use
the notation $\cong$ for the equality up to $\alpha$-dependent terms
i.e.
\begin{equation}
\langle\Psi|\mathbb{B}(\bar{t})\cong\bar{Z}_{K}^{N}(\bar{t}).
\end{equation}

\subsection{Untwisted case}

Let us derive the recursion for $t_{k}^{1}$-s in the untwisted case.
We start with the recurrence equation (\ref{eq:rec1})
\begin{equation}
\left\langle \Psi\right|\mathbb{B}(\{z,\bar{t}^{1}\},\left\{ \bar{t}^{\nu}\right\} _{\nu}^{N-1})\cong\left\langle \Psi\right|\frac{T_{1,2}(z)}{\lambda_{2}(z)f(\bar{t}^{2},z)}\mathbb{B}(\bar{t}).
\end{equation}
Now we can use the $KT$-relation (\ref{eq:KT_Utw})
\begin{equation}
K_{N,1}(u)\left\langle \Psi\right|T_{1,2}(u)=\sum_{j=1}^{N}\left\langle \Psi\right|T_{N,j}(-u)K_{j,2}(u)-\sum_{i=2}^{N}K_{N,i}(u)\left\langle \Psi\right|T_{i,2}(u)
\end{equation}
 to transform $T_{1,2}(z)$ into $T_{N,k}(-z)$ and $T_{j,2}(z)$
where $k=1,\dots,N$ and $j=2,\dots,N$. Substituting back, we obtain
that
\begin{equation}
\left\langle \Psi\right|\mathbb{B}(\{z,\bar{t}^{1}\},\left\{ \bar{t}^{\nu}\right\} _{\nu}^{N-1})\cong\frac{1}{f(\bar{t}^{2},z)}\left[\sum_{j=1}^{N}\frac{K_{j,2}(z)}{K_{N,1}(z)}\left\langle \Psi\right|\frac{T_{N,j}(-z)}{\lambda_{2}(z)}\mathbb{B}(\bar{t})-\sum_{i=2}^{N}\frac{K_{N,i}(z)}{K_{N,1}(z)}\left\langle \Psi\right|\frac{T_{i,2}(z)}{\lambda_{2}(z)}\mathbb{B}(\bar{t})\right].\label{eq:ttemp}
\end{equation}
Now we can use the action formula (\ref{eq:act}). Let us start with
$T_{N,N}(-z)$:
\begin{equation}
\left\langle \Psi\right|\frac{T_{N,N}(-z)}{\lambda_{2}(z)}\mathbb{B}(\bar{t})=\alpha_{1}(z)\sum_{\mathrm{part}}\left\langle \Psi\right|\mathbb{B}(\bar{\omega}_{\textsc{ii}})\prod_{s=1}^{N-1}\frac{f(\bar{\omega}_{\textsc{i}}^{s},\bar{\omega}_{\textsc{ii}}^{s})}{h(\bar{\omega}_{\textsc{i}}^{s},\bar{\omega}_{\textsc{i}}^{s-1})f(\bar{\omega}_{\textsc{i}}^{s},\bar{\omega}_{\textsc{ii}}^{s-1})},\label{eq:temp2}
\end{equation}
where we used the symmetry property (\ref{eq:constLamUTw}). The sum
goes over the partitions $\bar{\omega}=\bar{\omega}_{\textsc{i}}\cup\bar{\omega}_{\textsc{ii}}$
where $\bar{\omega}^{\nu}=\{-z,\bar{t}^{\nu}\}$ and $\#\bar{\omega}_{\textsc{i}}^{\nu}=1$.
The terms
\begin{equation}
\alpha_{1}(z)\left\langle \Psi\right|\mathbb{B}(\bar{\omega}_{\textsc{ii}})
\end{equation}
can be $\alpha$ independent only when $-z\in\bar{\omega}_{\textsc{ii}}^{N-1}$
for which
\begin{equation}
\alpha_{1}(z)\left\langle \Psi\right|\mathbb{B}(\bar{\omega}_{\textsc{ii}})\cong W_{K}^{N}(\emptyset^{\times N-2},\{-z\}|\{\bar{\omega}_{\textsc{ii}}^{\nu}\}_{\nu=1}^{N-2},\bar{t}_{\textsc{ii}}^{N-1}).
\end{equation}
Substituting back to (\ref{eq:temp2}) we obtain that
\begin{multline}
\left\langle \Psi\right|\frac{T_{N,N}(-z)}{\lambda_{2}(z)}\mathbb{B}(\bar{t})\cong\sum_{\mathrm{part}}Z_{K}^{N}(\emptyset^{\times N-2},\{-z\})\bar{Z}_{K}^{N}(\{\bar{\omega}_{\textsc{ii}}^{\nu}\}_{\nu=1}^{N-2},\bar{t}_{\textsc{ii}}^{N-1})\times\\
\prod_{s=1}^{N-2}\frac{f(\bar{\omega}_{\textsc{i}}^{s},\bar{\omega}_{\textsc{ii}}^{s})}{h(\bar{\omega}_{\textsc{i}}^{s},\bar{\omega}_{\textsc{i}}^{s-1})f(\bar{\omega}_{\textsc{i}}^{s},\bar{\omega}_{\textsc{ii}}^{s-1})}\frac{f(\bar{t}_{\textsc{i}}^{N-1},\bar{t}_{\textsc{ii}}^{N-1})f(\bar{t}^{N-1},-z)}{h(\bar{t}_{\textsc{i}}^{N-1},\bar{\omega}_{\textsc{i}}^{N-2})f(\bar{t}_{\textsc{i}}^{N-1},\bar{\omega}_{\textsc{ii}}^{N-2})}.
\end{multline}
where we used the identity (\ref{eq:WZZ}). This formula contains
the HC $Z(\emptyset^{\times N-2},\{-z\})$ which was already calculated
in appendix \ref{sec:Elementary} therefore
\begin{equation}
Z_{K}^{N}(\emptyset^{\times N-2},\{-z\})=-\frac{K_{N-1,1}(z)}{K_{N,1}(z)}.
\end{equation}
Let us continue with $T_{N,j}(-z)$ for $j<N$. The action formula
(\ref{eq:act}) explicitly contains $\alpha_{\nu}(\bar{\omega}_{\textsc{iii}}^{\nu})$
where $j\leq\nu<N$ therefore it is $\alpha_{\nu}(t_{k}^{\nu})$-independent
only when $\bar{\omega}_{\textsc{iii}}^{\nu}=\{-z\}$ therefore unwanted
terms $\alpha_{\nu}(-z)$ appears for $j<N-1$ which means
\begin{equation}
\left\langle \Psi\right|\frac{T_{N,j}(-z)}{\lambda_{2}(z)}\mathbb{B}(\bar{t})\cong0,
\end{equation}
for $j<N-1$. An $\alpha$-independent term comes form only $j=N-1$
which reads as
\begin{multline}
\left\langle \Psi\right|\frac{T_{N,N-1}(-z)}{\lambda_{2}(z)}\mathbb{B}(\bar{t})\cong\sum_{\mathrm{part}}\bar{Z}_{K}^{N}(\{\bar{\omega}_{\textsc{ii}}^{\nu}\}_{\nu=1}^{N-2},\bar{t}_{\textsc{ii}}^{N-1})\times\\
\prod_{s=1}^{N-2}\frac{f(\bar{\omega}_{\textsc{i}}^{s},\bar{\omega}_{\textsc{ii}}^{s})}{h(\bar{\omega}_{\textsc{i}}^{s},\bar{\omega}_{\textsc{i}}^{s-1})f(\bar{\omega}_{\textsc{i}}^{s},\bar{\omega}_{\textsc{ii}}^{s-1})}\frac{f(\bar{t}_{\textsc{i}}^{N-1},\bar{t}_{\textsc{ii}}^{N-1})f(\bar{t}^{N-1},-z)}{h(\bar{t}_{\textsc{i}}^{N-1},\bar{\omega}_{\textsc{i}}^{N-2})f(\bar{t}_{\textsc{i}}^{N-1},\bar{\omega}_{\textsc{ii}}^{N-2})}.
\end{multline}

Let us continue with the action of $T_{i,2}$
\begin{multline}
\frac{T_{i,2}(z)}{\lambda_{2}(z)}\mathbb{B}(\bar{t})=\frac{1}{\prod_{s=2}^{N-1}\alpha_{s}(z)}\sum_{\mathrm{part}}\mathbb{B}(\bar{w}_{\textsc{ii}})\frac{\prod_{s=2}^{i-1}f(\bar{w}_{\textsc{i}}^{s},\bar{w}_{\textsc{iii}}^{s})}{\prod_{s=2}^{i-2}f(\bar{w}_{\textsc{i}}^{s+1},\bar{w}_{\textsc{iii}}^{s})}\times\\
\prod_{s=1}^{i-1}\frac{f(\bar{w}_{\textsc{i}}^{s},\bar{w}_{\textsc{ii}}^{s})}{h(\bar{w}_{\textsc{i}}^{s},\bar{w}_{\textsc{i}}^{s-1})f(\bar{w}_{\textsc{i}}^{s},\bar{w}_{\textsc{ii}}^{s-1})}\prod_{s=2}^{N-1}\frac{\alpha_{s}(\bar{w}_{\textsc{iii}}^{s})f(\bar{w}_{\textsc{ii}}^{s},\bar{w}_{\textsc{iii}}^{s})}{h(\bar{w}_{\textsc{iii}}^{s+1},\bar{w}_{\textsc{iii}}^{s})f(\bar{w}_{\textsc{ii}}^{s+1},\bar{w}_{\textsc{iii}}^{s})},
\end{multline}
where the sum goes over the partitions $\bar{w}=\bar{w}_{\textsc{i}}\cup\bar{w}_{\textsc{ii}}\cup\bar{w}_{\textsc{iii}}$
where $\bar{w}^{\nu}=\{z,\bar{t}^{\nu}\}$ and $\#\bar{w}_{\textsc{iii}}^{\nu}=1$
for $\nu>1$ and $\#\bar{w}_{\textsc{iii}}^{1}=0$. This action formula
contains $\alpha_{s}(\bar{w}_{\textsc{iii}}^{s})$ for $s\geq2$ therefore
the terms are $\alpha_{s}(t_{k}^{s})$-independent only when $\bar{w}_{\textsc{iii}}^{\nu}=\{z\}$.
Applying this observation for the action formula we obtain that
\begin{multline}
\left\langle \Psi\right|\frac{T_{i,2}(z)}{\lambda_{2}(z)}\mathbb{B}(\bar{t})\cong\sum_{\mathrm{part}}\bar{Z}_{K}^{N}(\bar{w}_{\textsc{ii}}^{1},\{\bar{t}_{\textsc{ii}}^{\nu}\}_{\nu=2}^{i-1},\{\bar{t}^{\nu}\}_{\nu=i}^{N-1})\times\\
\frac{f(\bar{w}_{\textsc{i}}^{1},\bar{w}_{\textsc{ii}}^{1})}{h(\bar{w}_{\textsc{i}}^{1},z)}\frac{f(\bar{t}_{\textsc{i}}^{2},\bar{t}_{\textsc{ii}}^{2})f(\bar{t}^{2},z)}{h(\bar{t}_{\textsc{i}}^{2},\bar{w}_{\textsc{i}}^{1})f(\bar{t}_{\textsc{i}}^{2},\bar{w}_{\textsc{ii}}^{1})}\prod_{s=3}^{i-1}\frac{f(\bar{t}_{\textsc{i}}^{s},\bar{t}_{\textsc{ii}}^{s})}{h(\bar{t}_{\textsc{i}}^{s},\bar{t}_{\textsc{i}}^{s-1})f(\bar{t}_{\textsc{i}}^{s},\bar{t}_{\textsc{ii}}^{s-1})},
\end{multline}
where the sum goes over all the partitions of $\bar{w}^{1}=\bar{w}_{\textsc{i}}^{1}\cup\bar{w}_{\textsc{ii}}^{1}$
and $\bar{t}^{s}=\bar{t}_{\textsc{i}}^{s}\cup\bar{t}_{\textsc{ii}}^{s}$
where $\#\bar{w}_{\textsc{i}}^{1}=\#\bar{t}_{\textsc{i}}^{s}=1$ for
$s=2,\dots,i-1$. Substituting back to (\ref{eq:ttemp}) we obtain
a recurrence equation for the HC:

\begin{multline}
\bar{Z}_{K}^{N}(\{z,\bar{t}^{1}\},\left\{ \bar{t}^{\nu}\right\} _{\nu=2}^{N-1})=\\
F_{K}^{(1)}(z)\sum_{\mathrm{part}}\bar{Z}_{K}^{N}(\{\bar{\omega}_{\textsc{ii}}^{\nu}\}_{\nu=1}^{N-2},\bar{t}_{\textsc{ii}}^{N-1})\prod_{s=1}^{N-2}\frac{f(\omega_{\textsc{i}}^{s},\bar{\omega}_{\textsc{ii}}^{s})}{h(\omega_{\textsc{i}}^{s},\omega_{\textsc{i}}^{s-1})f(\omega_{\textsc{i}}^{s},\bar{\omega}_{\textsc{ii}}^{s-1})}\frac{f(\bar{t}_{\textsc{i}}^{N-1},\bar{t}_{\textsc{ii}}^{N-1})f(\bar{t}^{N-1},-z)}{h(\bar{t}_{\textsc{i}}^{N-1},\omega_{\textsc{i}}^{N-2})f(\bar{t}_{\textsc{i}}^{N-1},\bar{\omega}_{\textsc{ii}}^{N-2})f(\bar{t}^{2},z)}-\\
-\sum_{i=2}^{N}\frac{K_{N,i}(z)}{K_{N,1}(z)}\sum_{\mathrm{part}}\bar{Z}_{K}^{N}(\bar{w}_{\textsc{ii}}^{1},\{\bar{t}_{\textsc{ii}}^{\nu}\}_{\nu=2}^{i-1},\{\bar{t}^{\nu}\}_{\nu=i}^{N-1})\frac{f(\bar{w}_{\textsc{i}}^{1},\bar{w}_{\textsc{ii}}^{1})}{h(\bar{w}_{\textsc{i}}^{1},z)}\frac{f(\bar{t}_{\textsc{i}}^{2},\bar{t}_{\textsc{ii}}^{2})}{h(\bar{t}_{\textsc{i}}^{2},\bar{w}_{\textsc{i}}^{1})f(\bar{t}_{\textsc{i}}^{2},\bar{w}_{\textsc{ii}}^{1})}\prod_{s=3}^{i-1}\frac{f(\bar{t}_{\textsc{i}}^{s},\bar{t}_{\textsc{ii}}^{s})}{h(\bar{t}_{\textsc{i}}^{s},\bar{t}_{\textsc{i}}^{s-1})f(\bar{t}_{\textsc{i}}^{s},\bar{t}_{\textsc{ii}}^{s-1})},\label{eq:rect1-2}
\end{multline}
where
\begin{equation}
F_{K}^{(1)}(z):=\frac{K_{N-1,2}(z)}{K_{N,1}(z)}-\frac{K_{N,2}(z)}{K_{N,1}(z)}\frac{K_{N-1,1}(z)}{K_{N,1}(z)}.
\end{equation}
We just proved the equation (\ref{eq:recursionZ_UTw1}).

Let us calculate the recursion for the $(N-1)$-th Bethe roots. We
use the recurrence equation (\ref{eq:rec2})
\begin{equation}
\mathbb{B}(\left\{ \bar{t}^{s}\right\} _{s=1}^{N-2},\{z,\bar{t}^{N-1}\})=\sum_{k=1}^{N-1}\frac{T_{k,N}(z)}{\lambda_{N}(z)}\sum_{\mathrm{part}}\mathbb{B}(\left\{ \bar{t}^{s}\right\} _{s=1}^{k-1},\left\{ \bar{t}_{\textsc{ii}}^{s}\right\} _{s=k}^{N-2},\bar{t}^{N-1})\frac{\prod_{\nu=k}^{N-2}g(\bar{t}_{\textsc{i}}^{\nu+1},\bar{t}_{\textsc{i}}^{\nu})f(\bar{t}_{\textsc{i}}^{\nu},\bar{t}_{\textsc{ii}}^{\nu})}{\prod_{\nu=k}^{N-1}f(\bar{t}_{\textsc{i}}^{\nu},\bar{t}^{\nu-1})},
\end{equation}
and the $KT$-relation
\begin{equation}
\left\langle \Psi\right|T_{k,N}(z)=\sum_{i=1}^{N}\frac{K_{k,i}(-z)}{K_{N,1}(-z)}\left\langle \Psi\right|T_{i,1}(-z)-\sum_{j=1}^{N-1}\frac{K_{j,1}(-z)}{K_{N,1}(-z)}\left\langle \Psi\right|T_{k,j}(z).
\end{equation}
The action of the operators $T_{i,1}(-z)$ can be written as
\begin{multline}
\frac{T_{i,1}(-z)}{\lambda_{N}(z)}\mathbb{B}(\bar{t})=\prod_{s=1}^{N-1}\alpha_{s}(z)\sum_{\mathrm{part}}\mathbb{B}(\bar{\omega}_{\textsc{ii}})\frac{\prod_{s=1}^{i-1}f(\bar{\omega}_{\textsc{i}}^{s},\bar{\omega}_{\textsc{iii}}^{s})}{\prod_{s=1}^{i-2}f(\bar{\omega}_{\textsc{i}}^{s+1},\bar{\omega}_{\textsc{iii}}^{s})}\times\\
\prod_{s=1}^{i-1}\frac{f(\bar{\omega}_{\textsc{i}}^{s},\bar{\omega}_{\textsc{ii}}^{s})}{h(\bar{\omega}_{\textsc{i}}^{s},\bar{\omega}_{\textsc{i}}^{s-1})f(\bar{\omega}_{\textsc{i}}^{s},\bar{\omega}_{\textsc{ii}}^{s-1})}\prod_{s=1}^{N-1}\frac{\alpha_{s}(\bar{\omega}_{\textsc{iii}}^{s})f(\bar{\omega}_{\textsc{ii}}^{s},\bar{\omega}_{\textsc{iii}}^{s})}{h(\bar{\omega}_{\textsc{iii}}^{s+1},\bar{\omega}_{\textsc{iii}}^{s})f(\bar{\omega}_{\textsc{ii}}^{s+1},\bar{\omega}_{\textsc{iii}}^{s})},
\end{multline}
where we used the symmetry property (\ref{eq:constLamUTw}). The sum
goes over the partitions $\bar{\omega}=\bar{\omega}_{\textsc{i}}\cup\bar{\omega}_{\textsc{ii}}\cup\bar{\omega}_{\textsc{iii}}$
where $\bar{\omega}^{\nu}=\{-z,\bar{t}^{\nu}\}$ and $\#\bar{\omega}_{\textsc{i}}^{\nu}=\#\bar{\omega}_{\textsc{iii}}^{\mu}=1$
for $\mu=1,\dots,N$ and $\nu=0,\dots,i-1$. We can see that the $\alpha_{\nu}(\bar{t}_{k}^{\nu})$-independent
terms require that $\bar{\omega}_{\textsc{iii}}^{s}=\{-z\}$ for $s=1,\dots,N-1$.
Substituting back, the action formula simplifies as
\begin{equation}
\frac{T_{i,1}(-z)}{\lambda_{N}(z)}\mathbb{B}(\bar{t})\cong f(\bar{t}^{1},-z)\sum_{\mathrm{part}}\mathbb{B}(\left\{ \bar{t}_{\textsc{ii}}^{s}\right\} _{s=1}^{i-1},\left\{ \bar{t}^{s}\right\} _{s=i}^{N-1})\prod_{s=1}^{i-1}\frac{f(\bar{t}_{\textsc{i}}^{s},\bar{t}_{\textsc{ii}}^{s})}{h(\bar{t}_{\textsc{i}}^{s},\bar{t}_{\textsc{i}}^{s-1})f(\bar{t}_{\textsc{i}}^{s},\bar{t}_{\textsc{ii}}^{s-1})}.
\end{equation}
The sum goes over the partitions $\bar{t}^{s}=\bar{t}_{\textsc{i}}^{s}\cup\bar{t}_{\textsc{ii}}^{s}$
where $\#\bar{t}_{\textsc{i}}^{s}=1$ for $s=1,\dots,i-1$. The action
formula for $T_{k,j}(z)$ can be written as
\begin{multline}
\frac{T_{k,j}(z)}{\lambda_{N}(z)}\mathbb{B}(\bar{t})=\sum_{\mathrm{part}}\mathbb{B}(\bar{w}_{\textsc{ii}})\frac{\prod_{s=j}^{k-1}f(\bar{w}_{\textsc{i}}^{s},\bar{w}_{\textsc{iii}}^{s})}{\prod_{s=j}^{k-2}f(\bar{w}_{\textsc{i}}^{s+1},\bar{w}_{\textsc{iii}}^{s})}\times\\
\prod_{s=1}^{k-1}\frac{f(\bar{w}_{\textsc{i}}^{s},\bar{w}_{\textsc{ii}}^{s})}{h(\bar{w}_{\textsc{i}}^{s},\bar{w}_{\textsc{i}}^{s-1})f(\bar{w}_{\textsc{i}}^{s},\bar{w}_{\textsc{ii}}^{s-1})}\prod_{s=j}^{N-1}\frac{\alpha_{s}(\bar{w}_{\textsc{iii}}^{s})f(\bar{w}_{\textsc{ii}}^{s},\bar{w}_{\textsc{iii}}^{s})}{h(\bar{w}_{\textsc{iii}}^{s+1},\bar{w}_{\textsc{iii}}^{s})f(\bar{w}_{\textsc{ii}}^{s+1},\bar{w}_{\textsc{iii}}^{s})}.
\end{multline}
For $j<N$ all terms contains $\alpha_{N-1}(\bar{w}_{\textsc{iii}}^{N-1})$
therefore all terms are $\alpha$-dependent i.e.
\begin{equation}
\frac{T_{k,j}(z)}{\lambda_{N}(z)}\mathbb{B}(\bar{t})\cong0,
\end{equation}
for $j<N$. Substituting back to the recurrence equation we just obtained
an other recursion for the HC as
\begin{multline}
\bar{Z}_{K}^{N}(\left\{ \bar{t}^{s}\right\} _{s=1}^{N-2},\{z,\bar{t}^{N-1}\})=\sum_{i=1}^{N}\sum_{j=1}^{N-1}\frac{K_{j,i}(-z)}{K_{N,1}(-z)}f(\bar{t}^{1},-z)\sum_{\mathrm{part}}\bar{Z}_{K}^{N}(\left\{ \bar{t}_{\textsc{ii}}^{s}\right\} _{s=1}^{N-2},\bar{t}^{N-1})\times\\
\prod_{s=1}^{i-1}\frac{f(\bar{t}_{\textsc{i}}^{s},\bar{t}_{\textsc{ii}}^{s})}{h(\bar{t}_{\textsc{i}}^{s},\bar{t}_{\textsc{i}}^{s-1})f(\bar{t}_{\textsc{i}}^{s},\bar{t}_{\textsc{ii}}^{s-1})}\frac{\prod_{\nu=j}^{N-2}g(\bar{t}_{\textsc{iii}}^{\nu+1},\bar{t}_{\textsc{iii}}^{\nu})f(\bar{t}_{\textsc{iii}}^{\nu},\bar{t}_{\textsc{ii}}^{\nu})f(\bar{t}_{\textsc{iii}}^{\nu},\bar{t}_{\textsc{i}}^{\nu})}{\prod_{\nu=j}^{N-1}f(\bar{t}_{\textsc{iii}}^{\nu},\bar{t}^{\nu-1})},\label{eq:recUTWN}
\end{multline}
where we sum up to the partitions $\bar{t}^{\nu}=\bar{t}_{\textsc{i}}^{\nu}\cup\bar{t}_{\textsc{ii}}^{\nu}\cup\bar{t}_{\textsc{iii}}^{\nu}$
for $\nu=1,\dots,N-2$ where 
\begin{align*}
\#\bar{t}_{\textsc{i}}^{\nu} & =1,\quad\text{for \ensuremath{\nu<i}},\qquad\#\bar{t}_{\textsc{i}}^{\nu}=0,\quad\text{for \ensuremath{i\leq\nu}},\\
\#\bar{t}_{\textsc{iii}}^{\nu} & =0,\quad\text{for \ensuremath{\nu<j}},\qquad\#\bar{t}_{\textsc{iii}}^{\nu}=1,\quad\text{for \ensuremath{j\leq\nu}}
\end{align*}
and $\bar{t}_{\textsc{iii}}^{N-1}=\{z\},\bar{t}_{\textsc{ii}}^{N-1}=\bar{t}^{N-1}$
and $\bar{t}_{\textsc{i}}^{0}=\{-z\}$.

\subsection{Twisted case}

Let us derive the recursion for $t_{k}^{1}$-s in the twisted case.
We start with the recurrence equation (\ref{eq:rec1})

\begin{equation}
\left\langle \Psi\right|\mathbb{B}(\{z,\bar{t}^{1}\},\left\{ \bar{t}^{\nu}\right\} _{\nu}^{N-1})\cong\left\langle \Psi\right|\frac{T_{1,2}(z)}{\lambda_{2}(z)f(\bar{t}^{2},z)}\mathbb{B}(\bar{t}).
\end{equation}
Now we can use the twisted $KT$-relation (\ref{eq:KT_Tw})
\begin{equation}
K_{N,1}\left\langle \Psi\right|T_{1,2}(u)=\lambda_{0}(u)\sum_{j=1}^{N}\left\langle \Psi\right|\widehat{T}_{N,j}(-u)K_{j,2}-\sum_{i=2}^{N}K_{N,i}\left\langle \Psi\right|T_{i,2}(u).
\end{equation}
 The $T_{1,2}(z)$ transform into $\widehat{T}_{N,k}(-z)$ and $T_{j,2}(z)$
where $k=1,\dots,N$ and $j=2,\dots,N$ therefore
\begin{equation}
\left\langle \Psi\right|\mathbb{B}(\{z,\bar{t}^{1}\},\left\{ \bar{t}^{\nu}\right\} _{\nu=2}^{N-1})\cong\frac{1}{f(\bar{t}^{2},z)}\left[\sum_{j=1}^{N}\frac{K_{k,2}}{K_{N,1}}\left\langle \Psi\right|\frac{\lambda_{0}(z)\widehat{T}_{N,j}(-z)}{\lambda_{2}(z)}\mathbb{B}(\bar{t})-\sum_{i=2}^{N}\frac{K_{N,i}}{K_{N,1}}\left\langle \Psi\right|\frac{T_{i,2}(z)}{\lambda_{2}(z)}\mathbb{B}(\bar{t})\right].\label{eq:temp3}
\end{equation}
We can also use the action formula (\ref{eq:act}) of operator $\widehat{T}_{N,j}(-z)$.
Let us start with $\widehat{T}_{N,N}(-z)$:
\begin{multline}
\left\langle \Psi\right|\frac{\lambda_{0}(z)\widehat{T}_{N,N}(-z)}{\lambda_{2}(z)}\mathbb{B}(\bar{t})=\\
\alpha_{1}(z)\sum_{\mathrm{part}}\left\langle \Psi\right|\mathbb{B}(\bar{\omega}_{\textsc{ii}})\frac{\prod_{s=2}^{N-1}f(\bar{t}^{s-1}-c,\bar{t}^{s})}{\prod_{s=2}^{N-1}f(\bar{\omega}_{\textsc{ii}}^{s-1}-c,\bar{\omega}_{\textsc{ii}}^{s})}\prod_{s=1}^{N-1}\frac{f(\bar{\omega}_{\textsc{i}}^{s},\bar{\omega}_{\textsc{ii}}^{s})}{h(\bar{\omega}_{\textsc{i}}^{s},\bar{\omega}_{\textsc{i}}^{s+1}+c)f(\bar{\omega}_{\textsc{i}}^{s},\bar{\omega}_{\textsc{ii}}^{s+1}+c)}=\\
\alpha_{1}(z)\sum\left\langle \Psi\right|\mathbb{B}(\bar{\omega}_{\textsc{ii}})\prod_{s=1}^{N-1}\frac{f(\bar{\omega}_{\textsc{i}}^{s},\bar{\omega}_{\textsc{ii}}^{s})f(\bar{t}^{s},\bar{t}^{s+1}+c)}{h(\bar{\omega}_{\textsc{i}}^{s},\bar{\omega}_{\textsc{i}}^{s+1}+c)f(\bar{\omega}^{s},\bar{\omega}_{\textsc{ii}}^{s+1}+c)},
\end{multline}
where we used the symmetry property (\ref{eq:constLamTw}). The sum
goes over the partitions $\bar{\omega}=\bar{\omega}_{\textsc{i}}\cup\bar{\omega}_{\textsc{ii}}$
where $\bar{\omega}^{\nu}=\{-z-\nu c,\bar{t}^{\nu}\}$ and $\#\bar{\omega}_{\textsc{i}}^{\nu}=1$.
Since there is a $f(\bar{\omega}^{s},\bar{\omega}_{\textsc{ii}}^{s+1}+c)$
term in the denominator and $-z-sc\in\bar{\omega}^{s}$ therefore
the overlap is non-zero only if $-z-(s+1)c\notin\bar{\omega}_{\textsc{ii}}^{s+1}$
therefore $\bar{\omega}_{\textsc{i}}^{s+1}=-z-(s+1)c$ for $s>1$.
Using this observation, the action formula simplifies as
\begin{equation}
\left\langle \Psi\right|\frac{\lambda_{0}(z)\widehat{T}_{N,N}(-z)}{\lambda_{2}(z)}\mathbb{B}(\bar{t})=\alpha_{1}(z)\sum_{\mathrm{part}}\left\langle \Psi\right|\mathbb{B}(\bar{\omega}_{\textsc{ii}}^{1},\{\bar{t}^{\nu}\}_{\nu=2}^{N-1})\frac{f(\bar{\omega}_{\textsc{i}}^{1},\bar{\omega}_{\textsc{ii}}^{1})}{h(\bar{\omega}_{\textsc{i}}^{1},-z-c)}.
\end{equation}
 The term
\begin{equation}
\alpha_{1}(z)\left\langle \Psi\right|\mathbb{B}(\bar{\omega}_{\textsc{ii}}^{1},\{\bar{t}^{\nu}\}_{\nu=2}^{N-1})
\end{equation}
is $\alpha$-independent only when $-z-c\in\bar{\omega}_{\textsc{ii}}^{1}$.
These $\alpha$-independent terms read as
\begin{equation}
\alpha_{1}(z)\left\langle \Psi\right|\mathbb{B}(\bar{\omega}_{\textsc{ii}}^{1},\{\bar{t}^{\nu}\}_{\nu=2}^{N-1})\cong W_{K}^{N}(\{-z-c\},\emptyset^{\times N-2}|\bar{t}_{\textsc{ii}}^{1},\{\bar{t}^{\nu}\}_{\nu=2}^{N-1}).
\end{equation}
Substituting back, the action formula simplifies as
\begin{equation}
\left\langle \Psi\right|\frac{\lambda_{0}(z)\widehat{T}_{N,N}(-z)}{\lambda_{2}(z)}\mathbb{B}(\bar{t})\cong\sum_{\mathrm{part}}Z_{K}^{N}(\{-z-c\},\emptyset^{\times N-2})\bar{Z}_{K}^{N}(\bar{t}_{\textsc{ii}}^{1},\{\bar{t}^{\nu}\}_{\nu=2}^{N-1})\frac{f(\bar{t}^{1},-z-c)}{f(\bar{t}^{2},-z-c)}\frac{f(\bar{t}_{\textsc{i}}^{1},\bar{t}_{\textsc{ii}}^{1})}{h(\bar{t}_{\textsc{i}}^{1},-z-c)},
\end{equation}
where we used the identity (\ref{eq:WZZ}). This formula contains
the HC $Z(\{-z-c\},\emptyset^{\times N-2})$ which was already calculated
in appendix \ref{sec:Elementary}, therefore
\begin{equation}
Z_{K}^{N}(\{-z-c\},\emptyset^{\times N-2})=\frac{K_{N,2}}{K_{N,1}}.
\end{equation}
Let us continue with the action of $\widehat{T}_{N,j}(-z)$ for $j<N$.
The action formula (\ref{eq:actTw}) 
\begin{multline}
\frac{\lambda_{0}(z)\widehat{T}_{N,j}(-z)}{\lambda_{2}(z)}\mathbb{B}(\bar{t})=(-1)^{N-j}\alpha_{1}(z)\sum_{\mathrm{part}}\mathbb{B}(\bar{\omega}_{\textsc{ii}})\frac{\prod_{s=2}^{N-1}f(\bar{t}^{s-1}-c,\bar{t}^{s})}{\prod_{s=2}^{N-1}f(\bar{\omega}_{\textsc{ii}}^{s-1}-c,\bar{\omega}_{\textsc{ii}}^{s})}\frac{\prod_{s=1}^{N-j}f(\bar{\omega}_{\textsc{i}}^{s},\bar{\omega}_{\textsc{iii}}^{s})}{\prod_{s=2}^{N-j}f(\bar{\omega}_{\textsc{i}}^{s-1}-c,\bar{\omega}_{\textsc{iii}}^{s})}\times\\
\prod_{s=1}^{N-1}\frac{f(\bar{\omega}_{\textsc{i}}^{s},\bar{\omega}_{\textsc{ii}}^{s})}{h(\bar{\omega}_{\textsc{i}}^{s},\bar{\omega}_{\textsc{i}}^{s+1}+c)f(\bar{\omega}_{\textsc{i}}^{s},\bar{\omega}_{\textsc{ii}}^{s+1}+c)}\prod_{s=1}^{N-j}\frac{\alpha_{s}(\bar{\omega}_{\textsc{iii}}^{s})f(\bar{\omega}_{\textsc{ii}}^{s},\bar{\omega}_{\textsc{iii}}^{s})}{h(\bar{\omega}_{\textsc{iii}}^{s-1}-c,\bar{\omega}_{\textsc{iii}}^{s})f(\bar{\omega}_{\textsc{ii}}^{s-1}-c,\bar{\omega}_{\textsc{iii}}^{s})}
\end{multline}
explicitly contains $\alpha_{\nu}(\bar{\omega}_{\textsc{iii}}^{\nu})$
where $1\leq\nu<N-j$ therefore it can not be $\alpha$-independent
for $j<N-1$ i.e.
\begin{equation}
\left\langle \Psi\right|\frac{\lambda_{0}(z)\widehat{T}_{N,j}(-z)}{\lambda_{2}(z)}\mathbb{B}(\bar{t})\cong0,
\end{equation}
for $j<N-1$. The only $\alpha$-independent term comes form $j=N-1$,
and in this case $\bar{\omega}_{\textsc{iii}}^{N-1}=\{-z-(N-1)c\}$,
therefore
\begin{equation}
\left\langle \Psi\right|\frac{\lambda_{0}(z)\widehat{T}_{N,N-1}(-z)}{\lambda_{2}(z)}\mathbb{B}(\bar{t})\cong(-1)\frac{f(\bar{t}^{1},-z-c)}{f(\bar{t}^{2},-z-c)}\sum_{\mathrm{part}}\bar{Z}_{K}^{N}(\bar{t}_{\textsc{ii}}^{1},\{\bar{t}^{\nu}\}_{\nu=2}^{N-1})\frac{f(\bar{t}_{\textsc{i}}^{1},\bar{t}_{\textsc{ii}}^{1})}{h(\bar{t}_{\textsc{i}}^{1},-z-c)}.
\end{equation}

Let us continue with the action of $T_{i,2}$. The $\alpha$-independent
terms are completely the same as they were for the untwisted case
i.e.
\begin{multline}
\left\langle \Psi\right|\frac{T_{i,2}(z)}{\lambda_{2}(z)}\mathbb{B}(\bar{t})=\sum_{\mathrm{part}}\bar{Z}_{K}^{N}(\bar{w}_{\textsc{ii}}^{1},\{\bar{t}_{\textsc{ii}}^{\nu}\}_{\nu=2}^{i-1},\{\bar{t}^{\nu}\}_{\nu=i}^{N-1})\times\\
\frac{f(\bar{w}_{\textsc{i}}^{1},\bar{w}_{\textsc{ii}}^{1})}{h(\bar{w}_{\textsc{i}}^{1},z)}\frac{f(\bar{t}_{\textsc{i}}^{2},\bar{t}_{\textsc{ii}}^{2})f(\bar{t}^{2},z)}{h(\bar{t}_{\textsc{i}}^{2},\bar{w}_{\textsc{i}}^{1})f(\bar{t}_{\textsc{i}}^{2},\bar{w}_{\textsc{ii}}^{1})}\prod_{s=3}^{i-1}\frac{f(\bar{t}_{\textsc{i}}^{s},\bar{t}_{\textsc{ii}}^{s})}{h(\bar{t}_{\textsc{i}}^{s},\bar{t}_{\textsc{i}}^{s-1})f(\bar{t}_{\textsc{i}}^{s},\bar{t}_{\textsc{ii}}^{s-1})}.
\end{multline}
Substituting back to (\ref{eq:temp3}) we obtain a recurrence formula
for the HC:

\begin{multline}
\bar{Z}_{K}^{N}(\{z,\bar{t}^{1}\},\left\{ \bar{t}^{\nu}\right\} _{\nu=2}^{N-1})=F_{K}^{(1)}\frac{f(\bar{t}^{1},-z-c)}{f(\bar{t}^{2},-z-c)f(\bar{t}^{2},z)}\sum_{\mathrm{part}}\bar{Z}_{K}^{N}(\bar{t}_{\textsc{ii}}^{1},\{\bar{t}^{\nu}\}_{\nu=2}^{N-1})\frac{f(\bar{t}_{\textsc{i}}^{1},\bar{t}_{\textsc{ii}}^{1})}{h(\bar{t}_{\textsc{i}}^{1},-z-c)}-\\
-\sum_{i=2}^{N}\frac{K_{N,i}}{K_{N,1}}\sum\bar{Z}_{K}^{N}(\bar{w}_{\textsc{ii}}^{1},\{\bar{t}_{\textsc{ii}}^{\nu}\}_{\nu=2}^{i-1},\{\bar{t}^{\nu}\}_{\nu=i}^{N-1})\frac{f(\bar{w}_{\textsc{i}}^{1},\bar{w}_{\textsc{ii}}^{1})}{h(\bar{w}_{\textsc{i}}^{1},z)}\frac{f(\bar{t}_{\textsc{i}}^{2},\bar{t}_{\textsc{ii}}^{2})}{h(\bar{t}_{\textsc{i}}^{2},\bar{w}_{\textsc{i}}^{1})f(\bar{t}_{\textsc{i}}^{2},\bar{w}_{\textsc{ii}}^{1})}\prod_{s=3}^{i-1}\frac{f(\bar{t}_{\textsc{i}}^{s},\bar{t}_{\textsc{ii}}^{s})}{h(\bar{t}_{\textsc{i}}^{s},\bar{t}_{\textsc{i}}^{s-1})f(\bar{t}_{\textsc{i}}^{s},\bar{t}_{\textsc{ii}}^{s-1})},\label{eq:rect1-1-1}
\end{multline}
where
\begin{equation}
F_{K}^{(1)}:=\left(\frac{K_{N,2}}{K_{N,1}}\right)^{2}-\frac{K_{N-1,2}}{K_{N,1}}.
\end{equation}
We just proved the equation (\ref{eq:recursionZ_Tw}).

\section{Poles of the HC-s\label{sec:Poles}}

In this section we derive the equations (\ref{eq:pole_Utw}) and (\ref{eq:pole_Tw}).

\subsection{Twisted case}

Let us start with the twisted case. In the following we prove that
the HC has the following poles
\begin{equation}
\bar{Z}_{K}^{N}(\bar{t})=\frac{c}{t_{k}^{\nu}+t_{l}^{\nu}+\nu c}F_{K}^{(\nu)}\frac{f(\bar{t}_{k,l}^{\nu},t_{k}^{\nu})}{f(\bar{t}^{\nu+1},t_{k}^{\nu})}\frac{f(\bar{t}_{k,l}^{\nu},-t_{k}^{\nu}-\nu c)}{f(\bar{t}^{\nu+1},-t_{k}^{\nu}-\nu c)}\bar{Z}_{K}^{N}(\bar{t}\backslash\{t_{k}^{\nu},t_{l}^{\nu}\})+reg,\label{eq:poleTw}
\end{equation}
for $\nu=1,\dots,N-1$ where
\begin{equation}
F_{K}^{(k)}:=\left(\frac{K_{N+1-k,k+1}^{(k)}}{K_{N+1-k,k}^{(k)}}\right)^{2}-\frac{K_{N-k,k+1}^{(k)}}{K_{N+1-k,k}^{(k)}}.
\end{equation}
 We prove (\ref{eq:poleTw}) by induction. Let us assume that it is
true up to $N-1$. Using the embedding of the HC-s (\ref{eq:embedZ_Tw})
\begin{equation}
\bar{Z}_{K}^{N}(\emptyset,\left\{ \bar{t}^{k}\right\} _{k=2}^{N-1})=\bar{Z}_{K^{(2)}}^{N-1}(\left\{ \bar{t}^{k}+c/2\right\} _{k=2}^{N-1}),
\end{equation}
 we obtain that
\begin{equation}
\bar{Z}_{K}^{N}(\emptyset,\left\{ \bar{t}^{s}\right\} _{s=2}^{N-1})=\frac{c}{t_{k}^{\nu}+t_{l}^{\nu}+\nu c}F_{K}^{(\nu)}\frac{f(\bar{\tau}^{\nu},t_{k}^{\nu})}{f(\bar{t}^{\nu+1},t_{k}^{\nu})}\frac{f(\bar{\tau}^{\nu},-t_{k}^{\nu}-\nu c)}{f(\bar{t}^{\nu+1},-t_{k}^{\nu}-\nu c)}\bar{Z}_{K}^{N}(\emptyset,\left\{ \bar{\tau}^{s}\right\} _{s=2}^{N-1})+reg,\label{eq:ind_ass}
\end{equation}
for $\nu=2,\dots,N-1$ where $\bar{\tau}=\bar{t}\backslash\{t_{k}^{\nu},t_{l}^{\nu}\}$.
In the following we prove the equation (\ref{eq:poleTw}) for $\nu=1,2,\dots,N-1$
with general $r_{1}$.

\subsubsection*{Case $\nu>1$}

Let us start with the $\nu>1$ case of (\ref{eq:poleTw}). We do this
proof with an other inductions in $r_{1}$ therefore let us assume
that (\ref{eq:poleTw}) is satisfied for $r_{1}$ or less first type
of Bethe roots. Let us increase the number of the first roots and
apply the recurrence equation (\ref{eq:recursionZ_Tw})
\begin{multline}
\bar{Z}_{K}^{N}(\{z,\bar{t}^{1}\},\left\{ \bar{t}^{s}\right\} _{s=2}^{N-1})=F_{K}^{(1)}\frac{f(\bar{t}^{1},-z-c)}{f(\bar{t}^{2},-z-c)f(\bar{t}^{2},z)}\sum_{\mathrm{part}}\bar{Z}_{K}^{N}(\bar{t}_{\textsc{ii}}^{1},\{\bar{t}^{s}\}_{s=2}^{N-1})\frac{f(\bar{t}_{\textsc{i}}^{1},\bar{t}_{\textsc{ii}}^{1})}{h(\bar{t}_{\textsc{i}}^{1},-z-c)}-\\
-\sum_{i=2}^{N}\frac{K_{N,i}}{K_{N,1}}\sum_{\mathrm{part}}\bar{Z}_{K}^{N}(\bar{w}_{\textsc{ii}}^{1},\{\bar{t}_{\textsc{ii}}^{s}\}_{s=2}^{i-1},\{\bar{t}^{s}\}_{s=i}^{N-1})\frac{f(\bar{w}_{\textsc{i}}^{1},\bar{w}_{\textsc{ii}}^{1})}{h(\bar{w}_{\textsc{i}}^{1},z)}\frac{f(\bar{t}_{\textsc{i}}^{2},\bar{t}_{\textsc{ii}}^{2})}{h(\bar{t}_{\textsc{i}}^{2},\bar{w}_{\textsc{i}}^{1})f(\bar{t}_{\textsc{i}}^{2},\bar{w}_{\textsc{ii}}^{1})}\prod_{s=3}^{i-1}\frac{f(\bar{t}_{\textsc{i}}^{s},\bar{t}_{\textsc{ii}}^{s})}{h(\bar{t}_{\textsc{i}}^{s},\bar{t}_{\textsc{i}}^{s-1})f(\bar{t}_{\textsc{i}}^{s},\bar{t}_{\textsc{ii}}^{s-1})}.
\end{multline}
We can apply the induction assumption for the poles of the HC-s in
the r.h.s. since $\#\bar{w}_{\textsc{ii}}^{1}=r_{1}$ and $\#\bar{t}_{\textsc{ii}}^{1}=r_{1}-1$.
We can see that the poles at $t_{k}^{\nu}+t_{l}^{\nu}+\nu c=0$ appear
only in the HC-s. For the expression of the second line the poles
appear only when $t_{k}^{\nu},t_{l}^{\nu}\in\bar{t}_{\textsc{ii}}^{\nu}$
for $\nu<i$. Our strategy is that we use the induction assumption
for these poles and rearrange the expressions to get the following
form
\begin{equation}
\bar{Z}_{K}^{N}(\{z,\bar{t}^{1}\},\left\{ \bar{t}^{s}\right\} _{s=2}^{N-1})=\left(\frac{c}{t_{k}^{\nu}+t_{l}^{\nu}+\nu c}F_{K}^{(\nu)}\frac{f(\bar{\tau}^{\nu},t_{k}^{\nu})}{f(\bar{t}^{\nu+1},t_{k}^{\nu})}\frac{f(\bar{\tau}^{\nu},-t_{k}^{\nu}-\nu c)}{f(\bar{t}^{\nu+1},-t_{k}^{\nu}-\nu c)}\right)\mathcal{Q}(\{z,\bar{t}^{1}\},\left\{ \bar{\tau}^{\nu}\right\} _{\nu}^{N-1})+reg.
\end{equation}
where $\bar{\tau}=\bar{t}\backslash\{t_{k}^{\nu},t_{l}^{\nu}\}$.
Let us calculate the quotient $\mathcal{Q}$. We can use the induction
assumptions for the following HC-s 
\begin{equation}
\bar{Z}_{K}^{N}(\bar{t}_{\textsc{ii}}^{1},\{\bar{t}^{s}\}_{s=2}^{N-1})=\frac{c}{t_{k}^{\nu}+t_{l}^{\nu}+\nu c}F_{K}^{(\nu)}\frac{f(\bar{\tau}^{\nu},t_{k}^{\nu})}{f(\bar{t}^{\nu+1},t_{k}^{\nu})}\frac{f(\bar{\tau}^{\nu},-t_{k}^{\nu}-\nu c)}{f(\bar{t}^{\nu+1},-t_{k}^{\nu}-\nu c)}\bar{Z}_{K}^{N}(\bar{t}_{\textsc{ii}}^{1},\{\bar{\tau}^{s}\}_{s=2}^{N-1})+reg,
\end{equation}
and
\begin{multline}
\bar{Z}_{K}^{N}(\bar{w}_{\textsc{ii}}^{1},\{\bar{t}_{\textsc{ii}}^{s}\}_{s=2}^{i-1},\{\bar{t}^{s}\}_{s=i}^{N-1})=\\
\frac{c}{t_{k}^{\nu}+t_{l}^{\nu}+\nu c}F_{K}^{(\nu)}\frac{f(\bar{\tau}^{\nu},t_{k}^{\nu})}{f(\bar{t}^{\nu+1},t_{k}^{\nu})}\frac{f(\bar{\tau}^{\nu},-t_{k}^{\nu}-\nu c)}{f(\bar{t}^{\nu+1},-t_{k}^{\nu}-\nu c)}\bar{Z}_{K}^{N}(\bar{w}_{\textsc{ii}}^{1},\{\bar{\tau}_{\textsc{ii}}^{s}\}_{s=2}^{i-1},\{\bar{\tau}^{s}\}_{s=i}^{N-1})+reg,
\end{multline}
for $\nu\geq i$ and
\begin{multline}
\bar{Z}_{K}^{N}(\bar{w}_{\textsc{ii}}^{1},\{\bar{t}_{\textsc{ii}}^{s}\}_{s=2}^{i-1},\{\bar{t}^{s}\}_{s=i}^{N-1})=\\
\frac{c}{t_{k}^{\nu}+t_{l}^{\nu}+\nu c}F_{K}^{(\nu)}\frac{f(\bar{\tau}_{\textsc{ii}}^{\nu},t_{k}^{\nu})}{f(\bar{t}_{\textsc{ii}}^{\nu+1},t_{k}^{\nu})}\frac{f(\bar{\tau}_{\textsc{ii}}^{\nu},-t_{k}^{\nu}-\nu c)}{f(\bar{t}_{\textsc{ii}}^{\nu+1},-t_{k}^{\nu}-\nu c)}\bar{Z}_{K}^{N}(\bar{w}_{\textsc{ii}}^{1},\{\bar{\tau}_{\textsc{ii}}^{s}\}_{s=2}^{i-1},\{\bar{\tau}^{s}\}_{s=i}^{N-1})+reg,
\end{multline}
for $\nu<i-1$ and
\begin{multline}
\bar{Z}_{K}^{N}(\bar{w}_{\textsc{ii}}^{1},\{\bar{t}_{\textsc{ii}}^{s}\}_{s=2}^{i-1},\{\bar{t}^{s}\}_{s=i}^{N-1})=\\
\frac{c}{t_{k}^{\nu}+t_{l}^{\nu}+\nu c}F_{K}^{(\nu)}\frac{f(\bar{\tau}_{\textsc{ii}}^{\nu},t_{k}^{\nu})}{f(\bar{t}^{\nu+1},t_{k}^{\nu})}\frac{f(\bar{\tau}_{\textsc{ii}}^{\nu},-t_{k}^{\nu}-\nu c)}{f(\bar{t}^{\nu+1},-t_{k}^{\nu}-\nu c)}\bar{Z}_{K}^{N}(\bar{w}_{\textsc{ii}}^{1},\{\bar{\tau}_{\textsc{ii}}^{s}\}_{s=2}^{i-1},\{\bar{\tau}^{s}\}_{s=i}^{N-1})+reg,
\end{multline}
for $\nu=i-1$. Using the following identity
\begin{equation}
f(\bar{t}^{2},-z-c)f(\bar{t}^{2},z)=f(\bar{\tau}^{2},-z-c)f(\bar{\tau}^{2},z),
\end{equation}
after a straightforward calculation we can obtain that
\begin{multline}
\mathcal{Q}(\{z,\bar{t}^{1}\},\left\{ \bar{\tau}^{s}\right\} _{s=2}^{N-1})=F_{K}^{(1)}\frac{f(\bar{t}^{1},-z-c)}{f(\bar{\tau}^{2},-z-c)f(\bar{\tau}^{2},z)}\sum_{\mathrm{part}}\bar{Z}_{K}^{N}(\bar{t}_{\textsc{ii}}^{1},\{\bar{\tau}^{s}\}_{s=2}^{N-1})\frac{f(\bar{t}_{\textsc{i}}^{1},\bar{t}_{\textsc{ii}}^{1})}{h(\bar{t}_{\textsc{i}}^{1},-z-c)}-\\
-\sum_{i=2}^{N}\frac{K_{N,i}}{K_{N,1}}\sum_{\mathrm{part}}\bar{Z}_{K}^{N}(\bar{w}_{\textsc{ii}}^{1},\{\bar{\tau}_{\textsc{ii}}^{s}\}_{s=2}^{i-1},\{\bar{\tau}^{s}\}_{\nu=i}^{N-1})\frac{f(\bar{w}_{\textsc{i}}^{1},\bar{w}_{\textsc{ii}}^{1})}{h(\bar{w}_{\textsc{i}}^{1},z)}\frac{f(\bar{\tau}_{\textsc{i}}^{2},\bar{\tau}_{\textsc{ii}}^{2})}{h(\bar{\tau}_{\textsc{i}}^{2},\bar{w}_{\textsc{i}}^{1})f(\bar{\tau}_{\textsc{i}}^{2},\bar{w}_{\textsc{ii}}^{1})}\prod_{s=3}^{i-1}\frac{f(\bar{\tau}_{\textsc{i}}^{s},\bar{\tau}_{\textsc{ii}}^{s})}{h(\bar{\tau}_{\textsc{i}}^{s},\bar{\tau}_{\textsc{i}}^{s-1})f(\bar{\tau}_{\textsc{i}}^{s},\bar{\tau}_{\textsc{ii}}^{s-1})}.
\end{multline}
We can see that r.h.s is just the recurrence relation for $\bar{Z}_{K}(\{z,\bar{t}^{1}\},\left\{ \bar{\tau}^{s}\right\} _{s=2}^{N-1})$
(\ref{eq:recursionZ_Tw}) therefore we just proved (\ref{eq:poleTw})
for $\nu>1$.

\subsubsection*{Case $\nu=1$}

Let us continue with the case $\nu=1$. Now we also use an induction
in the number $r_{1}$. At first let us take $r_{1}=2$ and investigate
the pole at $t_{1}^{1}+t_{2}^{1}+c\to0$. For $r_{1}=2$ the HC reads
as (\ref{eq:recursionZ_Tw})
\begin{multline}
\bar{Z}_{K}^{N}(\{t_{1}^{1},t_{2}^{1}\},\left\{ \bar{t}^{s}\right\} _{s=2}^{N-1})=F_{K}^{(1)}\frac{f(t_{2}^{1},-t_{1}^{1}-c)}{f(\bar{t}^{2},-t_{1}^{1}-c)f(\bar{t}^{2},t_{1}^{1})}\bar{Z}_{K}^{N}(\emptyset,\{\bar{t}^{s}\}_{s=2}^{N-1})\frac{1}{h(t_{2}^{1},-t_{1}^{1}-c)}-\\
-\sum_{i=2}^{N}\frac{K_{N,i}}{K_{N,1}}\sum_{\mathrm{part}}\bar{Z}_{K}^{N}(\bar{t}_{\textsc{ii}}^{1},\{\bar{t}_{\textsc{ii}}^{s}\}_{s=2}^{i-1},\{\bar{t}^{s}\}_{s=i}^{N-1})\frac{f(\bar{t}_{\textsc{i}}^{1},\bar{t}_{\textsc{ii}}^{1})}{h(\bar{t}_{\textsc{i}}^{1},t_{1}^{1})}\frac{f(\bar{t}_{\textsc{i}}^{2},\bar{t}_{\textsc{ii}}^{2})}{h(\bar{t}_{\textsc{i}}^{2},\bar{t}_{\textsc{i}}^{1})f(\bar{t}_{\textsc{i}}^{2},\bar{t}_{\textsc{ii}}^{1})}\prod_{s=3}^{i-1}\frac{f(\bar{t}_{\textsc{i}}^{s},\bar{t}_{\textsc{ii}}^{s})}{h(\bar{t}_{\textsc{i}}^{s},\bar{t}_{\textsc{i}}^{s-1})f(\bar{t}_{\textsc{i}}^{s},\bar{t}_{\textsc{ii}}^{s-1})}.
\end{multline}
The pole at $t_{1}^{1}+t_{2}^{1}+c\to0$ reads as
\begin{equation}
\bar{Z}_{K}^{N}(\{t_{1}^{1},t_{2}^{1}\},\left\{ \bar{t}^{s}\right\} _{s=2}^{N-1})=\frac{c}{t_{1}^{1}+t_{2}^{1}+c}F_{K}^{(1)}\frac{1}{f(\bar{t}^{2},t_{1}^{1})f(\bar{t}^{2},-t_{1}^{1}-c)}\bar{Z}_{K}^{N}(\emptyset,\{\bar{t}^{s}\}_{s=2}^{N-1})+reg,
\end{equation}
therefore we just proved (\ref{eq:poleTw}) for $r_{1}=2$. In the
following let us assume that the equation (\ref{eq:poleTw}) holds
for $r_{1}$ or less first type of Bethe roots. Let us increase the
number of the first roots and apply the recurrence equation (\ref{eq:recursionZ_Tw})
\begin{multline}
\bar{Z}_{K}^{N}(\{z,\bar{t}^{1}\},\left\{ \bar{t}^{s}\right\} _{s=2}^{N-1})=F_{K}^{(1)}\frac{f(\bar{t}^{1},-z-c)}{f(\bar{t}^{2},-z-c)f(\bar{t}^{2},z)}\sum_{\mathrm{part}}\bar{Z}_{K}^{N}(\bar{t}_{\textsc{ii}}^{1},\{\bar{t}^{s}\}_{s=2}^{N-1})\frac{f(\bar{t}_{\textsc{i}}^{1},\bar{t}_{\textsc{ii}}^{1})}{h(\bar{t}_{\textsc{i}}^{1},-z-c)}-\\
-\sum_{i=2}^{N}\frac{K_{N,i}}{K_{N,1}}\sum_{\mathrm{part}}\bar{Z}_{K}^{N}(\bar{w}_{\textsc{ii}}^{1},\{\bar{t}_{\textsc{ii}}^{s}\}_{s=2}^{i-1},\{\bar{t}^{s}\}_{s=i}^{N-1})\frac{f(\bar{w}_{\textsc{i}}^{1},\bar{w}_{\textsc{ii}}^{1})}{h(\bar{w}_{\textsc{i}}^{1},z)}\frac{f(\bar{t}_{\textsc{i}}^{2},\bar{t}_{\textsc{ii}}^{2})}{h(\bar{t}_{\textsc{i}}^{2},\bar{w}_{\textsc{i}}^{1})f(\bar{t}_{\textsc{i}}^{2},\bar{w}_{\textsc{ii}}^{1})}\prod_{s=3}^{i-1}\frac{f(\bar{t}_{\textsc{i}}^{s},\bar{t}_{\textsc{ii}}^{s})}{h(\bar{t}_{\textsc{i}}^{s},\bar{t}_{\textsc{i}}^{s-1})f(\bar{t}_{\textsc{i}}^{s},\bar{t}_{\textsc{ii}}^{s-1})}.
\end{multline}
Our strategy is the same as before, i.e. using the induction assumption
we can express the poles of the HC-s at the r.h.s, and after that,
we rearrange the expressions to get the following form
\begin{equation}
\bar{Z}_{K}^{N}(\{z,\bar{t}^{1}\},\left\{ \bar{t}^{s}\right\} _{s=2}^{N-1})=\left(\frac{c}{t_{k}^{1}+t_{l}^{1}+c}F_{K}^{(1)}\frac{f(\bar{\mathrm{w}}^{1},t_{k}^{1})}{f(\bar{t}^{2},t_{k}^{1})}\frac{f(\bar{\mathrm{w}}^{1},-t_{k}^{1}-c)}{f(\bar{t}^{2},-t_{k}^{1}-c)}\right)\mathcal{Q}(\bar{\mathrm{w}}^{1},\left\{ \bar{t}^{s}\right\} _{s=2}^{N-1})+reg,
\end{equation}
where $\bar{\mathrm{w}}^{1}=\bar{w}^{1}\backslash\{t_{k}^{1},t_{l}^{1}\}$,
$\bar{w}^{1}=\{z,\bar{t}^{1}\}$. We will also use the notation $\bar{\mathrm{w}}_{\textsc{ii}}^{1}=\bar{w}_{\textsc{ii}}^{1}\backslash\{t_{k}^{1},t_{l}^{1}\}$.
We can use the induction assumptions for the following HC-s 
\begin{equation}
\bar{Z}_{K}^{N}(\bar{t}_{\textsc{ii}}^{1},\{\bar{t}^{s}\}_{s=2}^{N-1})=\frac{c}{t_{k}^{1}+t_{l}^{1}+c}F_{K}^{(1)}\frac{f(\bar{\tau}_{\textsc{ii}}^{1},t_{k}^{1})}{f(\bar{t}^{2},t_{k}^{1})}\frac{f(\bar{\tau}_{\textsc{ii}}^{1},-t_{k}^{1}-c)}{f(\bar{t}^{2},-t_{k}^{1}-c)}\bar{Z}_{K}^{N}(\bar{\tau}_{\textsc{ii}}^{1},\{\bar{t}^{s}\}_{s=2}^{N-1})+reg,
\end{equation}
and
\begin{equation}
\bar{Z}_{K}^{N}(\bar{w}_{\textsc{ii}}^{1},\{\bar{t}^{s}\}_{s=2}^{N-1})=\frac{c}{t_{k}^{1}+t_{l}^{1}+c}F_{K}^{(1)}\frac{f(\bar{\mathrm{w}}_{\textsc{ii}}^{1},t_{k}^{1})}{f(\bar{t}^{2},t_{k}^{1})}\frac{f(\bar{\mathrm{w}}_{\textsc{ii}}^{1},-t_{k}^{1}-c)}{f(\bar{t}^{2},-t_{k}^{1}-c)}\bar{Z}_{K}^{N}(\bar{\mathrm{w}}_{\textsc{ii}}^{1},\{\bar{t}^{s}\}_{s=2}^{N-1})+reg.
\end{equation}
Using the following identity
\begin{equation}
f(\bar{t}^{1},-z-c)=f(z,t_{k}^{1})f(z,-t_{k}^{1}-c)f(\bar{\tau}^{1},-z-c),
\end{equation}
after very straightforward calculation we can obtain that
\begin{multline}
\mathcal{Q}(\{z,\bar{\tau}^{1}\},\left\{ \bar{t}^{s}\right\} _{s=2}^{N-1})=F_{K}^{(1)}\frac{f(\bar{\tau}^{1},-z-c)}{f(\bar{t}^{2},-z-c)f(\bar{t}^{2},z)}\sum_{\mathrm{part}}\bar{Z}_{K}^{N}(\bar{\tau}_{\textsc{ii}}^{1},\{\bar{t}^{s}\}_{s=2}^{N-1})\frac{f(\bar{\tau}_{\textsc{i}}^{1},\bar{\tau}_{\textsc{ii}}^{1})}{h(\bar{\tau}_{\textsc{i}}^{1},-z-c)}-\\
-\sum_{i=2}^{N}\frac{K_{N,i}}{K_{N,1}}\sum_{\mathrm{part}}\bar{Z}_{K}^{N}(\bar{\mathrm{w}}_{\textsc{ii}}^{1},\{\bar{t}_{\textsc{ii}}^{s}\}_{s=2}^{i-1},\{\bar{t}^{s}\}_{s=i}^{N-1})\frac{f(\mathrm{\bar{\mathrm{w}}}_{\textsc{i}}^{1},\bar{\mathrm{w}}_{\textsc{ii}}^{1})}{h(\bar{\mathrm{w}}_{\textsc{i}}^{1},z)}\frac{f(\bar{t}_{\textsc{i}}^{2},\bar{t}_{\textsc{ii}}^{2})}{h(t_{\textsc{i}}^{2},\bar{\mathrm{w}}_{\textsc{i}}^{1})f(\bar{t}_{\textsc{i}}^{2},\bar{\mathrm{w}}_{\textsc{ii}}^{1})}\prod_{s=3}^{i-1}\frac{f(\bar{t}_{\textsc{i}}^{s},\bar{t}_{\textsc{ii}}^{s})}{h(\bar{t}_{\textsc{i}}^{s},\bar{t}_{\textsc{i}}^{s-1})f(\bar{t}_{\textsc{i}}^{s},\bar{t}_{\textsc{ii}}^{s-1})}.
\end{multline}
We can see that the r.h.s is just the recurrence relation for $\bar{Z}_{K}^{N}(\{z,\bar{\tau}^{1}\},\left\{ \bar{t}^{s}\right\} _{s=2}^{N-1})$.
Since the HC is symmetric under the permutations of the first type
of Bethe roots, we just proved the equation (\ref{eq:poleTw}) for
$\nu=1$.

\subsection{Untwisted case}

In this subsection we prove that the HC-s of the untwisted case have
the following poles
\begin{equation}
\bar{Z}_{K}^{N}(\bar{t})=\frac{c}{t_{k}^{\nu}+t_{l}^{N-\nu}}\tilde{F}_{K}^{(\nu)}(t_{k}^{\nu})\frac{f(\bar{\tau}^{\nu},t_{k}^{\nu})}{f(\bar{\tau}^{\nu+1},t_{k}^{\nu})}\frac{f(\bar{\tau}^{N-\nu},-t_{k}^{\nu})}{f(\bar{\tau}^{N-\nu+1},-t_{k}^{\nu})}\bar{Z}_{K}^{N}(\bar{\tau})+reg,\label{eq:pole-2-1}
\end{equation}
for $\nu=1,\dots,\left\lfloor N/2\right\rfloor $ where $\bar{\tau}=\bar{t}\backslash\{t_{k}^{\nu},t_{l}^{N-\nu}\}$,
$\tilde{F}_{K}^{(\nu)}(z)=F_{K}^{(\nu)}(z)$ for $\nu\neq\frac{N-1}{2}$
and
\begin{align}
\tilde{F}_{K}^{(\frac{N-1}{2})}(z) & =\frac{F_{K}^{(\frac{N-1}{2})}(z)}{f(-z,z)},
\end{align}
where

\begin{equation}
F_{K}^{(\nu)}(z):=\frac{K_{N-\nu,\nu+1}^{(\nu)}(z)}{K_{N+1-\nu,\nu}^{(\nu)}(z)}-\frac{K_{N+1-\nu,\nu+1}^{(\nu)}(z)}{K_{N+1-\nu,\nu}^{(\nu)}(z)}\frac{K_{N-\nu,\nu}^{(\nu)}(z)}{K_{N+1-\nu,\nu}^{(\nu)}(z)}.
\end{equation}
We follow the steps of the previous subsection therefore we prove
this statement by induction. Let us assume that it is true up to $\mathfrak{gl}(N-2)$
spin chain i.e. the HC-s
\begin{equation}
\bar{Z}_{K^{(2)}}^{N-2}(\left\{ \bar{t}^{s}\right\} _{s=2}^{N-2})=\bar{Z}_{K}^{N}(\emptyset,\left\{ \bar{t}^{s}\right\} _{s=2}^{N-2},\emptyset)
\end{equation}
satisfy the equation (\ref{eq:pole-2-1}) for $\nu=2,\dots,\left\lfloor N/2\right\rfloor $.
We have to prove the equation (\ref{eq:pole-2-1}) for $\nu=1,2,\dots,\left\lfloor N/2\right\rfloor $
with general $r_{1},r_{N-1}$ and we are done.

\subsubsection*{Case $\nu>1$}

Let us start with the $\nu>1$ case of (\ref{eq:pole-2-1}). We do
this proof with other inductions in $r_{1}$ and $r_{N-1}$. Let us
assume that (\ref{eq:pole-2-1}) is satisfied for $r_{1}$ or less
first type of Bethe roots and $r_{N-1}$ or less last type of Bethe
roots. In the following we show that we can increase $r_{1}$ and
$r_{N-1}$, separately. At first let us increase the number of $\bar{t}^{N-1}$.
We use the recurrence equation (\ref{eq:recursionZ_UTw2})

\begin{multline}
\bar{Z}_{K}^{N}(\left\{ \bar{t}^{s}\right\} _{s=1}^{N-2},\{z,\bar{t}^{N-1}\})=\sum_{i=1}^{N}\sum_{j=1}^{N-1}\frac{K_{j,i}(-z)}{K_{N,1}(-z)}f(\bar{t}^{1},-z)\sum_{\mathrm{part}}\bar{Z}_{K}^{N}(\left\{ \bar{t}_{\textsc{ii}}^{s}\right\} _{s=1}^{N-2},\bar{t}^{N-1})\times\\
\prod_{s=1}^{i-1}\frac{f(\bar{t}_{\textsc{i}}^{s},\bar{t}_{\textsc{ii}}^{s})}{h(\bar{t}_{\textsc{i}}^{s},\bar{t}_{\textsc{i}}^{s-1})f(\bar{t}_{\textsc{i}}^{s},\bar{t}_{\textsc{ii}}^{s-1})}\frac{\prod_{\nu=j}^{N-2}g(\bar{t}_{\textsc{iii}}^{\nu+1},\bar{t}_{\textsc{iii}}^{\nu})f(\bar{t}_{\textsc{iii}}^{\nu},\bar{t}_{\textsc{ii}}^{\nu})f(\bar{t}_{\textsc{iii}}^{\nu},\bar{t}_{\textsc{i}}^{\nu})}{\prod_{\nu=j}^{N-1}f(\bar{t}_{\textsc{iii}}^{\nu},\bar{t}^{\nu-1})}.\label{eq:temp-2}
\end{multline}
We follow the previous strategy i.e. we use the induction assumption
for the poles of HC in the r.h.s., and rearrange the expressions to
get the following form
\begin{multline}
\bar{Z}_{K}^{N}(\left\{ \bar{t}^{s}\right\} _{s=1}^{N-2},\{z,\bar{t}^{N-1}\})=\\
\left(\frac{c}{t_{k}^{\nu}+t_{l}^{N-\nu}}\tilde{F}_{K}^{(\nu)}(t_{k}^{\nu})\frac{f(\bar{\tau}^{\nu},t_{k}^{\nu})}{f(\bar{\tau}^{\nu+1},t_{k}^{\nu})}\frac{f(\bar{\tau}^{N-\nu},-t_{k}^{\nu})}{f(\bar{\tau}^{N-\nu+1},-t_{k}^{\nu})}\right)\mathcal{Q}(\left\{ \bar{\tau}^{s}\right\} _{s=1}^{N-2},\{z,\bar{t}^{N-1}\})+reg,
\end{multline}
where $\bar{\tau}=\bar{t}\backslash\{t_{k}^{\nu},t_{l}^{N-\nu}\}$.
Let us calculate the quotient which was denote as $\mathcal{Q}$.
We can use the induction assumptions for the HC-s in the r.h.s of
(\ref{eq:temp-2}). After a very straightforward calculation we can
obtain that
\begin{multline}
\mathcal{Q}(\left\{ \bar{\tau}^{s}\right\} _{s=1}^{N-2},\{z,\bar{t}^{N-1}\})=\sum_{i=1}^{N}\sum_{j=1}^{N-1}\frac{K_{j,i}(-z)}{K_{N,1}(-z)}f(\bar{\tau}^{1},-z)\sum_{\mathrm{part}}\bar{Z}_{K}^{N}(\left\{ \bar{\tau}_{\textsc{ii}}^{s}\right\} _{s=1}^{N-2},\bar{t}^{N-1})\times\\
\prod_{s=1}^{i-1}\frac{f(\bar{\tau}_{\textsc{i}}^{s},\bar{\tau}_{\textsc{ii}}^{s})}{h(\bar{\tau}_{\textsc{i}}^{s},\bar{\tau}_{\textsc{i}}^{s-1})f(\bar{\tau}_{\textsc{i}}^{s},\bar{\tau}_{\textsc{ii}}^{s-1})}\frac{\prod_{\nu=j}^{N-2}g(\bar{\tau}_{\textsc{iii}}^{\nu+1},\bar{\tau}_{\textsc{iii}}^{\nu})f(\bar{\tau}_{\textsc{iii}}^{\nu},\bar{\tau}_{\textsc{ii}}^{\nu})f(\bar{\tau}_{\textsc{iii}}^{\nu},\bar{\tau}_{\textsc{i}}^{\nu})}{\prod_{\nu=j}^{N-1}f(\bar{\tau}_{\textsc{iii}}^{\nu},\bar{\tau}^{\nu-1})}.\label{eq:Qfunc}
\end{multline}
We can see that the r.h.s is just the recurrence relation for $\bar{Z}_{K}(\left\{ \bar{\tau}^{\mu}\right\} _{\mu=1}^{N-2},\{z,\bar{t}^{N-1}\})$
(\ref{eq:recursionZ_UTw2}) therefore we just proved that if the equation
(\ref{eq:pole-2-1}) for $\nu>1$ is satisfied up to $\#\bar{t}^{1}=r_{1}$,
$\#\bar{t}^{N-1}=r_{N-1}$ then it is also satisfied for $\#\bar{t}^{1}=r_{1}$,
$\#\bar{t}^{N-1}=r_{N-1}+1$.

Let us continue with the recursion on $r_{1}$. Let us increase the
number of the first Bethe roots and use the recurrence equation (\ref{eq:recursionZ_UTw1})

\begin{multline}
\bar{Z}_{K}^{N}(\{z,\bar{t}^{1}\},\left\{ \bar{t}^{s}\right\} _{s=2}^{N-1})=\\
F_{K}^{(1)}(z)\sum_{\mathrm{part}}\bar{Z}_{K}^{N}(\{\bar{\omega}_{\textsc{ii}}^{s}\}_{s=1}^{N-2},\bar{t}_{\textsc{ii}}^{N-1})\prod_{s=1}^{N-2}\frac{f(\bar{\omega}_{\textsc{i}}^{s},\bar{\omega}_{\textsc{ii}}^{s})}{h(\bar{\omega}_{\textsc{i}}^{s},\bar{\omega}_{\textsc{i}}^{s-1})f(\bar{\omega}_{\textsc{i}}^{s},\bar{\omega}_{\textsc{ii}}^{s-1})}\frac{f(\bar{t}_{\textsc{i}}^{N-1},\bar{t}_{\textsc{ii}}^{N-1})f(\bar{t}^{N-1},-z)}{h(\bar{t}_{\textsc{i}}^{N-1},\bar{\omega}_{\textsc{i}}^{N-2})f(\bar{t}_{\textsc{i}}^{N-1},\bar{\omega}_{\textsc{ii}}^{N-2})f(\bar{t}^{2},z)}-\\
-\sum_{i=2}^{N}\frac{K_{N,i}(z)}{K_{N,1}(z)}\sum_{\mathrm{part}}\bar{Z}_{K}^{N}(\bar{w}_{\textsc{ii}}^{1},\{\bar{t}_{\textsc{ii}}^{s}\}_{s=2}^{i-1},\{\bar{t}^{s}\}_{s=i}^{N-1})\frac{f(\bar{w}_{\textsc{i}}^{1},\bar{w}_{\textsc{ii}}^{1})}{h(\bar{w}_{\textsc{i}}^{1},z)}\frac{f(\bar{t}_{\textsc{i}}^{2},\bar{t}_{\textsc{ii}}^{2})}{h(\bar{t}_{\textsc{i}}^{2},\bar{w}_{\textsc{i}}^{1})f(\bar{t}_{\textsc{i}}^{2},\bar{w}_{\textsc{ii}}^{1})}\prod_{s=3}^{i-1}\frac{f(\bar{t}_{\textsc{i}}^{s},\bar{t}_{\textsc{ii}}^{s})}{h(\bar{t}_{\textsc{i}}^{s},\bar{t}_{\textsc{i}}^{s-1})f(\bar{t}_{\textsc{i}}^{s},\bar{t}_{\textsc{ii}}^{s-1})}.\label{eq:poleproofUT}
\end{multline}
We follow the previous strategy i.e. we use the induction assumption
for the poles of HC in the r.h.s., and rearrange the expressions to
get the following form
\begin{equation}
\bar{Z}_{K}^{N}(\{z,\bar{t}^{1}\},\left\{ \bar{t}^{s}\right\} _{s=2}^{N-1})=\left(\frac{c}{t_{k}^{\nu}+t_{l}^{N-\nu}}\tilde{F}_{K}^{(\nu)}(t_{k}^{\nu})\frac{f(\bar{\tau}^{\nu},t_{k}^{\nu})}{f(\bar{\tau}^{\nu+1},t_{k}^{\nu})}\frac{f(\bar{\tau}^{N-\nu},-t_{k}^{\nu})}{f(\bar{\tau}^{N-\nu+1},-t_{k}^{\nu})}\right)\mathcal{Q}_{\nu}(\{z,\bar{t}^{1}\},\left\{ \bar{\tau}^{s}\right\} _{s=2}^{N-1})+reg,
\end{equation}
where $\bar{\tau}=\bar{t}\backslash\{t_{k}^{\nu},t_{l}^{N-\nu}\}$
for $\nu=2,\dots,\left\lfloor N/2\right\rfloor $. Using the induction
assumptions for the HC-s in the r.h.s of (\ref{eq:poleproofUT}),
after a straightforward calculation we can obtain that
\begin{multline}
\mathcal{Q}_{\nu}(\{z,\bar{t}^{1}\},\left\{ \bar{t}^{s}\right\} _{s=2}^{N-1})=\\
F_{K}^{(1)}(z)\sum_{\mathrm{part}}\bar{Z}_{K}^{N}(\{\bar{\mathtt{w}}_{\textsc{ii}}^{s}\}_{s=1}^{N-2},\bar{t}_{\textsc{ii}}^{N-1})\prod_{s=1}^{N-2}\frac{f(\bar{\mathtt{w}}_{\textsc{i}}^{s},\bar{\mathtt{w}}_{\textsc{ii}}^{s})}{h(\bar{\mathtt{w}}_{\textsc{i}}^{s},\bar{\mathtt{w}}_{\textsc{i}}^{s-1})f(\bar{\mathtt{w}}_{\textsc{i}}^{s},\bar{\mathtt{w}}_{\textsc{ii}}^{s-1})}\frac{f(\bar{t}_{\textsc{i}}^{N-1},\bar{t}_{\textsc{ii}}^{N-1})f(\bar{t}^{N-1},-z)}{h(\bar{t}_{\textsc{i}}^{N-1},\bar{\mathtt{w}}_{\textsc{i}}^{N-2})f(\bar{t}_{\textsc{i}}^{N-1},\bar{\mathtt{w}}_{\textsc{ii}}^{N-2})f(\bar{t}^{2},z)}-\\
-\sum_{i=2}^{N}\frac{K_{N,i}(z)}{K_{N,1}(z)}\sum_{\mathrm{part}}\bar{Z}_{K}^{N}(\bar{w}_{\textsc{ii}}^{1},\{\bar{\tau}_{\textsc{ii}}^{s}\}_{s=2}^{i-1},\{\bar{\tau}^{s}\}_{s=i}^{N-1})\frac{f(\bar{w}_{\textsc{i}}^{1},\bar{w}_{\textsc{ii}}^{1})}{h(\bar{w}_{\textsc{i}}^{1},z)}\frac{f(\bar{\tau}_{\textsc{i}}^{2},\bar{\tau}_{\textsc{ii}}^{2})}{h(\bar{\tau}_{\textsc{i}}^{2},\bar{w}_{\textsc{i}}^{1})f(\bar{\tau}_{\textsc{i}}^{2},\bar{w}_{\textsc{ii}}^{1})}\prod_{s=3}^{i-1}\frac{f(\bar{\tau}_{\textsc{i}}^{s},\bar{\tau}_{\textsc{ii}}^{s})}{h(\bar{\tau}_{\textsc{i}}^{s},\bar{\tau}_{\textsc{i}}^{s-1})f(\bar{\tau}_{\textsc{i}}^{s},\bar{\tau}_{\textsc{ii}}^{s-1})},
\end{multline}
where $\bar{\mathtt{w}}=\bar{\omega}\backslash\{t_{k}^{\nu},t_{l}^{N-\nu}\}$.
We can see that r.h.s is just the recurrence relation for $\bar{Z}_{K}(\{z,\bar{t}^{1}\},\left\{ \bar{t}^{s}\right\} _{s=2}^{N-1})$
(\ref{eq:recursionZ_UTw1}). We just proved that if the equation (\ref{eq:pole-2-1})
for $\nu>1$ is satisfied up to $\#\bar{t}^{1}=r_{1}$, $\#\bar{t}^{N-1}=r_{N-1}$
then it is also satisfied for $\#\bar{t}^{1}=r_{1}+1$, $\#\bar{t}^{N-1}=r_{N-1}$.
Previously, we also proved an induction for $r_{N-1}$, therefore
the proof of (\ref{eq:pole-2-1}) for $\nu>1$ is ready.

\subsubsection*{Case $\nu=1$}

To finish our proof we have to repeat the derivation for $\nu=1$.
At first we start with $\#\bar{t}^{1}=\#\bar{t}^{N-1}=1$ and take
the limit $t_{1}^{1}+t_{1}^{N-1}\to0$. The HC reads as (\ref{eq:recursionZ_UTw1})
\begin{multline}
\bar{Z}_{K}^{N}(\{t_{1}^{1}\},\left\{ \bar{t}^{s}\right\} _{s=2}^{N-2},\{t_{1}^{N-1}\})=\\
F_{K}^{(1)}(t_{1}^{1})\frac{f(t_{1}^{N-1},-t_{1}^{1})}{f(\bar{t}^{2},t_{1}^{1})}\sum_{\mathrm{part}}\frac{\bar{Z}_{K}^{N}(\emptyset,\{\bar{\omega}_{\textsc{ii}}^{s}\}_{s=2}^{N-2},\emptyset)}{h(t_{1}^{N-1},\bar{\omega}_{\textsc{i}}^{N-2})f(t_{1}^{N-1},\bar{\omega}_{\textsc{ii}}^{N-2})}\prod_{s=2}^{N-2}\frac{f(\bar{\omega}_{\textsc{i}}^{s},\bar{\omega}_{\textsc{ii}}^{s})}{h(\bar{\omega}_{\textsc{i}}^{s},\bar{\omega}_{\textsc{i}}^{s-1})f(\bar{\omega}_{\textsc{i}}^{s},\bar{\omega}_{\textsc{ii}}^{s-1})}\\
-\sum_{i=2}^{N}\frac{K_{N,i}(t_{1}^{1})}{K_{N,1}(t_{1}^{1})}\sum_{\mathrm{part}}\bar{Z}_{K}^{N}(\emptyset,\{\bar{t}_{\textsc{ii}}^{s}\}_{s=2}^{i-1},\{\bar{t}^{s}\}_{s=i}^{N-1})\frac{f(\bar{t}_{\textsc{i}}^{2},\bar{t}_{\textsc{ii}}^{2})}{h(\bar{t}_{\textsc{i}}^{2},z)}\prod_{s=3}^{i-1}\frac{f(\bar{t}_{\textsc{i}}^{s},\bar{t}_{\textsc{ii}}^{s})}{h(\bar{t}_{\textsc{i}}^{s},\bar{t}_{\textsc{i}}^{s-1})f(\bar{t}_{\textsc{i}}^{s},\bar{t}_{\textsc{ii}}^{s-1})}.
\end{multline}
We can see that the HC has a pole at the limit $t_{1}^{1}+t_{1}^{N-1}\to0$
which reads as
\begin{multline}
\bar{Z}_{K}^{N}(\{t_{1}^{1}\},\left\{ \bar{t}^{s}\right\} _{s=2}^{N-2},\{t_{1}^{N-1}\})=\\
\frac{c}{t^{1}+t^{N-1}}\frac{F_{K}^{(1)}(t_{1}^{1})}{f(\bar{t}^{2},t_{1}^{1})}\sum_{\mathrm{part}}\frac{\bar{Z}_{K}^{N}(\emptyset,\{\bar{\omega}_{\textsc{ii}}^{s}\}_{s=2}^{N-2},\emptyset)}{h(-t_{1}^{1},\omega_{\textsc{i}}^{N-2})f(-t_{1}^{1},\bar{\omega}_{\textsc{ii}}^{N-2})}\prod_{s=2}^{N-2}\frac{f(\bar{\omega}_{\textsc{i}}^{s},\bar{\omega}_{\textsc{ii}}^{s})}{h(\bar{\omega}_{\textsc{i}}^{s},\bar{\omega}_{\textsc{i}}^{s-1})f(\bar{\omega}_{\textsc{i}}^{s},\bar{\omega}_{\textsc{ii}}^{s-1})}+reg.
\end{multline}
In the denominator there is a factor $f(-t_{1}^{1},\bar{\omega}_{\textsc{ii}}^{N-2})$
therefore the residue is nonzero only when $-t_{1}^{1}\notin\bar{\omega}_{\textsc{ii}}^{N-2}\Rightarrow\bar{\omega}_{\textsc{i}}^{N-2}=\{-t_{1}^{1}\}$.
In an analogous way, the terms $f(\bar{\omega}_{\textsc{i}}^{s},\bar{\omega}_{\textsc{ii}}^{s-1})$
imply that $\bar{\omega}_{\textsc{i}}^{s}=\{-t_{1}^{1}\}$ for $s=2,\dots,N-3$
which means there is only one non-vanishing term in the sum
\begin{equation}
\bar{Z}_{K}^{N}(\{t_{1}^{1}\},\left\{ \bar{t}^{s}\right\} _{s=2}^{N-2},\{t_{1}^{N-1}\})=\frac{c}{t_{1}^{1}+t_{1}^{N-1}}\frac{F_{1}(t_{1}^{1})}{f(\bar{t}^{2},t_{1}^{1})}\bar{Z}_{K}^{N}(\emptyset,\{\bar{t}^{s}\}_{s=2}^{N-2},\emptyset)+reg.\label{eq:r1=00003D1}
\end{equation}
We can see that the equation (\ref{eq:pole-2-1}) is satisfied for
$r_{1}=r_{N}=1$.

Let us continue with general $r_{1}$ and $r_{N-1}$. Let assume that
(\ref{eq:pole-2-1}) is satisfied for $r_{1}$ or less first type
of Bethe roots and $r_{N-1}$ or less last type of Bethe roots. We
show that we can increase $r_{1}$ and $r_{N-1}$, separately. At
first let us increase the number of $\bar{t}^{N-1}$. We use the recurrence
relation (\ref{eq:recursionZ_UTw2})
\begin{multline}
\bar{Z}_{K}^{N}(\left\{ \bar{t}^{s}\right\} _{s=1}^{N-2},\{z,\bar{t}^{N-1}\})=\sum_{i=1}^{N}\sum_{j=1}^{N-1}\frac{K_{j,i}(-z)}{K_{N,1}(-z)}f(\bar{t}^{1},-z)\sum_{\mathrm{part}}\bar{Z}_{K}^{N}(\left\{ \bar{t}_{\textsc{ii}}^{s}\right\} _{s=1}^{N-2},\bar{t}^{N-1})\times\\
\prod_{s=1}^{i-1}\frac{f(\bar{t}_{\textsc{i}}^{s},\bar{t}_{\textsc{ii}}^{s})}{h(\bar{t}_{\textsc{i}}^{s},\bar{t}_{\textsc{i}}^{s-1})f(\bar{t}_{\textsc{i}}^{s},\bar{t}_{\textsc{ii}}^{s-1})}\frac{\prod_{\nu=j}^{N-2}g(\bar{t}_{\textsc{iii}}^{\nu+1},\bar{t}_{\textsc{iii}}^{\nu})f(\bar{t}_{\textsc{iii}}^{\nu},\bar{t}_{\textsc{ii}}^{\nu})f(\bar{t}_{\textsc{iii}}^{\nu},\bar{t}_{\textsc{i}}^{\nu})}{\prod_{\nu=j}^{N-1}f(\bar{t}_{\textsc{iii}}^{\nu},\bar{t}^{\nu-1})}.
\end{multline}
Now we can take the $t_{k}^{1}+t_{l}^{N-1}\to0$ limit. We can see
that this pole appears only in the HC-s for the partitions where $t_{k}^{1}\in\bar{t}_{\textsc{ii}}^{1}$.
Since $\#\bar{t}_{\textsc{ii}}^{1}\leq r_{1}$ and $\#\bar{t}^{N-1}=r_{N-1}$
we can use the induction assumption for the HC-s in the r.h.s. After
a rearrangement we can obtain the form
\begin{equation}
\bar{Z}_{K}^{N}(\left\{ \bar{t}^{s}\right\} _{s=1}^{N-2},\{z,\bar{t}^{N-1}\})=\left(\frac{c}{t_{k}^{1}+t_{l}^{N-1}}F_{K}^{(1)}(t_{k}^{1})\frac{f(\bar{\tau}^{1},t_{k}^{1})f(\bar{\mathrm{w}}^{N-1},-t_{k}^{1})}{f(\bar{t}^{2},t_{k}^{1})}\right)\mathcal{Q}(\left\{ \bar{\tau}^{s}\right\} _{s=1}^{N-2},\{z,\bar{\tau}^{N-1}\})+reg,
\end{equation}
where
\begin{multline}
\mathcal{Q}(\left\{ \bar{\tau}^{s}\right\} _{s=1}^{N-2},\{z,\bar{\tau}^{N-1}\})=\sum_{i=1}^{N}\sum_{j=1}^{N-1}\frac{K_{j,i}(-z)}{K_{N,1}(-z)}f(\bar{\tau}^{1},-z)\sum\bar{Z}_{K}^{N}(\left\{ \bar{\tau}_{\textsc{ii}}^{s}\right\} _{s=1}^{N-2},\bar{\tau}^{N-1})\times\\
\prod_{s=1}^{i-1}\frac{f(\bar{\tau}_{\textsc{i}}^{s},\bar{\tau}_{\textsc{ii}}^{s})}{h(\bar{\tau}_{\textsc{i}}^{s},\bar{\tau}_{\textsc{i}}^{s-1})f(\bar{\tau}_{\textsc{i}}^{s},\bar{\tau}_{\textsc{ii}}^{s-1})}\frac{\prod_{\nu=j}^{N-2}g(\bar{\tau}_{\textsc{iii}}^{\nu+1},\bar{\tau}_{\textsc{iii}}^{\nu})f(\bar{\tau}_{\textsc{iii}}^{\nu},\bar{\tau}_{\textsc{ii}}^{\nu})f(\bar{\tau}_{\textsc{iii}}^{\nu},\bar{\tau}_{\textsc{i}}^{\nu})}{\prod_{\nu=j}^{N-1}f(\bar{\tau}_{\textsc{iii}}^{\nu},\bar{\tau}^{\nu-1})}.
\end{multline}
We can see that r.h.s is just the recurrence equation (\ref{eq:recursionZ_UTw2})
for $\bar{Z}_{K}(\left\{ \bar{\tau}^{s}\right\} _{s=1}^{N-2},\{z,\bar{\tau}^{N-1}\})$.
Since the HC is symmetric under the permutations of the last type
of Bethe roots, we just proved that if the equation (\ref{eq:pole-2-1})
is satisfied up to $\#\bar{t}^{1}=r_{1}$, $\#\bar{t}^{N-1}=r_{N-1}$
then it is also satisfied for $\#\bar{t}^{1}=r_{1}$, $\#\bar{t}^{N-1}=r_{N-1}+1$.

Let us continue with the recursion in $r_{1}$. Let us take the $t_{k}^{1}+t_{l}^{N-1}\to0$
limit of the recurrence equation (\ref{eq:recursionZ_UTw1}). 
\begin{multline}
\bar{Z}_{K}^{N}(\{z,\bar{t}^{1}\},\left\{ \bar{t}^{s}\right\} _{s=2}^{N-1})=\\
F_{K}^{(1)}(z)\sum_{\mathrm{part}}\bar{Z}_{K}^{N}(\{\bar{\omega}_{\textsc{ii}}^{s}\}_{s=1}^{N-2},\bar{t}_{\textsc{ii}}^{N-1})\prod_{s=1}^{N-2}\frac{f(\bar{\omega}_{\textsc{i}}^{s},\bar{\omega}_{\textsc{ii}}^{s})}{h(\bar{\omega}_{\textsc{i}}^{s},\bar{\omega}_{\textsc{i}}^{s-1})f(\bar{\omega}_{\textsc{i}}^{s},\bar{\omega}_{\textsc{ii}}^{s-1})}\frac{f(\bar{t}_{\textsc{i}}^{N-1},\bar{t}_{\textsc{ii}}^{N-1})f(\bar{t}^{N-1},-z)}{h(\bar{t}_{\textsc{i}}^{N-1},\bar{\omega}_{\textsc{i}}^{N-2})f(\bar{t}_{\textsc{i}}^{N-1},\bar{\omega}_{\textsc{ii}}^{N-2})f(\bar{t}^{2},z)}-\\
-\sum_{i=2}^{N}\frac{K_{N,i}(z)}{K_{N,1}(z)}\sum_{\mathrm{part}}\bar{Z}_{K}^{N}(\bar{w}_{\textsc{ii}}^{1},\{\bar{t}_{\textsc{ii}}^{s}\}_{s=2}^{i-1},\{\bar{t}^{s}\}_{s=i}^{N-1})\frac{f(\bar{w}_{\textsc{i}}^{1},\bar{w}_{\textsc{ii}}^{1})}{h(\bar{w}_{\textsc{i}}^{1},z)}\frac{f(\bar{t}_{\textsc{i}}^{2},\bar{t}_{\textsc{ii}}^{2})}{h(\bar{t}_{\textsc{i}}^{2},\bar{w}_{\textsc{i}}^{1})f(\bar{t}_{\textsc{i}}^{2},\bar{w}_{\textsc{ii}}^{1})}\prod_{s=3}^{i-1}\frac{f(\bar{t}_{\textsc{i}}^{s},\bar{t}_{\textsc{ii}}^{s})}{h(\bar{t}_{\textsc{i}}^{s},\bar{t}_{\textsc{i}}^{s-1})f(\bar{t}_{\textsc{i}}^{s},\bar{t}_{\textsc{ii}}^{s-1})}.
\end{multline}
Since $\#\bar{\omega}_{\textsc{ii}}^{1}=r_{1}$, $\#\bar{t}_{\textsc{ii}}^{N-1}=r_{N-1}-1$,
$\#\bar{w}_{\textsc{ii}}^{1}=r_{1}$ and $\#\bar{t}^{N-1}=r_{N-1}$
we can use the induction assumption for all HC-s in the r.h.s. After
a rearrangement we can obtain the form
\begin{equation}
\bar{Z}_{K}^{N}(\{z,\bar{t}^{1}\},\left\{ \bar{t}^{s}\right\} _{s=2}^{N-1})=\left(\frac{c}{t_{k}^{1}+t_{l}^{N-1}}F_{K}^{(1)}(t_{k}^{1})\frac{f(\bar{\mathrm{w}}^{1},t_{k}^{1})f(\bar{\tau}^{N-1},-t_{k}^{1})}{f(\bar{t}^{2},t_{k}^{1})}\right)\mathcal{Q}(\{z,\bar{\tau}^{1}\},\left\{ \bar{\tau}^{\nu}\right\} _{\nu=2}^{N-1})+reg,
\end{equation}
where 
\begin{multline}
\mathcal{Q}(\{z,\bar{\tau}^{1}\},\left\{ \bar{\tau}^{\nu}\right\} _{\nu=2}^{N-1})=\\
F_{K}^{(1)}(z)\sum_{\mathrm{part}}\bar{Z}_{K}^{N}(\{\bar{\mathtt{w}}_{\textsc{ii}}^{\nu}\}_{\nu=1}^{N-2},\bar{\tau}_{\textsc{ii}}^{N-1})\prod_{s=1}^{N-2}\frac{f(\bar{\mathtt{w}}_{\textsc{i}}^{s},\bar{\mathtt{w}}_{\textsc{ii}}^{s})}{h(\bar{\mathtt{w}}_{\textsc{i}}^{s},\bar{\mathtt{w}}_{\textsc{i}}^{s-1})f(\bar{\mathtt{w}}_{\textsc{i}}^{s},\bar{\mathtt{w}}_{\textsc{ii}}^{s-1})}\frac{f(\bar{\tau}_{\textsc{i}}^{N-1},\bar{\tau}_{\textsc{ii}}^{N-1})f(\bar{\tau}^{N-1},-z)}{h(\bar{\tau}_{\textsc{i}}^{N-1},\bar{\mathtt{w}}_{\textsc{i}}^{N-2})f(\bar{\tau}_{\textsc{i}}^{N-1},\bar{\mathtt{w}}_{\textsc{ii}}^{N-2})f(\bar{\tau}^{2},z)}-\\
-\sum_{i=2}^{N}\frac{K_{N,i}(z)}{K_{N,1}(z)}\sum_{\mathrm{part}}\bar{Z}_{K}^{N}(\bar{\mathrm{w}}_{\textsc{ii}}^{1},\{\bar{\tau}_{\textsc{ii}}^{\nu}\}_{\nu=2}^{i-1},\{\bar{\tau}^{\nu}\}_{\nu=i}^{N-1})\frac{f(\bar{\mathrm{w}}_{\textsc{i}}^{1},\bar{\mathrm{w}}_{\textsc{ii}}^{1})}{h(\bar{\mathrm{w}}_{\textsc{i}}^{1},z)}\frac{f(\bar{\tau}_{\textsc{i}}^{2},\bar{\tau}_{\textsc{ii}}^{2})}{h(\bar{\tau}_{\textsc{i}}^{2},\bar{\mathrm{w}}_{\textsc{i}}^{1})f(\bar{\tau}_{\textsc{i}}^{2},\bar{\mathrm{w}}_{\textsc{ii}}^{1})}\prod_{s=3}^{i-1}\frac{f(\bar{\tau}_{\textsc{i}}^{s},\bar{\tau}_{\textsc{ii}}^{s})}{h(\bar{\tau}_{\textsc{i}}^{s},\bar{\tau}_{\textsc{i}}^{s-1})f(\bar{\tau}_{\textsc{i}}^{s},\bar{\tau}_{\textsc{ii}}^{s-1})},
\end{multline}
and $\bar{\mathrm{w}}=\bar{w}\backslash\{t_{k}^{1},t_{l}^{N-1}\}$,
$\bar{\mathtt{w}}=\bar{\omega}\backslash\{t_{k}^{1},t_{l}^{N-1}\}$.
We used the identity
\begin{equation}
\frac{f(\bar{\mathtt{w}}^{1},t_{k}^{1})f(-t_{k}^{1},-z)}{f(\bar{\omega}^{2},t_{k}^{1})}=\frac{f(\bar{\mathrm{w}}^{1},t_{k}^{1})}{f(\bar{t}^{2},t_{k}^{1})}.
\end{equation}
Applying the recurrence equation (\ref{eq:recursionZ_UTw1}) we obtain
that $\mathcal{Q}(\{z,\bar{\tau}^{1}\},\left\{ \bar{\tau}^{s}\right\} _{s=2}^{N-1})=\bar{Z}_{K}(\{z,\bar{\tau}^{1}\},\left\{ \bar{\tau}^{s}\right\} _{s=2}^{N-1})$.
Since the HC is symmetric under the permutations of the first type
of Bethe roots we just proved that if the equation (\ref{eq:pole-2-1})
is satisfied up to $\#\bar{t}^{1}=r_{1}$, $\#\bar{t}^{N-1}=r_{N-1}$
then it is also satisfied for $\#\bar{t}^{1}=r_{1}+1$, $\#\bar{t}^{N-1}=r_{N-1}$.
Previously, we also proved an induction for $r_{N-1}$, therefore
we finished the proof of (\ref{eq:pole-2-1}).

\section{Pair structure limit of the off-shell overlaps\label{sec:Pair-offshell}}

In this section we calculate the pair structure limits of the overlaps.
In the untwisted case we have achiral pair structure (see subsection
\ref{subsec:Achiral-pair-structure}) for which we have to take the
limit $t_{k}^{\nu}+t_{l}^{N-\nu}\to0$. In the twisted case we have
chiral pair structure (see subsection \ref{subsec:Chiral-pair-structure})
for which we have to take the limit $t_{k}^{\nu}+t_{l}^{\nu}+\nu c\to0$.

\subsection{Twisted case}

Let us start with the twisted case. Let us substitute the identities
(\ref{eq:WZZ}) and (\ref{eq:connHCTw}) to the sum formula (\ref{eq:sumFormula})
\begin{multline}
S_{K}^{N}(\bar{t})=(-1)^{\#\bar{t}_{\textsc{i}}}\sum_{\mathrm{part}(\bar{t})}\frac{\prod_{s=1}^{N-1}f(\bar{t}_{\textsc{ii}}^{s},\bar{t}_{\textsc{i}}^{s})}{\prod_{s=1}^{N-2}f(\bar{t}_{\textsc{ii}}^{s+1},\bar{t}_{\textsc{i}}^{s})f(\bar{t}_{\textsc{i}}^{s+1},\bar{t}_{\textsc{i}}^{s})}\bar{Z}_{K}^{N}(\pi^{c}(\bar{t}_{\textsc{i}}))\bar{Z}_{K}^{N}(\bar{t}_{\textsc{ii}})\prod_{\nu=1}^{N-1}\alpha_{\nu}(\bar{t}_{\textsc{i}}^{\nu}),\\
=(-1)^{\#\bar{t}_{\textsc{i}}}\sum\frac{\prod_{s=1}^{N-1}f(\bar{t}_{\textsc{ii}}^{s},\bar{t}_{\textsc{i}}^{s})}{\prod_{s=1}^{N-2}f(\bar{t}^{s+1},\bar{t}_{\textsc{i}}^{s})}\bar{Z}_{K}^{N}(\pi^{c}(\bar{t}_{\textsc{i}}))\bar{Z}_{K}^{N}(\bar{t}_{\textsc{ii}})\prod_{\nu=1}^{N-1}\alpha_{\nu}(\bar{t}_{\textsc{i}}^{\nu}).
\end{multline}
Let us take the $t_{k}^{\nu}+t_{l}^{\nu}+\nu c\to0$ limit. We can
see that the formal poles appear only for the partitions where $t_{k}^{\nu},t_{l}^{\nu}\in\bar{t}_{\textsc{ii}}$
or $t_{k}^{\nu},t_{l}^{\nu}\in\bar{t}_{\textsc{i}}$. At first let
us take a partition for which $t_{k}^{\nu},t_{l}^{\nu}\in\bar{t}_{\textsc{ii}}$.
The coefficient of $\prod_{\nu=1}^{N-1}\alpha_{\nu}(\bar{t}_{\textsc{i}}^{\nu})$
for this partition up to the factor $\bar{Z}_{K}^{N}(\pi^{c}(\bar{t}_{\textsc{i}}))\bar{Z}_{K}^{N}(\bar{t}_{\textsc{ii}})$
reads as 
\begin{equation}
\frac{(-1)^{\#\bar{t}_{\textsc{i}}}c}{t_{k}^{\nu}+t_{l}^{\nu}+\nu c}F_{K}^{(\nu)}\frac{f(\bar{\tau}_{\textsc{ii}}^{\nu},t_{k}^{\nu})}{f(\bar{\tau}_{\textsc{ii}}^{\nu+1},t_{k}^{\nu})}\frac{f(\bar{\tau}_{\textsc{ii}}^{\nu},-t_{k}^{\nu}-\nu c)}{f(\bar{\tau}_{\textsc{ii}}^{\nu+1},-t_{k}^{\nu}-\nu c)}\frac{f(t_{k}^{\nu},\bar{\tau}_{\textsc{i}}^{\nu})f(-t_{k}^{\nu}-\nu c,\bar{\tau}_{\textsc{i}}^{\nu})}{f(t_{k}^{\nu},\bar{\tau}_{\textsc{i}}^{\nu-1})f(-t_{k}^{\nu}-\nu c,\bar{\tau}_{\textsc{i}}^{\nu-1})}\frac{\prod_{s=1}^{N-1}f(\bar{\tau}_{\textsc{ii}}^{s},\bar{\tau}_{\textsc{i}}^{s})}{\prod_{s=1}^{N-2}f(\bar{\tau}^{s+1},\bar{\tau}_{\textsc{i}}^{s})}.
\end{equation}
For a partition for which $t_{k}^{\nu},t_{l}^{\nu}\in\bar{t}_{\textsc{i}}$
the coefficient of $\prod_{\nu=1}^{N-1}\alpha_{\nu}(\bar{t}_{\textsc{i}}^{\nu})$
up to the factor $\bar{Z}_{K}^{N}(\pi^{c}(\bar{t}_{\textsc{i}}))\bar{Z}_{K}^{N}(\bar{t}_{\textsc{ii}})$
reads as 
\begin{multline}
-\frac{(-1)^{\#\bar{t}_{\textsc{i}}}c}{t_{k}^{\nu}+t_{l}^{\nu}+\nu c}F_{K}^{(\nu)}\frac{f(t_{k}^{\nu},\bar{\tau}_{\textsc{i}}^{\nu})}{f(t_{k}^{\nu},\bar{\tau}_{\textsc{i}}^{\nu+1}+c)}\frac{f(-t_{k}^{\nu}-\nu c,\bar{\tau}_{\textsc{i}}^{\nu})}{f(-t_{k}^{\nu}-\nu c,\bar{\tau}_{\textsc{i}}^{\nu+1}+c)}\times\\
\frac{f(\bar{\tau}_{\textsc{ii}}^{\nu},t_{k}^{\nu})f(\bar{\tau}_{\textsc{ii}}^{\nu},-t_{k}^{\nu}-\nu c)}{f(\bar{\tau}^{\nu+1},t_{k}^{\nu})f(\bar{\tau}^{\nu+1},-t_{k}^{\nu}-\nu c)}\frac{1}{f(t_{k}^{\nu},\bar{\tau}_{\textsc{i}}^{\nu-1})f(-t_{k}^{\nu}-\nu c,\bar{\tau}_{\textsc{i}}^{\nu-1})}\frac{\prod_{s=1}^{N-1}f(\bar{\tau}_{\textsc{ii}}^{s},\bar{\tau}_{\textsc{i}}^{s})}{\prod_{s=1}^{N-2}f(\bar{\tau}^{s+1},\bar{\tau}_{\textsc{i}}^{s})}=\\
-\frac{(-1)^{\#\bar{t}_{\textsc{i}}}c}{t_{k}^{\nu}+t_{l}^{\nu}+\nu c}F_{K}^{(\nu)}\frac{f(t_{k}^{\nu},\bar{\tau}_{\textsc{i}}^{\nu})}{f(t_{k}^{\nu},\bar{\tau}_{\textsc{i}}^{\nu-1})}\frac{f(-t_{k}^{\nu}-\nu c,\bar{\tau}_{\textsc{i}}^{\nu})}{f(-t_{k}^{\nu}-\nu c,\bar{\tau}_{\textsc{i}}^{\nu-1})}\frac{f(\bar{\tau}_{\textsc{ii}}^{\nu},t_{k}^{\nu})f(\bar{\tau}_{\textsc{ii}}^{\nu},-t_{k}^{\nu}-\nu c)}{f(\bar{\tau}_{\textsc{ii}}^{\nu+1},t_{k}^{\nu})f(\bar{\tau}_{\textsc{ii}}^{\nu+1},-t_{k}^{\nu}-\nu c)}\frac{\prod_{s=1}^{N-1}f(\bar{\tau}_{\textsc{ii}}^{s},\bar{\tau}_{\textsc{i}}^{s})}{\prod_{s=1}^{N-2}f(\bar{\tau}^{s+1},\bar{\tau}_{\textsc{i}}^{s})},
\end{multline}
where we used the identity $f(u,v+c)f(v,u)=1$. Substituting back
to the sum formula we obtain that
\begin{multline}
S_{K}^{N}(\bar{t})\Biggr|_{t_{k}^{\nu}+t_{l}^{\nu}+\nu c\to0}=\\
\frac{(-1)^{\#\bar{t}_{\textsc{i}}}c}{t_{k}^{\nu}+t_{l}^{\nu}+\nu c}(1-\alpha_{\nu}(t_{k}^{\nu})\alpha_{\nu}(t_{l}^{\nu}))F_{K}^{(\nu)}\sum_{\mathrm{part}(\bar{\tau})}\frac{\prod_{s=1}^{N-1}f(\bar{\tau}_{\textsc{ii}}^{s},\bar{\tau}_{\textsc{i}}^{s})}{\prod_{s=1}^{N-2}f(\bar{\tau}^{s+1},\bar{\tau}_{\textsc{i}}^{s})}\bar{Z}_{K}^{N}(\pi^{c}(\bar{\tau}_{\textsc{i}}))\bar{Z}_{K}^{N}(\bar{\tau}_{\textsc{ii}})\times\\
\frac{f(t_{k}^{\nu},\bar{\tau}_{\textsc{i}}^{\nu})}{f(t_{k}^{\nu},\bar{\tau}_{\textsc{i}}^{\nu-1})}\frac{f(-t_{k}^{\nu}-\nu c,\bar{\tau}_{\textsc{i}}^{\nu})}{f(-t_{k}^{\nu}-\nu c,\bar{\tau}_{\textsc{i}}^{\nu-1})}\frac{f(\bar{\tau}_{\textsc{ii}}^{\nu},t_{k}^{\nu})f(\bar{\tau}_{\textsc{ii}}^{\nu},-t_{k}^{\nu}-\nu c)}{f(\bar{\tau}_{\textsc{ii}}^{\nu+1},t_{k}^{\nu})f(\bar{\tau}_{\textsc{ii}}^{\nu+1},-t_{k}^{\nu}-\nu c)}\prod_{s=1}^{N-1}\alpha_{s}(\bar{\tau}_{\textsc{i}}^{s})+\tilde{S}.
\end{multline}
This limit can be simplified as
\begin{equation}
S_{K}^{N}(\bar{t})\Biggr|_{t_{k}^{\nu}=-t_{l}^{\nu}-\nu c}\to\frac{f(\bar{\tau}^{\nu},t_{k}^{\nu})f(\bar{\tau}^{\nu},-t_{k}^{\nu}-\nu c)}{f(\bar{\tau}^{\nu+1},t_{k}^{\nu})f(\bar{\tau}^{\nu+1},-t_{k}^{\nu}-\nu c)}F_{K}^{(\nu)}X_{k}^{\nu}S_{K}^{N,mod}(\bar{\tau})+\tilde{S},\label{eq:Spairlimt-1}
\end{equation}
where
\begin{equation}
X_{k}^{\nu}=-c\frac{d}{du}\log\alpha_{\nu}(u)\Biggr|_{u=t_{k}^{\nu}}
\end{equation}
and $\tilde{S}$ depends only on the terms $\alpha_{\nu}(t_{k}^{\nu})$
but it does not depend on the derivative terms $X_{k}^{\nu}$. The
$S_{K}^{N,mod}$ contains the following modified $\alpha$-s
\begin{align}
\alpha_{\nu}^{(mod)}(u) & =\frac{f(t_{k}^{\nu},u)}{f(u,t_{k}^{\nu})}\frac{f(-t_{k}^{\nu}-\nu c,u)}{f(u,-t_{k}^{\nu}-\nu c)}\alpha_{\nu}(u),\nonumber \\
\alpha_{\nu+1}^{(mod)}(u) & =f(u,t_{k}^{\nu})f(u,-t_{k}^{\nu}-\nu c)\alpha_{\nu+1}(u),\\
\alpha_{\nu-1}^{(mod)}(u) & =\frac{1}{f(t_{k}^{\nu},u)f(-t_{k}^{\nu}-\nu c,u)}\alpha_{\nu-1}(u),\nonumber 
\end{align}
and $\alpha_{\mu}^{(mod)}(u)=\alpha_{\mu}(u)$ for $|\mu-\nu|>1$.

\subsection{Untwisted case}

Let us continue with the untwisted case. Let us substitute the identities
(\ref{eq:WZZ}) and (\ref{eq:connHCUTw}) to the sum formula (\ref{eq:sumFormula})

\begin{equation}
S_{K}^{N}(\bar{t})=\sum_{\mathrm{part}(\bar{t})}\frac{\prod_{s=1}^{N-1}f(\bar{t}_{\textsc{ii}}^{s},\bar{t}_{\textsc{i}}^{s})}{\prod_{s=1}^{N-2}f(\bar{t}_{\textsc{ii}}^{s+1},\bar{t}_{\textsc{i}}^{s})}\bar{Z}_{\Pi(K)}^{N}(\pi^{a}(\bar{t}_{\textsc{i}}))\bar{Z}_{K}^{N}(\bar{t}_{\textsc{ii}})\prod_{\nu=1}^{N-1}\alpha_{\nu}(\bar{t}_{\textsc{i}}^{\nu}).
\end{equation}
Let us take the $t_{k}^{\nu}+t_{l}^{N-\nu}\to0$ limit. We can see
that the poles of the HC-s appear only for the partitions where $t_{k}^{\nu},t_{l}^{N-\nu}\in\bar{t}_{\textsc{ii}}$
or $t_{k}^{\nu},t_{l}^{N-\nu}\in\bar{t}_{\textsc{i}}$. At first let
us take a partition for which $t_{k}^{\nu},t_{l}^{N-\nu}\in\bar{t}_{\textsc{ii}}$.
The coefficient of $\prod_{\nu=1}^{N-1}\alpha_{\nu}(\bar{t}_{\textsc{i}}^{\nu})$
for this partition up to the factor $\bar{Z}_{\Pi(K)}^{N}(\pi^{a}(\bar{t}_{\textsc{i}}))\bar{Z}_{K}^{N}(\bar{t}_{\textsc{ii}})$
reads as 
\begin{equation}
\frac{c}{t_{k}^{\nu}+t_{l}^{N-\nu}}\tilde{F}_{K}^{(\nu)}(t_{k}^{\nu})\frac{f(\bar{\tau}_{\textsc{ii}}^{\nu},t_{k}^{\nu})}{f(\bar{\tau}_{\textsc{ii}}^{\nu+1},t_{k}^{\nu})}\frac{f(\bar{\tau}_{\textsc{ii}}^{N-\nu},-t_{k}^{\nu})}{f(\bar{\tau}_{\textsc{ii}}^{N-\nu+1},-t_{k}^{\nu})}\frac{f(t_{k}^{\nu},\bar{\tau}_{\textsc{i}}^{\nu})f(-t_{k}^{\nu},\bar{\tau}_{\textsc{i}}^{N-\nu})}{f(t_{k}^{\nu},\bar{\tau}_{\textsc{i}}^{\nu-1})f(-t_{k}^{\nu},\bar{\tau}_{\textsc{i}}^{N-\nu-1})}\frac{\prod_{s=1}^{N-1}f(\bar{\tau}_{\textsc{ii}}^{s},\bar{\tau}_{\textsc{i}}^{s})}{\prod_{s=1}^{N-2}f(\bar{\tau}_{\textsc{ii}}^{s+1},\bar{\tau}_{\textsc{i}}^{s})}.
\end{equation}
For a partition for which $t_{k}^{\nu},t_{l}^{N-\nu}\in\bar{t}_{\textsc{i}}$
the coefficient of $\prod_{\nu=1}^{N-1}\alpha_{\nu}(\bar{t}_{\textsc{i}}^{\nu})$
up to the factor $\bar{Z}_{\Pi(K)}^{N}(\pi^{a}(\bar{t}_{\textsc{i}}))\bar{Z}_{K}^{N}(\bar{t}_{\textsc{ii}})$
reads as 
\begin{equation}
-\frac{c}{t_{k}^{\nu}+t_{l}^{N-\nu}}\tilde{F}_{\Pi\left(K\right)}^{(\nu)}(t_{k}^{\nu})\frac{f(t_{k}^{\nu},\bar{\tau}_{\textsc{i}}^{\nu})}{f(t_{k}^{\nu},\bar{\tau}_{\textsc{i}}^{\nu-1})}\frac{f(-t_{k}^{\nu},\bar{\tau}_{\textsc{i}}^{N-\nu})}{f(-t_{k}^{\nu},\bar{\tau}_{\textsc{i}}^{N-\nu-1})}\frac{f(\bar{\tau}_{\textsc{ii}}^{\nu},t_{k}^{\nu})f(\bar{\tau}_{\textsc{ii}}^{N-\nu},-t_{k}^{\nu})}{f(\bar{\tau}_{\textsc{ii}}^{\nu+1},t_{k}^{\nu})f(\bar{\tau}_{\textsc{ii}}^{N-\nu+1},-t_{k}^{\nu})}\frac{\prod_{s=1}^{N-1}f(\bar{\tau}_{\textsc{ii}}^{s},\bar{\tau}_{\textsc{i}}^{s})}{\prod_{s=1}^{N-2}f(\bar{\tau}_{\textsc{ii}}^{s+1},\bar{\tau}_{\textsc{i}}^{s})}.
\end{equation}
From the explicit form of the one-particle overlaps $\tilde{F}_{K}^{(\nu)}(z)$
(\ref{eq:onepartUTw}) we can see that the following identity is satisfied
\begin{equation}
\tilde{F}_{\Pi\left(K\right)}^{(\nu)}(u)=\tilde{F}_{K}^{(\nu)}(u).
\end{equation}
Substituting back to the sum formula we obtain that
\begin{multline}
S_{K}^{N}(\bar{t})\Biggr|_{t_{k}^{\nu}+t_{l}^{N-\nu}\to0}=\\
\frac{c}{t_{k}^{\nu}+t_{l}^{N-\nu}}(1-\alpha_{\nu}(t_{k}^{\nu})\alpha_{N-\nu}(t_{l}^{N-\nu}))\tilde{F}_{K}^{(\nu)}(t_{k}^{\nu})\sum_{\mathrm{part}(\bar{\tau})}\frac{\prod_{s=1}^{N-1}f(\bar{\tau}_{\textsc{ii}}^{s},\bar{\tau}_{\textsc{i}}^{s})}{\prod_{s=1}^{N-2}f(\bar{\tau}_{\textsc{ii}}^{s+1},\bar{\tau}_{\textsc{i}}^{s})}\bar{Z}_{\Pi(K)}^{N}(\pi^{a}(\bar{\tau}_{\textsc{i}}))\bar{Z}_{K}^{N}(\bar{\tau}_{\textsc{ii}})\times\\
\frac{f(t_{k}^{\nu},\bar{\tau}_{\textsc{i}}^{\nu})}{f(t_{k}^{\nu},\bar{\tau}_{\textsc{i}}^{\nu-1})}\frac{f(-t_{k}^{\nu},\bar{\tau}_{\textsc{i}}^{N-\nu})}{f(-t_{k}^{\nu},\bar{\tau}_{\textsc{i}}^{N-\nu-1})}\frac{f(\bar{\tau}_{\textsc{ii}}^{\nu},t_{k}^{\nu})f(\bar{\tau}_{\textsc{ii}}^{N-\nu},-t_{k}^{\nu})}{f(\bar{\tau}_{\textsc{ii}}^{\nu+1},t_{k}^{\nu})f(\bar{\tau}_{\textsc{ii}}^{N-\nu+1},-t_{k}^{\nu})}\prod_{s=1}^{N-1}\alpha_{s}(\bar{\tau}_{\textsc{i}}^{s})+\tilde{S}.
\end{multline}
This limit can be simplified as
\begin{equation}
S_{K}^{N}(\bar{t})\Biggr|_{t_{k}^{\nu}+t_{l}^{N-\nu}=0}\to\frac{f(\bar{\tau}^{\nu},t_{k}^{\nu})f(\bar{\tau}^{N-\nu},-t_{k}^{\nu})}{f(\bar{\tau}^{\nu+1},t_{k}^{\nu})f(\bar{\tau}^{N-\nu+1},-t_{k}^{\nu})}\tilde{F}_{K}^{(\nu)}(t_{k}^{\nu})X_{k}^{\nu}S_{K}^{N,mod}(\bar{\tau})+\tilde{S},\label{eq:Spairlimt}
\end{equation}
where $\tilde{S}$ depends only on the terms $\alpha_{\nu}(t_{k}^{\nu})$
but it does not depend on the derivative terms $X_{k}^{\nu}$. The
$S_{K}^{N,mod}$ contains the following modified $\alpha$-s
\begin{align}
\alpha_{\nu}^{(mod)}(u) & =\frac{f(t_{k}^{\nu},u)}{f(u,t_{k}^{\nu})}\alpha_{\nu}(u), & \alpha_{N-\nu}^{(mod)}(u) & =\frac{f(-t_{k}^{\nu},u)}{f(u,-t_{k}^{\nu})}\alpha_{N-\nu}(u),\nonumber \\
\alpha_{\nu+1}^{(mod)}(u) & =f(u,t_{k}^{\nu})\alpha_{\nu+1}(u), & \alpha_{N-\nu+1}^{(mod)}(u) & =f(u,-t_{k}^{\nu})\alpha_{N-\nu+1}(u),\\
\alpha_{\nu-1}^{(mod)}(u) & =\frac{1}{f(t_{k}^{\nu},u)}\alpha_{\nu-1}(u), & \alpha_{N-\nu-1}^{(mod)}(u) & =\frac{1}{f(-t_{k}^{\nu},u)}\alpha_{N-\nu-1}(u),\nonumber 
\end{align}
for $\nu<\frac{N-2}{2}$ and
\begin{equation}
\begin{split}\alpha_{\frac{N-1}{2}}^{(mod)}(u) & =\frac{f(t_{k}^{\frac{N-1}{2}},u)}{f(u,t_{k}^{\frac{N-1}{2}})f(-t_{k}^{\frac{N-1}{2}},u)}\alpha_{\frac{N-1}{2}}(u),\quad\alpha_{\frac{N+1}{2}}^{(mod)}(u)=\frac{f(-t_{k}^{\frac{N-1}{2}},u)f(u,t_{k}^{\frac{N-1}{2}})}{f(u,-t_{k}^{\frac{N-1}{2}})}\alpha_{\frac{N+1}{2}}(u),\\
\alpha_{\frac{N-3}{2}}^{(mod)}(u) & =\frac{1}{f(t_{k}^{\frac{N-1}{2}},u)}\alpha_{\frac{N-3}{2}}(u),\qquad\qquad\qquad\alpha_{\frac{N+3}{2}}^{(mod)}(u)=f(u,-t_{k}^{\frac{N-1}{2}})\alpha_{\frac{N+3}{2}}(u),
\end{split}
\end{equation}
for $\nu=\frac{N-1}{2}$ ($N$ is odd),
\begin{align}
\alpha_{N/2-1}^{(mod)}(u) & =\frac{f(t_{k}^{N/2-1},u)}{f(u,t_{k}^{N/2-1})}\alpha_{N/2-1}(u), & \alpha_{N/2+1}^{(mod)}(u) & =\frac{f(-t_{k}^{N/2-1},u)}{f(u,-t_{k}^{N/2-1})}\alpha_{N/2+1}(u),\nonumber \\
\alpha_{N/2-2}^{(mod)}(u) & =\frac{1}{f(t_{k}^{N/2-1},u)}\alpha_{N/2-2}(u), & \alpha_{N/2+2}^{(mod)}(u) & =f(u,-t_{k}^{N/2-1})\alpha_{N/2+2}(u),\\
\alpha_{N/2}^{(mod)}(u) & =\frac{f(u,t_{k}^{N/2-1})}{f(-t_{k}^{N/2-1},u)}\alpha_{N/2}(u),\nonumber 
\end{align}
for $\nu=N/2-1$ ($N$ is even),
\begin{align}
\alpha_{N/2}^{(mod)}(u) & =\frac{f(t_{k}^{N/2},u)}{f(u,t_{k}^{N/2})}\frac{f(-t_{k}^{N/2},u)}{f(u,-t_{k}^{N/2})}\alpha_{N/2}(u),\nonumber \\
\alpha_{N/2+1}^{(mod)}(u) & =f(u,t_{k}^{N/2})f(u,-t_{k}^{N/2})\alpha_{N/2+1}(u),\\
\alpha_{N/2-1}^{(mod)}(u) & =\frac{1}{f(t_{k}^{N/2},u)}\frac{1}{f(-t_{k}^{N/2},u)}\alpha_{N/2-1}(u),\nonumber 
\end{align}
for $\nu=N/2$ ($N$ is even). The rest of the $\alpha$-s are unmodified
i.e. $\alpha_{\mu}^{(mod)}(u)=\alpha_{\mu}(u)$ if $|\mu-\nu|>1$
and $|N-\mu-\nu|>1$.

\section{On-shell limit of the overlaps\label{sec:On-shell-limit}}

In appendix \ref{sec:Pair-offshell} we saw that, the pair structure
limit of the overlaps depends on the Bethe roots $t_{j}^{+,\nu}$,
the functional parameters $\alpha_{\nu}(t_{j}^{+,\nu})$ and the logarithmic
derivatives $X_{j}^{+,\nu}$. Taking the on-shell limit, the functional
parameters $\alpha_{\nu}(t_{j}^{+,\nu})$ can be expressed through
the Bethe roots $\bar{t}^{+}$ using the Bethe equations, therefore
the on-shell overlaps are the functions of $t_{j}^{+,\nu}$ and $X_{j}^{+,\nu}$.
Using the notations $\hat{\pi}\left(\bar{t}^{+,\nu}\right)=-\bar{t}^{+,\nu}$
or $\hat{\pi}\left(\bar{t}^{+,\nu}\right)=-\bar{t}^{+,\nu}-\nu c$
for the achiral or the chiral pair structures, the pair structure
limit reads as $\bar{t}^{-}\to\hat{\pi}\left(\bar{t}^{+}\right)$.

Let us define the re-normalized on-shell overlaps as
\begin{multline}
\mathbf{N}^{(\mathbf{r}^{+})}(\bar{X}^{+},\bar{t}^{+})=\\
\frac{1}{\prod_{\nu=1}^{N-1}\left(F_{K}^{(\nu)}\right)^{r_{\nu}/2}\prod_{k\neq l}f(t_{l}^{+,\nu},t_{k}^{+,\nu})\prod_{k<l}f(t_{l}^{+,\nu},-t_{k}^{+,\nu}-\nu c)f(-t_{k}^{+,\nu}-\nu c,t_{l}^{+,\nu})}\lim_{\bar{t}^{-,\mu}\to\hat{\pi}(\bar{t}^{+,\mu})}S_{K}^{N}(\bar{t})
\end{multline}
for the twisted case. For the untwisted case the re-normalization
reads as
\begin{equation}
\mathbf{N}^{(\mathbf{r}^{+})}(\bar{X}^{+},\bar{t}^{+})=\frac{1}{\prod_{\nu=1}^{\frac{N-1}{2}}\tilde{F}_{K}^{(\nu)}(\bar{t}^{+,\nu})}\frac{\prod_{\nu=1}^{\frac{N-3}{2}}f(\bar{t}^{+,\nu+1},\bar{t}^{+,\nu})\prod_{k<l}f(-t_{l}^{+,\frac{N-1}{2}},t_{k}^{+,\frac{N-1}{2}})}{\prod_{\nu=1}^{\frac{N-1}{2}}\prod_{k\neq l}f(t_{l}^{+,\nu},t_{k}^{+,\nu})}\lim_{\bar{t}^{-,\mu}\to\hat{\pi}(\bar{t}^{+,\mu})}S_{K}^{N}(\bar{t})
\end{equation}
for odd $N$ and
\begin{multline}
\mathbf{N}^{(\mathbf{r}^{+})}(\bar{X}^{+},\bar{t}^{+})=\\
\frac{1}{\prod_{\nu=1}^{\frac{N}{2}}F_{K}^{(\nu)}(\bar{t}^{+,\nu})}\frac{\prod_{\nu=1}^{\frac{N}{2}-1}f(\bar{t}^{+,\nu+1},\bar{t}^{+,\nu})f(-\bar{t}^{+,\frac{N}{2}},\bar{t}^{+,\frac{N}{2}-1})}{\prod_{\nu=1}^{\frac{N}{2}}\prod_{k\neq l}f(t_{l}^{+,\nu},t_{k}^{+,\nu})\prod_{k<l}f(-t_{l}^{+,\frac{N}{2}},t_{k}^{+,\frac{N}{2}})f(t_{l}^{+,\frac{N}{2}},-t_{k}^{+,\frac{N}{2}})}\lim_{\bar{t}^{-,\mu}\to\hat{\pi}(\bar{t}^{+,\mu})}S_{K}^{N}(\bar{t})
\end{multline}
for even $N$.

In the following we prove that these functions satisfy the Korepin
criteria (see subsection \ref{subsec:Gaudin-like-determinants}).
The properties \ref{enum:prop1},\ref{enum:prop2} are obvious. The
property \ref{enum:prop3} follows from a direct calculations of appendix
\ref{sec:Elementary}.

The property \ref{enum:prop4} follows from the equations (\ref{eq:Spairlimt-1})
and (\ref{eq:Spairlimt}). For the twisted case, from the equation
(\ref{eq:Spairlimt-1}) we obtain that
\begin{multline}
\frac{\partial}{\partial X_{k}^{+,\nu}}\lim_{\bar{t}^{-,\mu}\to\hat{\pi}(\bar{t}^{+,\mu})}S_{K}^{N}(\bar{t})=\\
f(\bar{\tau}^{+,\nu},t_{k}^{\nu})f(t_{k}^{\nu},\bar{\tau}^{+,\nu})f(-t_{k}^{\nu}-\nu c,\bar{\tau}^{+,\nu})f(\bar{\tau}^{+,\nu},-t_{k}^{\nu}-\nu c)F_{K}^{(\nu)}\lim_{\bar{\tau}^{-,\mu}\to\hat{\pi}(\bar{\tau}^{+,\mu})}S_{K}^{N,mod}(\bar{\tau}).
\end{multline}
For the untwisted case, from equation (\ref{eq:Spairlimt}) we obtain
that
\begin{equation}
\frac{\partial}{\partial X_{k}^{+,\nu}}\lim_{\bar{t}^{-,\mu}\to\hat{\pi}(\bar{t}^{+,\mu})}S_{K}^{N}(\bar{t})=\frac{f(\bar{\tau}^{+,\nu},t_{k}^{+,\nu})f(t_{k}^{+,\nu},\bar{\tau}^{+,\nu})}{f(\bar{\tau}^{+,\nu+1},t_{k}^{+,\nu})f(t_{k}^{+,\nu},\bar{\tau}^{+,\nu-1})}F_{K}^{(\nu)}(t_{k}^{\nu})\lim_{\bar{\tau}^{-,\mu}\to\hat{\pi}(\bar{\tau}^{+,\mu})}S_{K}^{N,mod}(\bar{\tau})
\end{equation}
for $\nu<\frac{N-1}{2}$. When $N$ is odd, for $\nu=\frac{N-1}{2}$
we obtain that 
\begin{multline}
\frac{\partial}{\partial X_{k}^{+,\frac{N-1}{2}}}\lim_{\bar{t}^{-,\mu}\to\hat{\pi}(\bar{t}^{+,\mu})}S_{K}^{N}(\bar{t})=\\
\frac{f(\bar{\tau}^{+,\frac{N-1}{2}},t_{k}^{+,\frac{N-1}{2}})f(t_{k}^{+,\frac{N-1}{2}},\bar{\tau}^{+,\frac{N-1}{2}})}{f(-\bar{\tau}^{+,\frac{N-1}{2}},t_{k}^{+,\frac{N-1}{2}})f(t_{k}^{+,\frac{N-1}{2}},\bar{\tau}^{+,\frac{N-3}{2}})}\tilde{F}_{K}^{(\frac{N-1}{2})}(t_{k}^{\frac{N-1}{2}})\lim_{\bar{\tau}^{-,\mu}\to\hat{\pi}(\bar{\tau}^{+,\mu})}S_{K}^{N,mod}(\bar{\tau}).
\end{multline}
When $N$ is even we obtain that
\begin{multline}
\frac{\partial}{\partial X_{k}^{+,\frac{N}{2}-1}}\lim_{\bar{t}^{-,\mu}\to\hat{\pi}(\bar{t}^{+,\mu})}S_{K}^{N}(\bar{t})=\\
\frac{f(\bar{\tau}^{+,\frac{N}{2}-1},t_{k}^{+,\frac{N}{2}-1})f(t_{k}^{+,\frac{N}{2}-1},\bar{\tau}^{+,\frac{N}{2}-1})}{f(\bar{\tau}^{+,\frac{N}{2}},t_{k}^{+,\frac{N}{2}-1})f(-\bar{\tau}^{+,\frac{N}{2}},t_{k}^{+,\frac{N}{2}-1})f(t_{k}^{+,\frac{N}{2}-1},\bar{\tau}^{+,\frac{N}{2}-2})}F_{K}^{(\frac{N}{2}-1)}(t_{k}^{\frac{N}{2}-1})\lim_{\bar{\tau}^{-,\mu}\to\hat{\pi}(\bar{\tau}^{+,\mu})}S_{K}^{N,mod}(\bar{\tau})
\end{multline}
for $\nu=\frac{N}{2}-1$ and

\begin{multline}
\frac{\partial}{\partial X_{k}^{+,\frac{N}{2}}}\lim_{\bar{t}^{-,\mu}\to\hat{\pi}(\bar{t}^{+,\mu})}S_{K}^{N}(\bar{t})=\\
\frac{f(\bar{\tau}^{+,\frac{N}{2}},t_{k}^{+,\frac{N}{2}})f(-\bar{\tau}^{+,\frac{N}{2}},t_{k}^{+,\frac{N}{2}})f(t_{k}^{+,\frac{N}{2}},\bar{\tau}^{+,\frac{N}{2}})f(t_{k}^{+,\frac{N}{2}},-\bar{\tau}^{+,\frac{N}{2}})}{f(-\bar{\tau}^{+,\frac{N}{2}-1},t_{k}^{+,\frac{N}{2}})f(t_{k}^{+,\frac{N}{2}},\bar{\tau}^{+,\frac{N}{2}-1})}F_{K}^{(\frac{N}{2})}(t_{k}^{\frac{N}{2}})\lim_{\bar{\tau}^{-,\mu}\to\hat{\pi}(\bar{\tau}^{+,\mu})}S_{K}^{N,mod}(\bar{\tau})
\end{multline}
for $\nu=\frac{N}{2}$. Since the modified $\alpha$-s in $S_{K}^{N,mod}$
(see appendix \ref{sec:Pair-offshell}) are in a complete agreement
with the modified $X$-s (see subsection \ref{subsec:modX}) of the
Korepin criteria, the property \ref{enum:prop4} is satisfied for
the normalized overlaps $\mathbf{N}^{(\mathbf{r}^{+})}(\bar{X}^{+},\bar{t}^{+})$.

Finally we prove the last property of the Korepin criteria for the
normalized overlap $\mathbf{N}^{(\mathbf{r}^{+})}(\bar{X}^{+},\bar{t}^{+})$.
Since the general on-shell overlaps vanish\footnote{In the subsections \ref{subsec:Achiral-pair-structure} and \ref{subsec:Chiral-pair-structure}
we showed that the non-vanishing on-shell overlaps requires certain
pair structures for the Bethe roots.}, we obtain an identically zero function if we replace the expression
$\alpha_{\nu}(\bar{t}_{k}^{\nu})$ with the r.h.s of the Bethe equation
(\ref{eq:bethe}) (since it is the on-shell limit), therefore we can
obtain that
\begin{multline}
S_{K}^{N}(\bar{t})\Biggr|_{\text{on-shell}}=\\
\sum_{\mathrm{part}(\bar{t})}\frac{\prod_{s=1}^{N-1}f(\bar{t}_{\textsc{ii}}^{s},\bar{t}_{\textsc{i}}^{s})}{\prod_{s=1}^{N-2}f(\bar{t}_{\textsc{ii}}^{s+1},\bar{t}_{\textsc{i}}^{s})}Z_{K}^{N}(\bar{t}_{\textsc{i}})\bar{Z}_{K}^{N}(\bar{t}_{\textsc{ii}})\prod_{\mu=1}^{N-1}\prod_{\bar{t}_{k}^{\mu}\in\bar{t}_{\textsc{i}}^{\mu}}\frac{f(t_{k}^{\mu},\bar{t}_{\textsc{i},k}^{\mu})}{f(\bar{t}_{\textsc{i},k}^{\mu},t_{k}^{\mu})}\frac{f(t_{k}^{\mu},\bar{t}_{\textsc{ii}}^{\mu})}{f(\bar{t}_{\textsc{ii}}^{\mu},t_{k}^{\mu})}\frac{f(\bar{t}_{\textsc{i}}^{\mu+1},t_{k}^{\mu})}{f(t_{k}^{\mu},\bar{t}_{\textsc{i}}^{\mu-1})}\frac{f(\bar{t}_{\textsc{ii}}^{\mu+1},t_{k}^{\mu})}{f(t_{k}^{\mu},\bar{t}_{\textsc{ii}}^{\mu-1})}=0.
\end{multline}
This equation simplifies as
\begin{equation}
\sum_{\mathrm{part}(\bar{t})}\frac{\prod_{s=1}^{N-1}f(\bar{t}_{\textsc{i}}^{s},\bar{t}_{\textsc{ii}}^{s})}{\prod_{s=1}^{N-2}f(\bar{t}_{\textsc{i}}^{s+1},\bar{t}_{\textsc{ii}}^{s})}Z_{K}^{N}(\bar{t}_{\textsc{i}})\bar{Z}_{K}^{N}(\bar{t}_{\textsc{ii}})=0.\label{eq:onshellVanish}
\end{equation}
Let us consider the pair structure limit $t_{k}^{-,\nu}\to-t_{k}^{+,\nu}$
or $t_{k}^{-,\nu}\to-t_{k}^{+,\nu}-\nu c$ in the untwisted or the
twisted cases. Let us start with the twisted case. Since all the poles
of the HC-s are simple, it is enough to develop the functions $\alpha_{\nu}(t_{k}^{-,\nu})$
up to the first order over $t_{k}^{-,\nu}+t_{k}^{+,\nu}+\nu c$:
\begin{equation}
\alpha_{\nu}(t_{k}^{-,\nu})=\frac{1}{\alpha_{\nu}(t_{k}^{+,\nu})}+(t_{k}^{-,\nu}+t_{k}^{+,\nu}+\nu c)\frac{\alpha'_{\nu}(t_{k}^{+,\nu})}{\alpha_{\nu}(t_{k}^{+,\nu})^{2}}+\mathcal{O}(t_{k}^{-,\nu}+t_{k}^{+,\nu}+\nu c)^{2}).
\end{equation}
If all $X_{k}^{+,\nu}=0$, then the derivatives of $\alpha_{\nu}$
vanish, and we can substitute $\alpha_{\nu}(t_{k}^{-,\nu})$ given
by (\ref{eq:bethe}) into the sum formula (\ref{eq:sumFormula}) in
the limit $t_{k}^{-,\nu}\to-t_{k}^{+,\nu}-\nu c$. This leads us to
\begin{equation}
\lim_{\bar{t}^{-,\mu}\to\hat{\pi}(\bar{t}^{+,\mu})}S_{K}^{N}(\bar{t})=\lim_{\bar{t}^{-,\mu}\to\hat{\pi}(\bar{t}^{+,\mu})}\sum_{\mathrm{part}(\bar{t})}\frac{\prod_{s=1}^{N-1}f(\bar{t}_{\textsc{i}}^{s},\bar{t}_{\textsc{ii}}^{s})}{\prod_{s=1}^{N-2}f(\bar{t}_{\textsc{i}}^{s+1},\bar{t}_{\textsc{ii}}^{s})}Z_{K}^{N}(\bar{t}_{\textsc{i}})\bar{Z}_{K}^{N}(\bar{t}_{\textsc{ii}}).\label{eq:parilim}
\end{equation}
However, due to (\ref{eq:onshellVanish}) the sum over partitions
in (\ref{eq:parilim}) vanishes for arbitrary set $\bar{t}$. In this
way we just showed that the property \ref{enum:prop5} is satisfied
for the normalized overlap $\mathbf{N}^{(\mathbf{r}^{+})}(\bar{X}^{+},\bar{t}^{+})$
in the twisted case. The proof for the untwisted case is completely
analog.

In summary we showed that the Korepin criteria is satisfied which
means
\begin{equation}
\mathbf{N}^{(\mathbf{r}^{+})}(\bar{X}^{+},\bar{t}^{+})=\det G^{+}.
\end{equation}

\bibliographystyle{elsarticle-num.bst}
\bibliography{ref.bib}

\end{document}